\documentclass[12pt, letterpaper, center, noupper]{uconnthesis}

\ThesisLineSpacing{1.5}
\usepackage[square,sort,comma,numbers]{natbib}
\usepackage[nointegrals]{wasysym}
\usepackage{fancyhdr}

\usepackage{epsfig}
\usepackage{algorithm}
\usepackage{algorithmic}
\newcommand{\ls}[1]
   {\dimen0=\fontdimen6\the\font
    \lineskip=#1\dimen0
    \advance\lineskip.5\fontdimen5\the\font
    \advance\lineskip-\dimen0
    \lineskiplimit=.9\lineskip
    \baselineskip=\lineskip
    \advance\baselineskip\dimen0
    \normallineskip\lineskip
    \normallineskiplimit\lineskiplimit
    \normalbaselineskip\baselineskip
    \ignorespaces
   }

\makeatletter 
\let\c@lofdepth\relax 
\let\c@lotdepth\relax 
\makeatother 
\usepackage{subfigure} 
\usepackage{breakcites}
\newcommand{\qea}{QUAD\xspace}

\newcommand{\mythesistitle}{Adaptive Bitrate Streaming Over Cellular Networks: Rate Adaptation and Data Savings Strategies}

\newcommand{\myname}{Yanyuan Qin}
\newcommand{\myyear}{2021}
\newcommand{\degreeone}{B.E. Nanjing University of Aeronautics and Astronautics, 2011}
\newcommand{\degreeoneshort}{B.E.}
\newcommand{\degreetwo}{M.E. Shanghai Jiao Tong University, 2014}
\newcommand{\degreetwoshort}{M.E.}
\newcommand{\degreethree}{M.S. University of Connecticut, 2019}

\newcommand{\majoradvisor}{\ \ \ \ \ Bing Wang}
\newcommand{\associateone}{Krishna R. Pattipati}
\newcommand{\associatetwo}{Song Han}
\newcommand{\associatethree}{Subhabrata Sen}

\usepackage{eqnarray,amsmath}
\usepackage{graphicx}
\usepackage[varg]{txfonts}
\usepackage{multirow}
\usepackage{color}
\usepackage{rotating}

\newcommand{\url}[1]{{#1}}
\newcommand{\fengc}[1]{}

\newcommand{\mysection}[1]{\vspace{-.03in}\section{#1}\vspace{-.00in}}
\newcommand{\mysubsection}[1]{\vspace{-.05in}\subsection{#1}\vspace{-.01in}}
\newcommand{\mysubsubsection}[1]{\vspace{-.05in}\subsubsection{#1}\vspace{-.00in}}
\newcommand*\SomeCommand[1]{\markboth{#1}{#1}}

\newcommand{\BULLET}{\vspace{+.00in} \noindent $\bullet$ \hspace{+.00in}}
\usepackage[hidelinks]{hyperref}

\begin{document}
% this is the abstraction part 
\pagestyle{empty}

%%%% Unless you have more than two associate advisors, or more than two previous degrees, there is no need to edit this file.  If you do fall into one of the latter categories, the set-up below should be transparent enough for suitable adjustments to be made.

%\frontmatter

%%%Title page 1
\thispagestyle{empty}
\begin{center}
{\LARGE
{\bf\mythesistitle}
}
\vspace{.6in}

\myname, Ph.D.\\
University of Connecticut, \myyear

\vspace{.6in}

{\large\bf ABSTRACT}
\end{center}

	\SomeCommand{Header 1 define here}

	\SomeCommand{Header 2 define here}

Adaptive bitrate streaming (ABR) has become the {\it de facto} technique for video streaming over the Internet. Despite a flurry of techniques, achieving high quality ABR streaming over cellular networks remains a tremendous challenge.  First, the design of an ABR scheme needs to balance conflicting Quality of Experience (QoE) metrics such as video quality, quality changes, stalls and startup performance, which is even harder under highly dynamic bandwidth in cellular network. Second, streaming providers have been moving towards using Variable Bitrate (VBR) encodings for the video content, which introduces new challenges for ABR streaming, whose nature and implications are little understood. Third, mobile video streaming consumes a lot of data. Although many video and network providers currently offer data saving options, the existing practices are suboptimal in QoE and resource usage. Last, when the audio and video tracks are stored separately, video and audio rate adaptation needs to be dynamically coordinated to achieve good overall streaming experience, which presents interesting challenges while, somewhat surprisingly, has received little attention by the research community. In this dissertation, we tackle each of the above four challenges.

Firstly, we design a framework called PIA (PID-control based ABR streaming) that strategically leverages PID control concepts and novel approaches to account for the various requirements of ABR streaming. The evaluation results demonstrate that PIA outperforms state-of-the-art schemes in providing high average bitrate with significantly lower bitrate changes and stalls, while incurring very small runtime overhead. We further design PIA-E (PIA Enhanced), which improves the performance of PIA in the important initial playback phase.

Secondly, we identify distinguishing characteristics of VBR encodings that impact user QoE and should be factored in any ABR adaptation decision and find that traditional ABR adaptation strategies designed for the Constatn Bitrate (CBR) encodings are not adequate for VBR. We develop novel best practice design principles to guide ABR rate adaptation for VBR encodings. As a proof of concept, we design a novel and practical control-theoretic rate adaptation scheme, CAVA (Control-theoretic Adaption for VBR-based ABR streaming), incorporating these concepts. Extensive evaluations show that CAVA substantially outperforms existing state-of-the-art adaptation techniques.

Thirdly, we analyze the underlying causes for suboptimal existing data saving practices and propose novel approaches to achieve better tradeoffs between video quality and data usage. The first approach is Chunk-Based Filtering (CBF), which can be retrofitted to any existing ABR scheme. The second approach is QUality-Aware Data-efficient streaming (QUAD), a holistic rate adaptation algorithm that is designed ground up. Our evaluations demonstrate that compared to the state of the art, the two proposed schemes achieve consistent video quality that is much closer to the user-specified target, lead to far more efficient data usage, and incur lower stalls.

For the fourth challenge, we shed light on a number of limitations in existing practices both in the protocols and the player implementations, which can cause undesirable behaviors such as stalls, selection of potentially undesirable combinations such as very low quality video with very high quality audio, etc. Based on our gained insights, we identify the underlying root causes of these issues, and propose a number of practical design best practices and principles  whose collective adoption will help avoid these issues and  lead to better QoE.

\newpage

%%%%Title page 2
\thispagestyle{plain}
\pagenumbering{roman} 
\setcounter{page}{1}
\begin{center}
{\LARGE
{\bf\mythesistitle}
}
\vspace{.6in}

\myname

\vspace{.6in}

\degreethree \\
\degreetwo \\
\degreeone 

\vfill
A Dissertation \\
Submitted in Partial Fulfillment of the \\
Requirements for the Degree of \\
Doctor of Philosophy \\
at the \\
University of Connecticut

\vspace{.25in}

\myyear

\end{center}
\newpage

%%%%%%%% Copyright Page
\thispagestyle{plain}
\phantom{skip}
\vspace{.5in}
\begin{center}
Copyright by

\vspace{1in}
\myname

\vspace{4.5in}
\myyear
\end{center}
\newpage

%%%%%%%%%%%%% Approval Page
\thispagestyle{plain}

\begin{center}
{\large\bf APPROVAL PAGE}

\vspace{.5in}
Doctor of Philosophy Dissertation

\vspace{.5in}
{\LARGE\bf\mythesistitle}

\vspace{.5in}
Presented by\\
\myname, \degreeoneshort, \degreetwoshort
\vspace{.75in}
\end{center}

\begin{center}
\begin{minipage}{4.5in}
Major Advisor \hfill$\underset{\hbox{\majoradvisor}}{\rule{3.25in}{1pt}}$\\[10pt]
Associate Advisor \hfill$\underset{\hbox{\associateone}}{\rule{3in}{1pt}}$\\[10pt]
Associate Advisor \hfill$\underset{\hbox{\associatetwo}}{\rule{3in}{1pt}}$\\[10pt]
Associate Advisor \hfill$\underset{\hbox{\associatethree}}{\rule{3in}{1pt}}$
\end{minipage}

\vspace{0.7in}
University of Connecticut\\
\myyear
\end{center}

\newpage

%%%%%%%%%%%%% Acknowledgments Page
\thispagestyle{plain}

\begin{center}
{\large\bf ACKNOWLEDGMENTS}
\vspace{1in}
\end{center}

%\doublespacing

 I would like to express my gratitude and appreciation for my major advisor, Professor Bing Wang, whose guidance, support and encouragement has been invaluable throughout this study. I feel extremely lucky being a member of the team, and having her as a constant source of courage and wisdom.  
 Her sharp insight and concise description of complicated research problems and her dedication have set an example of academic perfection, from which I will continue to benefit in my future career. 

I would  like to thank Dr. Shubho Sen and Dr. Shuai Hao from AT\&T Lab Research, Professor Krishna R. Pattipati from ECE department of UCONN, Professor Feng Qian from University of Minnesota-Twin City, for their guidance on my researches.

I would also like to thank, Professor Alexander Russel, Professor Peng Zhang, Professor Song Han, Professor Wei Wei,  Professor Suining He, Professor  Mohammad Maifi Hasan Khan, Professor Fei Miao and Professor Peter B. Luh  for their advice and broad knowledge. It was my great pleasure working with such intelligent and responsible professors.

I would like to extend my gratitude to my colleagues, Ruofan Jin,  Yuexin Mao, Levon Nazaryan, Sixia Chen, Lingyu Ren, Yan Li, Bing Yan,Reynaldo Morillo, Chaoqun Yue, Shweta Ware, Chinmaey Shende, Cheojin Park, Zefan Tang, Wenfeng Wan, Jiangwei Wang and Lizhi Wang for their discussions and comments in my research. My thanks also go to all my friends for their endless support whenever I needed. Last but not least, I give my deepest gratitude and love to my parents, who have been a constant source of happiness and motivation for me.

\pagestyle{plain}

\renewcommand\contentsname{Table of Contents}
\tableofcontents
\listoffigures
\listoftables
\thispagestyle{plain}
\mainmatter

 \section{Introduction}
\label{sec:pid_intro}

Video streaming has come to dominate mobile data consumption today. As per Cisco's 2016 Visual Network Index report~\cite{Cisco-feb-2016}, mobile video traffic now accounts for more than half of all mobile data traffic.
Despite much effort, achieving good quality video streaming over cellular networks remains a tremendous challenge.
%New technology (LTE) does not eliminate stalls in video streaming.
In fact, a report in late 2014  indicates significant stalling (10\% to 40\% depending on video quality) in both Europe and North America's cellular networks~\cite{Citrix-sept-2014}.

%Improving video viewing quality relies on innovative
Most video content are currently streamed using Adaptive Bit-Rate (ABR) streaming over HTTP, the \emph{de facto} technology adopted by industry.
In ABR streaming, a video is encoded into multiple resolutions/quality levels.
% (each with different network bandwidth requirements).
The encoding at  each quality level is  divided into small chunks, each containing data  for some time intervals' worth  of playback (e.g., several seconds).
A chunk at a higher quality level requires more  bits to encode and is therefore larger in size.
% than  the corresponding chunk encoded at a lower quality level,  for the same playback interval in the video.
During playback, the video player determines in real-time
% which video chunk (ie.,
which quality level to fetch according to its adaptation algorithm.
%, which bases its decision on factors including estimated available network bandwidth and the data already in the client's local buffer.

Various user engagement studies~\cite{dobrian2011understanding, krishnan2013video,Liu15:Deriving,ni2011flicker} indicate that satisfactory ABR streaming needs to achieve three conflicting goals simultaneously: (1) maximize the playback bitrate; (2) minimize the likelihood of stalls or rebuffering;
% that is, the client's playback buffer becoming empty, causing the video playback to freeze,
and (3) minimize the variability of the selected video bitrates for a smooth viewing experience. Reaching any of the three goals alone is relatively easy -- for instance, the player can simply stream at the highest bitrate to maximize the video quality; or it can stream at the lowest bitrate to minimize the stalls. The challenge lies in achieving all three goals simultaneously, especially over highly varying network conditions, typical of the last-mile scenarios in cellular networks.

Rate adaptation for ABR streaming can be naturally modeled as a control problem: the video player monitors the past network bandwidth and the amount of content in the playback buffer to decide the bitrate level for the current chunk; the decision will then affect the buffer level, which can be treated as feedback to adjust the decision for the next chunk. PID (named after its three correcting terms, namely ``proportional'', ``integral'', and ``derivative'' terms) is one of the most widely used feedback control technique in practice~\cite{astrom08:feedback}.
It is conceptually easy to understand and computationally simple.

There has been initial work on using PID control theory for ABR streaming. However, the  message so far  has been mixed and inconclusive. The studies~\cite{Cicco11:feedback,de2013elastic} directly apply the standard PID controller to ABR streaming with no modifications. As a result, video bitrate may fluctuate significantly as shown in the evaluation results. The studies~\cite{tian2012towards,yin2015control} conclude that PID control is not a suitable approach since the goal of PID control is misaligned with the goals of ABR streaming.

\subsection{Contributions}

In this paper, we adopt a contrarian perspective and take a fresh look at the potential of  PID control for ABR video streaming. We start by pointing out that a recent heuristic technique with commercial deployment, BBA~\cite{huang2014buffer}, in fact can be shown to be, in effect, using a simplified form of PID control.
%
% We start by pointing out that a recent heurstic  technique~\cite{huang2014buffer} that has been tested successfully in a large-scale video streaming service, in fact can be shown to be implicitly using a simplified form of PID control.
%
We then conduct an in-depth exploration of using PID control for ABR streaming.
We design \emph{PIA} (PID-control based ABR streaming), a novel control-theoretic video streaming scheme that strategically leverages PID control concepts as the base framework, and further incorporates domain knowledge of ABR streaming to improve the robustness and adaptiveness of video playback.
Our main contributions include the following.

\BULLET We take a fresh look at PID-based control for ABR streaming. We strategically leverage  PID control concepts  as the base framework for PIA. The {\em core controller}  in PIA differs from those in~\cite{Cicco11:feedback,de2013elastic} in that we define a control policy that makes the closed-loop control system linear, and easy to control and analyze. The core controller maintains the playback buffer to a target level, so as to reduce rebuffering.

\BULLET  We add domain-specific enhancements to further improve the robustness and adaptiveness of ABR streaming. Specifically, PIA addresses two key additional requirements of ABR streaming, i.e., maximizing playback bitrate and reducing frequent bitrate changes. It also incorporates strategies to accelerate initial ramp-up and protect the system from saturation.

\BULLET We explore parameter tuning. $K_p$ and $K_i$ (defined in Section~\ref{sec:background}) are the two fundamental and most critical parameters that guide the PIA controller's behavior. We develop a methodology that systematically examines a wide spectrum of network conditions and parameter settings to derive their ($K_p$, $K_i$) configurations that yield satisfactory quality of experience (QoE). A service provider can use this approach to appropriately tune ($K_p$, $K_i$) for their environment.

We conduct comprehensive evaluations of PIA
%using both simulation and emulation with a DASH player,
using a large number of real cellular network traces with diverse network variability characteristics. The traces were collected from two commercial LTE networks at diverse geographic locations. Our key findings include the following.

%\begin{itemize}
\BULLET PIA achieves comparable bitrates as two state-of-the-art schemes, BBA~\cite{huang2014buffer} and MPC~\cite{yin2015control}, while substantially reducing bitrate changes (49\% and 40\% lower respectively)  and rebuffering time (68\% and 85\%  lower respectively).
 %compared to the other two schemes.
 Overall, PIA achieves the best balance among the three QoE metrics.

\BULLET PIA has low computation overheads (e.g., comparable to BBA and only 0.5\% of MPC based on our simulation). Our emulation results also show that the execution time of PIA is less than 2 seconds for a 15-minute video.
%For a 20 minute video with 2~sec chunks, on a commodity laptop with Intel i5 CPU and 16GB RAM, the CPU time used by MPC, BBA and PIA were 36.04~s, 0.08~s and  0.17~s respectively.

\BULLET We also found that a  common set of ($K_p$, $K_i$) values  exist that have good performance across all the traces. This  is important as it suggests that a single appropriately selected ($K_p$, $K_i$)  pair can be used across a wide range of network conditions, facilitating easy deployment of our streaming scheme.
%\end{itemize}

%(\textbf{need more})

\subsection{Related Work}

Overall, our work differs from prior efforts that
%either implicitly borrow simple principles from control theory or
directly apply PID controller to video streaming without any adaptation. Instead, we show a somewhat surprising high-level finding: \emph{by applying control theories in an explicit and adaptive manner, our ABR streaming algorithm substantially outperforms state-of-the-art video streaming schemes.}
Besides the initial work on using control theory for ABR streaming~\cite{Cicco11:feedback,de2013elastic, tian2012towards,yin2015control} as described earlier,
we briefly review other recently proposed schemes for ABR streaming. %BBA~\cite{huang2014buffer} is a buffer based scheme.
BOLA~\cite{spiteri2016bola} improves BBA by selecting the bitrate to maximize a utility function considering both rebuffering and video quality. FESTIVE~\cite{jiang2012improving} and PANDA~\cite{li2014probe} both consider scenarios with multiple video flows.
%multi-player streaming scenarios.
piStream~\cite{Xie15:piStream} is designed specifically for LTE networks and uses PHY-layer information to improve the bandwidth prediction. As another related work, \cite{Chen13:Scheduling} proposes a network-based scheduling framework for adaptive video delivery over cellular
networks. The work~\cite{Zou15:video} demonstrates the benefits of knowing network bandwidth on the performance of ABR streaming. None of them focuses specifically on leveraging control theory for improving the video streaming QoE.

\chapter{Introduction}
\label{chapter:intro}

Video streaming has come to dominate mobile data consumption today. As per Cisco's  Visual Network Index report~\cite{Cisco-feb-2016}, mobile video traffic now accounts for more than half of all mobile data traffic.
Ensuring good viewing experience for this important application class is critical to content providers, content distribution networks, and mobile operators.
Despite much effort, achieving good quality video streaming over cellular networks remains a tremendous challenge~\cite{Citrix-sept-2014}.

\section{ABR streaming}
Most video contents are currently streamed using Adaptive Bit-Rate (ABR) streaming over HTTP  (specifically HLS~\cite{apple-hls} and DASH~\cite{dash-standard}), the \emph{de facto} technology adopted by industry.
In ABR streaming, a video is encoded into multiple resolutions/quality levels (or tracks).
The encoding at  each resolution/quality level is  divided into equal-duration chunks, each containing data for a short interval's worth of playback (e.g., several seconds).
A chunk at a higher resolutions/quality level requires more  bits to encode, and is therefore larger in size.
During playback, to fetch the content for a particular playpoint in the video, the video player dynamically determines what bitrate/quality chunk to download based on time-varying network conditions. The resulting play-back involves showing different portions of the video using chunks selected from different tracks.

Various user engagement studies~\cite{dobrian2011understanding, krishnan2013video,Liu15:Deriving,ni2011flicker} indicate that satisfactory ABR streaming needs to achieve three conflicting goals simultaneously: (1) maximize the playback bitrate; (2) minimize the likelihood of stalls or rebuffering;
and (3) minimize the variability of the selected video bitrates for a smooth viewing experience. Reaching any of the three goals alone is relatively easy -- for instance, the player can simply stream at the highest bitrate to maximize the video quality; or it can stream at the lowest bitrate to minimize the stalls. The challenge lies in achieving all three goals simultaneously, especially over highly varying network conditions, typical of the last-mile scenarios in cellular networks.

Control theories have been applied to design the ABR algorithm.
The studies~\cite{Cicco11:feedback,de2013elastic} directly use the standard PID controller, without adapting it to accommodate the special requirements of ABR streaming.
The study in~\cite{spiteri2016bola} shows that directly using PID control leads to worse performance than BOLA,
while our proposed schemes (PIA and PIA-E) carefully adapt PID control for ABR streaming and outperform BOLA.
The studies in~\cite{tian2012towards,yin2015control} conclude that PID control is not suitable for ABR streaming, and develop other control based approaches.
MPC~\cite{yin2015control} uses another branch of control theory, model predicative control, to solve a QoE optimization problem. It, however, requires accurate future network bandwidth estimation and incurs significant computation overhead. BBA~\cite{huang2014buffer} selects video bitrates purely based on buffer occupancy.  However, as we will discuss, BBA essentially uses a P-controller and does not consider the integral part, which affects its performance. BOLA~\cite{spiteri2016bola} and an improved version BOLA-E~\cite{spiteri2018theory}
select the bitrate based on maximizing a utility function, considering both rebuffering and video quality. We showed in Section~\ref{sec:pia_dash} that BOLA leads to significantly more rebuffering than our approach.

Another type of approach uses machine learning (e.g., reinforcement learning) to ``learn" an ABR scheme from data. The study in~\cite{Chiariotti2016OLA} proposes a tabular Q-Learning based reinforcement learning approach for ABR streaming. This approach, however, does not scale to large state/action
spaces. A more recent scheme, Pensieve~\cite{mao2017pensieve}, addresses the scalability issue using reinforcement learning based on neural networks.
In~\cite{mao2017pensieve},  Pensieve is realized as a server-side ABR algorithm -- the client feeds back information to the server and the server makes the decisions on bitrate choices.
For a given ABR algorithm, Oboe~\cite{akhtar2018oboe} pre-computes the best possible parameters
for different network conditions and dynamically adapts these parameters at run-time. The study in~\cite{qin2018abr} characterizes the VBR encoding characteristics, proposes design principles for VBR-based ABR streaming and a concrete scheme, CAVA, that instantiates these design principles. 

Our study takes a fresh look at using PID control for ABR streaming and shows a somewhat surprising high-level finding: \emph{by applying PID control in an explicit and adaptive manner, our ABR streaming algorithms substantially outperform the state-of-the-art video streaming schemes.} We will show that our proposed PIA~\cite{qin_tmc2020,sen2019methods} scheme incurs much lower computational overhead and achieves a significantly better balance among the three performance metrics than MPC; PIA also outperforms RobustMPC~\cite{yin2015control} and BBA~\cite{huang2014buffer}.

%Adaptive Bitrate (ABR) streaming (specifically HLS~\cite{apple-hls} and DASH~\cite{dash-standard}) has emerged as the \emph{de facto} video streaming technology in industry for dynamically adapting the video streaming quality based on varying network conditions. A video is compressed into multiple independent {\em streams} (or {\em tracks}), each specifying the same content but encoded with different bitrate/quality. A track is further divided into a series of {\em chunks}, each containing data for a few seconds' worth of playback. During streaming playback, to fetch the content  for a particular play point in the video, the adaptation logic at the client player dynamically determines what bitrate/quality chunk to download.
% based on time-varying network performance.
%The resulting playback involves showing different portions of the video using chunks from different tracks.

%%% CBR vs VBR
\section{ABR streaming for VBR encodings}

The video compression for a track can be (1) Constant Bitrate (CBR) --  encodes the entire video at a relatively fixed bitrate, allocating the same bit budget to  both \emph{simple scenes} (i.e., low-motion or low-complexity scenes) and \emph{complex scenes} (i.e., high-motion or high-complexity scenes), resulting in variable quality across scenes,
or (2) Variable Bitrate (VBR) -- encodes \emph{simple scenes}  with fewer bits and \emph{complex scenes}
with more bits, while maintaining a consistent quality throughout the track.
%%with more bits, while maintaining a consistent quality throughout the track.
VBR presents some key advantages over CBR~\cite{lakshman1998vbr}, such as the ability to realize better video quality for the same average bitrate, or the same quality with a lower bitrate encoding than CBR.
%As a result, VBR encodings allow more flexibility and use bits more efficiently.
Traditionally, streaming services mainly deployed only CBR,
partly due to various practical difficulties
%driven in part by various practical considerations
including the complex encoding pipelines for VBR,
as well as the demanding storage, retrieval, and transport challenges posed by the %inherent
multi-timescale bitrate burstiness of VBR videos.
%It is only fairly recently that
Most recently content providers have begun adopting VBR encoding~\cite{huang2014buffer,reed2016leaky,Xu17:Dissecting,sen2020apparatus,sen2020differential}, spurred by the promise of substantial improvements in the quality-to-bits ratio compared to CBR, and by technological advancements in VBR encoding pipelines.

Howerver, current commercial systems and ABR protocols do not fully make use of the characteristics of VBR videos; they typically assume a track's peak rate to be its bandwidth requirement  when making rate adaptation decisions~\cite{Xu17:Dissecting}.While content providers have recently begun adopting VBR encoding, their treatment to VBR videos is simplistic (e.g., by simply assuming that the bandwidth requirement of a chunk equals to the peak bitrate of the track)~\cite{Xu17:Dissecting}. As we will see in \S\ref{sec:cava_dash}, such simplistic treatment can lead to poor performance.
Of the large number of existing studies, only a few targeted VBR videos~\cite{li2014streaming,huang2014buffer,tong2017rbavbr};
most others were designed for and evaluated in the context of CBR encodings.
While several schemes  in the latter category can be applied for VBR videos by explicitly incorporating the actual bitrate of each chunk, they were not designed ground up to account for VBR characteristics, nor had their performance been evaluated for that case.
We evaluate the performance of schemes from both categories
(see~\S\ref{sec:cava_principles} and \S\ref{sec:cava_perf}), and show that
they can suffer from a large amount of rebuffering and/or significant viewing quality impairments when used for VBR videos.

There exists little prior understanding of the issues and challenges involved in  ABR streaming of VBR videos.
Our study differs from all the above works in several important aspects, including developing insights based on analyzing content complexity, encoding quality, and encoding bitrates, identifying general design principles for VBR streaming, and developing a concrete, practical ABR scheme for VBR streaming that significantly outperforms the state-of-the-art schemes.

\section{Data Usage for ABR Streaming}

Video streaming is data intensive. An important question is how can we effectively reduce the bandwidth consumption for mobile video streaming while minimizing the impact on users' quality of experience (QoE)?
This is a highly important and practical research problem since cellular network bandwidth is a relatively scarce resource.
The average data plan for a U.S. cellular customer is only 2.5 GB per month~\cite{ericsson-mobility-report}, while streaming just one-hour High Definition (HD) video on mobile Netflix can consume
3 GB data.
Therefore, the capability to make more efficient use of  data while still hitting quality targets is the key to enabling users to consume more content within their data budgets without adversely impacting QoE. In addition, downloading less data for a video session also translates to lower radio energy consumption, less thermal overhead on mobile devices, as well as
potentially better QoE for other users sharing the same cellular RAN {(radio access network)} or base station.

Existing ABR adaptation schemes (\S\ref{sec:related work}) focus mainly on maximizing the video quality and user QoE.
{While some schemes do conservatively utilize the bandwidth, their decisions are primarily driven by QoE impairment concerns such as the possibility of stalls caused by the potentially inaccurate bandwidth estimation.}
%While the schemes do  differ in their data usage, such behaviors are driven primarily by QoE impairment concerns, e.g., that the network bandwidth estimation at the client may not be accurate, thus selecting very high tracks can increase the chance of stalls.
For example, when selecting the next chunk, the default adaptation scheme in ExoPlayer~\cite{exoplayer} (a popular open-source player that is used in more than 10000 apps) only considers tracks whose declared bitrates\fengc{(typically set around the peak bitrate of that track)} are at least $25\%$ lower than the estimated network bandwidth.
%{While such a design does reduce the data usage, existing ABR schemes do not explicitly consider bandwidth efficiency as a key factor in making the track selection decisions.}
Such a design saves data just by being conservative. As we show later, it can lead to significantly lower quality/QoE (\S\ref{sec:exoplayer}).
Existing ABR schemes do not explicitly consider
bandwidth efficiency together with quality in making the track selection decisions.
%One challenge that existing ABR schemes do not explicitly consider
%bandwidth efficiency and quality together  as a key factor in making the track selection decisions.

Some mobile network operators and commercial video services  provide users with
certain ``data saver'' options.
%~(\S\ref{sec:current-practice-save-data}).
%the option of certain ``data saver'' modes~(\S\ref{sec:current-practice-save-data}).
{These options take either a service-based approach or a network-based approach.
	The former limits the highest level of quality/resolution/bitrate by, for example, streaming only  {standard definition} (SD) content over cellular networks.
	A network-based approach instead limits the network bandwidth.
}
%When activated, these options either (i)  limit the highest quality/screen resolution/bitrate track (e.g., stream only Standard Definition content on cellular) that the ABR logic will ever download~(service-based) or (ii)  limit the network bandwidth~(network-based),  which indirectly has a similar effect.
%These approaches primarily focus on data savings. Their implications for the delivered video {QoE} is little understood.
We survey the ``quality throttling'' mechanisms used by today's commercial video content providers for saving bandwidth for mobile devices, and pinpoint their inefficiencies~(\S\ref{sec:quality-bitrate-tradeoffs}-\S\ref{sec:reduce-data}).
Despite their simplicity, we find such approaches often achieve tradeoffs between video quality and bandwidth usage that are far from what viewers desire. A key reason is that for  state-of-the-art encoders (e.g., H.264~\cite{h264}, H.265~\cite{h265} and VP9~\cite{vp9}), the actual perceived picture quality exhibits significant variability across different chunks within the same track, for both Constant Bitrate (CBR) and Variable Bitrate~(VBR) encodings.
Therefore, a data-saving approach that
%uses a track-level filter
simply removes high tracks
%is bound to
leads to quality variations.\fengc{(e.g., streaming simple scenes at an unnecessarily high quality and complex scenes at a much lower quality).}
%, when there is ample network bandwidth available.
Such quality variability impairs user QoE, and makes suboptimal use of the network bandwidth.

% Whats missing is approaches that save data while not impacting quality.

%To address the drawbacks of the existing practices,  \emph{quality-aware data saving} strategy could be designed, which provides data savings while allowing users' direct control of the video quality. Specifically, a user selects from multiple quality options (e.g., good, better, best), and the player will map the selection to a \emph{target quality}.

\section{ABR Streaming with Separate Audio and Video Tracks
}

\begin{figure}[t]
	\centering
	\includegraphics[width=0.7\textwidth]{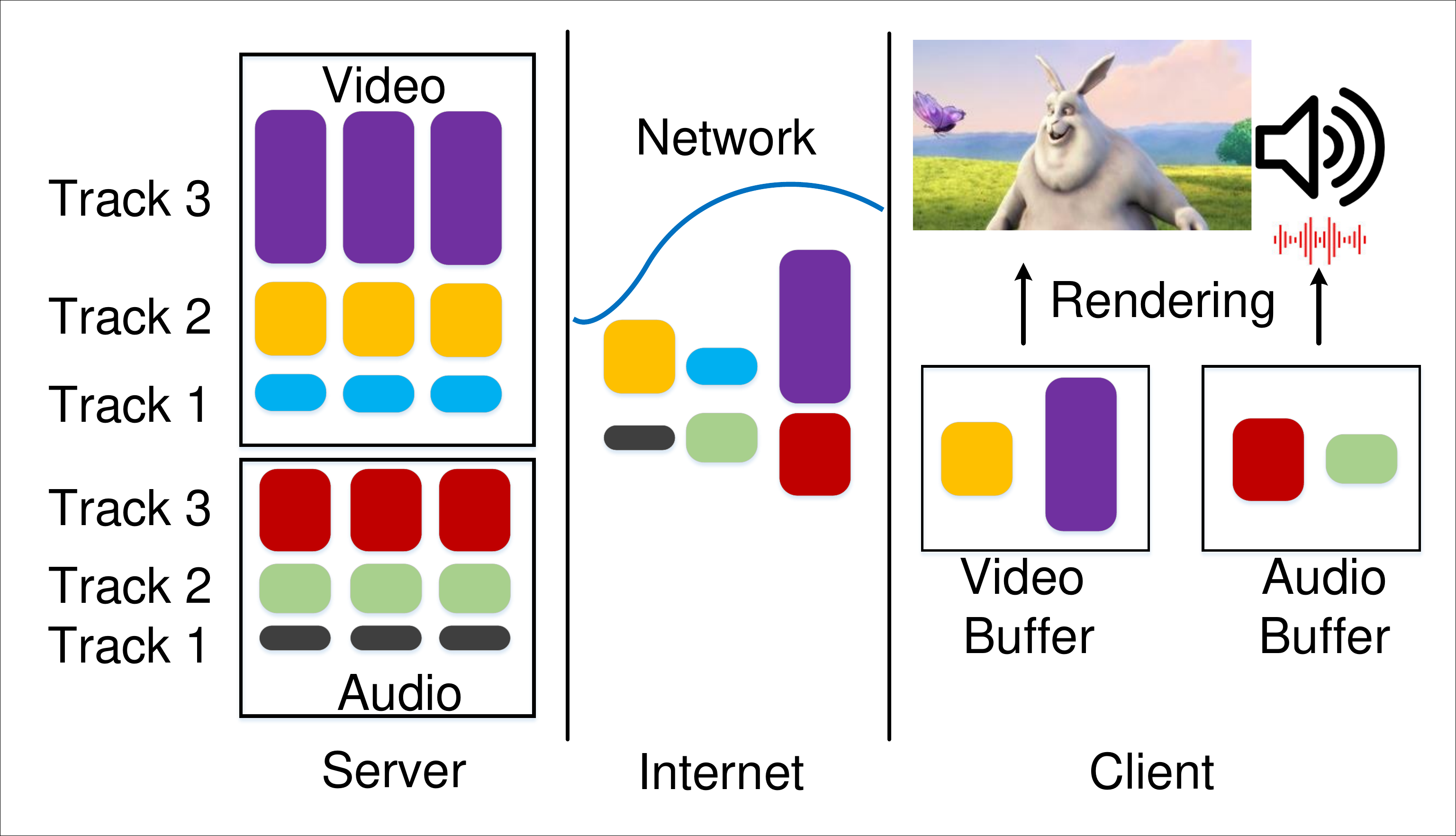}
	\caption{ABR streaming of demuxed audio and video {(the different colors represent the chunks from different tracks).} }
	\label{fig:audio_diagram}
\end{figure}

On the server for Adaptive bitrate (ABR) streaming, a video  track and its corresponding audio can be combined together as  a  single multiplexed  track (in {\em muxed} mode), where each chunk in the track contains the associated video and audio content.
Alternatively, the video and audio content can be maintained separately as  demultiplexed tracks (in {\em demuxed} mode)  on the server, as shown in Fig.~\ref{fig:audio_diagram}.
The  demuxed approach offers a number of advantages for services that need to have more than one audio variant -- e.g., to support multiple languages, or multiple audio quality levels or both.
First, it  requires significantly less storage at the origin server.
To see this,
consider a hypothetical service that uses $M$ video and $N$ audio tracks.
The server only needs to store $M$ video and $N$ audio tracks for the demuxed mode, while it has to store a much larger set of $M \times N$  muxed tracks for the muxed mode.
Second, {\em a subtle effect is that} the demuxed mode increases CDN cache hits. As an example, for a given chunk/playback position, suppose user $A$ requests video track variant $V1$ and audio track variant $A2$, and at a later time, user $B$ requests  the same video track variant $V1$ and a different audio track variant $A1$.
In muxed mode, $B$ needs to download both V1 and A1 from the original server,
while in demuxed mode, $B$ can get the video chunk for V1 from the CDN cache (which cached it due to  A's earlier request), and only needs to get the audio chunk for A1 from the original server.
Due to the above advantages, services are increasingly moving towards using demuxed audio and video tracks~\cite{Xu17:Dissecting}. We focus on ABR streaming for demuxed video and audio tracks in this work.

The literature on video rate adaptation is
extensive~\cite{jiang2012improving,huang2014buffer,yin2015control,Mao17:pensieve,spiteri2016bola,Akhtar18:oboe}.
A common assumption regarding audio is that its bitrate is significantly lower than that of the video, and hence the decision on audio track selection has little impact on the video track selection~\cite{Spiteri2018bolae}.
The reality, however, is that increasingly an audio track can be of {similar or even} much higher bitrate than some of the video tracks.
Based on the HLS authoring specification for Apple devices~\cite{hls2017}, the audio bitrate can be as high 384 Kbps
for mobile devices, which can be much higher than the typical encoding bitrates of lower-rung video tracks in real services (e.g., the two lowest video tracks have peak rates around 100  and 250 Kbps respectively for one popular service).
In addition, more and more devices
support Dolby Atmos audio~\cite{dolby}, for which the audio bitrate can be up to 768 Kbps~\cite{netflix-audio}.
As a result, audio can in reality  consume a significant fraction of the available network bandwidth,
and audio {track selection} can therefore significantly affect video selection and the overall viewing experience  (see \S\ref{sec:behaviors}).  Concurrently with the availability of higher bitrate audio tracks, the community has also realized the importance of selecting higher audio quality tracks when possible and appropriate. For example, Netflix very recently adopted audio adaptation at the player side
to boost the overall video streaming experience~\cite{netflix-audio,netflix-audio-blog}.

Little is known on how best to mesh together audio and video rate adaptation in ABR streaming. In fact, meshing together audio and video rate adaptation is a very challenging problem -- it involves many pieces (in ABR protocols, server and player designs)
and needs a holistic solution~\cite{qin_conext2019}.

\section{Contribution of This Dissertation}

The contributions of this dissertation are as follows.
\begin{itemize}

\item we take a fresh look at PID-based control for ABR streaming. We design a framework called PIA (PID-control based ABR streaming) that strategically leverages PID control concepts and incorporates several novel strategies to account for the various requirements of ABR streaming. We evaluate PIA using simulation based on real LTE network traces~\cite{yue2018}, as well as using real DASH implementation. The results demonstrate that PIA outperforms state-of-the-art schemes in providing high average bitrate with significantly lower bitrate changes (reduction up to 40\%) and stalls (reduction up to 85\%), while incurring very small runtime overhead. We further design PIA-E (PIA Enhanced), which improves the performance of PIA in the important initial playback phase. 

\item  We explore ABR streaming for VBR encodings across diverse video genres, encoding technologies, and platforms. We identify distinguishing characteristics of VBR encodings that impact user QoE and should be factored in any ABR adaptation decision.  Traditional ABR adaptation strategies designed for the CBR case are not adequate for VBR. We develop novel best practice design principles to guide ABR rate adaptation for VBR encodings. As a proof of concept, we design a novel and practical control-theoretic rate adaptation scheme, CAVA (Control-theoretic Adaption for VBR-based ABR streaming), incorporating these concepts. Extensive evaluations show that CAVA substantially outperforms existing state-of-the-art adaptation techniques, validating the importance of these design principles.

\item Our study shows that existing data saving practices for Adaptive Bitrate (ABR) videos are suboptimal: they often lead to highly variable video quality and do not make the most effective use of the network bandwidth~\cite{qin_mmsys2019}. We identify underlying causes for this and propose two novel approaches to achieve better tradeoffs between video quality and data usage.
The first approach is Chunk-Based Filtering (CBF), which can be retrofitted to any existing ABR scheme. The second approach is QUality-Aware Data-efficient  streaming~(QUAD), a holistic rate adaptation algorithm that is designed ground up.
We implement and integrate our solutions into two video player platforms (\texttt{dash.js} and ExoPlayer), and conduct thorough evaluations over emulated/commercial cellular networks using real videos. Our evaluations demonstrate that compared to the state of the art, the two proposed schemes achieve consistent video quality that is much closer to the user-specified target, lead to far more efficient data usage, and incur lower stalls.

\item We  examine the state of the art in the handling of demuxed audio and video tracks in predominant ABR protocols (DASH and HLS), as well as in real ABR client implementations in three popular players covering both browsers and mobile platforms. Combining experimental insights with code analysis, we shed light on a number of limitations in existing practices both in the protocols and the player implementations, which can cause undesirable behaviors such as stalls, selection of potentially undesirable combinations such as very low quality video with very high quality audio, etc. Based on our gained insights, we identify the underlying root causes of these issues, and propose a number of practical design best practices and principles  whose collective adoption will help avoid these issues and  lead to better QoE~\cite{qin_conext2019}. 
\end{itemize}

\section{Dissertation Roadmap}

The remainder of this dissertation is organized as follows.

In Chapter~\ref{chapter:pid}, we take a fresh look at PID-based bit rate adaptation and design novel ABR scheme that judiciously combines PID control and domain knowledge of ABR streaming.
In Chapter~\ref{chapter:cava}, we first characterize VBR encodings and develop design principles. Then we design novel ABR scheme for VBR videos using the principles.
In Chapter~\ref{chapter:quad}, {quality-aware data saving} strategies are designed, which provide data savings while allowing users' direct control of the video quality. 
In Chapter~\ref{chapter:audio}, we investigate the predominant ABR protocols and popular players and propose best practices for ABR Streaming with separate audio and video tracks.
We draw the conclusion and highlight the future work in Chapter~\ref{chapter:conclusion}.

\chapter{A Control Theoretic Approach to ABR Video Streaming: A Fresh Look at PID­-based Rate Adaptation}
\label{chapter:pid}

\section{Introduction}
\label{sec:pid_intro}

Rate adaptation for ABR streaming can be naturally modeled as a control problem: the video player monitors the past network bandwidth and the amount of content in the playback buffer to decide the bitrate level for the current chunk; the decision will then affect the buffer level, which can be treated as feedback to adjust the decision for the next chunk. PID (named after its three correcting terms, namely ``proportional'', ``integral'', and ``derivative'' terms) is one of the most widely used feedback control techniques in practice~\cite{astrom08:feedback}.
It is conceptually easy to understand and computationally simple.
There has been initial work on using PID control theory for ABR streaming. Specifically, the studies in~\cite{Cicco11:feedback,de2013elastic} directly apply the standard PID controller to ABR streaming with no modifications, which was shown to lead to significantly suboptimal performance~\cite{spiteri2016bola}.  The studies~\cite{tian2012towards,yin2015control} conclude that PID control is not suitable for ABR streaming.

In this section, we adopt a contrarian perspective and take a fresh look at the potential of  PID control for ABR video streaming. We start by pointing out that a recent heuristic technique,
 BBA~\cite{huang2014buffer}, can be shown to be, in effect, using a simplified form of PID control.
 We then conduct an in-depth study that explores using PID control for ABR streaming. Specifically, we
 design \emph{PIA} (PID-control based ABR streaming), a novel control-theoretic video streaming scheme that strategically incorporates PID control concepts and domain knowledge of ABR streaming.
Our main contributions include the following.

\begin{itemize}

\item We take a fresh look at PID-based control for ABR streaming, and strategically leverage  PID control concepts as the base framework for PIA. Specifically, the {\em core controller}  in PIA differs from those in~\cite{Cicco11:feedback,de2013elastic} in that we define a control policy that makes the closed-loop control system linear, and easy to control and analyze. The core controller maintains the playback buffer to a target level, so as to reduce rebuffering.

\item  We add domain-specific enhancements to further improve the robustness and adaptiveness of ABR streaming. Specifically, PIA addresses two key additional requirements of ABR streaming, i.e., maximizing playback bitrate and reducing frequent bitrate changes. It also incorporates strategies to accelerate initial ramp-up and protect the system from saturation. We further develop \emph{PIA-E} (PIA Enhanced) that improves the performance of PIA in the  important initial playback phase, by dynamically adjusting the parameters used in the control loop.

\item We explore parameter tuning. Specifically,
the proportional gain $K_p$ and the integral gain $K_i$ 
are the two fundamental and most critical parameters that guide the PIA controller's behavior. We develop a methodology that systematically examines a wide spectrum of network conditions and parameter settings to derive their ($K_p$, $K_i$) configurations that yield satisfactory quality of experience (QoE). 
Our results demonstrate that a  common set of ($K_p$, $K_i$) values exist that have good performance across a wide range of network settings, indicating our schemes can be easily deployed in practice. 

\end{itemize}

We conduct comprehensive evaluations of PIA using a large number of real cellular network traces with diverse network variability characteristics. The traces were collected from two commercial LTE networks at diverse  geographic locations, with a range of mobility conditions. Our key findings include the following.

\begin{itemize}
\item PIA achieves comparable bitrates as two state-of-the-art schemes, BBA~\cite{huang2014buffer} and MPC~\cite{yin2015control}, while substantially reducing bitrate changes (49\% and 40\% lower, respectively)  and rebuffering time (68\% and 85\%  lower, respectively).
 Overall, PIA achieves the best balance among the three QoE metrics.
 {Compared to PIA, the enhanced version, PIA-E, achieves higher bitrate for the beginning part of the video; for the entire video, PIA-E has similar rebuffering, average bitrate and bitrate changes as PIA.}

\item PIA and PIA-E have low computation overhead (e.g., comparable to BBA and only 0.5\% of MPC based on our simulations). Our emulation results also show that their execution time is less than 2 seconds for a 15-minute video.

\end{itemize}

The rest of this chapter is organized as follows.
Section~\ref{sec:background} summarizes the background and motivations.
Section~\ref{sec:PID} presents PIA,  the PID based controller for ABR streaming.
Section~\ref{sec:performance} evaluates the performance of PIA under a wide range of settings.
Section~\ref{sec:startup-phase} presents PIA-E, designed to improve the startup performance of PIA.
Section~\ref{sec:pia_dash} presents the implementation and evaluation of PIA and PIA-E using a real video player.
%Section~\ref{sec:pia_related} briefly describes the related work.
Finally, Section~\ref{sec:pia_concl} summarize this section.

\section{Motivation and Background} \label{sec:background}

\begin{table}
    \centering
    \begin{tabular}[h]{l|l} %\hline
        $C_t$ & Network bandwidth at time $t$          \\ \hline
        $\widehat{C}_t$ & Estimated network bandwidth at time $t$          \\ \hline
        $x_t$   & Buffer level at time $t$   (in seconds)               \\ \hline
        $x_r$     & Target buffer level  (in seconds)        \\ \hline
	$R_t$   & Selected video bitrate for time $t$ \\ \hline
	$\Delta$ &   Video chunk duration  (in seconds)      \\ \hline
	$\delta$ &   Startup latency  (in seconds)     \\ \hline
        $u_t$   & PID controller output      \\  \hline
   $K_p, K_i, K_d$ & PID controller parameters \\ \hline
        $\zeta$     & Damping ratio   \\ \hline
        $\omega_n$     & Natural frequency  \\ \hline
        $\beta$     & Setpoint weighting parameter \\ 
    \end{tabular}
    \caption{Key notation.}
    \vspace{-0.0in}
    \label{tab:notations}
\end{table}

ABR streaming has been used in many commercial systems~\cite{ms-smoothstreaming,apple-hls,adobe-hds}.
For a satisfactory user-perceived QoE, ABR streaming needs to optimize several conflicting goals, including maximizing the average playback rate, minimizing stalls (or rebuffering), and reducing sudden and frequent quality variations~\cite{dobrian2011understanding, krishnan2013video,Liu15:Deriving,ni2011flicker}.
We next formulate ABR streaming as a control problem and describe the motivation for our study. 
Table~\ref{tab:notations} summarizes the main notation used in this chapter.

\subsection{ABR streaming as a control problem}
Deciding which level to choose in ABR streaming can be modeled as a control problem. Specifically, let $x_t$ be the buffer level (in seconds) of the video player at time $t$, $C_t$ the real-time network bandwidth at time $t$, and $R_t$ the  bitrate of the video chunk that is being downloaded at time $t$. Further, let $\Delta$ denote the {video chunk duration} (i.e., the duration of a chunk's playback time), $\delta$ denote the startup delay, i.e., how long it will take for the player to start playing. Then the player's buffer dynamics can be written as
\begin{eqnarray}
\label{eq:dynamics}
\dot x_t &=&
\begin{cases}
     \frac{C_t}{R_t},   & \text{if } t \leq \delta\\
     \frac{C_t}{R_t} - \textbf{1}(x_t - \Delta),      & \text{otherwise}
\end{cases}
\end{eqnarray}
where $\mathbf{1}(x_t - \Delta) =   1$ if  $x_t\geq \Delta$; otherwise, $\mathbf{1}(x_t - \Delta) =   0$
{since in ABR streaming, a chunk has to be downloaded completely before any part of it can be played back.}

In Eq. (\ref{eq:dynamics}), $\dot x_t$ is the rate of change of the buffer at time $t$. Here, ${C_t}/{{R_t}}$ models the relative buffer filling rate.
If $C_t > R_t$, i.e., the actual network bandwidth is larger than the bitrate of the video chunk being downloaded, the buffer level will increase. Otherwise, the buffer level will be at the same level (if $C_t = R_t$) or decrease (if $C_t < R_t$).
\begin{figure}[t]
	\centering
	\subfigure[ \label{subfig:openloop}]{%
		\includegraphics[width=0.8\textwidth]{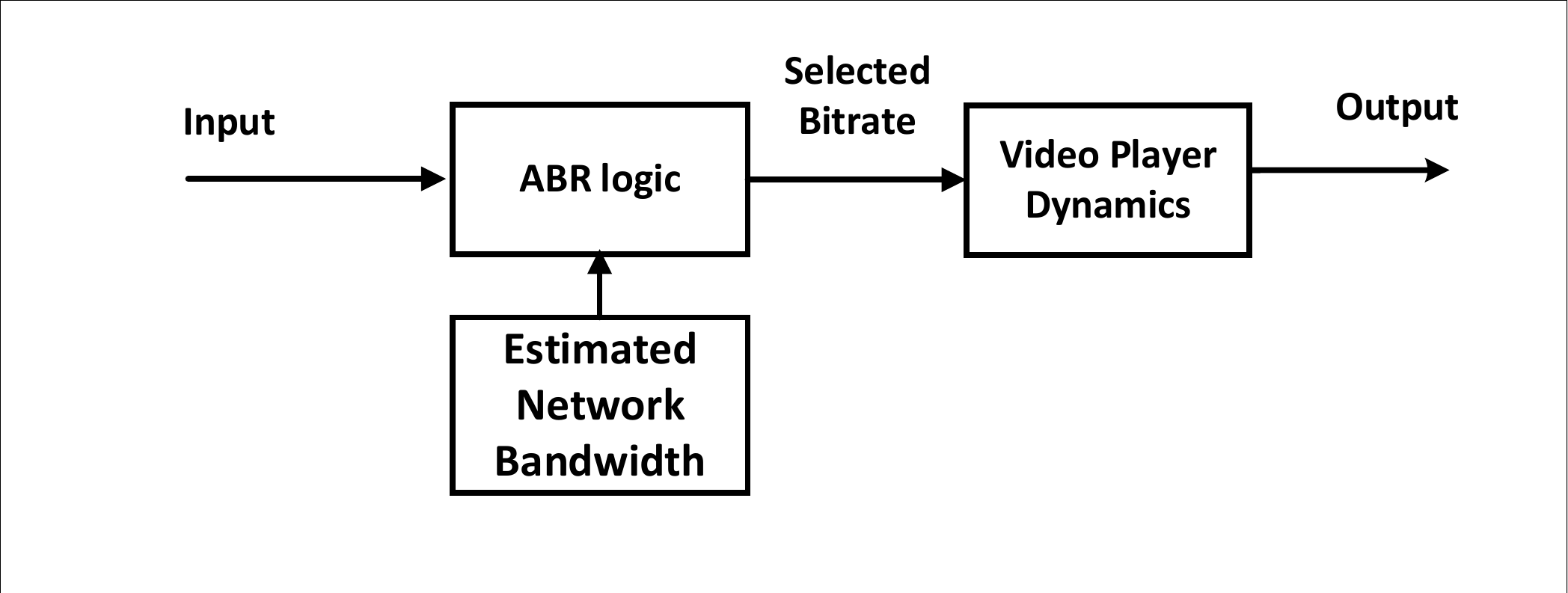}
	}\\
	\subfigure[ \label{subfig:closeloop}]{%
		\includegraphics[width=0.8\textwidth]{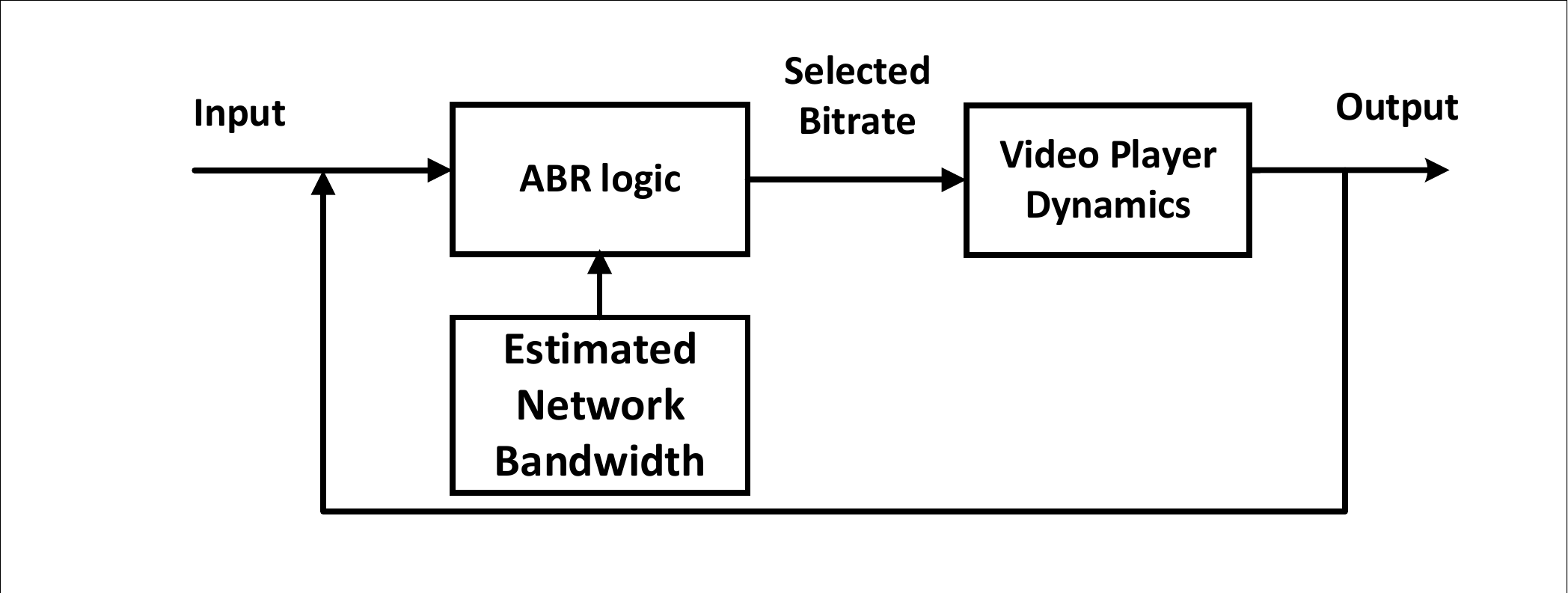}
	}
	\caption{ABR streaming based on open-loop control (a) and closed-loop control (b).}
	\label{fig:open-closeloop-cmp}
\end{figure}

One simple control strategy is to select the video bitrate for each chunk based on the prediction of real-time link bandwidth, $\widehat{C}_t$. Specifically, it simply chooses the highest bitrate that is less than $\widehat{C}_t$.
This is an open-loop control strategy (illustrated in Fig.~\ref{fig:open-closeloop-cmp}(a)), since the output (e.g., the buffer level) is not fed back to the system to assist the decision making.
It is not robust against network link bandwidth estimation errors. As an example, it may choose a high video bitrate if the estimated bandwidth, $\widehat{C}_t$, is high, even if the current playback buffer level is very low. If it turns out that $\widehat{C}_t$ is an overestimate of the actual network bandwidth, the buffer can be even further drained and become empty, causing stalls. Closed-loop or feedback control, as illustrated Fig.~\ref{fig:open-closeloop-cmp}(b), is more effective in dealing with network link bandwidth estimation errors. We therefore focus on closed-loop/feedback control in this chapter.

\subsection{PID control}
 As mentioned earlier, PID control is by far the most common way of using feedback in engineering systems.
 A PID controller works by continuously monitoring an ``error value", defined as the difference between the setpoint and measured process variable~\cite{astrom08:feedback}. Specifically, let $u_t$ represent the control signal, and $e_t$ the error feedback at time $t$. Then
\begin{eqnarray}
u_t = {K_p}e_t + {K_i}\int_0^t e_\tau d\tau  + K_d \frac{de_t}{dt},
\label{eq:PID}
\end{eqnarray}
where the three parameters $K_p$, $K_i$ and $K_d$ are all non-negative, and denote the coefficients for the proportional, integral and derivative terms, respectively. As defined above, a PID controller takes account of the present, past and future values of the errors through the three terms, respectively.
Some applications may require using only one or two terms to provide the appropriate system control. This is achieved by setting the other parameters to zero. A PID controller is called a PI, PD, P or I controller in the absence of the respective control actions~\cite{astrom08:feedback}.

In the video streaming scenario, the real-time buffer level is the measured process variable, and the reference/target buffer level is the setpoint.
We next show that a recent state-of-the-art buffer based scheme, BBA~\cite{huang2014buffer}, can be mapped to a P-controller (though the paper does not claim any control-theoretic underpinnings).
 In BBA, the video player maintains a buffer level, and empirically sets two thresholds, $\theta_\text{high} > \theta_\text{low}$.
If the buffer level is below $\theta_\text{low}$, the video player always picks the lowest bitrate, $R_{\text{min}}$; if the buffer level is above $\theta_\text{high}$, the video player picks the highest bitrate, $R_{\text{max}}$; otherwise, the video player picks the video bitrate proportionally to buffer level. The selected bitrate, $R_t$, can therefore be represented as
\begin{eqnarray*}
R_t &=&
\begin{cases}
R_{\text{min}}, & x_t < \theta_\text{low}, \\
\frac{R_{\text{max}}-R_{\text{min}}}{\theta_\text{high} - \theta_\text{low}}\left(x_t-\theta_\text{low}\right)+R_{\text{min}}, & \theta_\text{low} \le x_t \le \theta_\text{high} \\
R_{\text{max}}, & x_t > \theta_\text{high}
\end{cases}
\label{eq:bba}
\end{eqnarray*}
Comparing the above with Eq. (\ref{eq:PID}), we see that it is equivalent to a P-controller when $x_t \in [\theta_\text{low},\theta_\text{high}]$ with $K_p=\frac{R_{\text{max}}-R_{\text{min}}}{\theta_\text{high} - \theta_\text{low}}$, $K_i=0$ and $K_d=0$.

BBA has been tested successfully in a large-scale deployment~\cite{huang2014buffer}, indicating that a PID-type control framework has the potential for ABR streaming.
On the other hand, P-controller only considers the present error (i.e., the proportional term), and ignores the other two terms. It is well known that the absence of an integral term in a system may prevent the system from reaching its target value~\cite{astrom08:feedback}. This is especially true for video streaming, where inaccurate network bandwidth estimation may cause the error to accumulate over time.
Therefore, including the integral term can potentially further improve the performance of BBA.
PID's ability to address accumulative errors is advantageous compared to model predictive control (MPC) based approach in~\cite{yin2015control}, which does not consider accumulative errors (unless new state variables are added) and also requires accurate network bandwidth prediction. In addition, MPC is much more computation-intensive than PID (see Section~\ref{sec:perf-comp-overhead}).

We investigate PID-based control for ABR streaming in this chapter, motivated by the widespread adoption of PID control in {various domains beyond video streaming (e.g., industrial control, process control)} and BBA. Using PID for ABR streaming, however, has several challenges.
First, the goal of PID control is to maintain a target buffer level that is only indirectly related to QoE. Indeed, while maintaining the buffer at a target level can help in preventing rebuffering, it does not help with the other two metrics on playback quality and bitrate variation. Second, PID is often used in continuous time and state space, while video streaming is a discrete-time system, where the decisions are made at chunk boundaries and the video bitrate levels are discrete. Finally, while PID is conceptually simple, the parameters ($K_p$, $K_i$ and $K_d$) need to be tuned carefully. Here important questions are how to choose these parameters, and to determine whether there exists a parameter set that is applicable to a wide range of network and video settings. Some of the above challenges have been pointed out in~\cite{tian2012towards,yin2015control}, which take the position that PID is not suitable for ABR streaming. As we shall show, none of the above challenges is a fundamental hurdle for using PID-based control for ABR streaming.

\section{Adapting PID Control for ABR Streaming} \label{sec:PID}

\begin{figure}[t]
	\centering
	\includegraphics[width=0.7\textwidth]{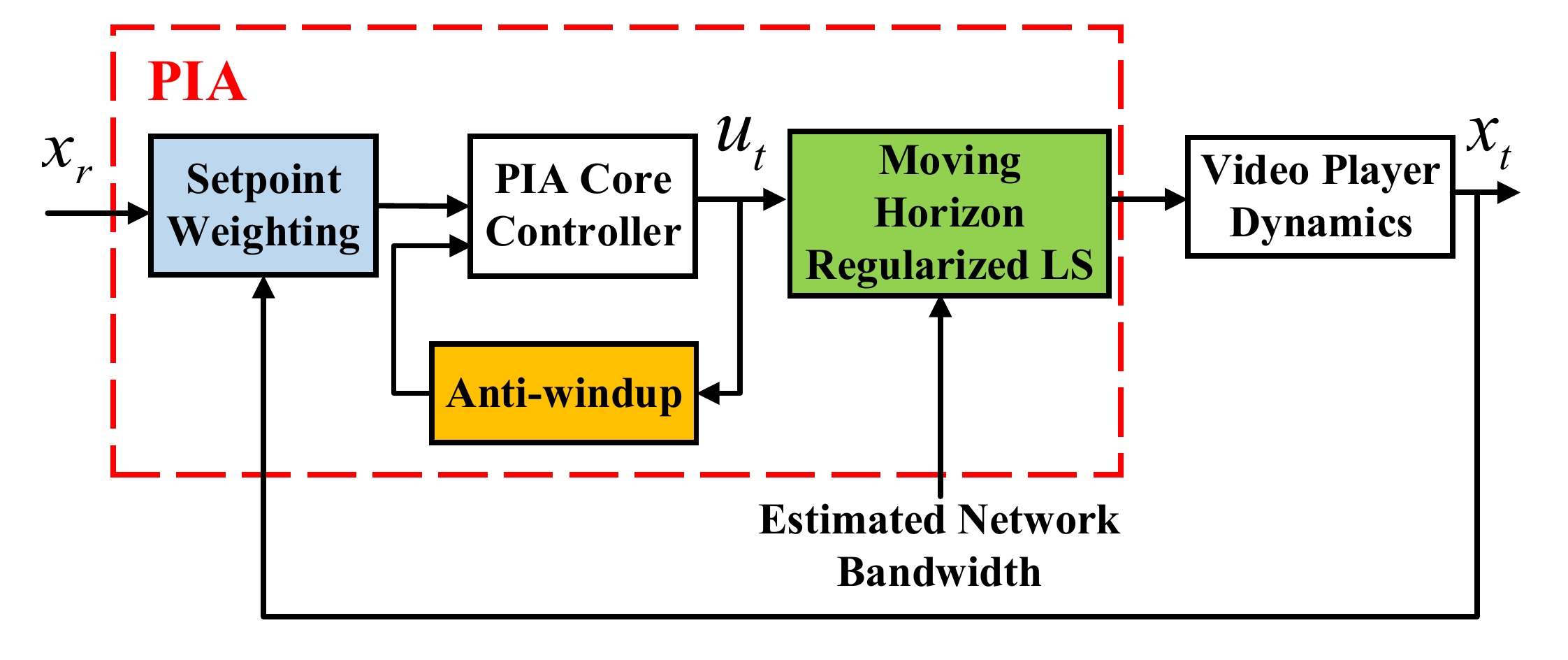}
	\caption{PIA main components.} 
	\label{fig:pia_diagram}
\end{figure}

We propose PIA, a PID based rate adaption algorithm for ABR streaming. As shown in Fig.~\ref{fig:pia_diagram}, it contains a PI-based core control block as well as three mechanisms to address specific requirements for ABR streaming. We first describe the core component, and then the three performance enhancing mechanisms.

%\bigsinglefig{PIA-E/figure/video_PID_control1}{PIA main components.}{diagram}

\subsection{PIA core component}
The core component of PIA adjusts the standard PID control policy in Eq. (\ref{eq:PID}) so that the resultant closed-loop system is linear, and hence easier to control and analyze. We next define the controller output, analyze the system behavior, and  provide insights into how to choose the various parameters.

Recall the dynamic video streaming model in Eq. (\ref{eq:dynamics}), where $x_t$ is the video player buffer level at $t$, $C_t$ is the network bandwidth at time $t$, and $R_t$ is the video bitrate chosen for time $t$.
We define the controller output, $u_t$, as
%The relative buffer filling rate is
\begin{equation}
\label{eq:pia_pid_ut}
u_t = \frac{C_t}{R_t},
\end{equation}
and set the control policy as
\begin{eqnarray}
u_t = {K_p}({x_r} - x_t) + {K_i}\int_0^t {(x_r - x_\tau)d\tau} + \textbf{1}(x_t - \Delta)
\label{eq:PID-video-streaming}
\end{eqnarray}
 where $K_p$ and $K_i$ denote, respectively, the parameters for proportional and integral control, $x_r$ denotes target buffer level, and $\Delta$ is the chunk duration.
The choice of $x_r$ depends on system constraints, a point we will come back to in Section~\ref{sec:performance}.

 The above control policy differs from the standard PID control policy in Eq. (\ref{eq:PID}) in the last term $\textbf{1}(x_t - \Delta)$, which is a novel aspect of our design. As we shall see, it provides linearity, making the closed-loop control system easier to control and analyze. In our control policy, the parameter for derivative control $K_d=0$ (hence strictly speaking, our controller is a PI controller). This is because derivative action is sensitive to measurement noise~\cite{astrom08:feedback} and measuring network bandwidth in our context is prone to noise.

Intuitively, $u_t$ defined in Eq. (\ref{eq:pia_pid_ut}) is a unitless quantity representing the relative buffer filling rate.
With $u_t$ selected, based on Eq. (\ref{eq:pia_pid_ut}), the player can select the corresponding bitrate as
\begin{equation}
\label{eq:pid_bitrate}
R_t = \frac{\widehat{C}_t}{u_t},
\end{equation}
where $\widehat{C}_t$  is the estimated link bandwidth at time $t$. Since video bitrate levels are discrete, we can choose the bitrate to be the highest that is below ${\widehat{C}_t}/{u_t}$.
This choice of $R_t$ can increase, decrease or maintain the buffer level.

We next analyze the system to provide insights into its behavior as well as providing guidelines in choosing the controller parameters.
Combining equations (\ref{eq:dynamics}) and (\ref{eq:PID-video-streaming}) yields
\vspace{-.0in}
\begin{equation}
\dot x_t = u_t - \textbf{1}(x_t - \Delta) = {K_p}({x_r} - x_t) + {K_i}\int\limits_0^t (x_r - x_\tau)d\tau,
\label{eq:system_eq}
\end{equation}
when the video starts playback (i.e., when $t \ge \delta$). We see that it is a linear system.
Taking Laplace transform on both sides of Eq. (\ref{eq:system_eq}) yields
\begin{eqnarray}
\label{eq:laplace-form}
sx(s) = {K_p}({x_r}(s) - x(s)) + \frac{K_i}{s}\left({x_r}(s) - x(s)\right),
\end{eqnarray}
where $s$ is a complex Laplace transform variable. Let $T(s)$ be the system transfer function, which describes the relationship of the input and output of a linear time-invariant system. From Eq. (\ref{eq:laplace-form}), we have
\begin{eqnarray}
T(s) = \frac{x(s)}{x_r(s)} = \frac{K_ps + K_i}{s^2 + K_p s + K_i}, \label{eq:Ts-1}
\end{eqnarray}
which is a second-order system. In the above transfer function, since $T(0)=1$, the system can track step changes in the target buffer level, $x_r$, with zero error in the steady state~\cite{Stefani02:feedback-control}, indicating that our system can maintain the preset target buffer level.
From Eq. (\ref{eq:Ts-1}), we have
\begin{eqnarray}
2{\zeta}{\omega_n} = K_p, \quad {\omega_n^2} = K_i \label{eq:zeta-omega}\,,
\end{eqnarray}
where $\zeta$ and $\omega_n$ are the \emph{damping ratio} and the \emph{natural frequency}, respectively, two important properties of the system.
Solving the above two equations, we have
\begin{eqnarray}
\zeta  = \frac{K_p}{2\sqrt {K_i}}, \quad \omega_n = \sqrt {K_i}\,.
\label{eq:zeta}
\end{eqnarray}
Damping ratio represents the system's ability for reducing its oscillations. In our context, it measures how the buffer will oscillate around the target buffer level -- small damping will cause the buffer to change rapidly, while large damping will cause the buffer to change slowly in a sluggish manner. 
Natural frequency represents the frequency at which a system tends to oscillate in the absence of any driving or damping force.
Empirically it has been found that $\zeta$ in the range $[0.6,0.8]$ yields a very good system performance~\cite{ogata2010modern}. As a result, $K_p$ and $K_i$  should be chosen so that $\zeta \in [0.6,0.8]$.
We will discuss how to tune $K_p$ and $K_i$ in Section~\ref{subsec:kp-ki}.

\subsection{PIA performance enhancing techniques} \label{subsec:PID-enhanced}

Based on the core component of PIA, we now add three domain-specific enhancements to PIA to further improve its robustness and adaptiveness for ABR streaming.

\medskip
\noindent{\textbf{Accelerating initial ramp-up.}}  At the beginning of the video playback, the buffer level $x_t$ can be much smaller than the target buffer level $x_r$. In this case, observe from Eq. (\ref{eq:PID-video-streaming}) that $u_t$ will be large. A large $u_t$ will result in low video bitrate, $R_t$, and therefore a low quality at the beginning, which can adveresley impact initial user experience. To address this issue, we include a setpoint weighting parameter~\cite{astrom08:feedback},  $\beta \in (0,1]$, into the control policy as
\vspace{-.00in}
\begin{eqnarray}
%u_t &=& {K_p}({\beta x_r} - x_t) + {K_i}\int\limits_0^t {(x_r - x_\tau)d\tau} \nonumber \\
u_t = {K_p}({\beta x_r} - x_t) + {K_i}\int_0^t {(x_r - x_\tau)d\tau} + \textbf{1}(x_t - \Delta).\
\label{eq:PID-video-streaming-improved}
\end{eqnarray}
Note that $\beta$ is only included in the proportional term; it does not affect the steady-state behavior of the control system~\cite{astrom08:feedback}. When $\beta=1$, the above control policy reduces to (\ref{eq:PID-video-streaming}). When $\beta < 1$, it can lead to smaller $u_t$, and hence faster initial ramp-up in video bitrate.
 However, very small $\beta$ can lead to aggressive choice of video bitrate, and hence increase the chance of  buffer emptying, causing rebuffering at the beginning of the playback. We explore how to set $\beta$ in Section~\ref{subsec:kp-ki}.
The system transfer function corresponding to the above control policy is
\begin{eqnarray}
T(s) = \frac{x(s)}{x_r(s)} = \frac{\beta K_ps + K_i}{s^2 + K_p s + K_i}= \frac{\beta K_ps + K_i}{s^2 + 2{\zeta}{\omega_n} s + \omega_n^2}\,. \label{eq:Ts-with-beta}
\end{eqnarray}
Note that both damping ratio, $\zeta$, and natural frequency, $\omega_n$ remain the same as those in Eq. (\ref{eq:zeta}).

\medskip
\noindent{\textbf{Minimizing bitrate fluctuations.}} The simple choice of $R_t$ in Eq. (\ref{eq:pid_bitrate}) mainly tracks the network bandwidth.
{It, however,   may lead to frequent and/or abrupt bitrate (and hence quality) changes, adversely impacting viewing quality.}
To address the above issue, we develop a regularized least squares (LS) formulation that considers both video bitrate and the changes in video bitrate to achieve a balance between both of these metrics.
Specifically, it minimizes the following objective function
\begin{eqnarray}
\label{eq:lms}
J({\ell_t}) = \sum\limits_{k = t}^{t + N-1} \left({u_k}{R(\ell_t)} - \widehat C_{k}\right)^2 + \eta \left(R(\ell_t) - R(\ell_{t - 1})\right)^2,
\end{eqnarray}
where $\ell_t$ and $\ell_{t-1}$ represent the tracks selected for chunks $t$ (the current chunk) and $t-1$ (i.e., the previous chunk), respectively\footnote{Here we slightly abuse notation by using $t$ to represent the index of the chunk chosen at time $t$, and using $t-1$ to represent the index of the previous chunk.}, $u_k$ is the controller output for the $k$-th chunk,
 $\widehat C_k$ is the estimated link bandwidth for the $k$-th chunk, $\eta$ is the weight factor for bitrate changes, and $R(\ell)$ represents the bitrate corresponding to track $\ell$.

Let $\mathcal L$ denote the set of all possible track levels.  For every $\ell_t \in \mathcal L$, the formulation in Eq. (\ref{eq:lms}) considers a moving horizon of $N$ chunks in the future (represented as the sum of $N$ terms, one for each of the $N$ future chunks).
The first term in the sum aims to minimize the difference between $u_k R(\ell_t)$ and the estimated network bandwidth $\widehat C_{k}$ so as to
maximize $R(\ell_t)$ (and hence quality) under the bandwidth constraint and the selected $u_k$.
The second term aims to minimize the variability in bitrate for two adjacent chunks (i.e., the current and previous chunks) for a smooth viewing experience.
The weight factor $\eta$ can be set to reflect the relative importance of these two terms. We use $\eta=1$ (i.e., equal importance) in the rest of the chapter since maximizing the video quality and reducing quality variation are both important for user QoE.
To reduce the number of video bitrate changes, in Eq. (\ref{eq:lms}), we assume that the same track is chosen for the next $N$ chunks  (this assumption is used only for determining the track/bitrate for chunk $t$; the actual track/bitrate for the future chunks will be decided at later times, independent of the assumption made for the decision of chunk $t$).
{
	For Constant Bitrate (CBR) videos (which is the focus of this chapter), the chunks in the same track have the same bitrate. Therefore, in Eq. (\ref{eq:lms}), the bitrate of the next $N$ chunks are all equal to $R(\ell_t)$. For Variable Bitrate (VBR) videos, the formulation can be modified in a straightforward manner\footnote{For VBR videos, even the chunks in the same track have variable bitrate. In that case, we can modify (\ref{eq:lms}) as follows. In the first term of (\ref{eq:lms}), we replace $R(\ell_t)$ with $R_k(\ell_t)$ to reflect that the bitrate is time varying; in the second term, we replace $R(\ell_t)$ and  $R(\ell_{t-1})$ with  the average bitrate of the corresponding tracks,  denoted as $\bar{R}(\ell_t)$ and  $\bar{R}(\ell_{t-1})$, respectively, so that the penalty for quality variation is zero as long as the two adjacent chunks are in the same track.}.
}	
Last, in Eq. (\ref{eq:lms}), the control output, $u_k$, is updated according to the control policy in Eq. (\ref{eq:PID-video-streaming-improved}), based on the estimated buffer size, $x_k$, over the moving horizon.  The estimated buffer size, $x_k$, is updated through Eq. (\ref{eq:dynamics}) using $R(\ell_t)$ as the video bitrate and the estimated network bandwidth $\widehat C_{k}$.

The optimal solution of Eq. (\ref{eq:lms}) is
\begin{eqnarray}
\ell^*_t = \arg \min \limits_{\ell_t \in \mathcal L}  J(\ell_t),
\end{eqnarray}
where $\mathcal L$ denotes the set of all possible track levels.
We can find $\ell^*_t$ easily by plugging in all possible values of $\ell_t$, $\ell_t \in \mathcal L$, into Eq. (\ref{eq:lms}), and select the value that provides the minimum objective function value in Eq. (\ref{eq:lms}).

The complexity of the above formulation is as follows. For every $\ell_t \in \mathcal L$, obtaining $J(\ell_t)$ requires computation of $N$ steps. Therefore, the total computational overhead is $O(|{\mathcal L}| N)$, significantly lower than the complexity of $O(|{\mathcal L}|^N)$ in~\cite{yin2015control}.

\medskip
\noindent{\textbf{Dealing with bitrate saturation.}} Following the control policy, $u_t$ may become negative (e.g., when the current buffer level exceeds the target buffer level).  In this case, solving Eq. (\ref{eq:lms}) will lead $R_t$ to be the minimum bitrate.
%a rational choice is to set video bitrate, $R_t$, to be the maximum value to quickly reduce the buffer level.
During this time period, if we continue using the integral term, $I_{\text{out}} = {K_i}\int_0^t {(x_r - x_\tau)d\tau}$, $u_t$ may remain negative for an extended period of time, causing $R_t$ to stay at the minimum bitrate level for an extended period of time (so called system saturation~\cite{astrom08:feedback}), and causing the buffer level to continue to grow.

Such system saturation is undesirable since it will cause the client to select
low bitrate (and hence low quality) chunks that adversely impact user  QoE, even though  the network bandwidth is able to support higher quality streaming.
To deal with the above scenario, we incorporate an anti-windup technique (to deal with integral windup, i.e., integral term accumulates a significant error) for negative $u_t$, or more specifically, when $u_t \le \epsilon$, $0 <\epsilon \ll 1$.
Many anti-windup techniques have been proposed in the literature~\cite{astrom08:feedback}.
We adopt a simple technique, which sets $u_t$ to $\epsilon$, chooses $R_t$ as the maximum bitrate, and does not change $I_{\text{out}}$ when $u_t \le \epsilon$. This corresponds to turning off the integral control when $u_t$ is below $\epsilon$.  We set $\epsilon$ to a small positive value, $10^{-10}$, in the rest of this chapter.

\subsection{PIA parameter tuning} \label{subsec:PIA-tuning}
Three important parameters in PIA are $K_p$, $K_i$ and $\beta$, where $K_p$ and $K_i$ determine the system behavior and $\beta$ is used for faster initial ramp-up.
For a given network setting (e.g., cellular networks), since $\beta$ does not affect the steady-state behavior~\cite{astrom08:feedback}, we can first assume a fixed $\beta$ (e.g., $\beta=1$) and tune $K_p$ and $K_i$ to achieve a desirable steady-state behavior (i.e., jointly maximize the three metrics in QoE). Once $K_p$ and $K_i$ are fixed, we then tune $\beta$ for the initial stage of the video playback. The values of $K_p$ and $K_i$ need to be tuned so that  the resultant system behavior is compatible with the network setting. Taking cellular networks as an example, since the bandwidth is highly dynamic, it is reasonable to tune the system so that the buffer level does not fluctuate drastically. Otherwise, the buffer can suddenly become very low, making the system vulnerable to stalls.
We describe this approach using a set of network traces from commercial cellular networks in Section~\ref{subsec:kp-ki}.

\subsection{Putting it all together} \label{subsec:PIA-summary}

We now summarize the workflow of PIA depicted in Fig.~\ref{fig:pia_diagram}.  PIA takes the target buffer level $x_r$, the current buffer level $x_t$, and the estimated network bandwidth as input, and computes the selected track level $\ell^*_t$, which is then fed into the Video Player Dynamics block to
update the buffer level. PIA considers both present and past estimation errors, as well as incorporates all the three QoE metrics in
the control loop. PIA also includes an anti-windup mechanism to deal with bitrate saturation, and a setpoint weighting
technique to provide faster initial ramp-up.

\section{Performance Evaluation} \label{sec:performance}
In this section, we evaluate the performance of PIA using simulation; evaluation through real implementation on a video player is deferred to Section~\ref{sec:pia_dash}.
Simulation allows us to evaluate a large set of parameters in a scalable manner, while real implementation provides insights under various system constraints.
In both cases, the network conditions are driven by
a set of traces captured from commercial LTE networks that allow reproducible runs, as well as apple-to-apple comparison of different schemes.
We first describe the evaluation setup, choice of parameters for PIA, and then compare PIA against several state-of-the-art schemes.

\subsection{Evaluation setup} \label{sec:eval-setup}

\noindent{\textbf{Network bandwidth traces.}} We focus on LTE networks that dominate today's cellular access technology. For evaluation under realistic LTE network environments, we collected 50 network bandwidth traces from two large commercial LTE networks in the US. These traces were collected under a wide range of settings, including different times of day, different locations (in three U.S. states,
 CT, NJ and NY),  and different movement speed (stationary, walking, local driving, and highway driving).

\begin{figure}[t]
	\centering
	\subfigure[ Throughput.]{%
		\includegraphics[width=0.4\textwidth]{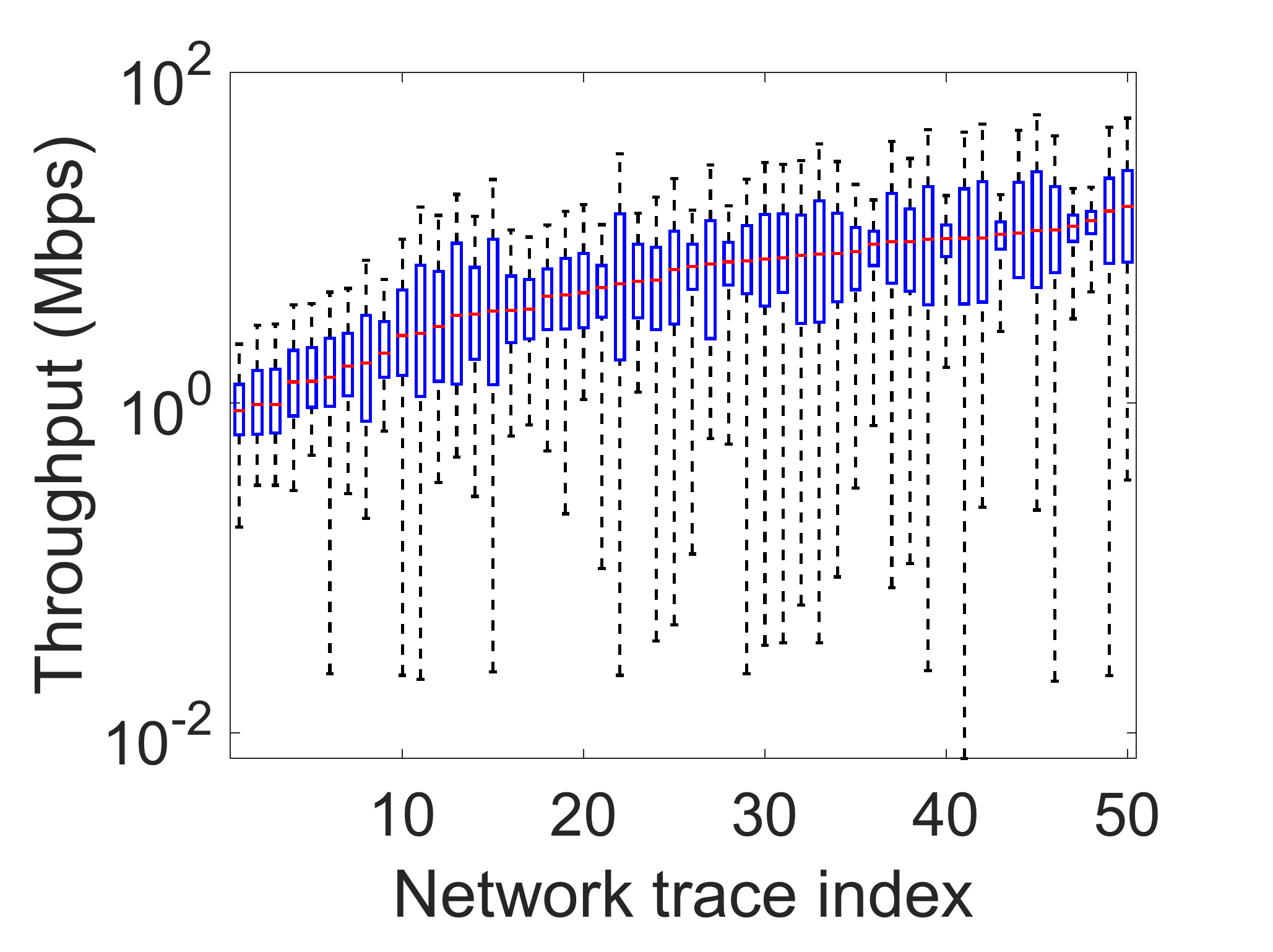}
		%\caption{}
	}
	\subfigure[ Prediction error.]{%
		\includegraphics[width=0.4\textwidth]{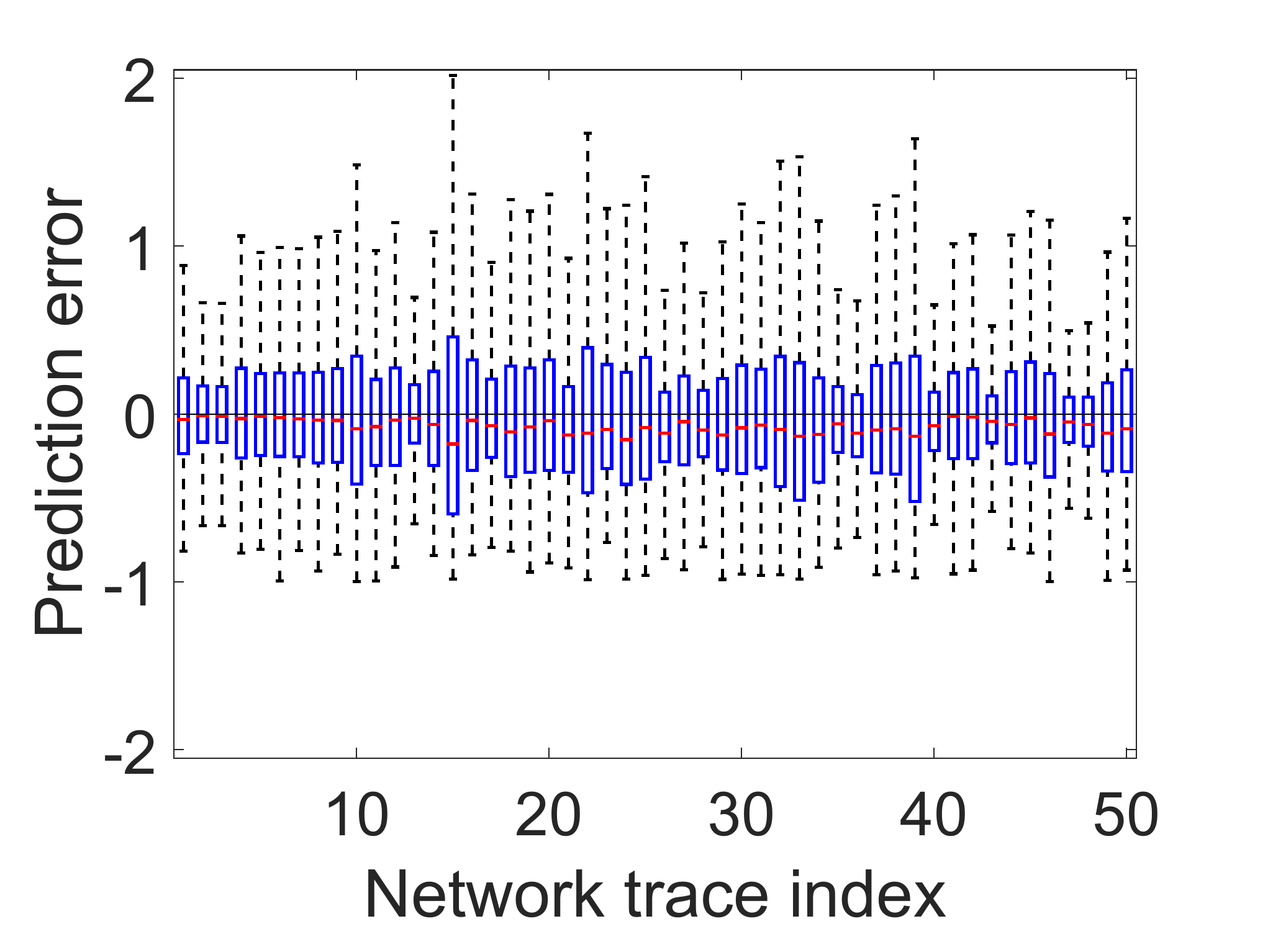}
		%\caption{Prediction error.}
	}
	\caption{Characteristics of the network bandwidth traces that are used in performance evaluation.}
	\label{fig:trace-charact}
\end{figure}

Each trace contains 30 minutes of one-second measurement of network bandwidth.
The bandwidth was measured  on a mobile device, as the throughput of a large file downloading from a well provisioned server to the device.
Fig.~\ref{fig:trace-charact}(a) is a boxplot that shows the minimum, first quartile, median, third quartile, and maximum bandwidth of each trace, where the traces are sorted by the median bandwidth.
We see that the network bandwidth is indeed highly dynamic. For some traces, the maximum bandwidth is tens of Mbps, while the minimum bandwidth is less than 10 Kbps.

These network traces, by capturing the bandwidth variability over time, accurately
reflect the impact of lower level network
characteristics (e.g., signal strength, loss, and RTT) on the network bandwidth perceived by an application.
Using the network traces is sufficient when evaluating ABR schemes;
there is no need to explicitly incorporate lower level network characteristics
 since ABR adaptation operates at the application level, using
application-level estimation of the network bandwidth.

%\tydubfigsingle{PIA-E/figure/trace_characteristics_box}{Throughput.}{PIA-E/figure/trace_characteristics_predicterror_box}{Prediction error.}{Characteristics of the network bandwidth traces that are used in performance evaluation.}{trace-charact}

\smallskip
\noindent{\textbf{Video parameters.}}
We use three video bitrate sets, all being Constant Bitrate  (CBR) videos: $\mathcal R_1=[0.35, 0.6, 1, 2, 3]$ Mbps, $\mathcal R_2=[0.35, 0.6, 1, 2, 3, 5]$ Mbps and $\mathcal R_3=[0.2, 0.4, 0.6, 1.2, 3.5, 5, 6.5, 8.5]$ Mbps. The first set is based on the reference for YouTube video bitrate levels (corresponding to 240p, 360p, 480p, 720p and 1080p respectively)~\cite{youtube-bitrate-levels}. The second set adds a higher bitrate level of 5 Mbps to the first set. The third set is based on Apple's HTTP Live Streaming standard~\cite{apple-bitrate-levels}. For each bitrate set, we further consider three variants with chunk duration of 2, 4, and 8 s.
%, and video length as 5, 10 or 20 minutes, respectively.

\smallskip
\noindent{\textbf{ABR Schemes.}}
We compare PIA against four other schemes; in Section~\ref{sec:pia_dash}, we further compare the performance of PIA and BOLA~\cite{spiteri2016bola}, another state-of-the-art ABR scheme, using DASH implementation.
\begin{itemize}
  \item {RB}: The bitrate is picked as the maximum possible bitrate that is below the predicted network bandwidth. This is a simple open-loop controller (see Section~\ref{sec:background}), serving as a baseline.

  \item {BBA~\cite{huang2014buffer}}: It is a state-of-the-art buffer based scheme. We use BBA-0, which is the BBA variant for CBR streaming, the focus of this work.
      We set the lower and upper buffer thresholds as $\theta_\text{low}=10$ s and $\theta_\text{high}=60$ s, respectively,
      and empirically verified that the above thresholds work well on our dataset.

  \item {MPC and RobustMPC~\cite{yin2015control}}: Both are state-of-the-art ABR schemes based on model predictive control. RobustMPC is more conservative in estimating network bandwidth, and has been shown to outperform MPC~\cite{mao2017pensieve} in more dynamic network settings (e.g., cellular networks). Both schemes use a look-ahead horizon of 5 chunks (as suggested by the paper).

  \item PIA: Unless otherwise stated, the target buffer level $x_r=60$ s that is compatible with the setting of BBA. In Section~\ref{subsec:perf-comparison}, we vary the target buffer level and explore its impact on performance. The look-ahead horizon is set to 5 chunks, i.e., $N=5$ in Eq. (\ref{eq:lms}).
\end{itemize}

For all the schemes, unless otherwise stated, the startup playback latency (i.e., the latency from when requesting the first chunk of the video to starting the playback of the video), $\delta$, is set to $5$, $10$ or $15$ s.
For BBA and MPC, their parameters are either selected based on the original papers, or configured by us based on the properties of the videos (e.g., chunk duration and encoding rates) as justified above.

\smallskip
\noindent{\textbf{Network bandwidth prediction.}}
For the schemes that require network bandwidth estimation, it is set as the harmonic mean of the network bandwidth of the past $20$ s. Harmonic mean has been shown to be robust to measurement outliers~\cite{jiang2012improving}. Fig.~\ref{fig:trace-charact}(b) shows the boxplots of the bandwidth prediction errors (the difference of the predicted and actual bandwidths divided by the actual bandwidth) of the network bandwidth traces used in our evaluations.
Each box in the plot corresponds to the distribution of all prediction instances within a particular trace.
The prediction is at the beginning of every second. As shown, the median prediction error is 20\% to 40\%, highlighting the challenges of accurate bandwidth prediction in LTE networks.
As we shall show later in Section~\ref{subsec:perf-comparison}, our PIA scheme is  very robust to such levels of inaccuracy in network bandwidth estimation.

\subsection{PIA: Choice of parameters} \label{subsec:kp-ki}

Following the methodology outlined in Section~\ref{subsec:PIA-tuning}, we first tune $K_p$ and $K_i$, and then tune $\beta$ for PIA.
One question we aim to answer is whether there exists a set of $K_p$ and $K_i$ values that works well in a wide range of settings.
This is an important issue related to the practicality of PIA -- because if the choice of $K_p$ and $K_i$ were too sensitive to the settings,
then tuning and/or adapting them for different settings would require more effort.

\subsubsection{Tuning $K_p$ and $K_i$}
We use a single combined performance metric when tuning $K_p$ and $K_i$. This is because, while as described earlier, the QoE is affected by three metrics (average video bitrate, the amount of bitrate changes and rebuffering) jointly, comparing the QoE under different choices of $K_p$ and $K_i$ is much simpler when using a single combined metric. Currently there is no consensus in the field around the form of such a metric. One approach is using a weighted sum of the three metrics as in~\cite{yin2015control}.
 %Following the approach in~\cite{yin2015control}, we use a weighted sum of the three metrics as QoE.
 Specifically, for a video of $M$ chunks,
\begin{eqnarray}
\label{eq:qoe}
\text{QoE}=\sum_{t=1}^{M}R_t -\mu \sum_{t=1}^{M-1}\left|R_{t+1}-R_t\right| - \lambda \sum_{t=1}^{M} S_t.
\end{eqnarray}
where
$R_t$ is the bitrate of the $t$-th chunk and $S_t$ is the amount of stalls for the $t$-th chunk, and $\mu$ and $\lambda$ are weights that represent, respectively, the importance of the middle and last terms (i.e., bitrate changes and rebuffering) relative to the first term (i.e., average bitrate) in the sum.
There is no well agreed-upon settings for $\mu$ and $\lambda$; we therefore vary $\mu$ and $\lambda$ over multiple values to assess sensitivity.
%for sensitivity test.

\begin{figure}[t]
	\centering
	\subfigure[ Heatmap.]{%
		\includegraphics[width=0.4\textwidth]{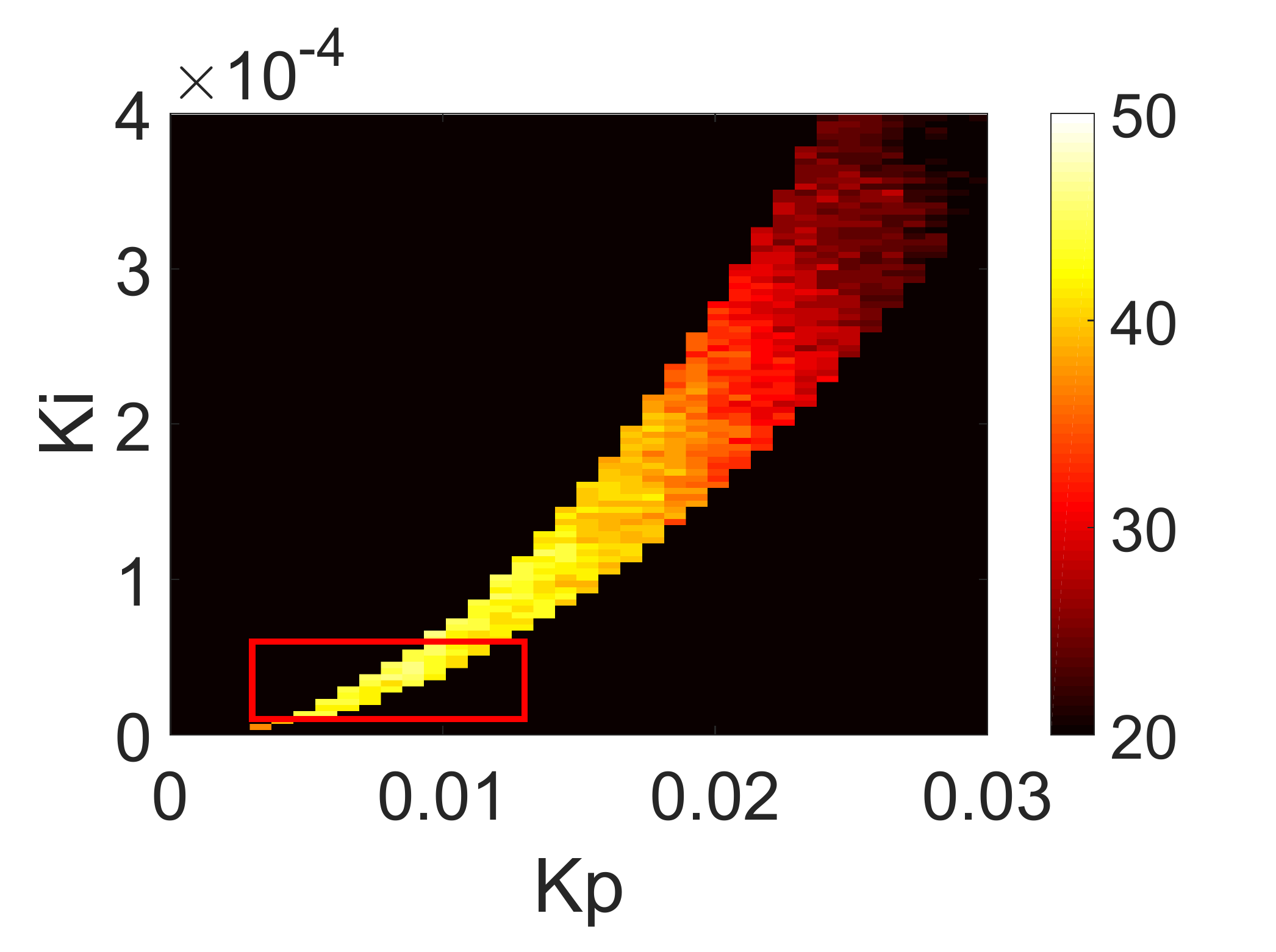}
		%\caption{}
	}
	\subfigure[ Histogram of ``heat" values.]{%
		\includegraphics[width=0.4\textwidth]{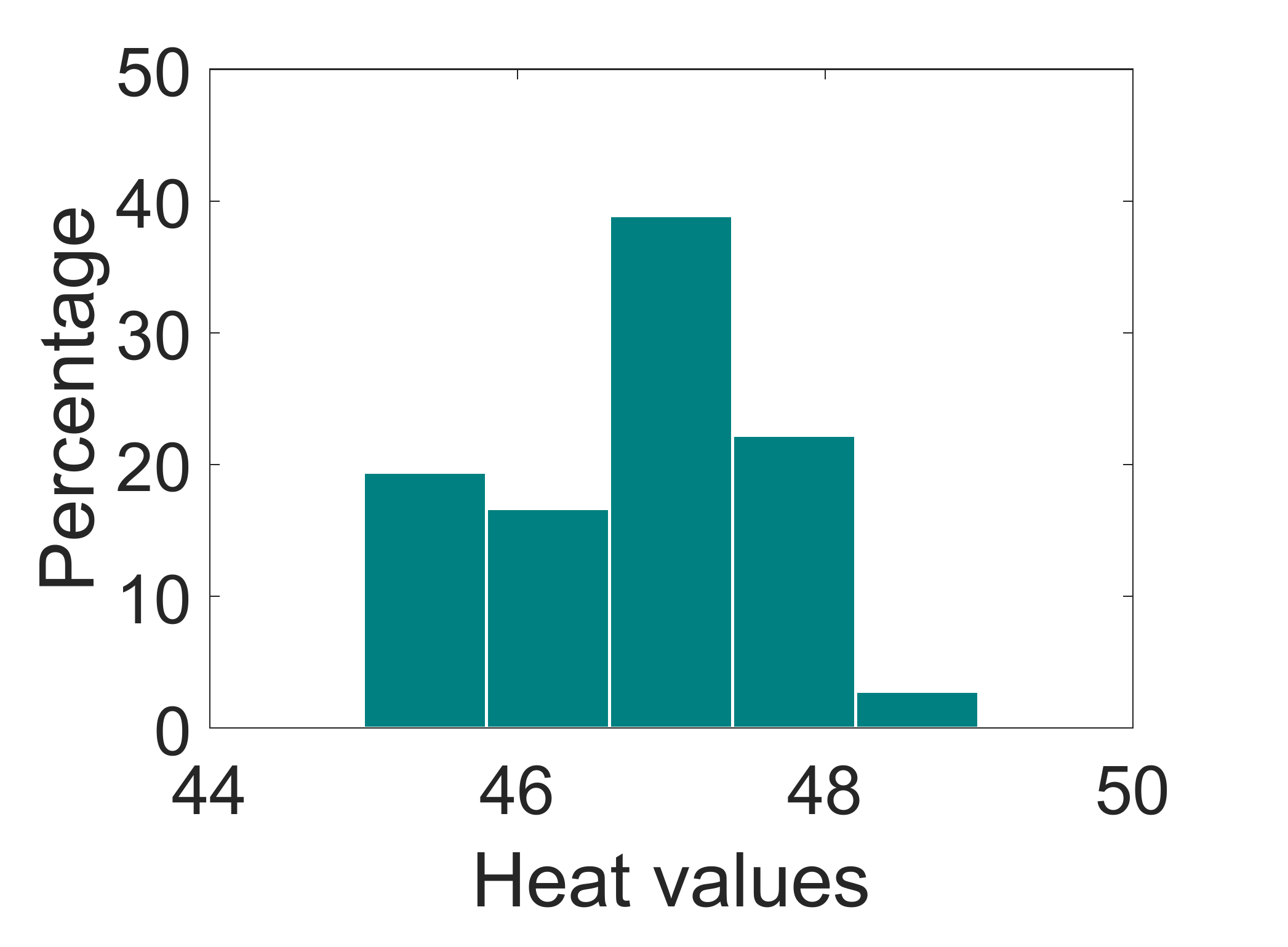}
		%\caption{Prediction error.}
	}
	\caption{Region of $K_p$ and $K_i$ and the corresponding ``heat" values in one setting (video bitrate set $\mathcal R_3$, chunk duration 2 s, video length 20 min,  startup latency 10 s, $\mu=1$, $\lambda=8.5$ (which is the same as the maximum bitrate (in Mbps) in $\mathcal R_3$).}
	\label{fig:region-kpki}
\end{figure}

%\tydubfigsingle{PIA-E/figure/heatmap.pdf}{Heatmap.}{PIA-E/figure/heatmap_hist.pdf}{Histogram of ``heat" values.}{Region of $K_p$ and $K_i$ and the corresponding ``heat" values in one setting (video bitrate set $\mathcal R_3$, chunk duration 2 s, video length 20 min,  startup latency 10 s, $\mu=1$, $\lambda=8.5$ (which is the same as the maximum bitrate (in Mbps) in $\mathcal R_3$).}{region-kpki}

The trace-driven simulation allows us to consider a very wide range of settings by varying a  number of parameters: the video bitrate set, video length, chunk size, startup latency, and $\mu$ and $\lambda$ in (\ref{eq:qoe}). Specifically, the video bitrate set is either $\mathcal R_1$, $\mathcal R_2$, or $\mathcal R_3$ (see Section~\ref{sec:eval-setup}), video length is 5, 10 or 20 minutes, chunk duration is 2, 4 or 8 s, startup latency is 5, 10 or 15 s, and $\mu$ is 1 or 2, and $\lambda$ is the maximum bitrate level of a video (e.g., 3 Mbps in $\mathcal R_1$) or twice as much. The choice of of $\mu$ and $\lambda$ is based on the settings in~\cite{yin2015control}. In each setting (i.e., after fixing the above parameters), we consider each of the 50 network bandwidth traces individually.
For the $k$-th network trace, we vary the values of $K_p$ and $K_i$ in a wide range to find a pair of $K_p$ and $K_i$ that maximizes the QoE (note that as described in Section~\ref{sec:PID}, we only consider valid combinations of $K_p$ and $K_i$ values, i.e., those so that the damping ratio is in $[0.6,0.8]$).
Once the maximum QoE, denoted as $Q^*_k$, is determined, the QoE under each valid ($K_p$, $K_i$) pair is compared to $Q^*_k$ to see whether it is within $90\%$ of $Q^*_k$. Specifically, we define a binary function $f_k(K_p,K_i)$ for the $k$-th network bandwidth trace, where $f_k(K_p,K_i)=1$ if the resulting QoE under $K_p$ and $K_i$ is within $90\%$ of $Q^*_k$;
otherwise, $f_k(K_p,K_i)=0$. We then consider all the network bandwidth traces, and create a heat map with the ``heat" for each valid pair of $K_p$ and $K_i$ values as $\sum_{k} f_k(K_p,K_i)$. Clearly, a larger ``heat" value for a $K_p$ and $K_i$ pair means that it leads to good performance for more network bandwidth traces.

Fig.~\ref{fig:region-kpki}(a) shows an example heat map for one setting (details of the setting described in the caption of the figure).
The black region represents invalid $K_p$ and $K_i$ pairs (i.e., those causing the damping ratio out of the desired range $[0.6,0.8]$). For the valid $K_p$ and $K_i$ pairs, the ``heat" value varies, with the highest values in the bottom left region, marked by the rectangle (brighter color represents higher ``heat" values). Fig.~\ref{fig:region-kpki}(b) is  the histogram of the ``heat" values in
the rectangle area (excluding those corresponding to invalid $K_p$ and $K_i$ pairs). It shows that majority of the values are close to 50 (i.e., the maximum ``heat"), indicating that the valid $K_p$ and $K_i$ pairs marked by the rectangle provide good performance across almost all network traces.

We repeat the above procedure for all the settings, and find the following region of $K_p$ and $K_i$ values leads to good performance for all the settings
\begin{equation}
\begin{split}
{K_p} \in \left[ {1 \times {{10}^{ - 3}}, 14 \times {{10}^{ - 3}}} \right]\\
{K_i} \in \left[ {1 \times {{10}^{ - 5}}, 6 \times {{10}^{ - 5}}} \right]\\
\text{s.t.}\,\,\zeta \left( {{K_p},{K_i}} \right) \in \left[0.6, 0.8\right].
\end{split}
\label{eq:good-Kp-Ki}
\end{equation}
Specifically, under the above range of values, the average ``heat" for the different settings varies from 31 to 50, and the standard deviation varies from 0.25 to 3.45.
The results in the rest of the chapter use $K_p=8.8\times 10^{-3}$, approximately the middle of the range of $K_p$ in (\ref{eq:good-Kp-Ki}), and $K_i=3.6\times10^{-5}$ so that the damping ratio is ${1}/\sqrt{2}$, a widely recommended value for damping ratio~\cite{ogata2010modern,McClamroch80:State-Models}.

The finding that a set of $K_p$ and $K_i$ values works well under a wide range of settings is encouraging. 
Considering that the network traces were collected under a wide range of settings and that they exhibit significantly different characteristics (see Fig.~\ref{fig:trace-charact}), our results show that $K_p$ and $K_i$ can be tuned to accommodate the large variations among individual traces. The above results indicate that we can find a range of $K_p$ and $K_i$ values to make the system capable of dealing with the rapid bandwidth variations (one of the predominant characteristics of cellular networks), despite the differences across individual network conditions.

%\tydubfigsingle{PIA-E/figure/step_response.pdf}{Step response.}{PIA-E/figure/avg-QoE-pid-R3.pdf}{QoE.}{Choosing $\beta$ in PIA. In (b), the setting is the same as that used for Fig.~\ref{fig:region-kpki}. }{system-characteristics}
%\vspace{-1cm}

\begin{figure}[t]
	\centering
	\subfigure[ Step response.]{%
		\includegraphics[width=0.4\textwidth]{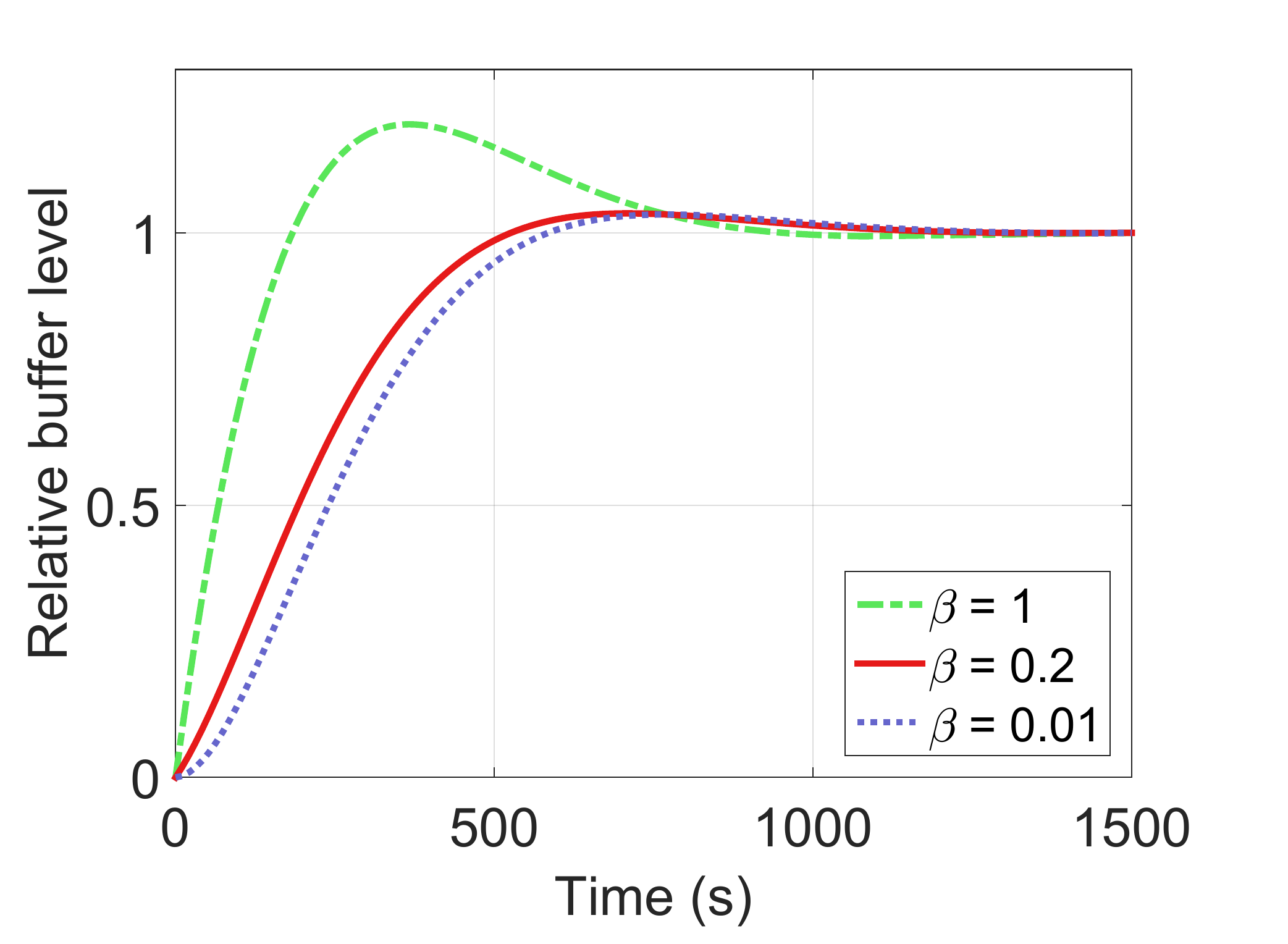}
		%\caption{}
	}
	\subfigure[ QoE.]{%
		\includegraphics[width=0.4\textwidth]{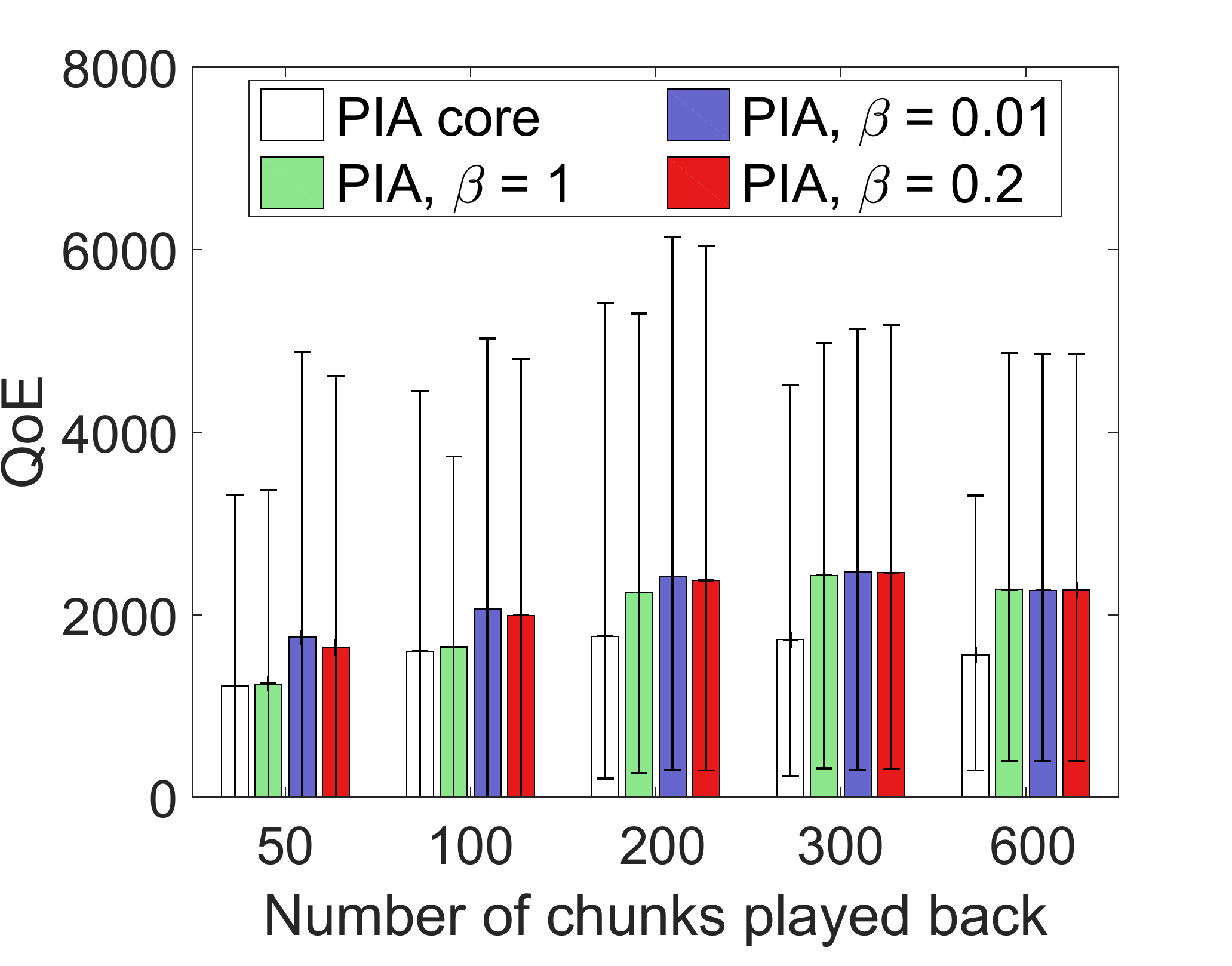}
		%\caption{Prediction error.}
	}
	\caption{Choosing $\beta$ in PIA. In (b), the setting is the same as that used for Fig.~\ref{fig:region-kpki}.}
	\label{fig:system-characteristics}
\end{figure}

\subsubsection{Tuning $\beta$}
Once $K_p$ and $K_i$ are determined, we tune $\beta$ for the initial stage of the video playback.
Specifically, we set $\beta$ to 0.01, 0.2, 0.4, 0.6, 0.8, and 1.0.
Fig.~\ref{fig:system-characteristics}(a) shows the step response of the control policy (only the results for $\beta=0.01$, 0.2 and 1 are shown for better clarity). When $\beta = 1$, the buffer becomes full much more quickly than when $\beta=0.2$ and 0.01. This is because, as explained in Section~\ref{subsec:PID-enhanced}, lower bitrate tends to be selected when $\beta=1$, causing the buffer to fill up more quickly. 

To examine the three QoE metrics jointly, Fig.~\ref{fig:system-characteristics}(b) plots the QoE when playing up to the $i$-th chunk of a video of 600 chunks when $\beta=0.01$, 0.2 or 1. The results are averaged over all the network traces; the 95\% confidence intervals are plotted in the figure as well. 
We see that $\beta$ indeed affects the QoE for the initial playback, and $\beta=1$ leads to lower QoE compared to $\beta=0.01$ and 0.2. Further investigation reveals that $\beta=0.01$ leads to more rebuffering than $\beta=0.2$. The above results are for the setting used for Fig.~\ref{fig:region-kpki}. Results in other settings show similar trends. Since rebuffering has very detrimental effects on viewing quality, we use $\beta=0.2$ in the rest of the chapter.

Last, the results of PIA core (i.e., without the three enhancing techniques) are also shown in Fig.~\ref{fig:system-characteristics}(b). We see that PIA core indeed leads to lower QoE compared to the full-fledged PIA, demonstrating the benefits of our three enhancing techniques.

\subsection{Performance comparison} \label{subsec:perf-comparison}

%\triplefig{PIA-E/figure/indx-cs-2-v2-sf-3-avgbit-cdf-robustMPC}{}{PIA-E/figure/indx-cs-2-v2-sf-3-bitchange-cdf-robustMPC}{}{PIA-E/figure/indx-cs-2-v2-sf-3-rebuffer-cdf-robustMPC}{}
%{Performance comparison in the default setting (chunk duration 2 s, video bitrate set $\mathcal R_2$, video length 20 minutes, startup latency 10 s).}{video-performance}

\begin{figure}[t]
	\centering
	\subfigure[ ]{%
		\includegraphics[width=0.3\textwidth]{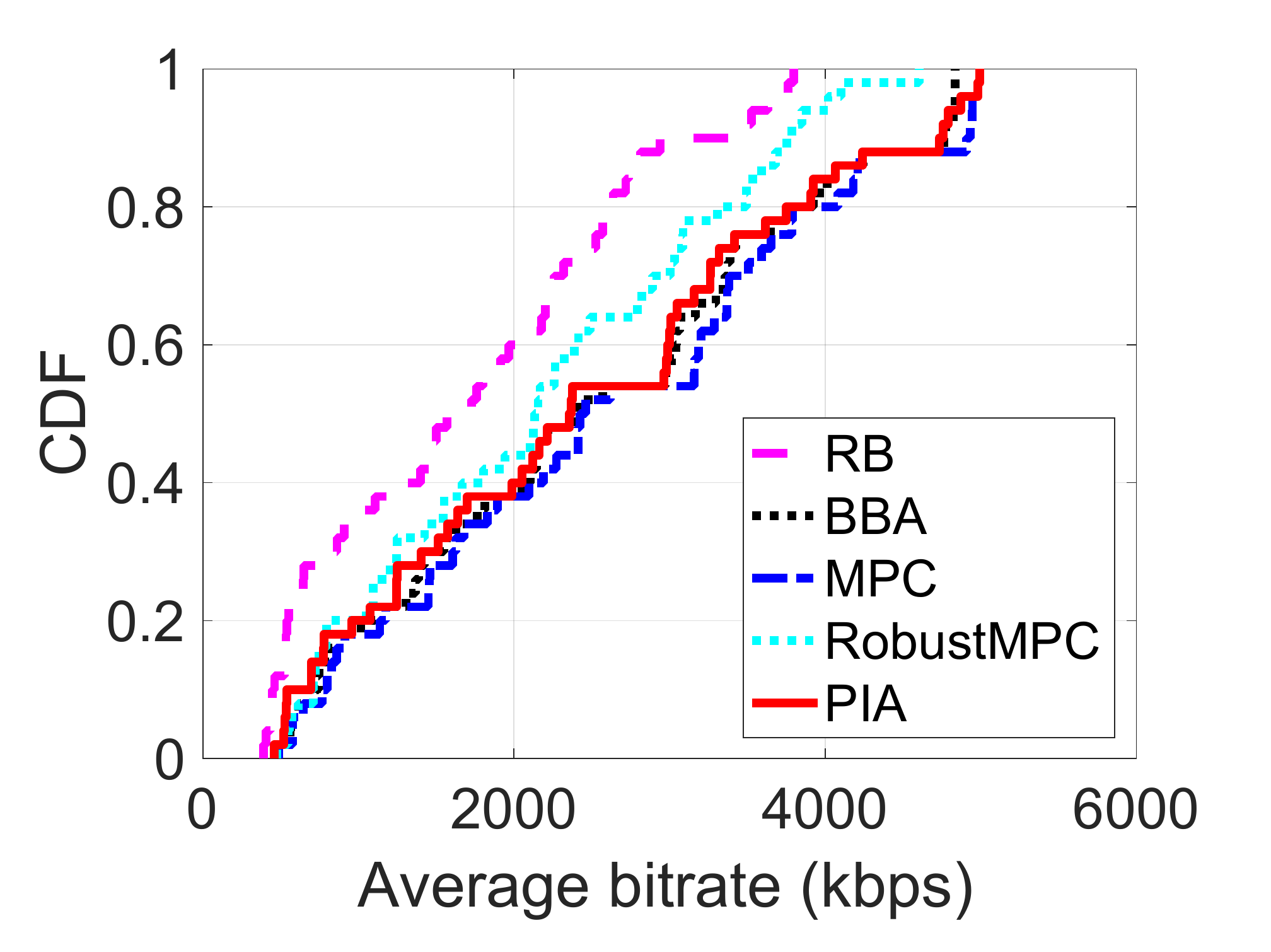}
		%\caption{}
	}
	\subfigure[ ]{%
		\includegraphics[width=0.3\textwidth]{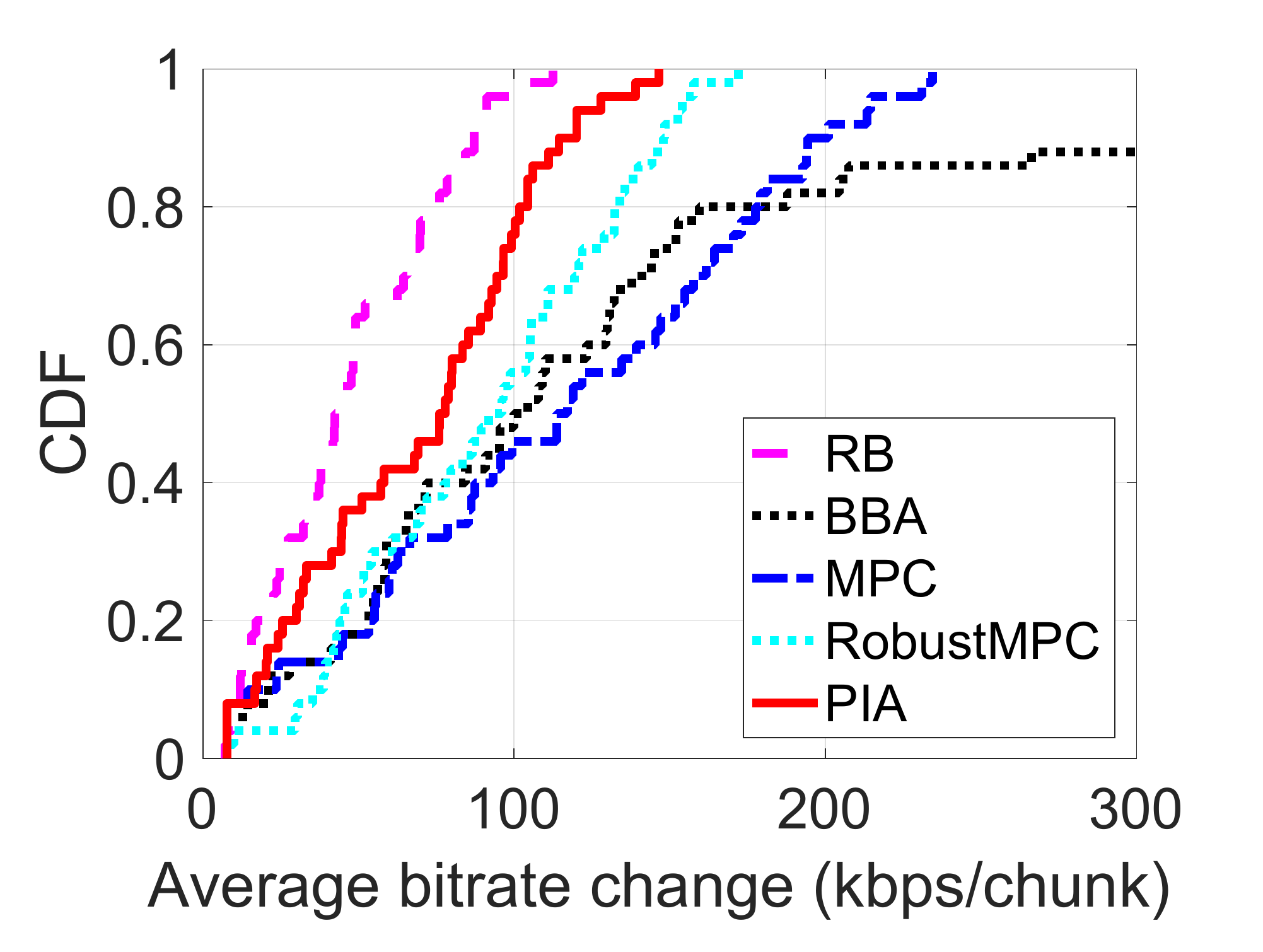}
		%\caption{Prediction error.}
	}
\subfigure[]{%
	\includegraphics[width=0.3\textwidth]{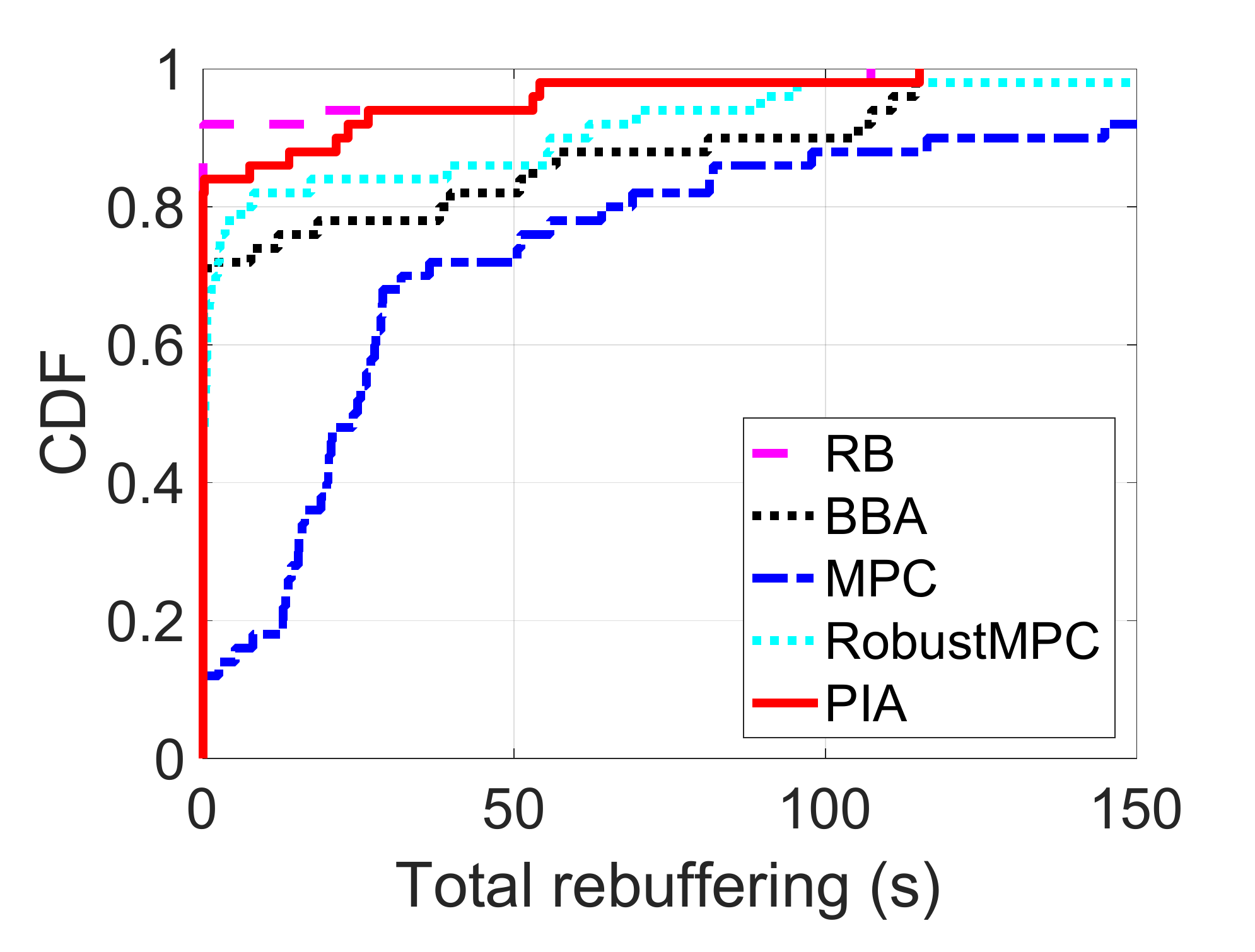}
	%\caption{Prediction error.}
}
	\caption{Performance comparison in the default setting (chunk duration 2 s, video bitrate set $\mathcal R_2$, video length 20 minutes, startup latency 10 s).}
	\label{fig:video-performance}
\end{figure}

In the following, we first present the performance of PIA in the default setting, i.e., chunk duration of 2 s, video bitrate set $\mathcal R_2$,
video length of 20 minutes,
and startup latency of 10 s. After that, we evaluate the impact of the various parameters on the performance of PIA.

Fig.~\ref{fig:video-performance} plots the CDF of the three QoE metrics
over all network bandwidth traces in the default setting.
The performance of four schemes, RB, BBA, MPC, RobustMPC and PIA, are plotted. We see that, while the amount of bitrate change and rebuffering is low under RB, its average bitrate is significantly lower than those of the other schemes.
PIA achieves comparable average bitrate as BBA and MPC, with significantly less bitrate changes and rebuffering. Specifically, the average bitrate of PIA is 98\% and 96\% of that of BBA and MPC, respectively, while the average amount of bitrate change is 49\% and 40\% lower, and the average amount of rebuffering is 68\% and 85\% lower than BBA and MPC, respectively. 
RobustMPC has significantly lower rebuffering than MPC, but its rebuffering is still higher than that under PIA. In addition,
RobustMPC leads to significantly lower average bitrate than MPC, PIA and BBA.

Overall, \emph{PIA achieves the best balance among the three conflicting QoE metrics}.
As described earlier, the inferior performance of RB is because it uses an open-loop control without any feedback. The superior performance of PIA compared to BBA is because BBA implicitly uses one form of P-control (Section~\ref{sec:background}) that only takes the present error into account, while PIA considers both the present and past errors.
PIA's approach of applying PID in an explicit and adaptive manner further facilitates the design and improves the performance.
The performance of MPC is sensitive to network bandwidth estimation errors~\cite{yin2015control}: it solves a discrete optimization problem at each step; when network bandwidth estimation is inaccurate, the input to the optimization problem is correspondingly inaccurate, leading to suboptimal performance. RobustMPC leads to lower rebuffering than MPC due to its more conservative network bandwidth estimation. 
It, however, also leads to significantly lower bitrate choices.

\begin{figure}[h]
    \centering
        \includegraphics[height=2.5in]{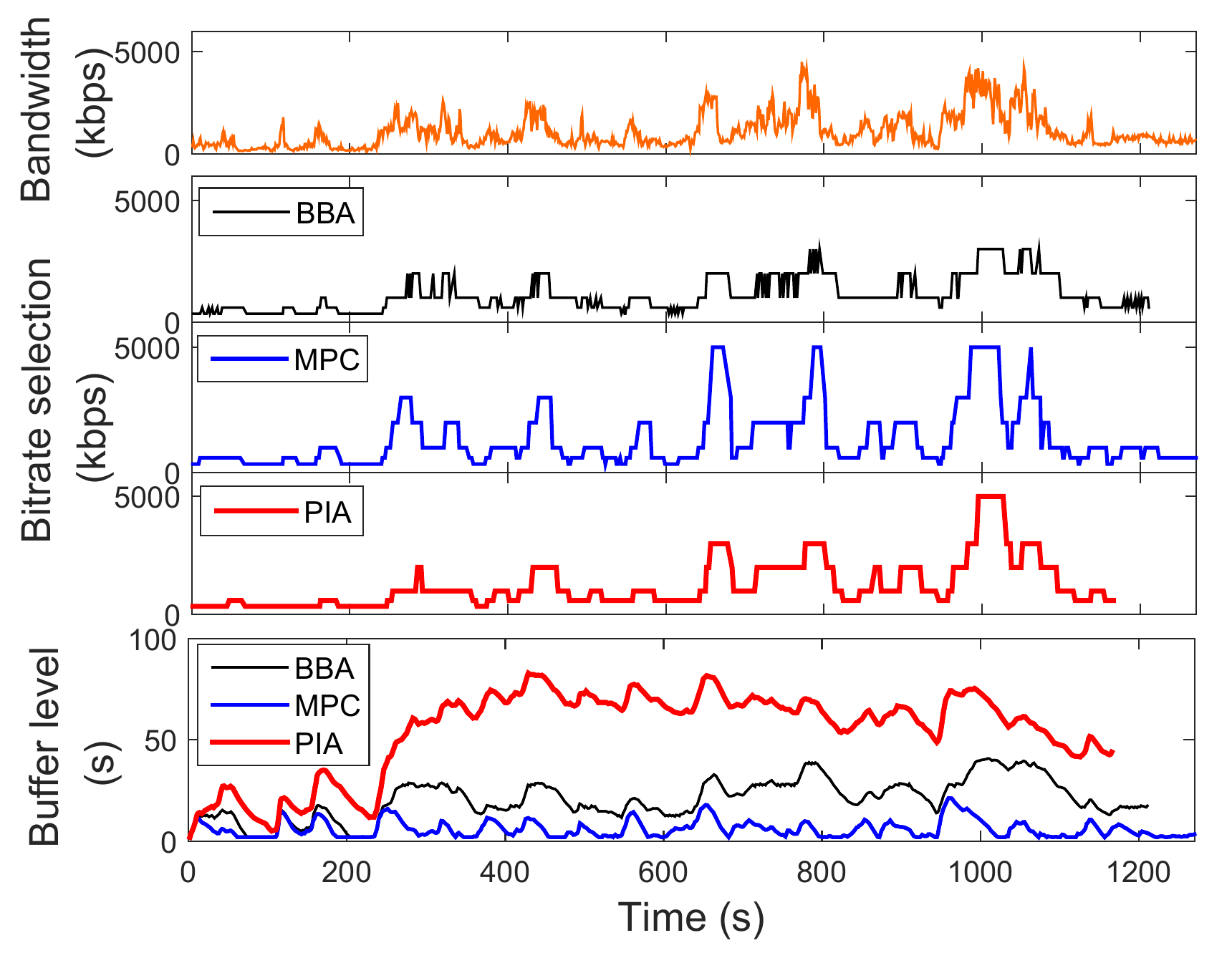}
        \caption{Comparison of different schemes for one trace under the default setting (chunk duration 2 s, video bitrate set $\mathcal R_2$, video length 20 minutes, startup latency 10 s). {The plot on buffer level shows the results until the end of the downloading.}}
\label{fig:individual_cmp}
\end{figure}

To provide further insights, Fig.~\ref{fig:individual_cmp} plots the bitrate selection and the buffer level over time for BBA, MPC and PIA when using one network trace. For reference, it also plots the network bandwidth of the trace.  We clearly see that BBA has significantly more bitrate changes; MPC tends to be more aggressive in choosing higher bitrates, which can lead to excessive rebuffering. The bitrate selection under PIA matches well with the network bandwidth without frequent bitrate changes.
In terms of the buffer level/occupancy (i.e., the duration of the video content that has been brought in and has not yet been played back) shown in the bottom plot in Fig.~\ref{fig:individual_cmp}, the buffer level of MPC is lower than that of BBA and PIA due to its aggressive choice of bitrate; the buffer level of PIA reaches the target level  of 60 s at around 300 s, and then stays around the target level; the buffer level of BBA is in between that of MPC and PIA.

\begin{figure}[t]
	\centering
	\subfigure[ Bitrate.]{%
		\includegraphics[width=0.4\textwidth]{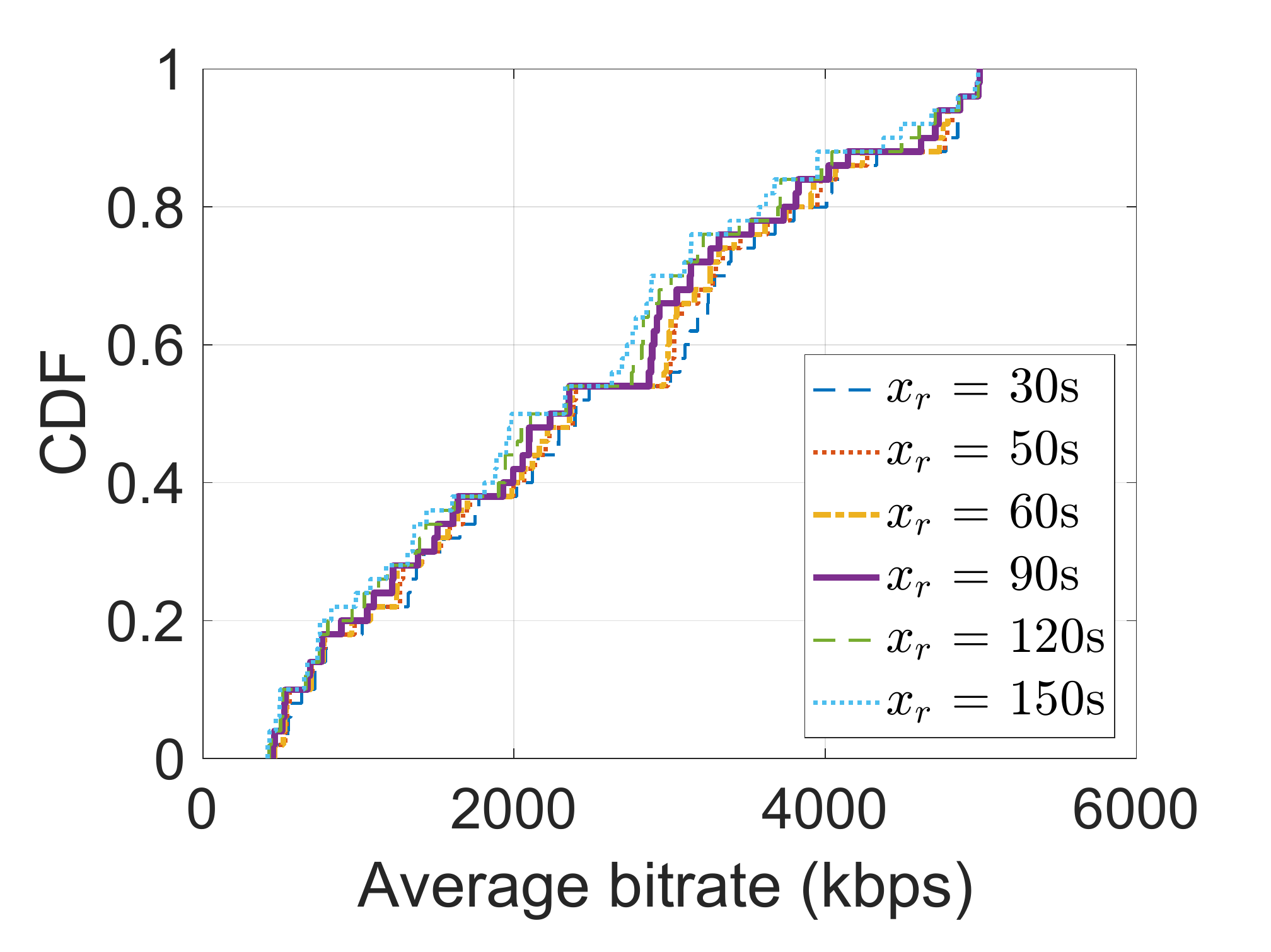}
		%\caption{}
	}
	\subfigure[ Rebuffering.]{%
		\includegraphics[width=0.4\textwidth]{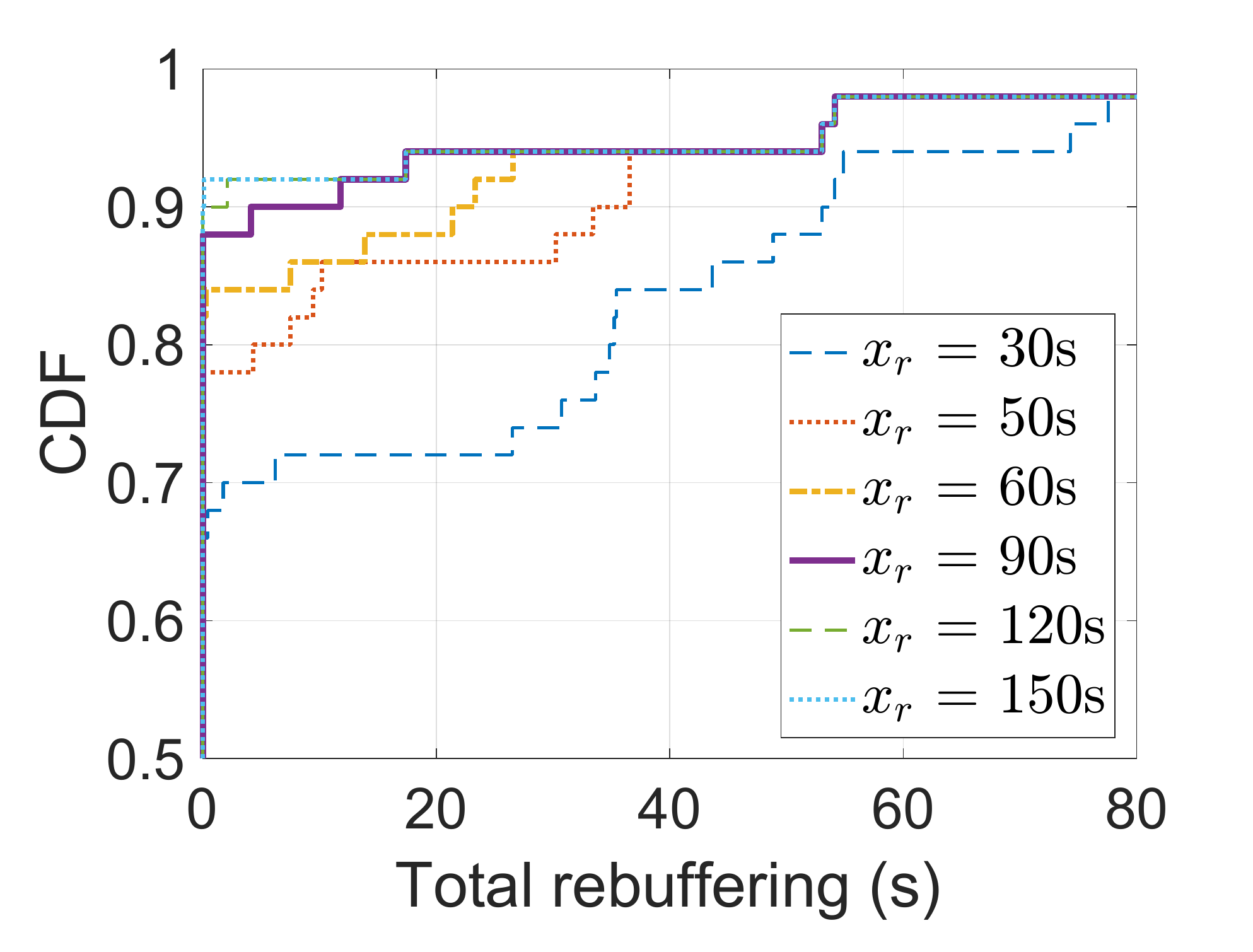}
		%\caption{Prediction error.}
	}
	\caption{Impact of the target buffer level on performance (video bitrate set $\mathcal R_2$, chunk duration 2 s, video length 20 minutes, startup latency 10 s).}
	\label{fig:impact-xr}
\end{figure}

\medskip
\noindent{\textbf{Impact of target buffer level.}} We vary the target buffer level, $x_r$, from 30 to 200 seconds, and evaluate its impact on the various performance metrics. We observe that larger target buffer levels lead to lower bitrate choices (to reach the larger target buffer levels during buffer ramp-up periods, e.g., at the beginning of the playback or after stall events).
%This is not surprising since lower bitrate tends to be chosen in order to maintain the larger target buffer levels. 
A large target buffer level also has the drawback that it may lead to more waste of resources when a user abandons watching a video in the middle of the playback. Using a small target buffer level, however, can lead to  higher rebuffering. Fig.~\ref{fig:impact-xr} plots the average bitrate  and the amount of rebuffering for different $x_r$ values. We observe noticeably lower bitrate when $x_r=150$ seconds, and noticeably higher rebuffering when $x_r=30$ seconds. Setting $x_r$ to 50 to 120 seconds leads to similarly good performance in all three performance metrics (the average bitrate change between two consecutive chunks across the $x_r$ values is similar, and is not shown in the figure).

%\tydubfigsingle{PIA-E/figure/target-buffer-bitrate-cdf}{Bitrate. }{PIA-E/figure/target-buffer-rebuffer-cdf}{Rebuffering.}{Impact of the target buffer level on performance (video bitrate set $\mathcal R_2$, chunk duration 2 s, video length 20 minutes, startup latency 10 s).}{impact-xr}

\medskip
\noindent{\textbf{Impact of video length.}}
The above results are for video length of 20 mins. We vary the ending time of the video to investigate PIA's performance for shorter videos.
When the ending time is larger than 5 mins (i.e., video length longer than 5 mins), we observe similar results as before; for much shorter videos, PIA has lower average bitrate compared to MPC (the average bitrate of PIA is 81\% of MPC when the video length is 2 minutes, and 90\% of MPC when the video length is 5 minutes), but still outperforms BBA and MPC on the other two metrics. We investigate how to improve the bitrate of PIA in the startup phase in Section~\ref{sec:startup-phase}.

\medskip
\noindent{\textbf{Impact of video bitrate sets.}} Recall that the video bitrate set $\mathcal R_2$ has one higher bitrate level of 5 Mbps compared to $\mathcal R_1$.
We further investigate two more video bitrate sets $\mathcal R_4=[0.2, 0.35, 0.6, 1, 2, 3]$ Mbps, which has one lower bitrate of 0.2 Mbps compared to $\mathcal R_1$; and $\mathcal R_5=[0.2, 0.35, 0.6, 1, 2, 3, 5]$ Mbps, which has one lower and one higher video bitrate levels (of 0.2 and 5 Mbps) compared to $\mathcal R_1$.
We observe consistent trend for all the schemes under the above three video bitrate sets.
Comparing the results under $\mathcal R_1$ and $\mathcal R_2$, we see that adding one higher bitrate level leads to higher average video bitrate, more bitrate changes, and more rebuffering;
comparing the results under $\mathcal R_1$ and $\mathcal R_4$, we see that adding one lower bitrate level maintains the average video bitrate while reducing bitrate changes and rebuffering;
comparing $\mathcal R_1$ and $\mathcal R_5$, we see that adding both one lower and higher bitrate levels increases the average video bitrate and bitrate changes, while reducing the rebuffering.
In general, adding more bitrate levels helps improve at least one of the three metrics. Across the settings,
PIA has the lowest bitrate switches, the lowest rebuffering, and similar average bitrate compared to BBA and MPC.

\medskip
\noindent{\textbf{Impact of video chunk duration.}}
We vary the video chunk duration by setting it to 2, 4, or 8 s.
We found that, for all chunk durations, PIA consistently outperforms MPC and BBA in balancing the tradeoffs incurred by the three metrics. For example, for chunk duration of 8 s, PIA's average playback bitrate differs from BBA and MPC only by 0.1\% and 3.2\%, respectively, while PIA reduces the rebuffering duration by 67\% and 68\% compared to those of BBA and MPC, respectively.

\begin{figure}[h]
    \centering
        \vspace{0in}
        \includegraphics[height=1.5in]{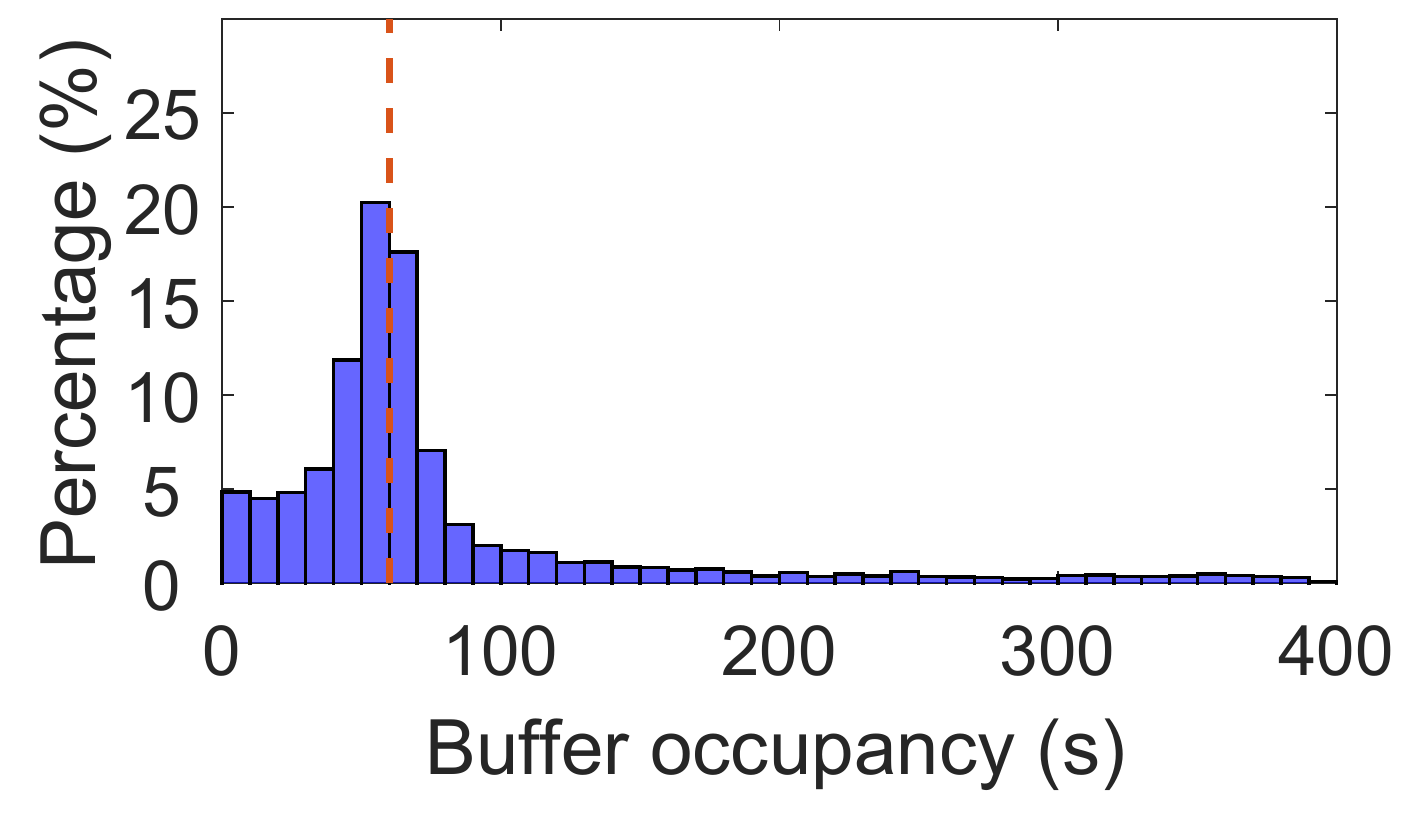}
        \caption{The distribution of buffer occupancy of PIA (under the default setting, i.e., chunk duration 2 s, video bitrate set $\mathcal R_2$, video length 20 minutes, startup latency 10 s).}
        \vspace{0in}
\label{fig:buffer_occupancy_dist}
\end{figure}

\medskip
\noindent{\textbf{Buffer occupancy and impact of maximum buffer size.}}
In the evaluations so far, we assume that all data  downloaded ahead of the current playback point
is stored in the client buffer until it is played back.
Fig.~\ref{fig:buffer_occupancy_dist} plots the distribution of the buffer occupancy
%(the amount of content in the buffer
(i.e., all the chunks brought in that have not yet been played back) under PIA in the default setting across all the network traces, where we record the buffer occupancy after downloading each chunk. We observe that $52$\% of the time, the buffer occupancy is below the target buffer size (60 s), $85$\% of the time it is below 100 s, and $95$\% of the time it is below 200 s. The above results indicate that, while there is no explicit constraint on buffer size, the amount of video stored at the client under PIA is not large (200 s of video corresponds to at most 125 MB even if we consider buffering the maximum bitrate track in $\mathcal R_2$).

In practice, a player may impose a maximum buffer size, $B_{\max}$, so that the amount of video downloaded before its playback time does not exceed this limit. We explore a simple strategy for this case. Specifically, the client stops downloading video when the buffer is full, and resumes downloading when there is space in the buffer. We set $B_{\max}=90$, 120, 150, 180 or 210 s, motivated by the practice of commercial players, which set the maximum buffer limit to tens to hundreds of seconds~\cite{dash-js,exoplayer,huang2014buffer,Xu17:Dissecting}. In the following, in the interest of space, we only report the results when $B_{\max}=90$ or 210 s.

\begin{figure}[h]
    \centering
        \vspace{0in}
        \includegraphics[height=1.5in]{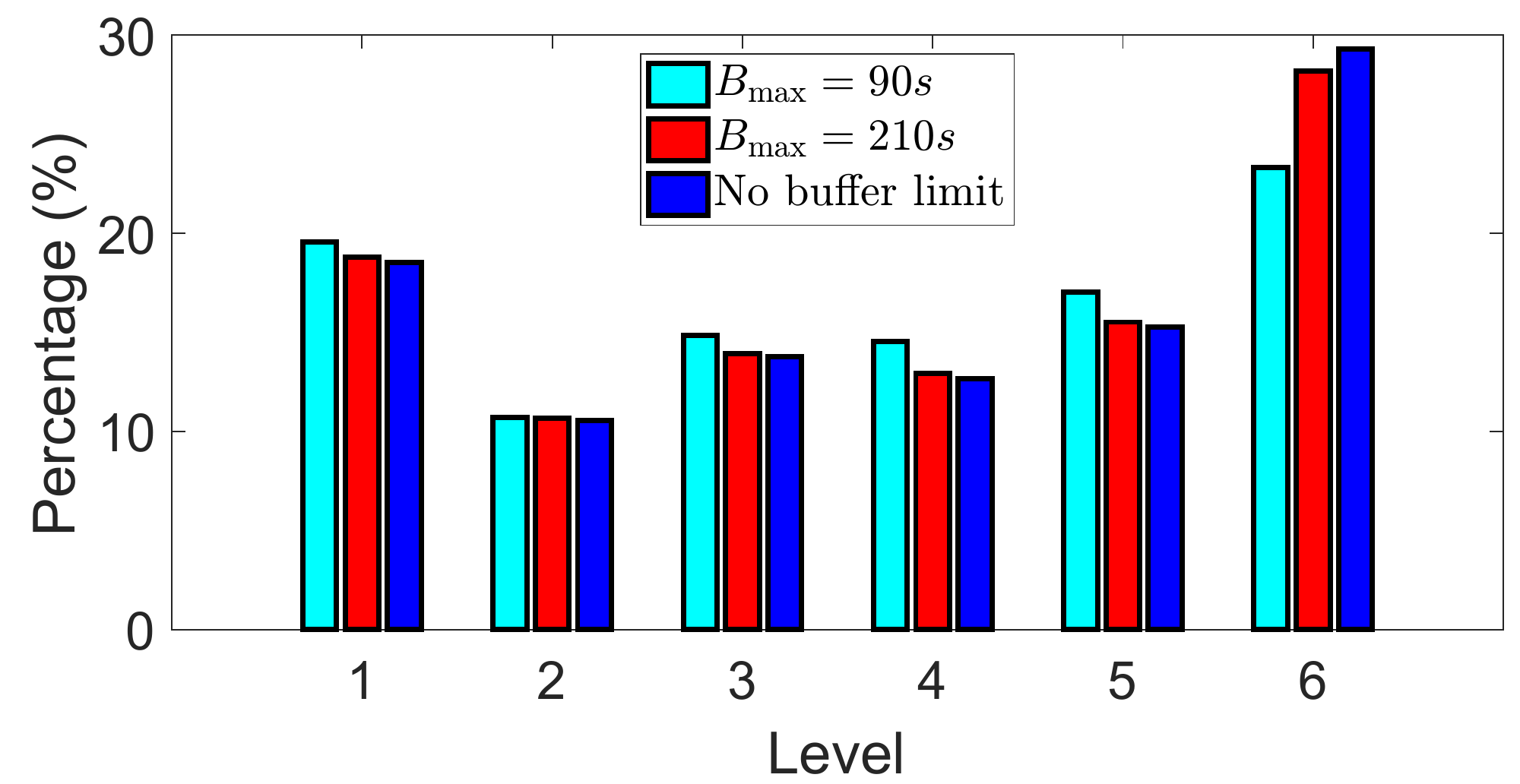}
        \caption{Impact of the maximum buffer size on the bitrate adaptation of PIA (under the default setting, i.e., chunk duration 2 s, video bitrate set $\mathcal R_2$, video length 20 minutes, startup latency 10 s). }
        \vspace{0in}
\label{fig:bufferlimit_cmp}
\end{figure}

\begin{figure}[h]
    \centering
        \vspace{0in}
        \includegraphics[height=2.2in]{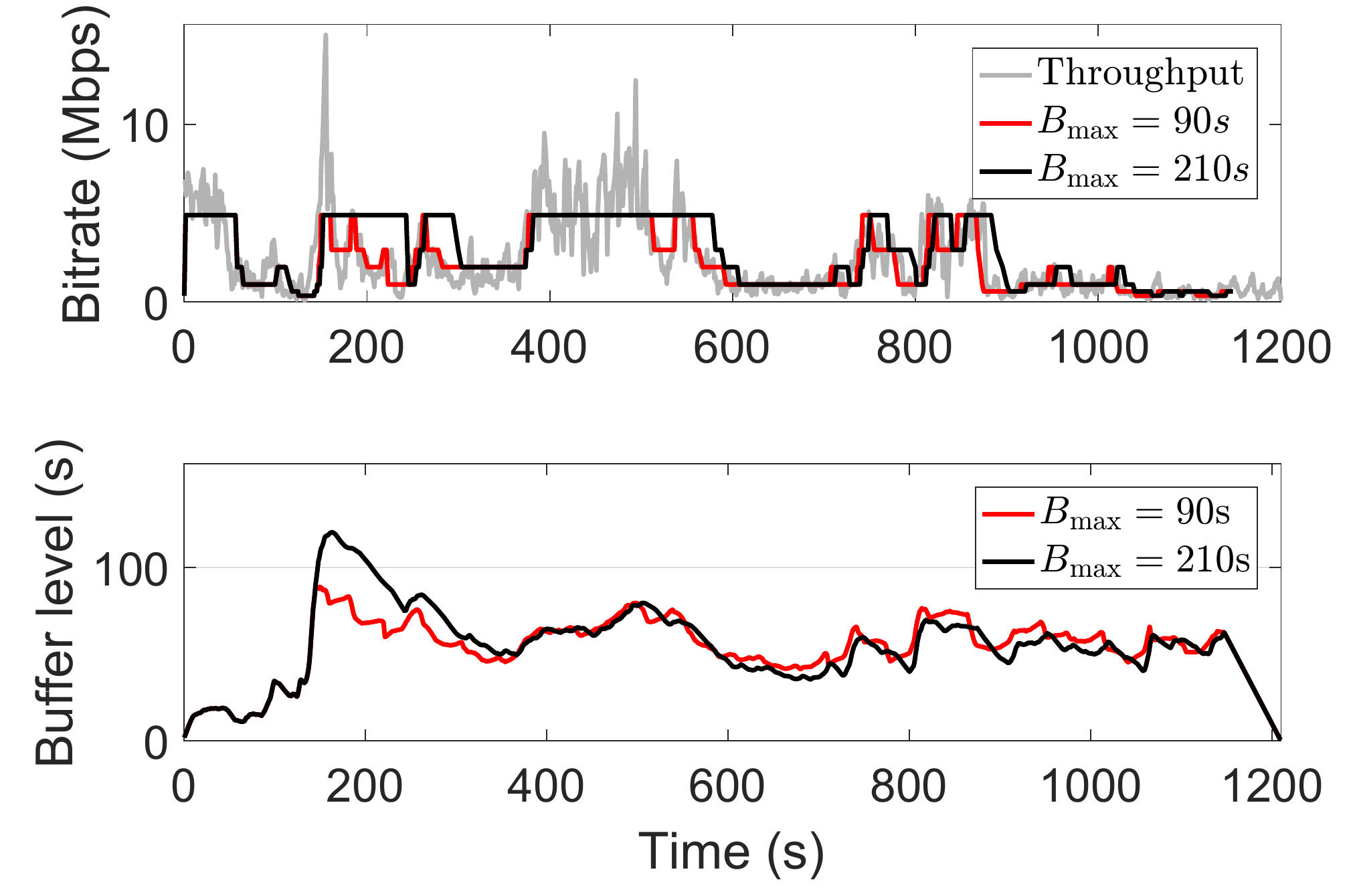}
        \caption{An example that illustrates the impact of the maximum buffer size on the bitrate choice of PIA (under the default setting, i.e., chunk duration 2 s, video bitrate set $\mathcal R_2$, video length 20 minutes, startup latency 10 s).}
        \vspace{0in}
\label{fig:bufferlimit_cmp_indv}
\end{figure}

When imposing the maximum buffer limit, not surprisingly, the average bitrate of all the schemes is reduced. On the other hand, 
the reduction under PIA  when $B_{\max}=90$ s is only slightly more than that $B_{\max}=200$ s (consistent with the observation in Fig.~\ref{fig:buffer_occupancy_dist} that 82\% and 96\% of the time, the buffer size is less than 90 s and 210 s, respectively).  PIA still achieves the best balance among the three conflicting performance metrics. 
Fig.~\ref{fig:bufferlimit_cmp} plots the distribution of the tracks selected by PIA under the default setting across all the network traces (the video bitrate set $\mathcal R_2$ has six tracks with increasing bitrate). 
For comparison, the bitrate choices when there is no maximum buffer limit are also plotted in the figure. We observe that, when imposing a maximum buffer limit, the probability of choosing the highest track is reduced (particularly when $B_{\max}=90$ s), while the probability of choosing the lower tracks is increased.
This is because, with the maximum buffer limit, the amount of video in the buffer is limited (and hence tends to be less than the amount when there is no buffer limit); the player thus has less cushion for preventing buffer underruns, and is less likely to choose the highest track (which takes longer to download compared to lower tracks, causing more buffer drainage).
Fig.~\ref{fig:bufferlimit_cmp_indv} shows an example. We see from 150 to 260 seconds, the buffer level is significantly lower under $B_{\max}=90$ s than that when $B_{\max}=210$ s. As a result, the player chooses lower tracks for the former, while choosing the highest track for the latter.

In the above, we use a simple strategy that stops downloading when the buffer is full, which does not fully leverage the network bandwidth. This simple strategy can be improved in two directions. The first direction is segment replacement, i.e., the player can leverage the bandwidth to preempt some earlier downloaded chunks that are of lower bitrate (i.e., replace them with chunks with higher bitrate and hence quality). While segment replacement has been used in commercial systems, the performance of existing commercial implementation is not satisfactory~\cite{Xu17:Dissecting}.
The key decision of segment replacement is to select which chunks to preempt and which bitrate levels to replace them with, which we will investigate in future studies. The second direction is designing bitrate selection strategies to download higher bitrate chunks proactively, instead of filling in the buffer with lower bitrate chunks. Further exploration along this direction is also left as future work.

\medskip
\noindent{\textbf{Impact of startup delay.}}
So far we have set the startup delay to be 10 s, i.e., waiting for 10 s after requesting the first chunk of the video before starting the playback of the video.
We next vary the startup delay.
Specifically, we assume the chunk duration is 2 or 6 s and the startup delay is 6 or 12 s, equivalent to 3 or 6 chunks for chunk duration of 2 s, and 1 or 2 chunks for chunk duration of 6 s.
Our results show that the amount of rebuffering depends more on the number of chunks that is downloaded during the startup delay, and is less sensitive to the length of the startup delay. For the same amount of startup delay (in seconds), using a smaller chunk duration leads to less rebuffering than using a larger chunk duration. In addition, having at least 2 or 3 chunks in buffer before playback starts leads to significantly less rebuffering than having a single chunk. The above findings are consistent with those in~\cite{Xu17:Dissecting}.

\subsection{Computational overhead} \label{sec:perf-comp-overhead}
As described earlier, the computational overhead of PIA is much lower than that of MPC: for $|{\mathcal L}|$ bitrate levels and horizon $N$, the complexity of MPC is $O(|{\mathcal L}| ^N)$, while the complexity of PIA is $O(|{\mathcal L}| N)$.
Table~\ref{table:cpu-compare} compares the average execution time of running MPC, BBA and PIA for a 600-chunk video (2-second chunk with bitrate set $\mathcal R_2$) on a commodity laptop with Intel i5 2.6 GHz CPU and 16 GB RAM.
The CPU time for PIA is $0.17$ s,
comparable to that of BBA.
The CPU time for MPC is more than 200 times higher than that for PIA.

\begin{table}[ht]
\caption{Comparison of computational overhead.} % title of Table
\centering
\begin{tabular}{c c c c} % centered columns (5 columns)
\hline\hline %inserts double horizontal lines
&MPC & BBA  & PIA \\ [0ex]
\hline
CPU time (s)  &  36.04      &  0.08        & 0.17\\[0ex]
\hline %inserts single line
\end{tabular}
\label{table:cpu-compare} % is used to refer this table in the text
\end{table}

\section{Improving Startup Performance} \label{sec:startup-phase}

\begin{figure*}[ht]
 \centering
 \subfigure[\label{subfig:fig:videolengthAnalysisBitrate}]{%
      \includegraphics[width=0.3\textwidth]{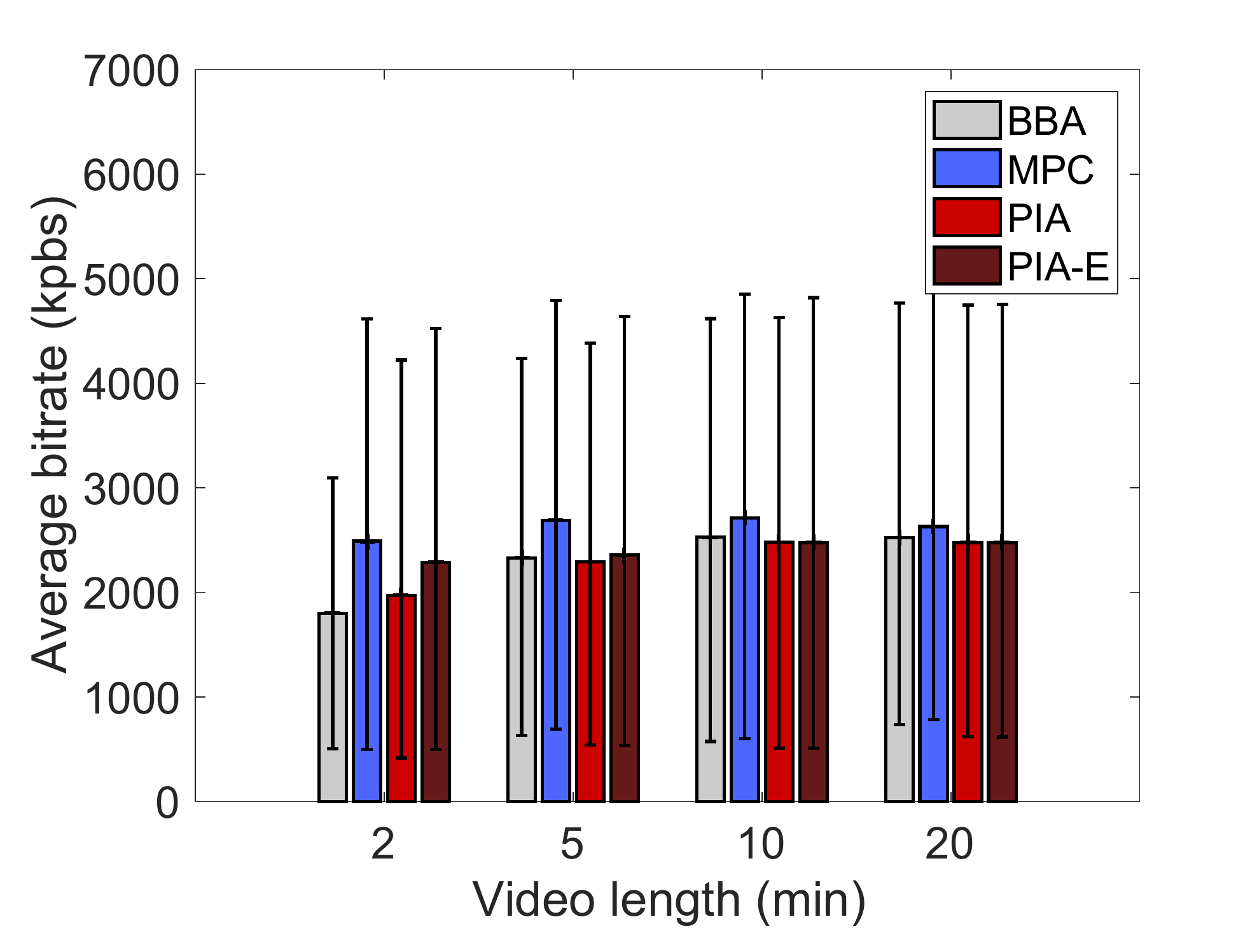}
    }
    \subfigure[\label{subfig:videolengthAnalysisBitrateChange}]{%
      \includegraphics[width=0.3\textwidth]{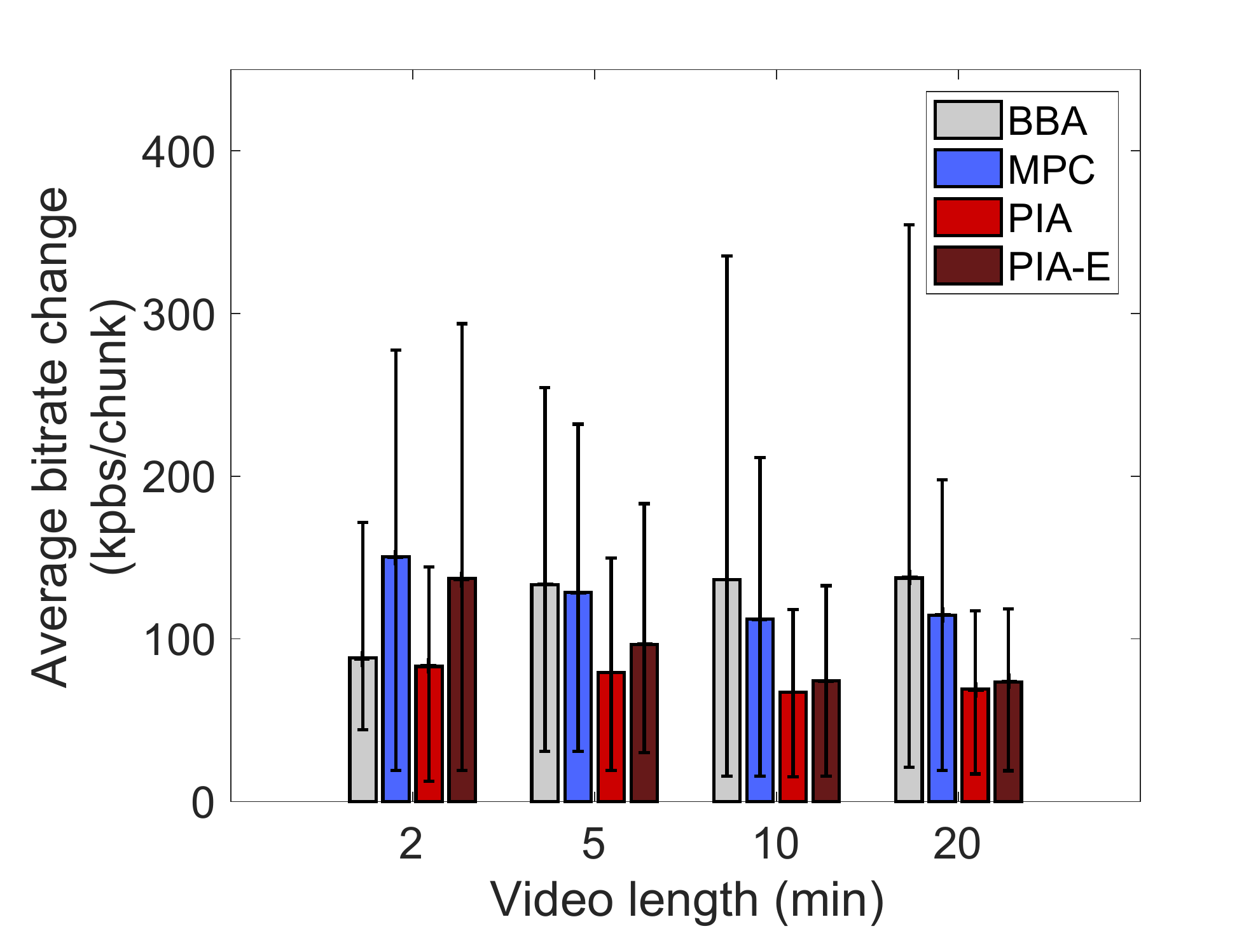}
    }
    \subfigure[\label{subfig:videolengthAnalysisRebuffer}]{%
      \includegraphics[width=0.3\textwidth]{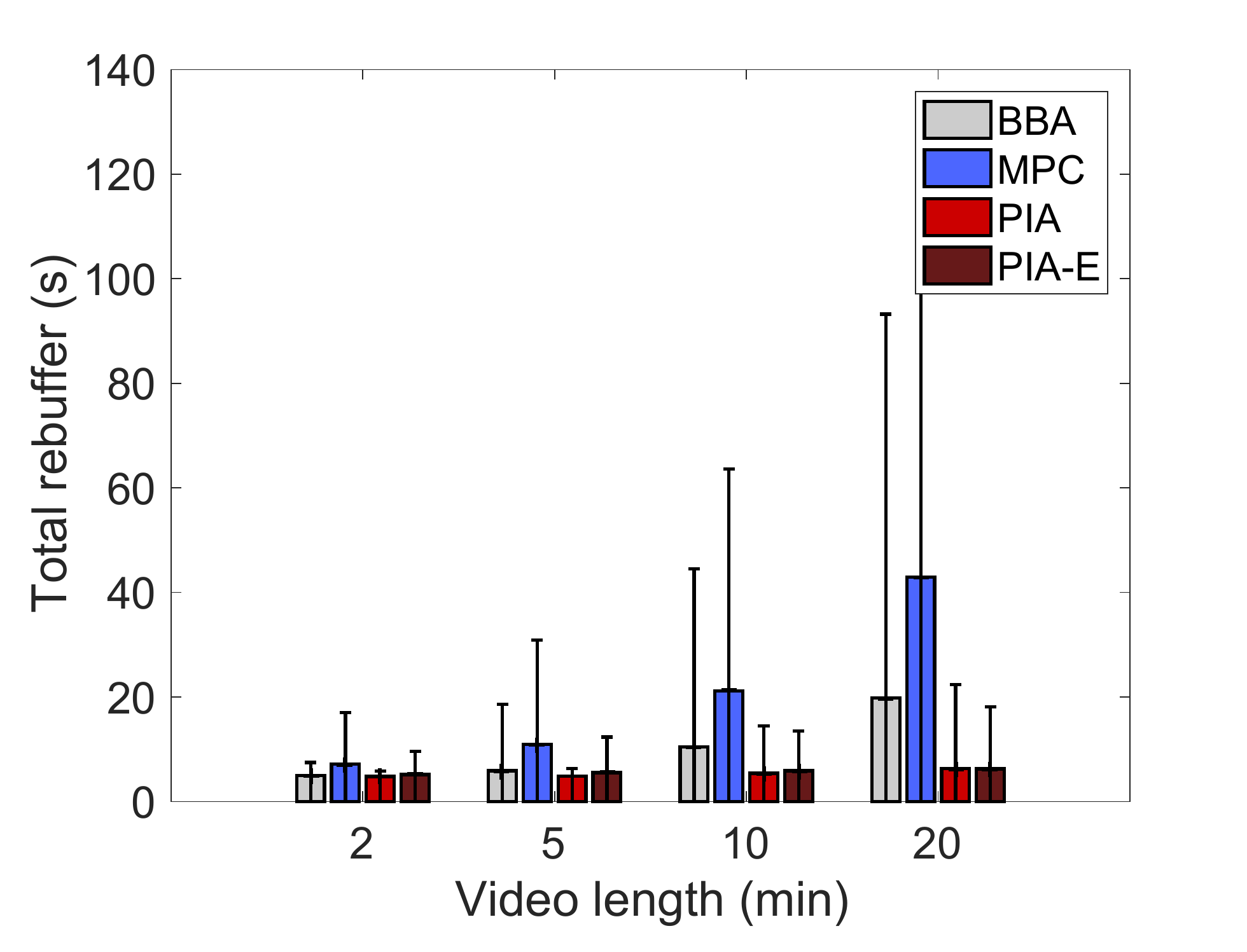}
    }
    \caption{Performance of PIA-E versus other schemes (chunk duration 2 s, video bitrate set $\mathcal R_2$, startup latency 10 s).}
    \label{fig:videolength-analysis}
\end{figure*}

As shown in Section~\ref{sec:performance}, the average bitrate under PIA in the startup phase (up to 5 minutes into the playback) can be 20\% lower than that of MPC. We next present a variant of PIA, PIA-E (PIA Enhanced), that improves PIA's startup performance. The bitrate selection during the startup phase needs to balance two aspects: the quality of the video and the accumulation of the video content. The quality of the early part of the video (the chunks downloaded during the startup phase) is  important, since low quality may prompt a user to stop watching the video. On the other hand, the startup phase also plays an important role in building up the video content in the buffer to reduce the likelihood of rebuffering in the future; choosing lower bitrate chunks helps to accumulate more content in the buffer. A good strategy for the startup phase thus needs to account for these two conflicting aspects.

Our goal of designing PIA-E is to increase the bitrate for the early part of the video, without increasing the amount of rebuffering during the later part of the playback.
We achieve this goal by dynamically adjusting a selective set of parameters over time.
Specifically, observe from Eq. (\ref{eq:lms}) that the bitrate for the early part of the video can be increased by decreasing the controller output, $u_t$,
which can be achieved by adjusting the parameters, $K_p$, $\beta$, $x_r$, and $K_i$, based on Eq. (\ref{eq:PID-video-streaming-improved}).
Let $K_p(t)$, $\beta(t)$, $x_r(t)$, and $K_i(t)$ denote the values of these parameters at time $t$.
We next describe one design of PIA-E that adjusts $K_p(t)$ and $x_r(t)$ overtime. At the end of this section, we discuss other design options.

\textbf{Adjusting $K_p(t)$ and $x_r(t)$.} This design keeps $\beta(t)=\beta$ and $K_i(t)=K_i$, and dynamically adjusts $K_p(t)$ and $x_r(t)$ over time,
starting with initial values that are selected for the startup phase, and ending with the values that have been tuned for the steady state.
Specifically, we set $K_p(t)$ as a decreasing function of $t$ that decreases from an initial value $\alpha  K_p$, $\alpha > 1$, to $K_p$ over a time interval, and set $x_r(t)$ as an increasing function of $t$ that increases from a small value to $x_r$ over a time interval. As a result, the first term in  Eq. (\ref{eq:PID-video-streaming-improved}) is a non-negligible negative value at the beginning of the playback, leading to lower $u_t$ and hence larger bitrate choices. 
Specifically, let $\tau$ denote the time interval. We set
\begin{eqnarray}
K_p(t) &=&
\begin{cases}
\alpha K_p -   \frac{(\alpha K_p - K_p)t}{\tau}\,, & 0 \leq t \leq \tau \\
K_p\,, & t > \tau
\end{cases}
\end{eqnarray}
Similarly, we set $x_r(t)$ as a linear function with the minimum value of $2\Delta$ initially (which is twice of the chunk duration; the target buffer level cannot be zero, and the value $2\Delta$ is chosen empirically).
\begin{eqnarray}
x_r(t) &=&
\begin{cases}
\max\left(2\Delta, \frac{x_r t}{\tau}\right)\,, & 0 \leq t \leq \tau \\
x_r\,, & t > \tau
\end{cases}
\end{eqnarray}

In the above design, $x_r(t)$ follows a ramp change. For such a ramp change, the system in Eq. (\ref{eq:PID-video-streaming-improved}) can track $x_r$ with the error proportional to the inverse of the so-called velocity constant $K_v$ in the steady state~\cite{Stefani02:feedback-control}.
From the transfer function Eq. (\ref{eq:Ts-with-beta}), we derive $K_v$ as $1/K_v = K_p(1-\beta)/K_i$. Evidently, when $\beta=1$, the steady sate error is zero, and is non-zero when $\beta<1$. Therefore, we choose $\beta=1$ for the above choice of $x_r(t)$.

We explored the choice of $\alpha$ (as 2 or 4) and $\tau$ (as 2, 5, or 10 minutes).
Fig.~\ref{fig:videolength-analysis} plots the performance of PIA-E versus other schemes, $\alpha=4$ and $\tau=5$ minutes (which achieves the best tradeoff). The results when playing up to 2, 5, 10 or 20 minutes of the video are shown in the figure.  We observe that  for the first 2 minutes, PIA-E indeed leads to significantly higher bitrate: the average bitrate is 14\% higher than that of PIA, 27\% higher than that of BBA and only 8\% lower than that of MPC. On the other hand, the average bitrate change for PIA-E is higher than that of PIA, but still 9\% less compared to MPC on average. The amount of rebuffering of PIA-E is similar to that of PIA.
 When the ending time is larger (i.e., 5, 10 or 20 minutes), the performance of PIA-E is very close to PIA, confirming that the more aggressive bitrate choice of PIA-E for the early part of the video does not adversely affect the performance for the later part of the video.

\textbf{Other designs.} In the above design, we set $K_p(t)$ and $x_r(t)$ as piece-wise linear functions. We also explored setting them as exponential functions, and observed that the resultant design can achieve similar performance.
 In addition, we have explored dynamically adjusting $K_p(t)$ and $\beta(t)$, and observed similar performance\footnote{Note that in Eq. (\ref{eq:PID-video-streaming-improved}), since $u_t$ is  affected by the product of $\beta$ and $x_r$, it is sufficient to vary either $\beta(t)$ or $x_r(t)$; there is no need to vary them simultaneously.}. Last, we explored dynamically adjusting a single parameter (e.g., $x_r(t)$), and observed that the resultant performance is inferior to that when adjusting two parameters
 simultaneously.

\section{Evaluation using DASH Implementation} \label{sec:pia_dash}
\begin{figure*}[ht]
 \centering
 \vspace{-0in}
 \subfigure[\label{subfig:dashBitrate}]{%
      \includegraphics[width=0.4\textwidth]{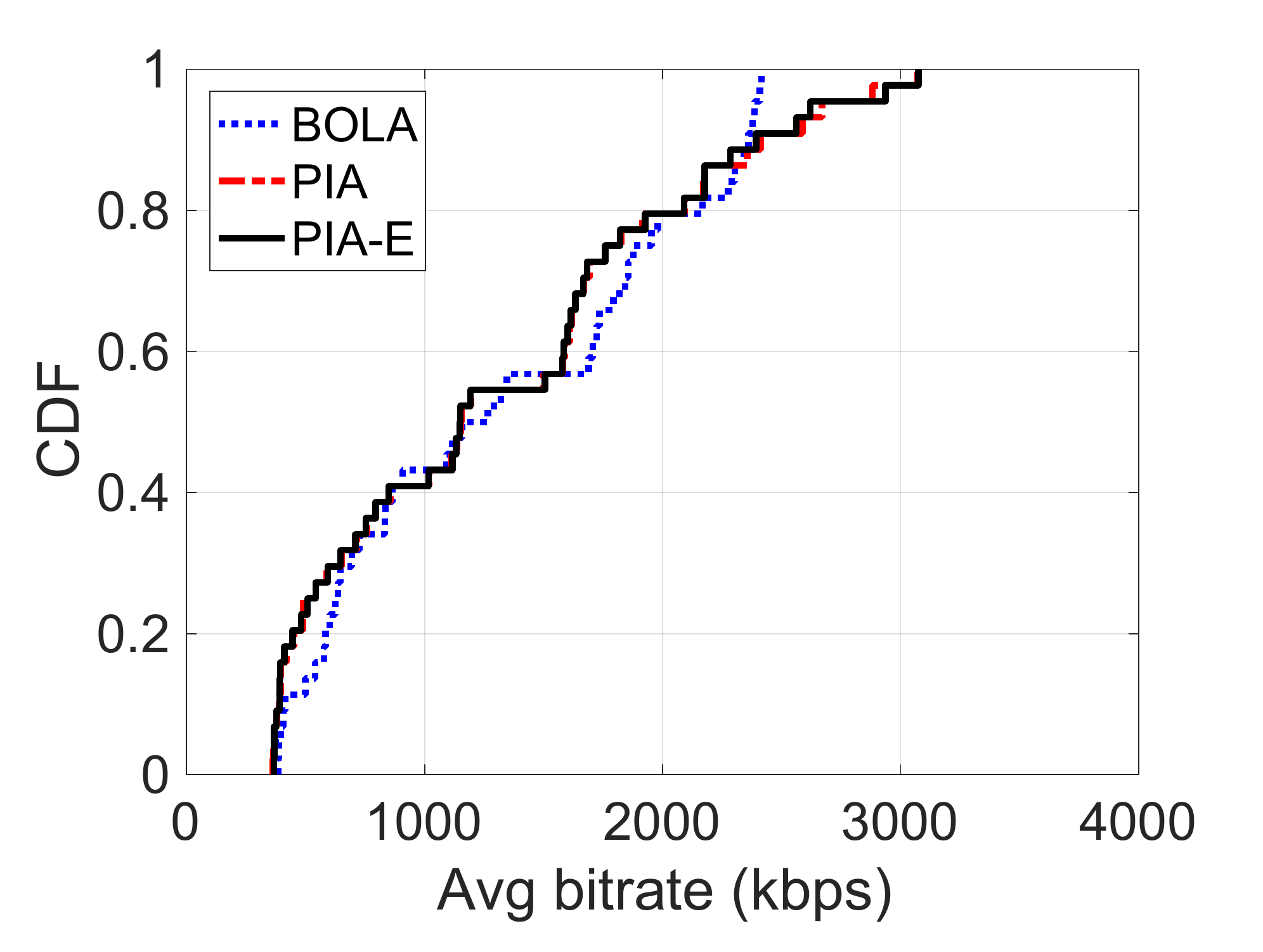}
    }
    \subfigure[\label{subfig:dashBitrateChange}]{%
      \includegraphics[width=0.4\textwidth]{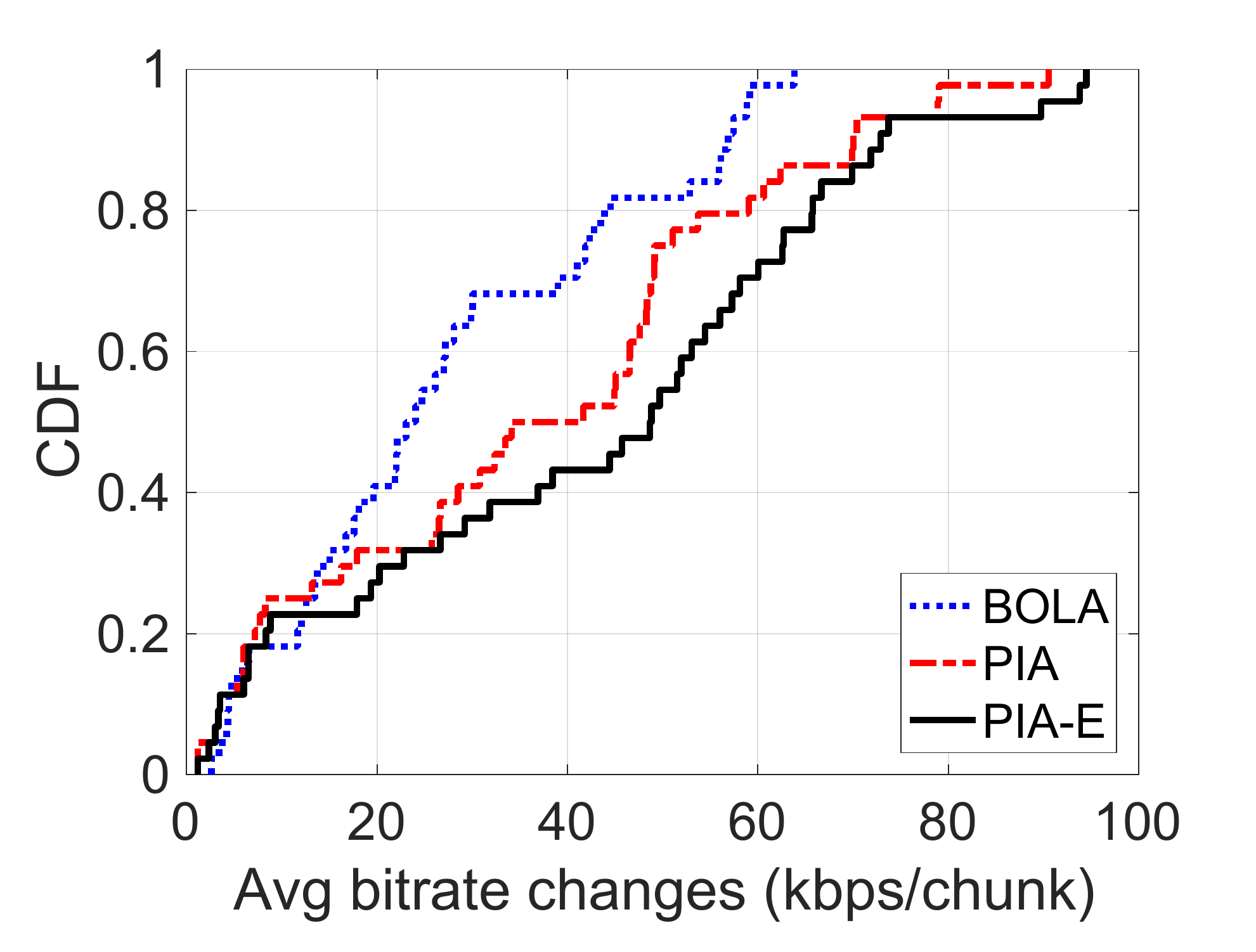}
    }
    \subfigure[\label{subfig:dashRebuffer}]{%
      \includegraphics[width=0.4\textwidth]{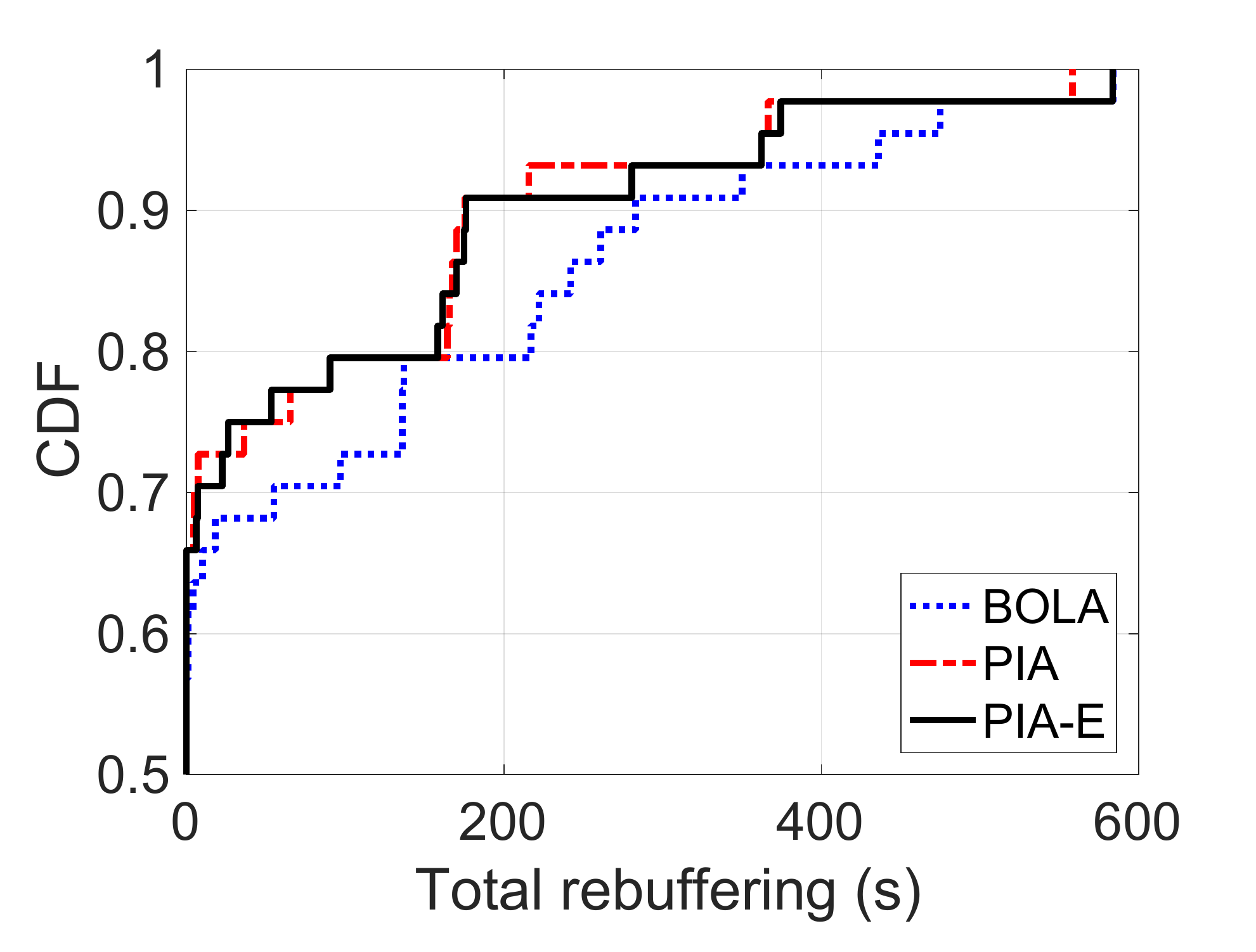}
    }
    \subfigure[\label{subfig:dashBitrate-init}]{%
      \includegraphics[width=0.4\textwidth]{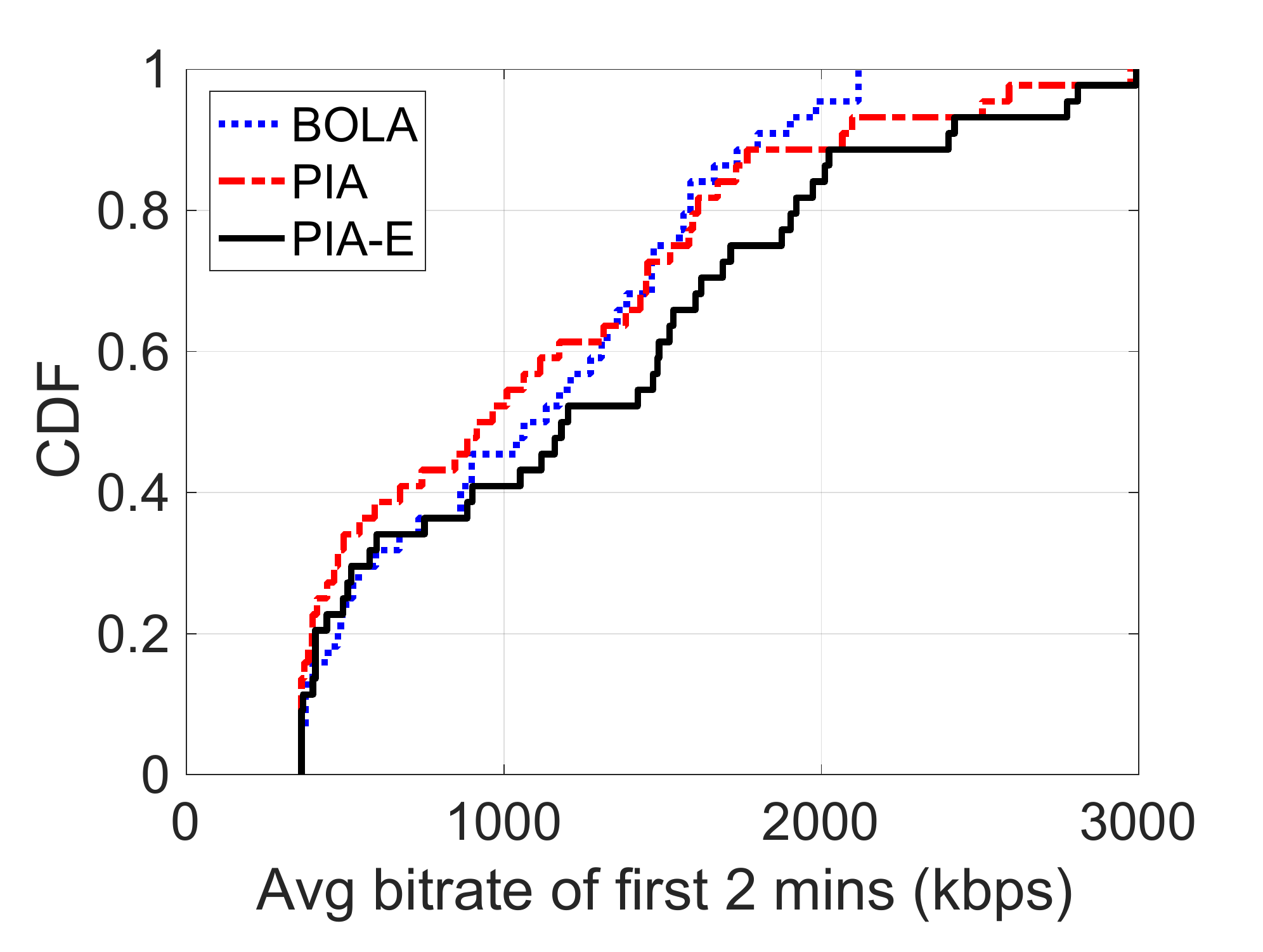}
    }
    \caption{Performance comparison in DASH (music show, chunk duration 2 s, video length 15 minutes, startup latency 10 s).}
    \label{fig:dash-analysis}
\end{figure*}

\begin{figure*}[ht]
	\centering
	\subfigure[\label{subfig:dashBitrate}]{%
		\includegraphics[width=0.4\textwidth]{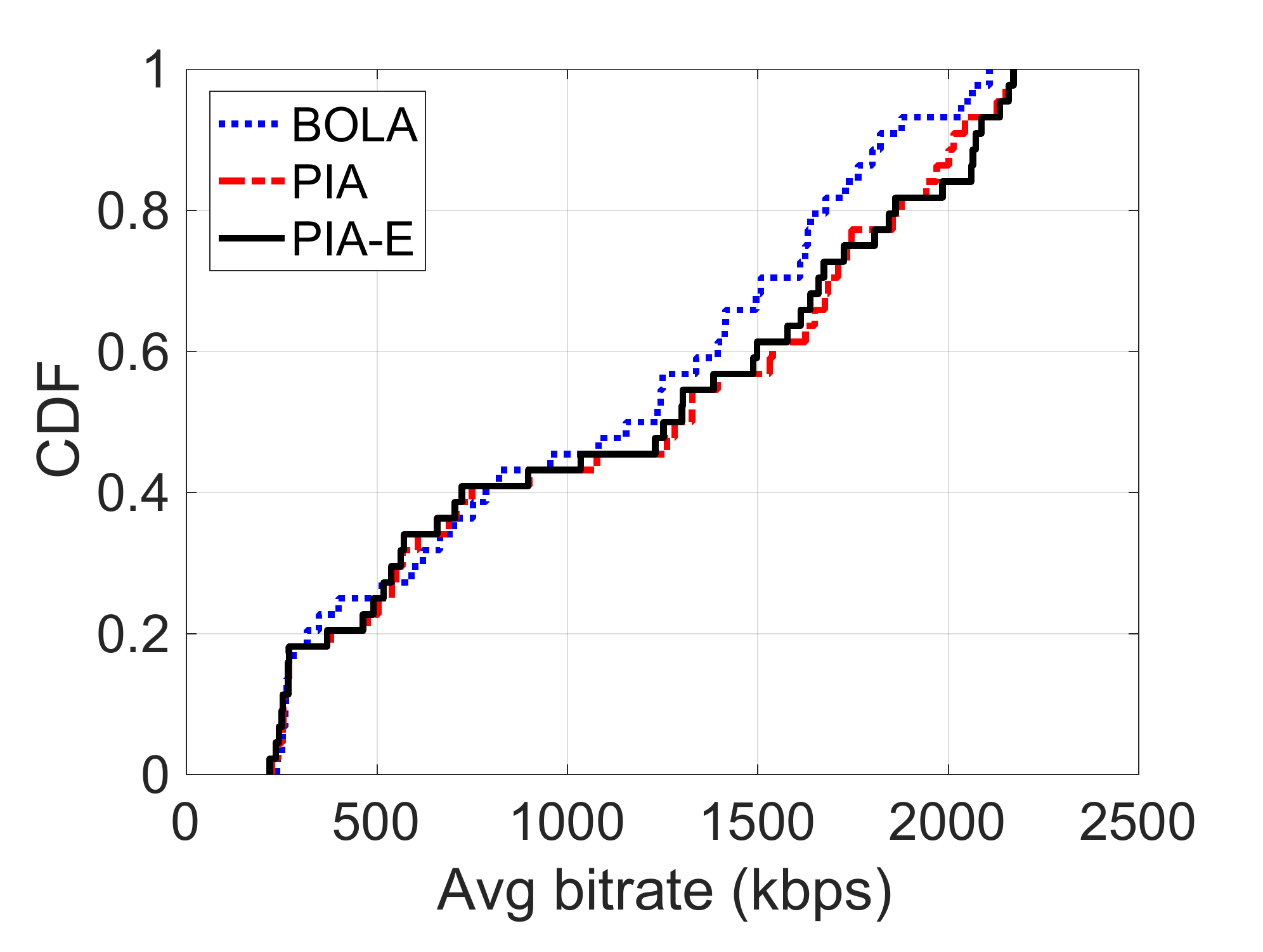}
	}
	\subfigure[\label{subfig:dashBitrateChange}]{%
		\includegraphics[width=0.4\textwidth]{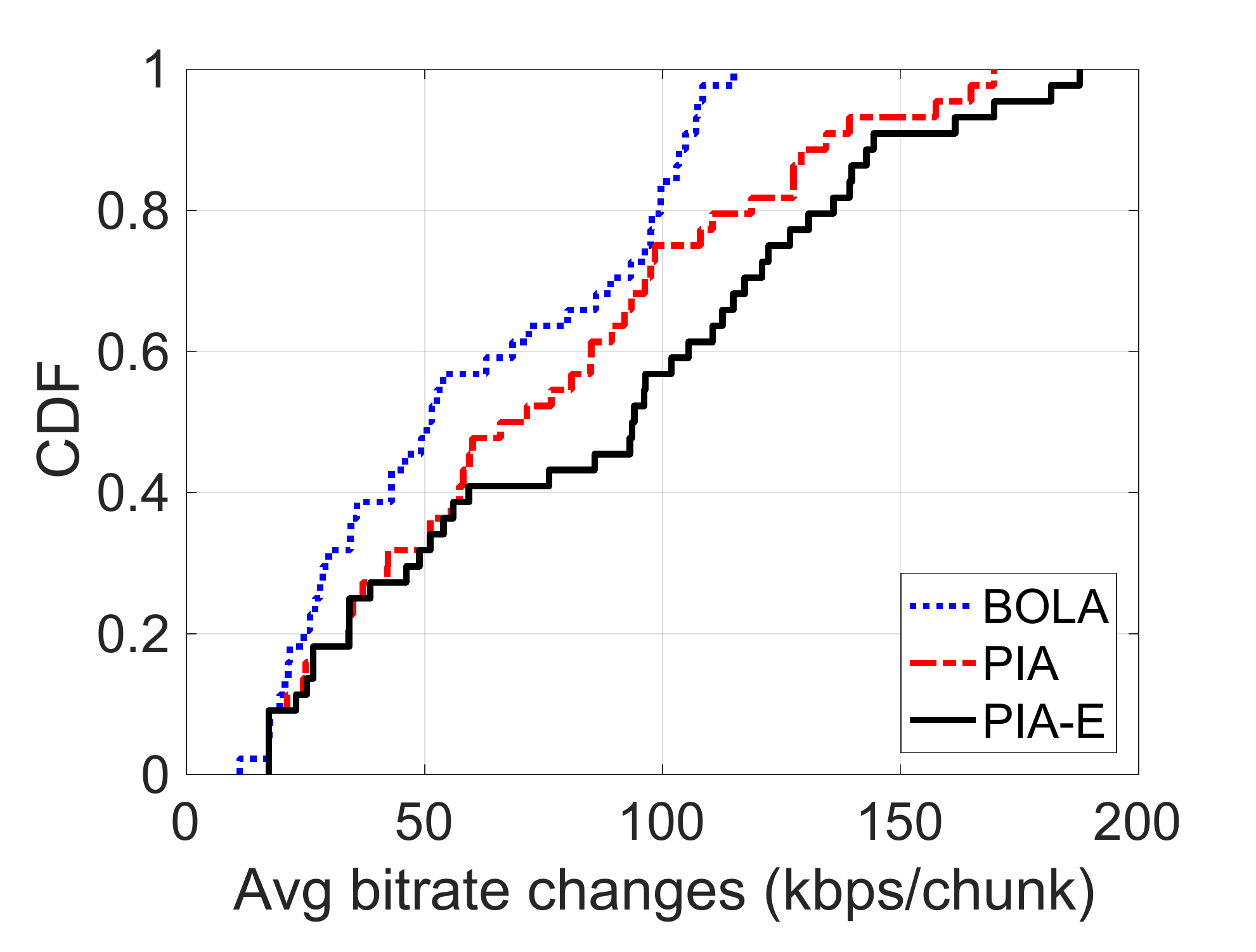}
	}
	\subfigure[\label{subfig:dashRebuffer}]{%
		\includegraphics[width=0.4\textwidth]{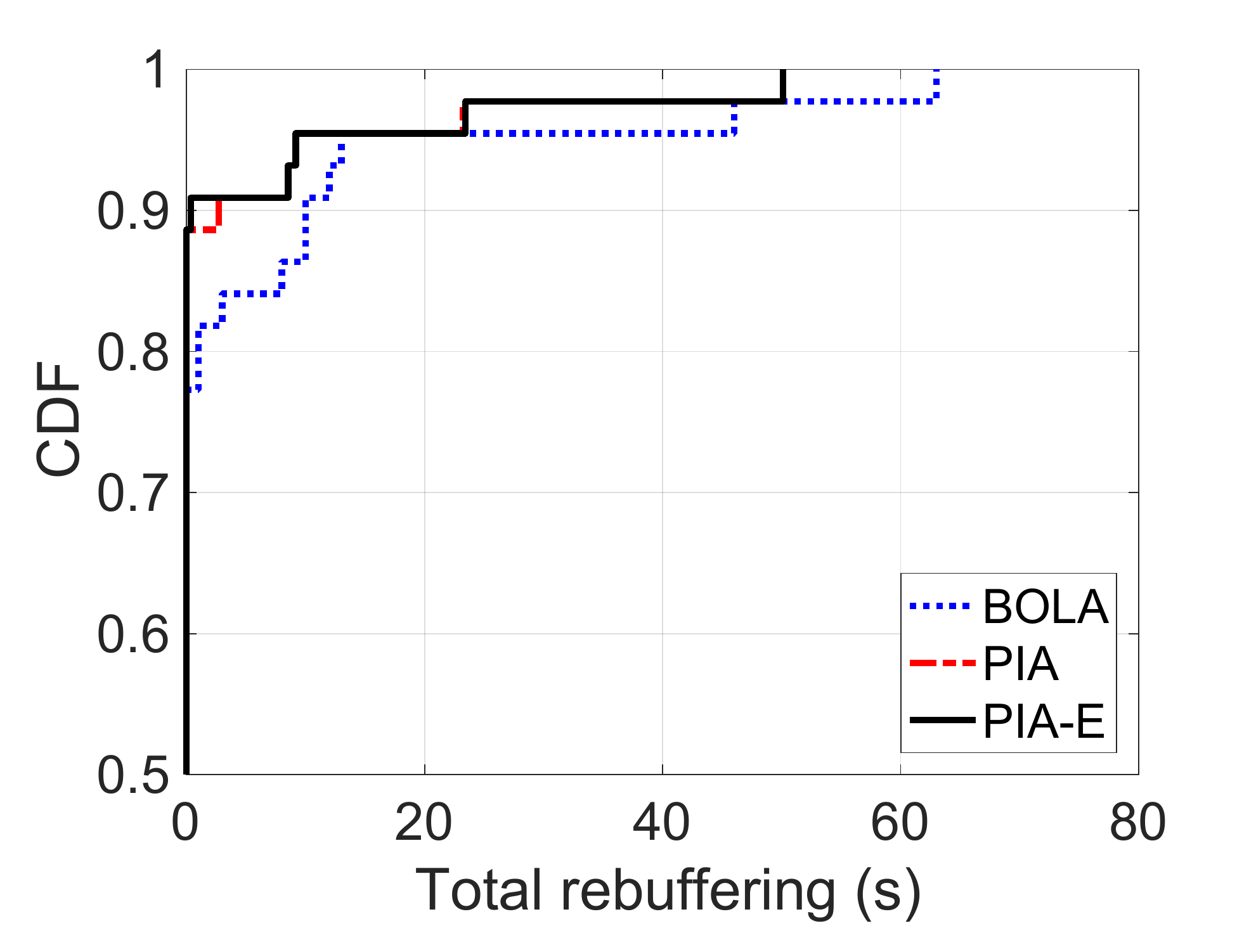}
	}
	\subfigure[\label{subfig:dashBitrate-init}]{%
		\includegraphics[width=0.4\textwidth]{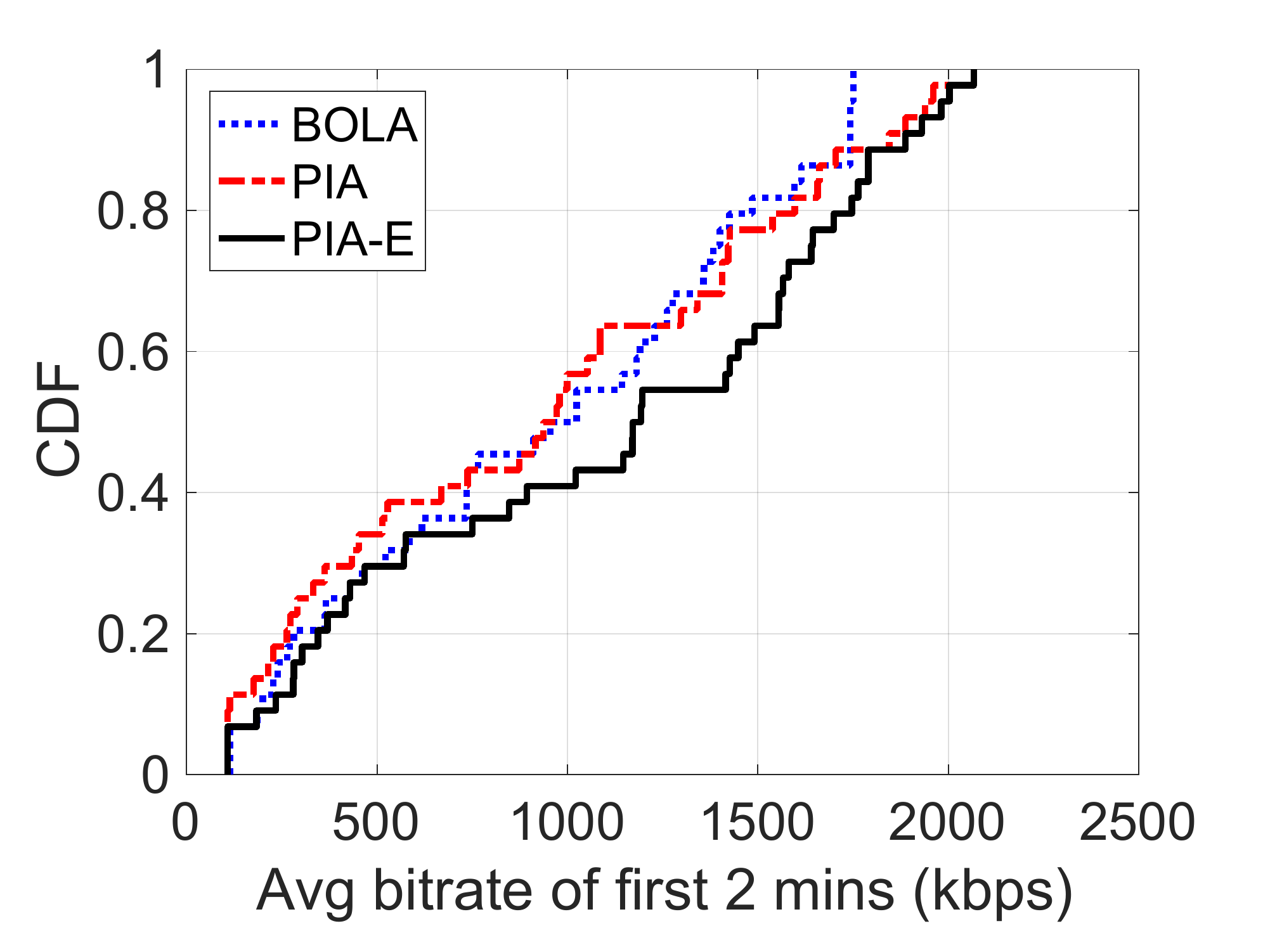}
	}
	\caption{Performance comparison in DASH (BBB, chunk duration 5 s, video length 10 minutes, startup latency 10 s).}
	\label{fig:dash-analysis-BBB}
\end{figure*}

We have implemented PIA and PIA-E using \texttt{dash.js} (version 2.2.0)~\cite{dash-js}, a production quality open source framework provided by DASH Industry Forum~\cite{sodagar2011mpeg}.
To evaluate the performance of PIA and PIA-E under realistic network settings, we create an emulation environment as follows.
We use a Linux machine running Apache httpd as the video server and a Windows laptop (with i7-5700HQ 3.50 GHz CPU and 16 GB memory) as the client.
The server and client are connected by a 100 Mbps Ethernet link.
We apply the Linux \texttt{tc} tool at the server to emulate the downlink bandwidth using the LTE bandwidth traces that we collected. The latency between the server and client is set to 70 ms as it is the average latency reported by OpenSignal's latency report. The client uses Chrome browser to run \texttt{dash.js}.

\textbf{Implementation of PIA and PIA-E in \texttt{dash.js}.} We implemented two new ABR streaming rules (each about 400 LoC) in \texttt{dash.js} to realize PIA and PIA-E. The parameters used for PIA and PIA-E are as those used in Sections~\ref{sec:performance} and~\ref{sec:startup-phase}, respectively. The bandwidth estimation requires knowing the past throughput per second (we use the harmonic mean of the throughput of the past 20 seconds).
We therefore further developed a bandwidth estimation module that uses
progress events in \texttt{dash.js} to obtain the past throughput per second.
Specifically, when receiving one progress event, we calculate the number of bytes downloaded since the last event.

\textbf{Videos.} We use three CBR videos encoded using \texttt{FFmpeg}~\cite{FFmpeg}. The first video is a music show, around 15 minutes long, with five tracks of resolution 240p, 360p, 480p, 720p and 1080p, respectively (the bitrate of the tracks varies from 0.36 to 3.09 Mbps); the chunk duration is 2 s. The other two videos are Big Buck Bunny (BBB) and Tears of Steel (ToS),
both around 10 minutes long, with chunk duration of 5 s. ToS has five tracks, with the same resolutions and slightly lower bitrate (0.32 to 2.84 Mbps across the tracks) compared to the music show. BBB has one additional lower track (144p resolution) and the bitrate of the six tracks varies from 0.11 to 2.20 Mbps.

\textbf{Comparing simulation and implementation results.} We compare the results obtained from our \texttt{dash.js} implementation with those from the simulations, and confirm that the results are consistent.
Specifically, for the music show, under PIA, 90\% of the relative differences between implementation and simulation results are within 6.5\% for average bitrate;
for bitrate changes and rebuffering duration, 90\% of the absolute differences are within 15 Kbps/chunk and 3 s, respectively.
The results for PIA-E are similar. The performance differences between implementation and simulation results are due to multiple reasons. First, the simulation assumes a perfect CBR video where all the chunks in the same track have exactly the same bitrate; the video used in the implementation, while encoded as CBR, has bitrate variability across the chunks.
Second, the implementation results are affected by various practical factors (e.g., server response time, client computational and response time, network RTT and TCP window size), which are not accounted for in the simulations.

\textbf{Performance comparison.}
We compare the performance of PIA and PIA-E with a state-of-the-art scheme, BOLA~\cite{spiteri2016bola}, that was implemented in \texttt{dash.js} version 2.2.0. BOLA selects the bitrate to maximize a utility function considering both rebuffering and delivered bitrate.
The upper threshold of BOLA is set to 60 s (the default value is 30 s) to be compatible with the target buffer level of 60 s in PIA.
Fig.~\ref{fig:dash-analysis} plots the QoE metrics of PIA, PIA-E and BOLA for the music show video.
We observe that PIA-E achieves higher average bitrate than PIA and BOLA for the first 2 minutes of the video. For the entire video, PIA-E leads to comparable performance as PIA: PIA-E leads to slightly more rebuffering, and similar average bitrate and bitrate changes. 
BOLA has higher rebuffering and lower bitrate changes than PIA and PIA-E.
Fig.~\ref{fig:dash-analysis-BBB} plots the results for the BBB video. We observe similar results as those for the music show except that (i) the average bitrate change per chunk for all the three schemes is larger, which is due to the larger chunk duration in BBB (5 versus 2 s), and (ii) the amount of rebuffering is significantly lower due to the much lower bitrate of the lowest track in BBB (0.11 versus 0.36 Mbps).  For ToS, the average bitrate change per chunk is similar to that for BBB due to the same chunk duration of 5 s of these two videos, and the amount of rebuffering is similar to that of the music show due to the similar bitrate of
the lowest track of these two videos (the figures are omitted in the interest of space).

\textbf{Runtime overhead.}
We record the CPU execution time of the ABR logic in the JavaScript code when a video is being played. The execution time of the default ABR logic in the \texttt{dash.js} player is 1.2 s for the entire 15-min music show video. The execution time of our PIA logic is 1.9 s, only slightly larger than that of the default ABR logic. For PIA-E, the execution time is 2.0 s (compared to PIA, it has additional calculation of $K_p(t)$ and $\beta(t)$). The execution time PIA-E and for the other two videos is similar.  The above results indicate that PIA incurs very small runtime overhead, despite its non-trivial decision process shown in Fig.~\ref{fig:pia_diagram}.

\section{Conclusion and Future work}
\label{sec:pia_concl}

In this chapter, we have explored using feedback control theory for creating the adaptation logic component critical to  ABR video streaming.
By strategically applying a PID controller in an explicit and adaptive manner, PIA considerably outperforms the state-of-the-art video streaming schemes
%such as BBA and MPC
in balancing the complex tradeoffs associated with the key QoE metrics, as demonstrated by extensive evaluations. PIA is also lightweight and easy to deploy.
We believe the same high-level principle can be applied to other multimedia applications with content quality adaptation, such as live video conferencing.
In our future work, we plan to port our implementation to mobile devices to better assess PIA's performance in the wild, and also  conduct deeper exploration of other network and streaming settings (e.g., live streaming).

%\documentclass[9pt,sigconf]{acmart}
%\usepackage{booktabs} % For formal tables

%\usepackage{citesort}

%\usepackage[english]{babel}
%\usepackage{blindtext}

%Conference Info

%\input{bing-command}
%\usepackage{amsmath}
%\usepackage{subfig}
%\usepackage{graphicx}
%\usepackage{multirow}
%\usepackage{color}
%\usepackage[normalem]{ulem}

%%%%%%%%%%%%%%%%%%%%%%%%%% COLORS
%\definecolor{brown}{cmyk}{0,0.81,1,0.60}
%\definecolor{magenta}{rgb}{0.4,0.7,0}
%\definecolor{magenta(dye)}{rgb}{0.79, 0.08, 0.48}
%\definecolor{lightcarminepink}{rgb}{0.9, 0.4, 0.38}
%\definecolor{gray}{rgb}{0.5,0.5,0.5}
%\definecolor{red}{rgb}{1,0,0}
%\definecolor{green}{rgb}{0,0.5,0}
%\definecolor{blue}{rgb}{0,0,1}

%\newcommand{\feng}[1]{{\color{red}[FQ:#1]}}
%\newcommand{\fengc}[1]{{\color{red}\sout{[FQ:#1]}}}

%\newcommand{\shuai}[1]{{\color{blue}[SH: #1]}}
%\newcommand{\shuaic}[1]{{\color{blue}\sout{[SH: #1]}}}

%\newcommand{\sen}[1]{{\color{blue}[SS: #1]}}
%\newcommand{\senc}[1]{{\color{blue}\sout{[SS: #1]}}}

%\newcommand{\bing}[1]{{\color{red}#1}}

%\newcommand{\bingc}[1]{{\color{red}[BW:#1]}}
%\newcommand{\BULLET}{\vspace{+.00in} \noindent $\bullet$ \hspace{+.00in}}

\newcommand{\RN}[1]{%
	\textup{\uppercase\expandafter{\romannumeral#1}}%
}

%\newcommand{\yanyuan}[1]{{\color{green}[YQ:#1]}}

%%wb, define macro for the basic and enhanced version of the schemes
%\usepackage{xspace}

%\newcommand*{\affaddr}[1]{#1} % No op here. Customize it for different styles.
%\newcommand*{\affmark}[1][*]{\textsuperscript{#1}}

% correct bad hyphenation here
\hyphenation{net-works semi-conduc-tor}
\chapter{ABR Streaming of VBR-encoded Videos: Characterization, Challenges, and Solutions} \label{chapter:cava}

	%\begin{abstract}
	%\blindtext
	% \input{CAVA/abstract}
	
\section{Introduction}

Adaptive Bitrate (ABR) streaming (specifically HLS~\cite{apple-hls} and DASH~\cite{dash-standard}) has emerged as the \emph{de facto} video streaming technology in industry for dynamically adapting the video streaming quality based on varying network conditions. A video is compressed into multiple independent {\em streams} (or {\em tracks}), each specifying the same content but encoded with different bitrate/quality. A track is further divided into a series of {\em chunks}, each containing data for a few seconds' worth of playback. During streaming playback, to fetch the content  for a particular play point in the video,
the adaptation logic at the client player dynamically determines what bitrate/quality chunk to download.
% based on time-varying network performance.
%The resulting playback involves showing different portions of the video using chunks from different tracks.

%%% CBR vs VBR

The video compression for a track can be (1) Constant Bitrate (CBR) --  encodes the entire video at a relatively fixed bitrate, allocating the same bit budget to  both \emph{simple scenes} (i.e., low-motion or low-complexity scenes) and \emph{complex scenes} (i.e., high-motion or high-complexity scenes), resulting in variable quality across scenes,
or (2) Variable Bitrate (VBR) -- encodes \emph{simple scenes}  with fewer bits and \emph{complex scenes}
with more bits, while maintaining a consistent quality throughout the track.

%%with more bits, while maintaining a consistent quality throughout the track.
VBR presents some key advantages over CBR~\cite{lakshman1998vbr}, such as the ability to realize better video quality for the same average bitrate, or the same quality with a lower bitrate encoding than CBR.
%As a result, VBR encodings allow more flexibility and use bits more efficiently.
Traditionally, streaming services mainly deployed only CBR,
partly due to various practical difficulties
%driven in part by various practical considerations
including the complex encoding pipelines for VBR,
as well as the demanding storage, retrieval, and transport challenges posed by the %inherent
multi-timescale bitrate burstiness of VBR videos.
%It is only fairly recently that
Most recently content providers have begun adopting VBR encoding~\cite{huang2014buffer,reed2016leaky,Xu17:Dissecting}, spurred by the promise of substantial improvements in the quality-to-bits ratio compared to CBR, and by technological advancements in VBR encoding pipelines.
Current commercial systems and ABR protocols do not fully make use of the characteristics of VBR videos; they typically assume a track's peak rate to be its bandwidth requirement  when making rate adaptation decisions~\cite{Xu17:Dissecting}.
Existing ABR streaming research
centered mostly on CBR encoding, and very little on the VBR case (see \S\ref{sec:cava_related}).
Therefore there exists little prior understanding of the issues and challenges involved in  ABR streaming of VBR videos.

In this chapter, we  explore a number of  important open research questions around ABR streaming for VBR-encoded video.
We make the following main contributions:

\BULLET  We analyze the characteristics of VBR videos (\S\ref{sec:cava_characteristics})  using a rich set of videos spanning diverse content genres, predominant encoding technologies  (H264~\cite{h264} and HEVC (H.265)~\cite{h265}) and platforms (\S\ref{sec:cava_video-dataset}). Jointly examining content complexity, encoding quality, and encoding bitrates, we identify distinguishing characteristics of VBR encodings that impact user QoE, and their implications for ABR rate adaptation and QoE metrics.
% (e.g., the need for differential treatment of  complex scenes).
 We find that, for real-world VBR encoding settings, complex scenes, although encoded with more bits,
have inferior encoding quality compared to other scenes in the same track, across a range of video quality metrics.
We also find that different chunk sizes across the video can be used to identify relative scene complexity with high accuracy.  These two key observations (i) suggest the need to provide complex scenes with  {\em differential treatment}
 %\sen{whats more accurate - is it "differential" or "preferential" treatment?}
 during streaming, as improving the quality of complex scenes brings more QoE enhancement compared to improving simple scenes~\cite{li2014streaming}, and (ii) provide us a practical pathway towards achieving this differential treatment in
 the context of today's \emph{de facto} ABR protocols where the scene complexity or quality information is typically unavailable to the ABR logic.

%\smallskip
\BULLET Based on  our analysis, we enunciate three key design principles ({\em non-myopic}, {\em differential treatment}, and {\em proactive}) for ABR rate adaptation for VBR videos (\S\ref{sec:cava_principles}). These principles explicitly account for the characteristics of VBR videos (including scene complexity, quality, and  bitrate variability across the video).
  %and avoid the shortcomings of  state-of-the-art ABR streaming techniques that do not incorporate these principles (examined in \S\ref{subsec:cbr-schemes-for-vbr}-\ref{sec:use-per-chunk-info})
Specifically, the {\em non-myopic} principle dictates considering multiple future chunks when making rate adaptation decisions, leading to better usage of the available network bandwidth;
%more elastic resource allocation
%\sen{what does more elastic resource allocation mean ?};
the {\em differential treatment} principle strategically favors complex scenes over simple scenes for better user QoE;  the last principle proactively reacts to chunks' variable sizes/bitrates by, for example, pre-adjusting the target buffer level, thus further facilitating the smoothness and adaptiveness of VBR streaming.

%\smallskip
%\vspace{0.2in}
\BULLET We design CAVA~(Control-theoretic Adaption for VBR-based ABR streaming), a novel, practical ABR rate adaptation algorithm based on concrete instantiations of these principles~(\S\ref{sec:cava_design}). CAVA uses control theory that has shown promise in adapting to dynamic network bandwidth such as that in cellular networks~\cite{yin2015control,qin2017control}. Its design involves  two tightly coupled controller loops that work in synergy using easy-to-obtain information such as VBR chunk sizes and historically observed network bandwidth.
%\senc{CAVA instantiates the three aforementioned principles of VBR streaming.}
 Specifically, CAVA improves the QoE by (i) using Proportional-Integral-Derivative~(PID) control concepts~\cite{astrom08:feedback} to maintain a dynamic target buffer level to limit stalls, (ii) judiciously selecting track levels to maximize the quality while minimizing quality changes, and (iii)
strategically favoring chunks for complex scenes by saving bandwidth for such chunks.

%\smallskip
\BULLET  We evaluate the performance of CAVA for a wide range of real-world network scenarios, covering both LTE and broadband~(\S\ref{sec:cava_perf}).
%network bandwidth traces collected from two large commercial LTE cellular networks
Our results show that CAVA considerably outperforms existing state-of-the-art ABR schemes across several important dimensions.
It provides significantly enhanced quality for complex scenes while not sacrificing the quality of other scenes. Depending on the video, the percentage of low-quality chunks under CAVA is up to 75\% lower than that of other schemes. CAVA also leads to substantially lower rebuffering (up to $95$\%) and quality variation (up to $48$\%).
%$29$\%-$95$\% under LTE traces depending on the video)
%and $23$\%-$48$\% reduction in quality variation.
The network data usage under CAVA is in the same ballpark or slightly lower than most of the other schemes. The above results demonstrate that CAVA achieves a much better balance in the multiple-dimension design space than other schemes.

The performance improvements hold across a wide spectrum of settings, indicating that the three design  principles hold broadly.
Our implementation of CAVA in the open source \texttt{dash.js}~\cite{dash-js} player
further demonstrates that CAVA is lightweight and practical.

	%talk about the diverse set of videos, how they are encoded, why

\section{The VBR Video Dataset}\label{sec:cava_video-dataset}

%As shown in Table~\ref{tab:videoset},
Our video dataset includes 16 videos: 8 encoded by YouTube, and 8 encoded by us using
% a popular open source encoder
\texttt{FFmpeg}~\cite{FFmpeg}, following specifications 
%and recommendations 
in~\cite{de2016large}.
%popular video services~\cite{de2016large}.
%
Each video is around 10 minutes, and includes 6 tracks/levels (we use the terms \emph{track} and \emph{level} interchangeably) for ABR streaming,
with resolutions of 144p, 240p, 360p, 480p, 720p, and 1080p, respectively.

Our use of YouTube and \texttt{FFmpeg} encodings complement each other.
YouTube represents the state-of-the-art practice of VBR encoding from a popular commercial streaming service.  With \texttt{FFmpeg}, we
explore state-of-the-art  VBR encoding recommendations from another  commercial streaming service, Netflix~\cite{de2016large}.
We also use \texttt{FFmpeg}
% to broaden our exploration
to explore a variety of other settings (e.g., 4x-capped video (\S\ref{sec:cava_video-discussion}), H.265 encoding) beyond  the above cases.
We next explain in detail how we obtain our dataset.

\textbf{Videos Encoded by \texttt{FFmpeg}.}
We select 4 publicly available \emph{raw} videos from~\cite{xiph}.
%each consisting a sequence of unprocessed images captured by a camera
These videos include Elephant Dream (ED), Big Buck Bunny (BBB), Tears of Steel (ToS), and Sintel,  in the categories of animation and science fiction.
We encode these 4 raw videos into 4 H.264 and 4 H.265 videos using \texttt{FFmpeg} (8 encoded videos in total).
%We encode the four publicly available raw videos using \texttt{FFmpeg}~\cite{FFmpeg} as follows.
For each video, we choose the aforementioned six tracks (144p to 1080p),
% which are consistent with the recommendation from
consistent with~\cite{netflix:per-title}.
%with resolutions of 240p, 360p, 480p, 720p, and 1080p, as recommended by YouTube~\cite{youtube-bitrate-levels} and Netflix~\cite{netflix:per-title}.
To encode each track, we follow the per-title ``three-pass'' encoding procedure from Netflix~\cite{de2016large}.
In the first pass, a video clip is encoded using a fixed constant rate factor (CRF) value, which accounts for motion and has been shown to outperform other methods such as constant quantization parameter (QP)~\cite{crf-guide}.
In the second and third passes, we use the output bitrates of the first step to drive a two-pass VBR encoder to
avoid over-allocating bits for simple scenes and under-allocating bits for complex scenes.
In practice, {\em capped VBR} (i.e., the chunk bitrate is capped at a certain high limit) is often used to limit the amount of bitrate variability and enable the player to estimate the incoming chunk sizes~\cite{Xu17:Dissecting}.
Following the latest recommendations from HLS~\cite{hls2017},
 we encode the videos to be 2$\times$ capped (i.e., the peak bitrate is limited to 200\% of the average) using \texttt{-maxrate} and \texttt{-bufsize}, two options in \texttt{FFmpeg}. 
After a track is encoded, we segment it into 2-sec chunks.

Our encoding uses both the widely adopted H.264, and a more recent and efficient codec, H.265~\cite{h265}.
We use CRF value of 25, which provides good viewing quality~\cite{crf-guide,de2016large}.

For an encoded video, we assess its quality relative to a \emph{reference video} (\S\ref{sec:VBR-characteristics}), which is its corresponding raw video footage.

\textbf{Videos Encoded by YouTube.}
We upload the aforementioned 4 raw videos to YouTube and download the encoded videos using \texttt{youtube-dl} ~\cite{youtube-dl}.
% In addition, to make  video dataset more diverse,
We also downloaded 4 other videos from YouTube, in the categories of sports, animal, nature and action movies.

The highest quality track of these 4 videos has a resolution of 2160p,
with average bitrate of $14-18$~Mbps, significantly higher than lower tracks.
To be consistent with the other videos which were 6-track with top track at 1080p,
for these 4 videos, we use the bottom six tracks between 144p and 1080p as the set of tracks for ABR adaptation.
We use the top track (2160p) as the reference video to measure the video quality of the lower tracks, as we do not have the raw source for these videos.
All YouTube videos are encoded using H.264 codec~\cite{h264}, with chunk duration around 5 seconds, consistent with that in~\cite{lin2015multipass}.
%

%The shaded vertical bars represent Q4 chunks.
%%for space
%\iffalse
\begin{figure}[t]
%\vspace{-.3in}
 \centering
     \includegraphics[width=0.7\textwidth]{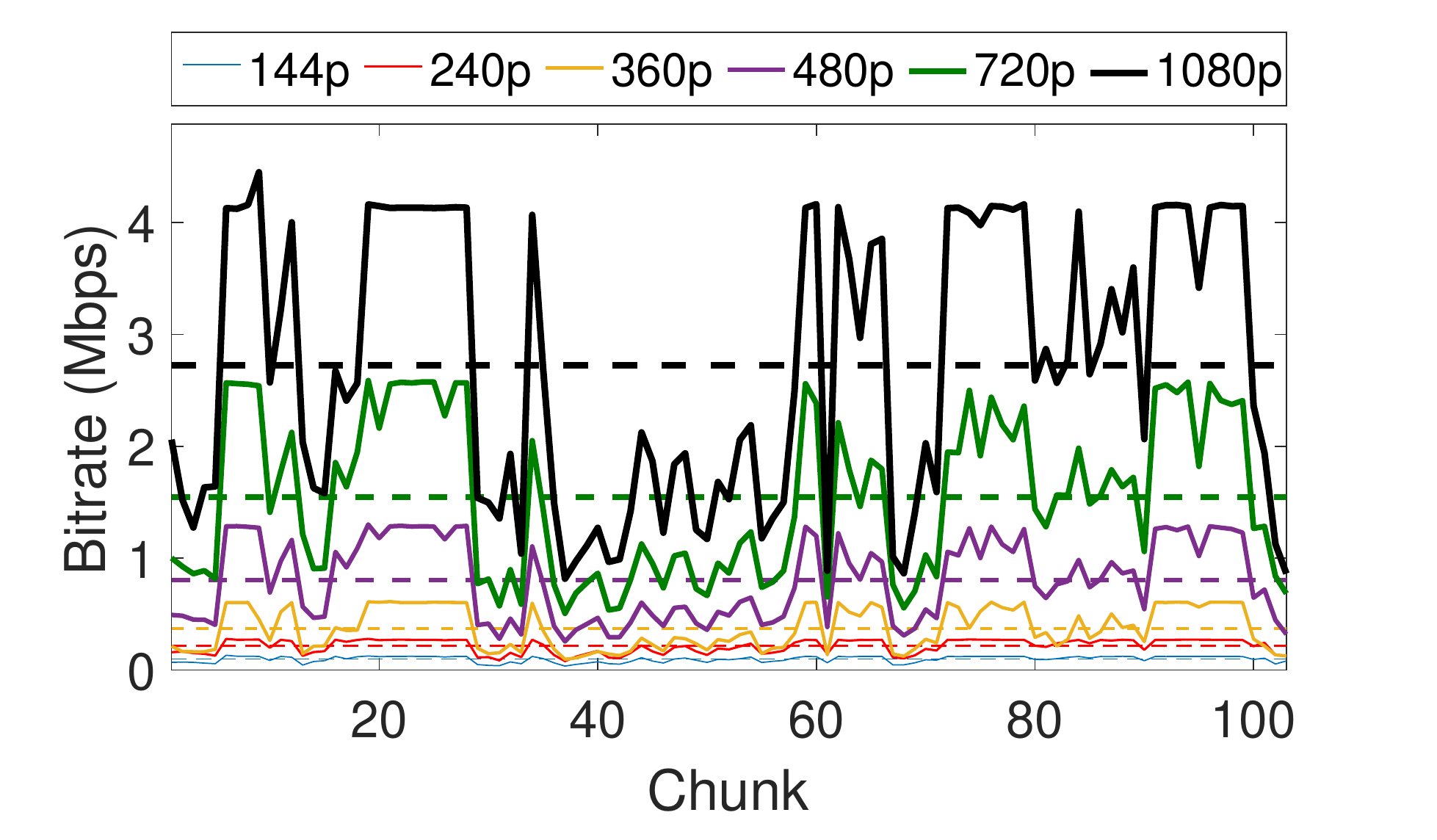}
     \vspace{-.2in}
    \caption{Bitrate of the chunks of a VBR video (Elephant Dream, H.264,  YouTube encoded). }
    \label{fig:cs-dist}
    \vspace{-0.15in}
\end{figure}
%\fi

\textbf{Chunk Duration and Bitrate Variability.}
Overall, our dataset includes two chunk durations: 2 (\texttt{FFmpeg} encodings) and 5 seconds (YouTube encodings). 
They are consistent with the range of chunk durations (2 to 10 seconds) that are used in commercial services~\cite{Xu17:Dissecting}, and allow us to investigate the impact of chunk duration on the performance of ABR streaming in \S\ref{sec:cava_perf}.
All encoded videos exhibit significant bitrate variability: the coefficient of variation of the bitrate in a track varies from 0.3 to 0.6.
Fig.~\ref{fig:cs-dist} shows an example YouTube video; the six horizontal dashed lines mark the average bitrate of the six tracks.
All encodings are capped VBR:
for YouTube videos, the maximum bitrate of a track ranges between 1.1$\times$ to 2.3$\times$ the average bitrate; for the \texttt{FFmpeg} videos, the corresponding range is 1.4$\times$ to 2.4$\times$.
Note that although we set the cap to 2$\times$ explicitly for \texttt{FFmpeg}, the resulting videos can exceed the cap slightly to achieve the specified quality. For all videos, the two lowest tracks have the lowest variability, since the low bitrate limits the amount of variability that can be introduced in VBR encoding; for all the other tracks, the ratio is 1.4 to 2.4.

\section{VBR Streaming: Challenges} \label{sec:cava_characteristics}

We now analyze our video dataset described in \S\ref{sec:cava_video-dataset}:
we first present the characteristics of VBR encoding, and then show its implications for ABR streaming, highlighting challenges of streaming VBR videos.

\subsection{VBR Video Characteristics} \label{sec:VBR-characteristics}
\vspace{+.05in}
\subsubsection{Scene Complexity and Chunk Size}

%Complex scenes in a video play a particularly important role in viewing quality~\cite{li2014streaming}.
Classifying chunks by scene complexity is useful for ABR streaming since the knowledge of scene complexity can be used for rate adaptation.
%since scenes of different complexities can be treated differently.
For instance, since complex scenes in a video play a particularly important role in viewing quality~\cite{li2014streaming},
the player may choose higher tracks for complex scenes to bring more enhancement to viewing experience.

%Identification of complex scenes can be achieved
One way of determining scene complexity
%They can be identified using
is through Spatial information (SI) and Temporal Information (TI)~\cite{itu2008}, two commonly used metrics for quantifying the spatial and temporal complexity of the scenes. This approach, however, requires computation-heavy content-level analysis.
More importantly, such scene complexity information is not available in the complex ABR streaming pipeline in today's commercial services, and would involve non-trivial changes to realize, including resources
 %at the server side
 to compute the complexity information and new features to the ABR streaming standard specifications.
Below, we propose
%a method that uses relative chunk size to quantify the scene complexity, which is simple, and more importantly, can be directly used for ABR streaming (\S\ref{sec:implication}).
a simple and lightweight method that uses relative chunk size to approximate scene complexity, and can be readily used for ABR streaming (\S\ref{sec:implication}).

Our approach is motivated by the following two properties.
%based on the encoding principles used in VBR.
(1) Since the VBR encoding principles dictate that simple scenes are encoded with fewer bits and complex scenes with more bits~\cite{effelsberg1996high},
we expect that the relative size of a chunk (in bytes) can be used as a proxy for the relative scene complexity.
 %specifically, complex scenes which are important for perceptual quality can be identified as
%as we shall see later in Section~\ref{sec:implication}, has important implications for ABR streaming of VBR videos. We  visually verified this property.
%
%By manually watching the videos in our dataset, we confirmed that large chunks indeed contain more complex scenes (e.g., fighting scenes with a lot of fast scene changes), while small chunks capture simple scenes (e.g., mostly static scenes with people talking, preview title, rolling credits).
%
%We visually verified that this is indeed the case.
%Specifically, we select four videos in our dataset. For each video, we select a set of chunks with large sizes and a set of chunks with small sizes. We then watched each chunk, and confirmed that the large chunks indeed contain more complex scenes (e.g., fighting scenes with a lot of fast scene changes) while the small chunks capture simple scenes (e.g., mostly static scenes with people talking, preview title, rolling credits).
(2) Since the scene complexity is a function of the content itself (e.g., motion, details in the scene), we expect that the scene complexity at a given playback point in a video is consistent among different tracks.
%(different tracks are encodings of the same raw video)
Combining with the property that the relative size of a chunk is correlated with the scene complexity, we expect that a chunk that is relatively large/small in one track (compared to other chunks in that track) is also relatively large/small in another track.

\begin{figure}[ht]
	\centering
	%\subfloat[H.264, YouTube-encoded\label{subfig:si-ti-264}]{%
	\includegraphics[width=0.43\textwidth]{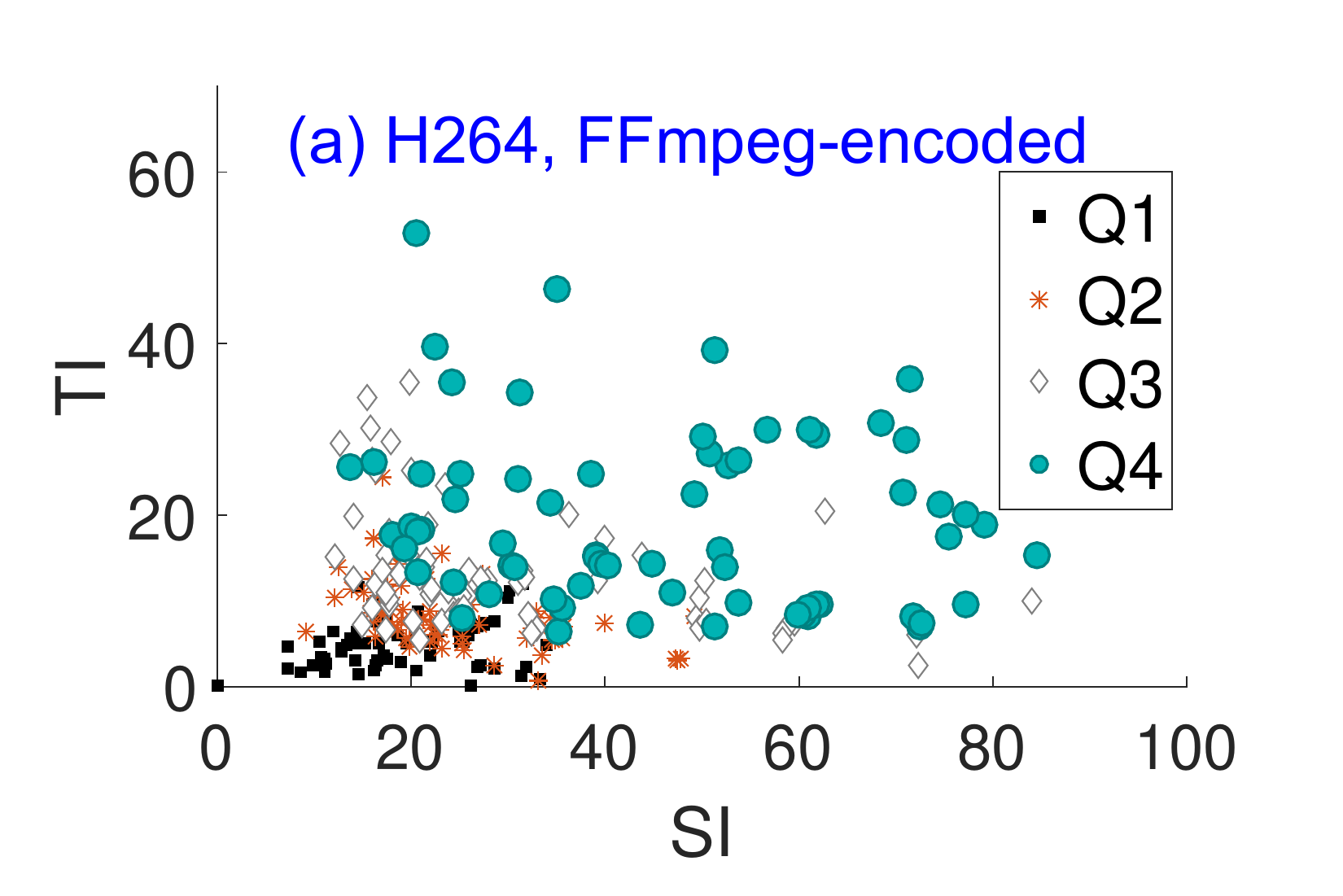}
	%}
	%\subfloat[H.265, \texttt{FFmpeg}-encoded\label{subfig:si-ti-265}]{%
	\includegraphics[width=0.43\textwidth]{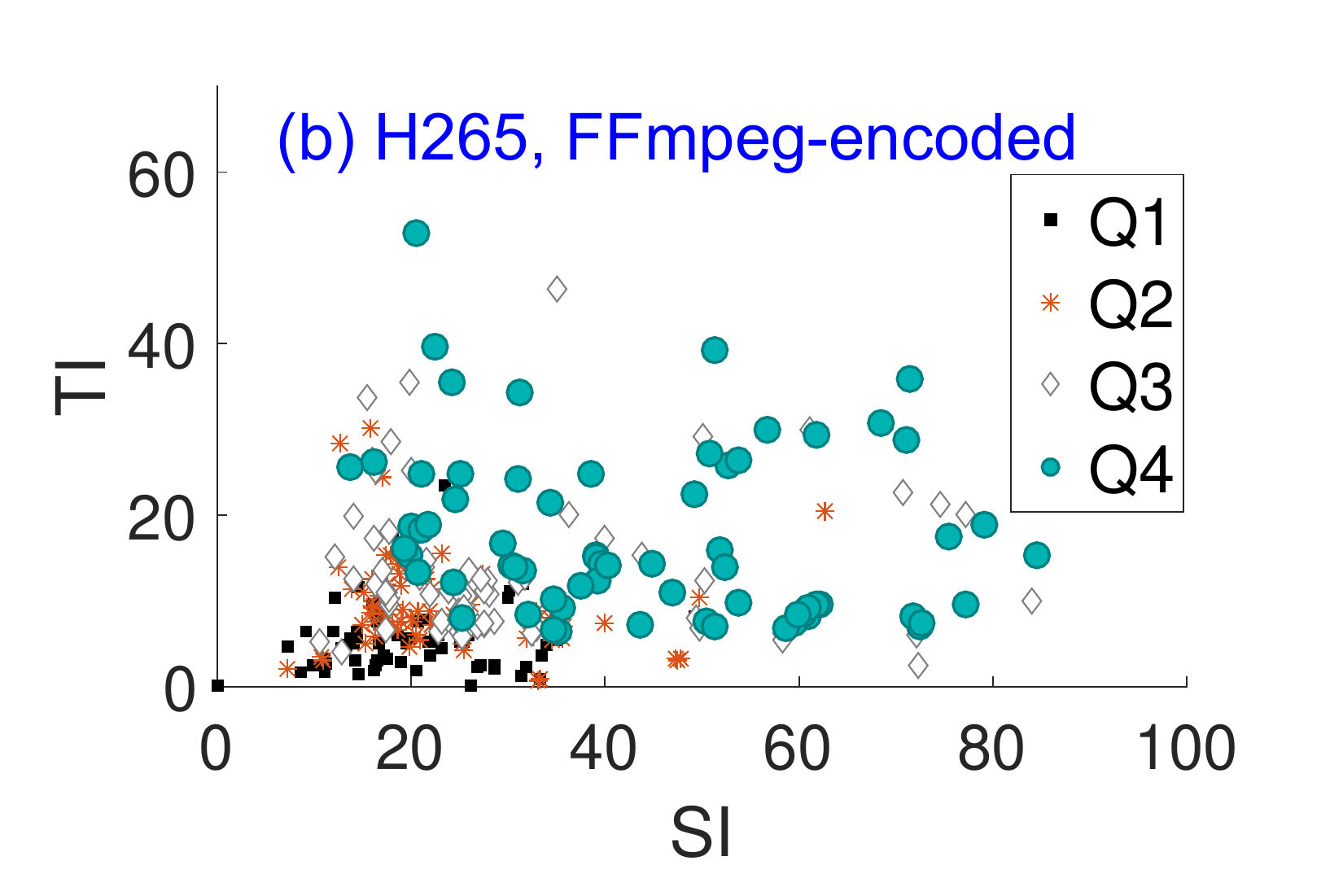}
	%}
	\caption{Chunk SI \& TI (Elephant Dream, Track 3).}
	\label{fig:si-ti}
\end{figure}

%\bing{We verify the above two properties as follows. For each track, we classify the chunks into four categories, based on their size distribution. Specifically, a chunk with a size falling into the first quartile is called a Q1 chunk, and a chunk with a size falling into the second quartile is called a Q2 chunk, and so on.  To verify Property (1), we calculate the SI and TI (recall that SI and TI are two commonly used complexity metrics) for the chunks in each category. Specifically, we obtain SI and TI values from the raw video, which is not affected by any encoding distortion and hence the SI and TI values thus obtained reflect the inherent scene complexity more accurately. We observe that indeed chunks in lower quartiles have lower SI and TI values, while the Q4 chunks tend to have the largest SI and TI values.

We verify the above two properties as follows. To verify Property (1), we calculate the SI and TI values (recall that SI and TI are two commonly used complexity metrics) for each chunk in a video. Specifically, we obtain SI and TI values from the raw video, which is not affected by encoding distortion, and hence the SI and TI values thus obtained reflect the inherent scene complexity more accurately. We then consider an encoded video, for 
each track, we classify the chunks into four categories, based on their size distribution. Specifically, a chunk with a size falling into the first quartile is called a Q1 chunk, and a chunk with a size falling into the second quartile is called a Q2 chunk, and so on.  We observe that indeed chunks in lower quartiles have lower SI and TI values, while the Q4 chunks tend to have the largest SI and TI values.
Fig.~\ref{fig:si-ti} shows an example.
For the H.264 encoding, Fig.~\ref{fig:si-ti}(a) shows that
$78$\% of Q4 chunks have SI and TI larger than 25 and 7, respectively, while only 11\% of Q1 chunks and 14\% of Q2 chunks have SI and TI larger than these thresholds.
%Fig.~\ref{fig:si-ti}(b) shows similar results for H.265 encoding.
For  H.265 (Fig.~\ref{fig:si-ti}(b)), $75$\% of Q4 chunks have SI and TI larger than 25 and 7, respectively, while only 5\% of Q1 chunks and 14\% of Q2 chunks have SI and TI larger than these thresholds.
%		
%	For the H.264 encoding, Fig.~\ref{fig:si-ti}(a) shows that
%	$93$\% of Q4 chunks have SI and TI larger than 20 and 7, respectively, while only 13\% of Q1 chunks and 27\% of Q2 chunks have SI and TI larger than these thresholds.
%	%Fig.~\ref{fig:si-ti}(b) shows similar results for H.265 encoding.
%	For  H.265 (Fig.~\ref{fig:si-ti}(b)), $90$\% of Q4 chunks have SI and TI larger than 20 and 7, respectively, while only 10\% of Q1 chunks and 28\% of Q2 chunks have SI and TI larger than these thresholds.
%
%
To verify Property (2), 	
%suppose there are $n$ chunks in each track, and 
let $\{c_{\ell,1},\ldots,c_{\ell,n}\}$ denote the sequence of the chunk categories for track/level $\ell$, where $c_{\ell,i} \in \{1,2,3,4\}$ represents the category of the $i$th chunk in track $\ell$, and $n$ is the number of chunks in a track.
We then calculate the correlation between each pair of tracks.
%Table~\ref{corr-matrix-table} shows the correlation matrix for one video; the results for other videos show similar trend.
Our results confirm that all the correlation values are close to 1, indicating that
%the chunks with the same playback positions
chunks with the same index (i.e., at the same playback position) are indeed in consistent categories across tracks.
%
%In addition, the correlation between a middle track (track 2, 3, 4 or 5) and all the other tracks is high; the correlation between the lowest and highest track (tracks 1 and 6) is slightly lower, which is perhaps not surprising because they differ the most in quality/resolution.
%
%these two tracks have the lowest and highest quality/resolution, respectively.

%%wb, 1/28/2018, for space
\iffalse
\begin{table}[]
\centering
\caption{Correlation Matrix (Elephant Dream, H.264, encoded by us). \bing{To change. There are 6 tracks instead of 5 tracks.}}
\small{
\begin{tabular}{|c|c|c|c|c|c|}
\hline
        & Track 1 & Track 2 & Track 3 & Track 4 & Track 5 \\ \hline
Track 1 & 1.00    & 0.96    & 0.95    & 0.92    & 0.89    \\ \hline
Track 2 & 0.96    & 1.00    & 0.97    & 0.95    & 0.92    \\ \hline
Track 3 & 0.95    & 0.97    & 1.00    & 0.97    & 0.94    \\ \hline
Track4  & 0.92    & 0.95    & 0.97    & 1.00    & 0.97    \\ \hline
Track 5 & 0.89    & 0.92    & 0.94    & 0.97    & 1.00    \\ \hline
\end{tabular}
}
\label{corr-matrix-table}
\end{table}
\fi

The above two properties together suggest that we can classify chunks into different categories of scene complexity simply based on chunk size. For consistent classification across tracks, we can leverage the consistent relative sizes across tracks, use one track as the reference track, and classify chunks at different playback positions based on their chunk sizes in that track (so that chunks with the same index will be in the same class).
Specifically, we choose a reference track (e.g., a middle track), and classify the chunks at different playback positions into four categories, Q1, Q2, Q3, and Q4 chunks, based on which quartile that the size of a chunk falls into.
%corresponding respectively to the first to the fourth quartile size chunks in the track.
After that, all chunks in the same playback positions are placed into the same category, regardless of the tracks.
%belong to the same category, regardless of the tracks. We refer to these four categories as Q1, Q2, Q3, and Q4 chunks.
%In the rest of the paper, these four categories are referred as Q1, Q2, Q3, and Q4 chunks.
%
%Since chunk size reflects scene complexity, the above four categories correspond to increasingly more complex scenes, with the Q1 chunks containing the simplest scenes, and Q4 chunks containing the most complex scenes.

%
The above classification method is based on quartiles. Other methods can also be used (e.g., using five classes instead of four classes); our design principles (\S\ref{sec:cava_principles}) and rate adaptation scheme (\S\ref{sec:cava_design}) are independent of this specific classification method.

\mysubsubsection{Scene Complexity and Encoding Quality} \label{sec:complexity-quality}
After classifying chunks into the four categories as above, we now analyze the encoding quality of the chunks in each category.
VBR encoding allocates fewer bits to simple scenes and more bits to complex scenes so as to maintain a constant quality across the scenes~\cite{effelsberg1996high}.
We show below, however,
the quality across the scenes is not  constant in a specific track.
%In the following, we demonstrate the above in both objective quality and perceptual quality.
Specifically, we use three quality metrics, Peak Signal-to-Noise Ratio (PSNR), Structural Similarity Index (SSIM)~\cite{Wang04:SSIM}, and Video Multimethod Assessment Fusion (VMAF)~\cite{vmaf,Liu13:Visual,Lin14:fusion-based}.
PSNR is a traditional image quality metric.
%for space
%\fengc{While there is a consensus that PSNR does not {always} correlate well with perceptual quality, there is no single quality metric that can replace PSNR~\cite{de2016large}, and PSNR is still a widely used quality metric.}
The PSNR of a chunk is represented as the median PSNR of all frames in that chunk. SSIM is a widely used perceptual quality metric; it utilizes local luminance,
contrast and structure measures to predict quality. VMAF is a recently proposed perceptual quality metric that correlates quality strongly with subjective scores and has been validated independently in~\cite{Rassool17:VMAF-rep}.
%for space
%\fengc{It is derived using a large and diverse set of video clips; a recent effort~\cite{Rassool17:VMAF-rep} has validated it independently.
%Specifically, VMAF uses a machine learning approach that takes a series of high-performance metrics as features and a pre-trained prediction model.
%%(obtained by asking subjects to evaluate the quality of a variety of short video clips)
%to predict per-frame quality.}
%
There are two VMAF models: \emph{TV model} for larger screens (TV, laptop, tablet),
and \emph{phone model} for smaller screens (phones).
% one for larger screens (TV, laptop, tablet), referred to as VMAF TV model, and the other for small screens (phones), referred to as VMAF phone model.
%henceforth referred to as VMAF TV model and VMAF phone model, respectively.
%one using a prediction model trained for larger screens (TV, laptop, tablet), and the other using a prediction model trained for small screens (phones).
The aggregate VMAF of a chunk can be calculated in multiple ways~\cite{VMAF-aggr}. We use the median VMAF value of all frames in a chunk as the VMAF value for the chunk (using mean leads to similar values for the videos in our dataset).
%We use the model for phones below.

%%present both PSNR, SSIM and VMAF
%%wb, 1/31/2018, only kept the results for 480p
\iffalse
\begin{figure}[t]
	\centering
	\begin{minipage}{1\textwidth}
		\raisebox{-\height}{\includegraphics[width=0.25\textwidth]{figs/quartile-quality/YTpsnrED264_480p}}
		\raisebox{-\height}{\includegraphics[width=0.25\textwidth]{figs/quartile-quality/YTssimED264_480p}}
		\vspace{.6ex}
	\end{minipage}
	\begin{minipage}{1\textwidth}
		\raisebox{-\height}{\includegraphics[width=0.25\textwidth]{figs/quartile-quality/YTvmafED264_480p}}
		\raisebox{-\height}{\includegraphics[width=0.25\textwidth]{figs/quartile-quality/YTvmafphoneED264_480p}}
		%\caption{caption of the four small images\\second line\\third line}
	\end{minipage}
    \vspace{-.1in}
	\caption{Quality of chunks in a VBR video (Elephant Dream, YouTube encoded, H.264).}
	\label{fig:Q4-quality}
	\vspace{-.2in}
\end{figure}
\fi

\begin{figure}[t]
	\centering
	\begin{minipage}{1\textwidth}
		\raisebox{-\height}{\includegraphics[width=0.44\textwidth]{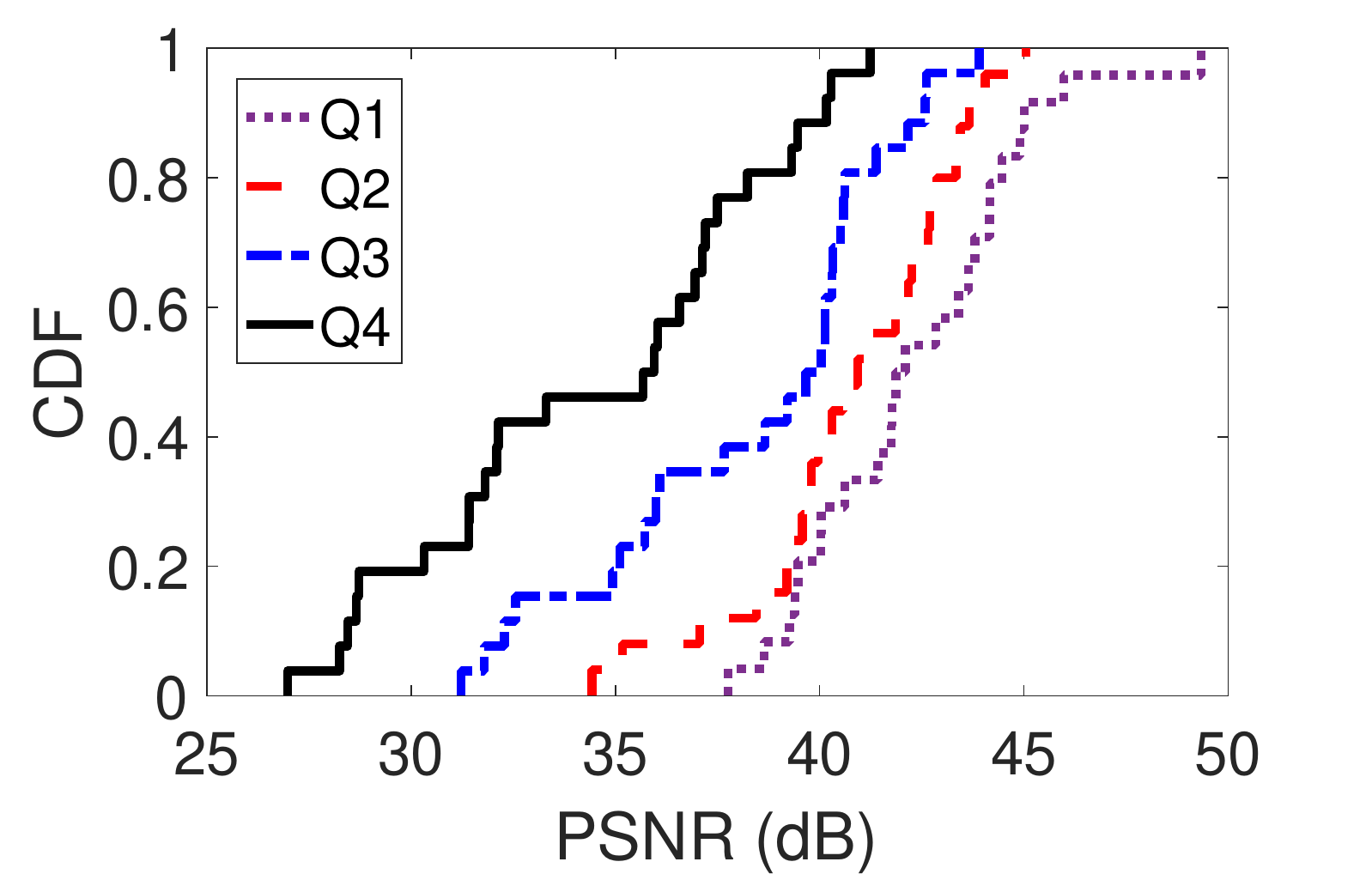}}
		\raisebox{-\height}{\includegraphics[width=0.44\textwidth]{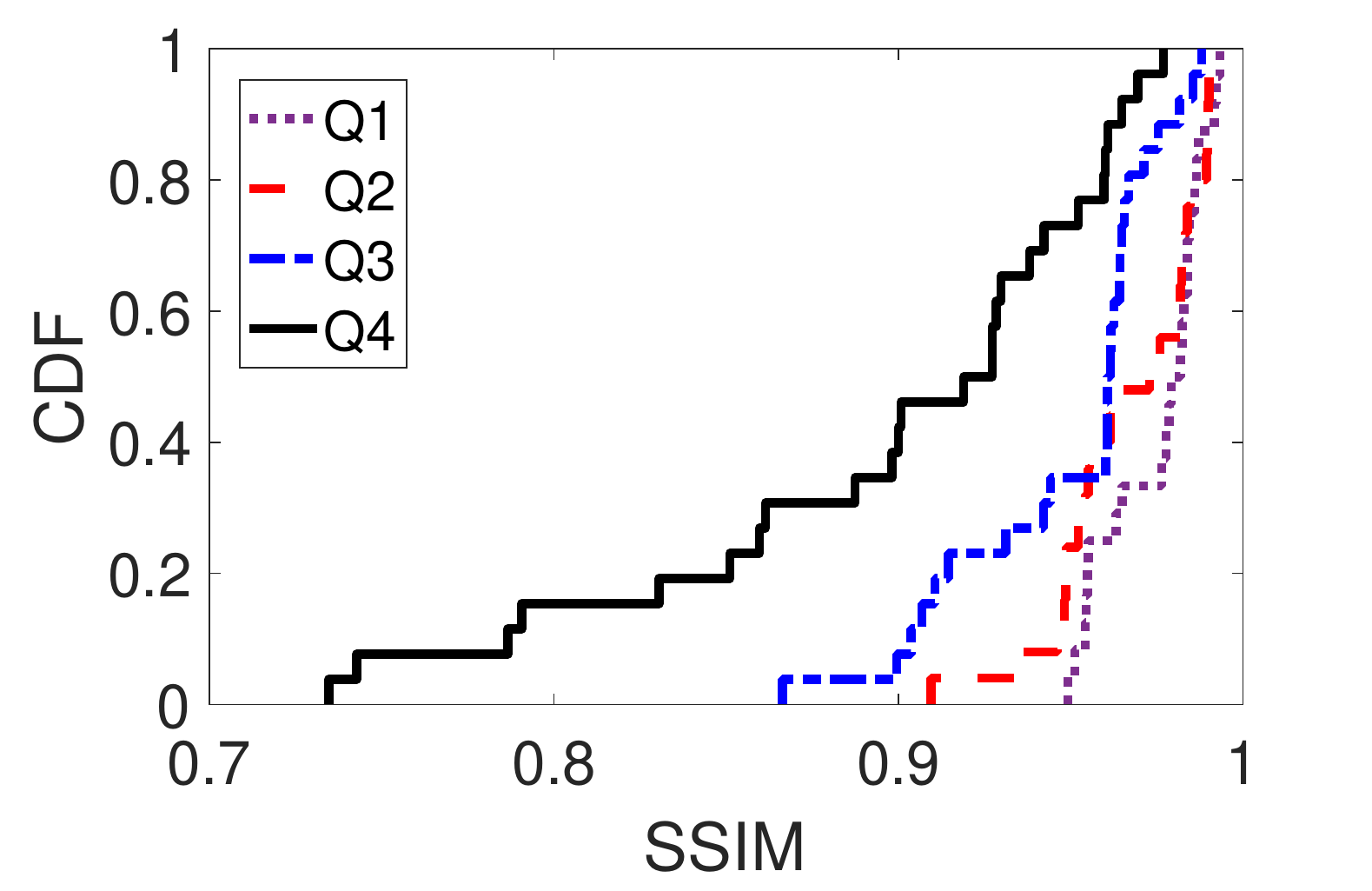}}
		\vspace{-.1ex}
	\end{minipage}
	\begin{minipage}{1\textwidth}
		\raisebox{-\height}{\includegraphics[width=0.44\textwidth]{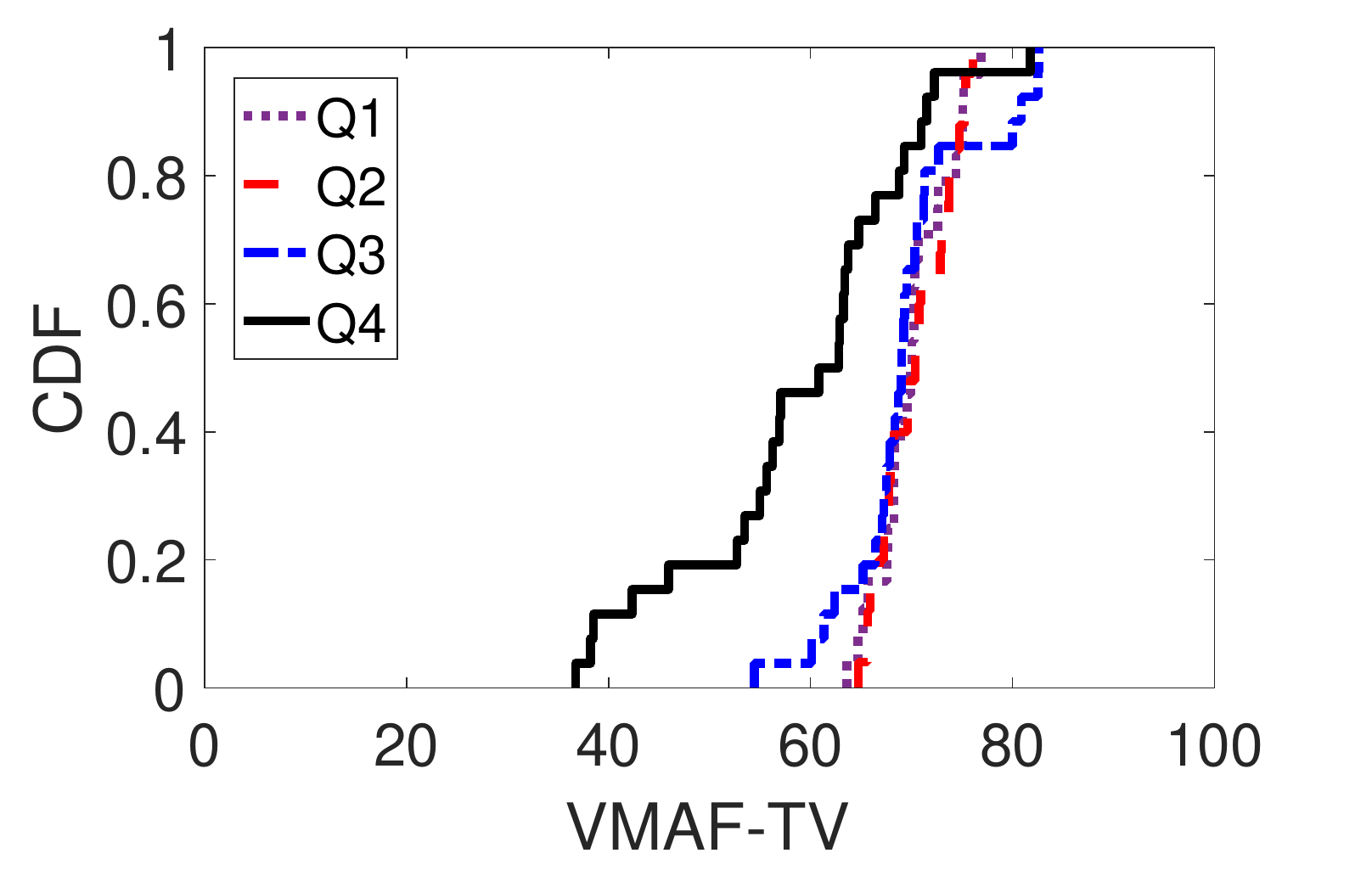}}
		\raisebox{-\height}{\includegraphics[width=0.44\textwidth]{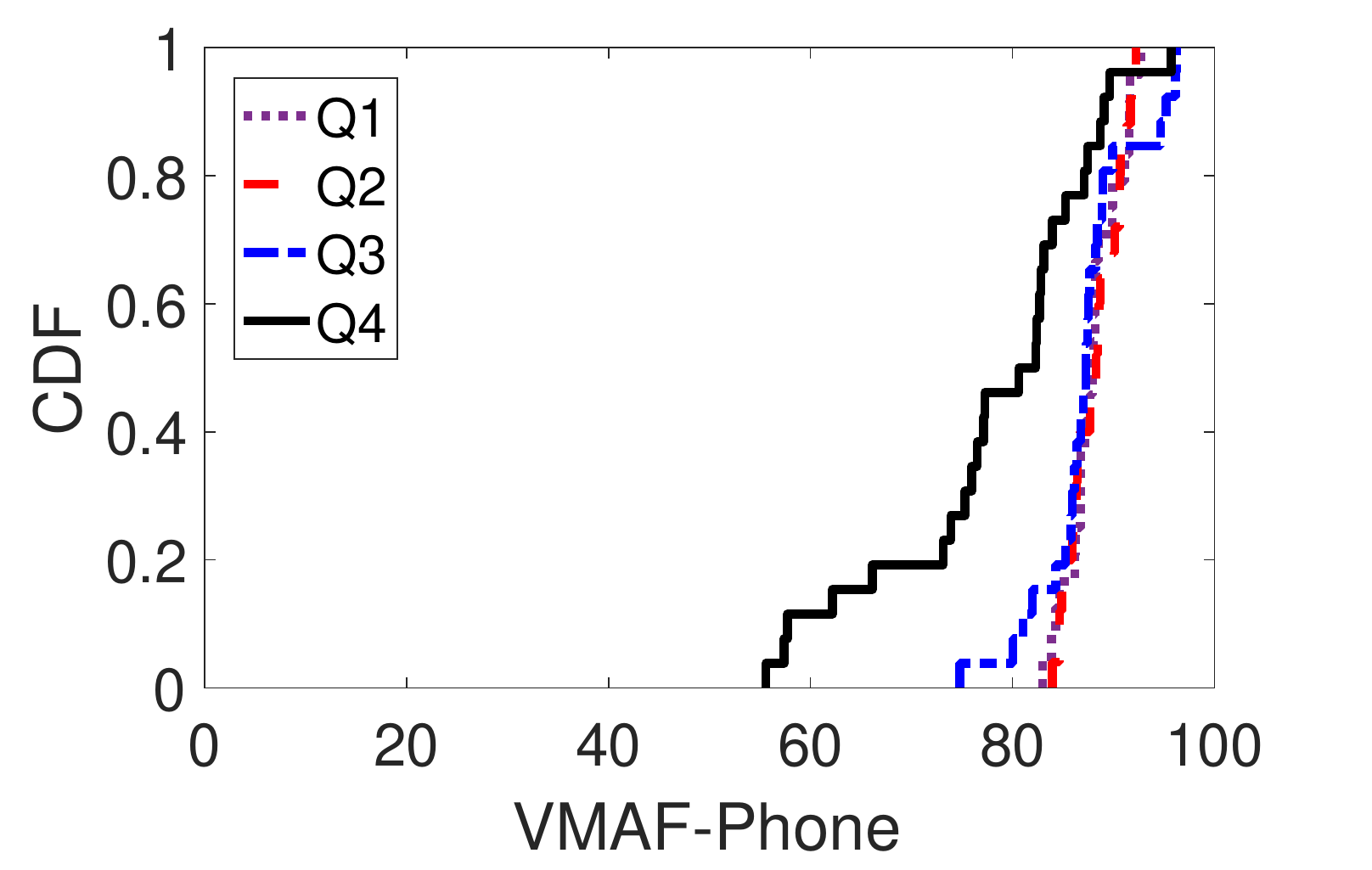}}
		%\caption{caption of the four small images\\second line\\third line}
	\end{minipage}
	\vspace{-.1in}
	\caption{Quality of chunks in a VBR video (Elephant Dream, YouTube encoded, H.264).}
	\label{fig:Q4-quality}
	\vspace{-.2in}
\end{figure}

Fig.~\ref{fig:Q4-quality} plots the CDF of the quality, measured in PSNR, SSIM, VMAF TV, and VMAF phone model, for a YouTube-encoded video. For each quality metric, the results of
a middle track (480p)
%two tracks (240p and 480p in the left and right columns, respectively)
are shown in the figure.
We observe that, while Q1 to Q4 chunks have increasing sizes (in bits), they have decreasing quality. In addition, the quality gap between Q4 and Q1--Q3 chunks is particularly large, despite that Q4 chunks have the most bits.
This is because it is challenging to encode complex scenes to have the same quality of simple scenes.
% This may be due to the limitations of the current encoding techniques -- it is challenging to encode complex scenes to have the same quality of simple scenes.
%for space
%\fengc{The gap between Q4 chunks and other chunks is even larger for lower tracks (figure omitted), which may be due to the limited number of bits in such tracks.}
 %; as a result, even Q4 chunks have relatively more bits, they are not sufficient to bring up the quality of these chunks.
% The above video has 2$\times$ cap; we show the same trend even for videos with a much larger cap in \S\ref{sec:video-discussion}.
%\feng
We make similar observations for videos encoded using \texttt{FFmpeg} and  H.265 encoded videos (figures omitted).

\mysubsection{Implications for Rate Adaptation} \label{sec:implication}
As described above, we have identified two important characteristics of VBR encoding: \emph{(1) the relative size of a chunk (relative to the other chunks in the same track) is a good indication of the scene complexity, and (2) complex scenes, despite being encoded with more bits, have inferior qualities compared to other scenes in the same track, across a range of quality metrics}. They both have important implications on rate adaption during VBR streaming process.

In particular, the second observation is opposite to what one would like to achieve during playback, where having higher quality for complex scenes is more important~\cite{li2014streaming}.
% since it brings more enhancement to the viewing experience, as indicated in~\cite{li2014streaming}.
It indicates that complex scenes need to receive differential treatment to achieve a higher quality,

%
%This differential treatment
which requires (i) inferring the scene complexity, and (ii) carefully designing ABR streaming algorithms.
 %to can be achieved for VBR videos.
 %This is because, as indicated by the first observation,
 The first requirement can be achieved based on our first observation,
 %, i.e., relative chunk sizes can be used to infer the complexity of the scenes.
 %
 by using chunk sizes to infer scene complexity.
 It has two advantages compared to using those based on video content (e.g., SI and TI). First, it is simple to compute.
  %requiring no computation-heavy content-level analysis.
 Second, chunk sizes are already known at the client in current dominant ABR standards\footnote{Chunk size information is included in the manifest file sent from server to client in DASH, and more recently HLS has added this feature~\cite{Xu17:Dissecting}.}.
 
%The second requirement,
Achieving the second requirement is very challenging.

Given that complex scenes are already encoded with more bits than other scenes in the same track, choosing higher tracks for complex scenes indicates that a significant amount of network bandwidth needs to be allocated to stream complex scenes,  which may come at the cost of lower tracks for simple scenes. We therefore need to design ABR streaming schemes to judiciously use network bandwidth, favoring Q4 chunks while not degrading the quality of other chunks too drastically (\S\ref{sec:cava_principles}, \S\ref{sec:cava_design}).
%for space

%discuss when the cap of VBR is more relaxed
\mysubsection{VBR with Larger Cap} \label{sec:cava_video-discussion}
In this chapter, we mainly use 2$\times$ capped VBR videos. The bitrate cap, while limits the amount of bitrate variability and makes it easier to stream VBR videos over the Internet, also limits the quality of the videos, particularly for complex scenes. The recommendation on VBR encoding has been evolving. The original recommendation was to set the cap to  1.1$\times$~\cite{apple-hls}, and has now evolved to 2$\times$~\cite{hls2017}. It is conceivable that the cap will be even larger in the future. For this reason, we further encode one of the publicly available videos to be 4$\times$ capped.

We observe that the characteristics described in \S\ref{sec:cava_characteristics} still hold for this video. Specifically, Q4 chunks, even under
%uncapped encoding or
4$\times$ cap, are still of significantly lower quality than other chunks. 

As an example, for the middle track (480p), the median VMAF value (phone model) for the Q4 chunks is 79, significantly lower than the values of 88, 88, and 85 for Q1 to Q3 chunks.
This might be because it is inherently very difficult to encode complex scenes to reach the same quality as simple scenes~\cite{chen2017encoding}.
Therefore,  for VBR videos of larger caps, we still need to incorporate these characteristics when designing ABR schemes (\S\ref{sec:perf-large-cap}).

	\section{Design Principles}\label{sec:cava_principles}

We believe that a good rate adaptation scheme for VBR videos must (i) take the per-chunk bitrate information into consideration (also suggested by~\cite{Xu17:Dissecting}), which is available to the clients in dominant ABR
standards (see \S\ref{sec:implication}), and (ii) use per-chunk bitrate information properly, by incorporating the characteristics of VBR videos. We next propose three design principles for VBR video rate adaptation.

\begin{figure}[t]
	%\vspace{-.3in}
	\centering
	\includegraphics[width=0.85\textwidth]{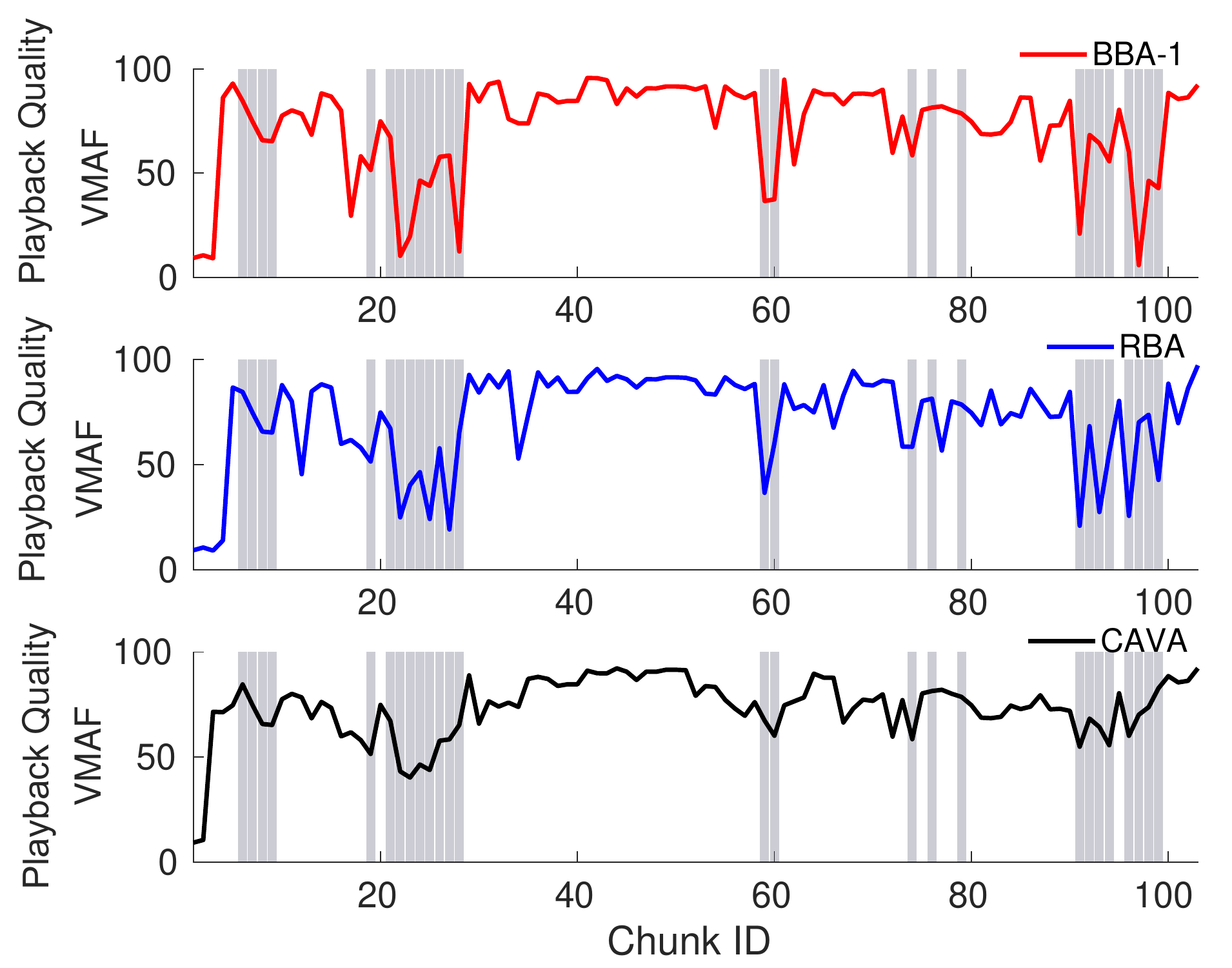}
	\vspace{-.1in}
	\caption{Two myopic schemes and CAVA.}
	\label{fig:indv-trace}
	\vspace{-.3in}
\end{figure}

%The first principle is to be
\textbf{Be Non-myopic (P1).} This requires considering not only the bitrate of the immediate next chunk,
%but also several future chunks.
but also the bitrate of multiple future chunks.
{A myopic scheme only considers the bitrate of the immediate next chunk, leading to level selections that only match the current resource (in terms of network bandwidth or player buffer level).
As a result,
a myopic scheme oftentimes mechanically selects very high (low) levels for chunks with small (large) sizes -- exactly the opposite to what is desirable for VBR videos (\S\ref{sec:cava_characteristics}).}
%A non-myopic strategy can consider multiple future chunks as a whole, more adaptively and elastically allocating the resources.
We illustrate this using
two myopic schemes:
(i)~BBA-1~\cite{huang2014buffer}, a buffer-based scheme,
%that selects the maximum allowable size for the next chunk based on the current buffer level;
and (ii) a rate-based scheme~\cite{tong2017rbavbr},
%the scheme in~\cite{tong2017rbavbr}, which is essentially a rate-based scheme,
referred to as RBA henceforth.
%BBA-1 is derived from a CBR scheme, BBA-0~\cite{huang2014buffer}.
Both schemes are myopic: when selecting the track for the next chunk, BBA-1 selects the highest track
%so that the corresponding chunk size is allowed by the
based on a \emph{chunk map}, which defines the allowed chunk sizes as a range from the average chunk size of the lowest track to that of the highest track;
RBA selects the highest track so that after downloading the corresponding chunk,
the player buffer will still contain at least four chunks, where the downloading time of a chunk is obtained as its size divided by the estimated network bandwidth.
Fig.~\ref{fig:indv-trace} shows an example that illustrates the performance of BBA-1 and RBA (the top two plots), where the shaded bars mark the playback positions of Q4 chunks.
We observe that indeed BBA-1 and RBA tend to lead to low qualities for Q4 chunks  (recall that these tend to correspond to complex scenes)  and high qualities for Q1-Q3 chunks (with simpler scenes).
%%wb, 6/19/2018, this is too detailed; omit
%
%
For comparison, we also plot our scheme CAVA  (to be described in~\S\ref{sec:cava_design}) in Fig.~\ref{fig:indv-trace}. In this example,
for BBA-1 and RBA, the average VMAF quality of Q4 chunks is 49 and 52, and the amount of rebuffering  is 6s and 4s, respectively; both are significantly worse than our scheme CAVA, under which the average VMAF quality of Q4 chunks is 65 with no rebuffering.

\textbf{Offer Differential Treatment (P2).} This principle favors more complex scenes by explicitly accounting for the chunk size and quality characteristics of VBR videos (\S\ref{sec:implication}). The differential treatment can be achieved by using resources saved from simple scenes for complex scenes. While this suggests potentially lower quality for simple scenes, a good VBR rate adaptation scheme needs to achieve a balance: improve the quality of complex scenes while not degrading too much the quality of simple scenes.
This principle is particularly important when complex scenes have lower quality than simple scenes in the same track, as we have observed in \S\ref{sec:complexity-quality}, since in this case, it is preferable to choose a higher track for complex scenes to achieve comparable quality as simple scenes. Even in an ideal VBR encoding, where complex scenes have similar quality as simple scenes in the same track, this principle is still important, since complex scenes, being of larger sizes, naturally need more resources than simple scenes in the same track.
%, and need more resources to achieve comparable quality as simple scenes.
Offering differential treatment thus helps the VBR algorithm make judicious rate adaptation decisions when the network bandwidth is limited.

\textbf{Be Proactive (P3).} Since VBR videos have high bitrate variability and the size/bitrate of future chunks
%each chunk
is known beforehand, it is desirable to account for the variability proactively.
This is especially true for complex scenes corresponding to large chunks.
Suppose when making a decision at time $t$,
the player notices that consecutive positions after $t$ contain complex scenes.
If the playback buffer is low and the network bandwidth is low, then rebuffering will inevitably occur.
Fortunately, with the known chunk size information, the rate adaptation algorithm
can, instead, react to the cluster of large chunks \emph{early}---before time $t$ instead of right at $t$---by, for example, pre-adjusting the target buffer level for buffer-related approaches.
%By taking actions before it is too late, this proactive principle further enhances the adaptiveness of rate adaptation algorithms.

\textbf{Discussion.} Several existing schemes~\cite{yin2015control,li2014streaming,qin2017control} use a window-based approach that considers multiple future chunks in bitrate adaptation, which differs from our non-myopic principle that is motivated specifically by the dynamic chunk bitrates/sizes of VBR videos (\S\ref{sec:inner}).
%Several existing schemes~\cite{yin2015control,li2014streaming,qin2017control} consider multiple future chunks in bitrate adaptation. Their notion of  ``non-myopic" differs from our non-myopic principle, which is motivated specifically by the dynamic chunk bitrates/sizes of VBR videos (see \S\ref{sec:inner}).
%
Note that \textbf{P1} and \textbf{P3} share in common that they both require considering future chunks.
But they differ mainly in the timescale:
the non-myopic principle considers a small number of future chunks to account for the variability of VBR videos in the short timescale, in order to facilitate the \emph{current} decision; the proactive principle, on the other hand, considers chunks further away in future to benefit \emph{future} decisions in a proactive manner.

	\section{CAVA Design and Implementation} \label{sec:cava_design}
We now instantiate the three principles (\S\ref{sec:cava_principles}) and design a control-theoretic rate adaptation scheme, CAVA, for VBR streaming.
The reason for adopting a control-theoretic approach is because it has shown promise in adapting to dynamic network bandwidth~\cite{yin2015control,qin2017control}.
%\fengc{, particularly in cellular networks}~\cite{yin2015control,qin2017control}.
The design of CAVA is compatible with the current dominant ABR standards (DASH and HLS), and can be readily used in practice.

\begin{figure}[t]
%\vspace{-.3in}
 \centering
     \includegraphics[width=0.8\textwidth]{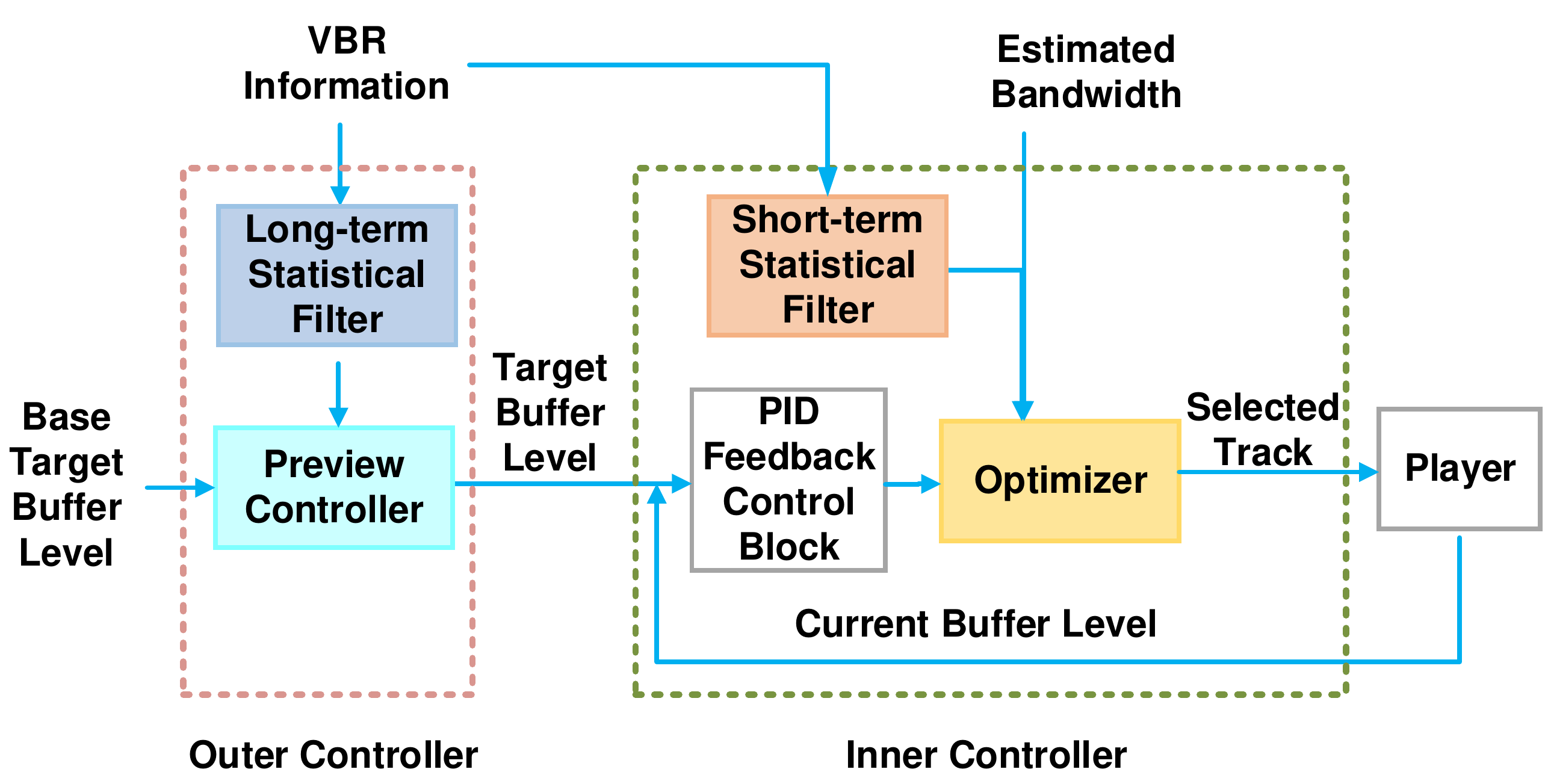}
     \vspace{-.15in}
    \caption{CAVA design diagram.}
    \label{fig:PIB-diagram}
    \vspace{-.2in}
\end{figure}

\subsection{Overview of CAVA}
	%Leveraging all the three design principles for VBR streaming, 
	We design CAVA to address both chunk size variability, an inherent feature of VBR videos, and quality variability in VBR videos, a characteristic that we identified in today's VBR encodings (\S\ref{sec:VBR-characteristics}). Even under 
	%theoretically idealized 
	an ideal VBR encoding where complex scenes have similar quality as other scenes in the same track, CAVA still has its value in addressing the inherent chunk size variability of VBR videos.

Fig.~\ref{fig:PIB-diagram} shows the design diagram for CAVA, which builds on the basic feedback control framework~\cite{qin2017control}.
%, an ABR scheme for CBR videos.
CAVA introduces significant innovations by
(i) generalizing the control framework
% in~\cite{qin2017control} that was designed for
from plain CBR to VBR,
(ii) designing an inner controller that reacts in short timescale, and incorporates the \emph{non-myopic} and \emph{differential treatment principles}, and (iii) designing an outer controller that reacts over a longer timescale, and incorporates the \emph{proactive principle}.
%The inner controller incorporates the \emph{non-myopic} and \emph{differential treatment principles}; the outer controller incorporates the \emph{proactive principle} (\S\ref{sec:principles_details}).

The control framework uses PID control~\cite{astrom08:feedback}, a widely used feedback control technique.
The reason for using PID control is its good performance and robustness~\cite{qin2017control}.
%for space
%
As shown in Fig.~\ref{fig:PIB-diagram}, CAVA consists of two controllers that work in synergy to realize the aforementioned three principles.

\BULLET The \emph{inner controller} is responsible for selecting the appropriate track level. It employs a PID feedback control block that maintains the target buffer level by properly adjusting the control signals (\S\ref{sec:basic-control}). Then based on the controller's output, together with the estimated bandwidth and a short-term average of future chunks' bitrate, CAVA invokes an optimization algorithm that incorporates
non-myopic (\textbf{P1}) and differential treatment (\textbf{P2}) principles
% described in~\S\ref{sec:principles}
to perform VBR-aware track level selection (\S\ref{sec:inner}).

\BULLET The \emph{outer controller}'s job is to determine the target buffer level that is used by the inner controller. When doing so, CAVA exercises proactive principle (\textbf{P3}) to adjust the target buffer level proactively based on dynamics of future chunks (\S\ref{sec:outer}).

%%wb, 1/26/2018, comment for now
\iffalse
\begin{table}
    \centering
    \caption{Key notation in \piv.}
    \small{
    \begin{tabular}[h]{l|l} %\hline
 %   Symbol & Meaning   \\ \hline\hline
        $\widehat{C}_t$ & Estimated network bandwidth at time $t$          \\ \hline
        $x_t$   & Buffer level at time $t$ (second)                 \\ \hline
        $x_r$     & Base target buffer level  (second)        \\ \hline
        $x_r(t)$     & Target buffer level at time $t$ (second)        \\ \hline
	$\ell_t$   & Selected track/level at time $t$ \\ \hline
	$R_t(\ell_t)$   & Bitrate of the chunk being selected at time $t$ \\ \hline
%	$B_t(\ell)$   & Bitrate of chunk for time $t$ for level $\ell$ \\ \hline
%	$\bar{B_t}(\ell)$   & Short-term average corresponding to $B_t(\ell)$ \\ \hline
	$r(\ell)$   & Average bitrate of track/level $\ell$ \\ \hline
%	$\Delta$ &   Video chunk duration (second)      \\ \hline
%	$\delta$ &   Startup latency (second)      \\ \hline
        $u_t$   & Inner controller output at time $t$     \\  \hline
%   $K_p, K_i $ & Parameters for proportional and integral control \\ \hline
   $W, W'$ & Window size used in inner and outer controllers \\ \hline
%        $\beta$     & Setpoint weighting parameter \\ %\hline
    \end{tabular}
    }
    \vspace{-.2in}
    \label{tab:notations}
\end{table}
\fi

\subsection{PID Feedback Control Block} \label{sec:basic-control}
The PID feedback control continuously monitors an ``error value'',
i.e., the difference between the target and current buffer levels of the player, and adjusts the control signal to maintain the target buffer level.
To account for the characteristics of VBR videos, it considers per-chunk size information and uses a dynamic target buffer level.
This differs from the PID controller design for CBR~\cite{qin2017control}, which uses a fixed target buffer level and a fixed average bitrate for each track.
%unlike that in~\cite{qin2017control}, which is for CBR, and uses per-track average bitrate and a fixed target buffer size.

Specifically, let $C_t$ denote the network bandwidth at time $t$.
Let $\ell_t$ denote the track number selected at time $t$, with the corresponding bitrate of the chunk denoted as $R_t(\ell_t)$. We use $R_t(\ell_t)$ instead of $R(\ell_t)$ to reflect that it is a function of time $t$ since for VBR videos, the bitrate varies significantly even inside a track.
%(this notation reflects that the bitrate is both a function of $\ell_t$ and time $t$ since for VBR videos, the bitrate varies significantly even inside a track).
The controller output, $u_t$, is defined as
\vspace{-.05in}
\begin{equation}
\label{eq:cava_pid_ut}
u_t = \frac{C_t}{R_t(\ell_t)},
\vspace{-.06in}
\end{equation}
which is a unitless quantity representing
the relative buffer filling rate. The control policy is defined as
\begin{eqnarray}
u_t = {K_p}({x_r(t)} - x_t) + {K_i}\int_0^t {(x_r(t) - x_\tau)d\tau} + \textbf{1}(x_t - \Delta)
\label{eq:cava_PIA-control}
\vspace{-.06in}
\end{eqnarray}
where $K_p$ and $K_i$ denote respectively the parameters for proportional and integral control (two key parameters in PID control), $x_t$ and $x_r(t)$ denote respectively the current and target buffer level at time $t$ (both in seconds), $\Delta$ denotes the playback duration of a chunk, and the last term, $\mathbf{1}(x_t - \Delta)$, is an indicator function (it is 1 when  $x_t\geq \Delta$ and 0 otherwise), which makes the feedback control system linear, and hence easier to control and analyze.
%controller a linear controller, which is easier to analyze and set parameters~\cite{qin2017control}.
The target buffer level is dynamic, set by the outer controller (\S\ref{sec:outer}).

\begin{figure}[t]
 \centering
 \vspace{-.15in}
    %\subfloat[\label{subfig:illus-inner}]{%
      \includegraphics[width=0.43\textwidth]{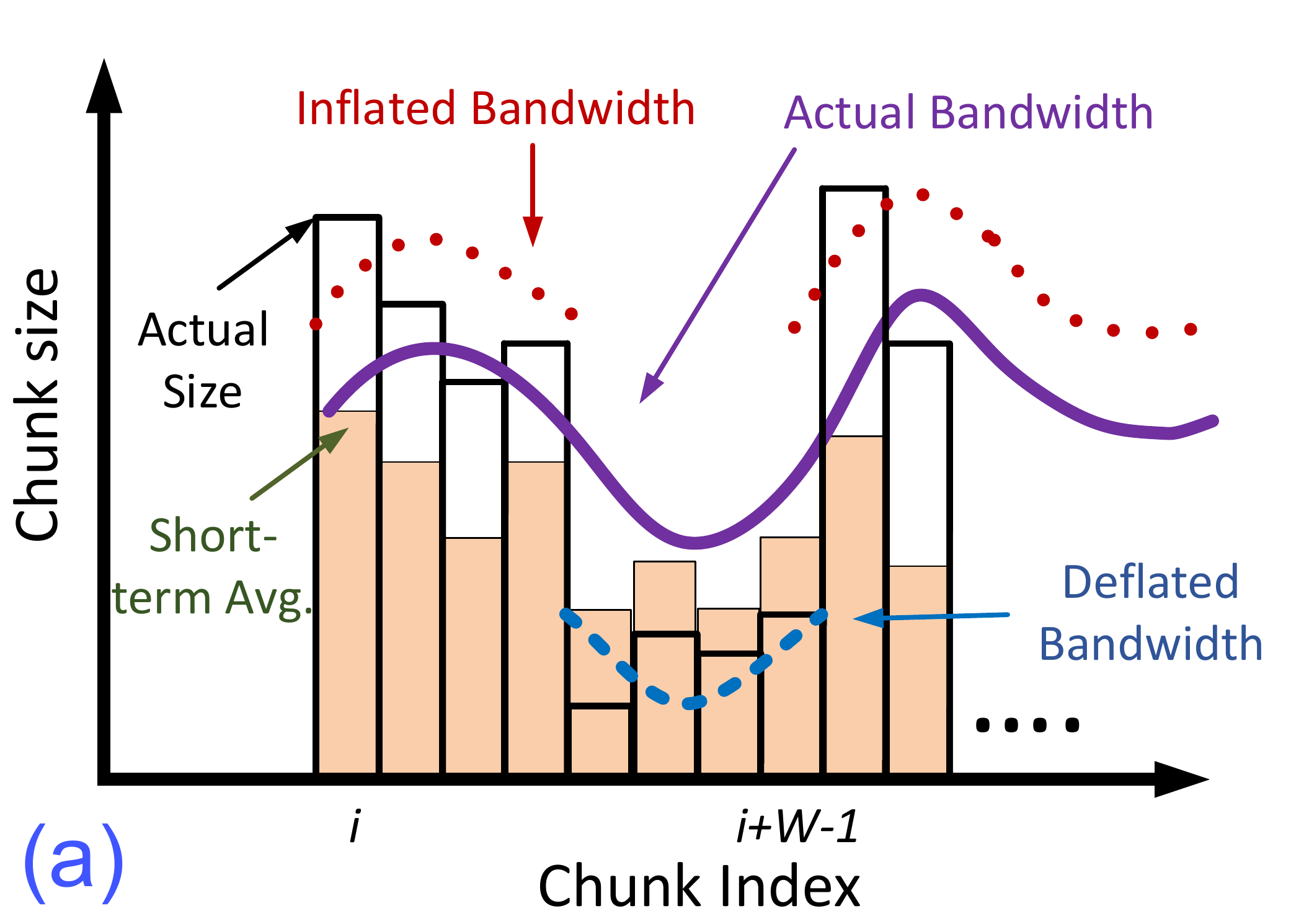}
    %}
    %\subfloat[\label{subfig:illus-outer}]{%
      \includegraphics[width=0.43\textwidth]{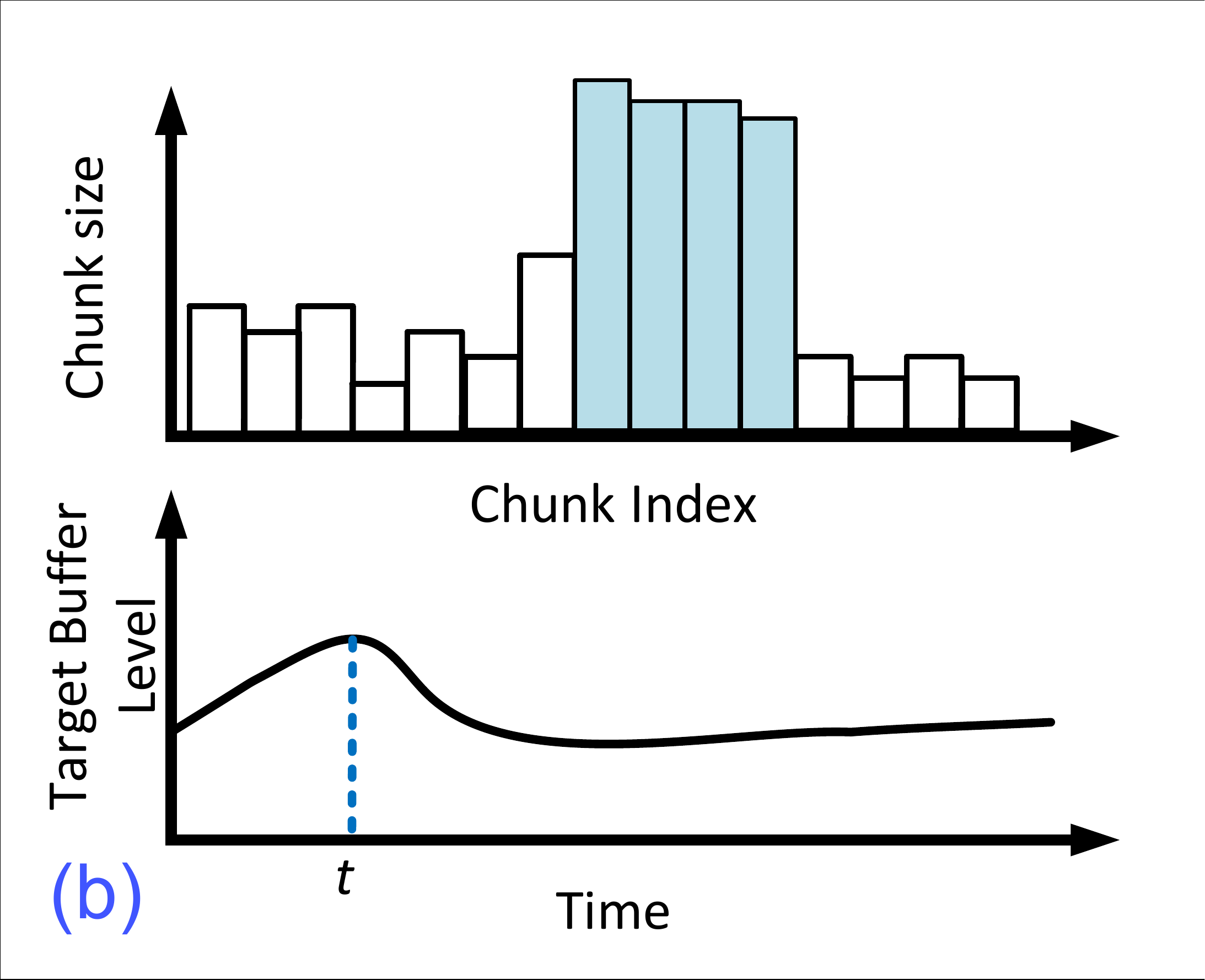}
    %}
    \vspace{-.15in}
    \caption{Illustration of (a) track selection, and (b) setting target buffer level.}
    \label{fig:design-illus}
    \vspace{-.15in}
  \end{figure}

\mysubsection{The Inner Controller} \label{sec:inner}
The inner controller determines $\ell_t$, the track at time $t$.
Based on the definition in Eq. (\ref{eq:cava_pid_ut}), once the controller output $u_t$ is determined using Eq. (\ref{eq:cava_PIA-control}), we can simply determine  $\ell_t$ so that the corresponding bitrate $R_t(\ell_t)$ is the maximum bitrate that is below ${\widehat{C}_t}/{u_t}$,
where $\widehat{C}_t$ is the estimated network bandwidth at time $t$.
This approach is, however, myopic (only considering the next chunk), and is thus undesirable for VBR videos.

We next formulate an optimization problem that accounts for both the non-myopic and differential treatment principles (\S\ref{sec:cava_principles}) to determine $\ell_t$.
Fig.~\ref{fig:design-illus}(a) illustrates the approach. Suppose at time $t$, we need to select the track number for chunk $i$. Following the {non-myopic} principle (\textbf{P1}), we consider a window of $W$ chunks, from $i$ to $i+W-1$, and use the average bitrate of these $W$ chunks to represent the bandwidth requirement of chunk $i$.
% (represented as the shaded region).
{This is handled by the ``Short-term Statistical Filter'' in Fig.~\ref{fig:PIB-diagram}.}
%\feng{[[The shaded region actually goes beyond $i+W-1$]]}
Using the average bitrate of $W$ chunks (instead of the bitrate of chunk $i$) leads to smoother bandwidth requirement from the chunks, avoiding mechanically choosing very high (low) levels for chunks with small (large) bitrates.
% (\S\ref{sec:principles}).

To provide {differential treatment} (\textbf{P2}) to chunks with different complexities, we assume either \emph{inflated} or \emph{deflated} network bandwidth, depending on whether chunk $i$ contains complex or simple scenes. For complex scenes, we assume inflated network bandwidth, allowing a higher track to be chosen, while for simple scenes, we assume deflated bandwidth so as to save bandwidth for complex scenes.
The track selection algorithm maximizes the quality level to match the assumed network bandwidth, while minimizing the number of quality level changes between two adjacent chunks.
In the example in Fig.~\ref{fig:design-illus}(a), chunk $i$ is a Q4 chunk, and hence we assume inflated bandwidth, which allows a higher level track to be selected.

\smallskip
\noindent{\textbf{Optimization Formulation.}}
%\noindent{\textbf{Quality Representation.}}
We next describe our optimization formulation. Our current design uses track level to approximate the quality since this is currently the only available quality information in predominant ABR standards; using such information only allows CAVA to be easily deployable. In our performance evaluation (\S\ref{sec:cava_perf}), we use a perceptual quality metric (VMAF) to evaluate CAVA, and show that it achieves good perceptual quality although the formulation does not explicitly use a perceptual quality metric.

%\smallskip
The optimization aims to minimize a weighted sum of two penalty terms: the first term represents the deviation of the selected bitrate from the estimated network bandwidth, and the second  term represents the track change relative to the previous chunk. 
	Other formulations 
	%of the optimization objectives 
	might also be possible; our empirical results show that this formulation works well in practice.
Specifically, we aim to minimize
\begin{eqnarray}
\label{eq:lms-vbr}
Q(\ell_t) = \sum\limits_{k = t}^{t+N-1} \left(u_{k}\bar{R_t}(\ell_t) - \alpha_t \widehat C_{k}\right)^2 + \eta_t \left(r({\ell_t}) - r({\ell_{t-1}})\right)^2,
\vspace{-.1in}
\end{eqnarray}
where $u_k$ and $\widehat C_{k}$ are respectively the controller output and  the estimated link bandwidth for the $k$-th chunk,  $\eta_t \ge 0$ is a parameter that represents the weight of the second term, $\bar{R_t}(\ell_t)$ is the average bitrate over a window of $W$ chunks (according to \textbf{P1}), from $t$ to $t+W$, at track $\ell_t$ (here we slightly abuse the notation to use $t$ to represent the chunk index for time $t$), and $r(\ell)$ denotes the average bitrate of track $\ell$.

\textbf{Penalty on deviation from estimated bandwidth.}
The first term in Eq. (\ref{eq:lms-vbr}) represents the penalty based on how much the required bandwidth deviates from the estimated network bandwidth. Specifically,
minimizing this penalty term aims at selecting the largest $\ell_t$ (and hence the highest quality) so that the difference between the required bandwidth, $u_{k}\bar{R_t}(\ell_t)$, and the assumed bandwidth, $\alpha_t \widehat C_{k}$, is minimized ($\alpha_t$ is a factor that deflates or inflates the estimated bandwidth; see details later on).
Following the {non-myopic} principle (\textbf{P1}), we use $\bar{R_t}(\ell_t)$ (instead of  $R_t(\ell_t)$), which 
%considers $W$ future chunks and  
is the average bitrate of $W$ future chunks, to represent the bandwidth requirement of chunk $t$.
%,  where $W \ge N$, thus ensuring that $\bar{R_t}(\ell_t)$ is stable over the finite horizon.
We refer to $W$ as the \emph{inner controller window size} and explore its configuration in \S\ref{sec:perf-parameter}.

%As shown in Eq. (\ref{eq:lms-vbr}), the penalty term considers the network bandwidth needed to download a finite horizon of $N$ future chunks since when operating online, \piv only has a limited view towards future network conditions.
When operating online, CAVA only has a limited view towards future network conditions. 
Therefore, the penalty term on deviation from estimated bandwidth considers the bandwidth needed to download a finite horizon of $N$ future chunks.
% (the first term in Eq. (\ref{eq:lms-vbr}) is the sum of $N$ terms).
We  assume that these $N$ chunks
are all from the same track, $\ell_t$, to reduce the number of quality level changes.
In the rest of the chapter, we set $N$ to 5 chunks, a relatively short horizon to accommodate network bandwidth dynamics.

%While several existing schemes~\cite{yin2015control,li2014streaming,qin2017control} also consider a finite horizon of $N$ chunks, none of them adopts our non-myopic principle in computing the bandwidth requirement from $W$ future chunks for rate adaptation.

Note that several existing schemes~\cite{yin2015control,li2014streaming,qin2017control} also examine a finite horizon of chunks in their {online} algorithms (controlled by $N$ in CAVA). But none of them utilizes concepts like our non-myopic principle in computing the bandwidth requirement from multiple future chunks (controlled by $W$ in CAVA), whose sizes may vary significantly in VBR, for rate adaptation.

\textbf{Penalty on track change.} The second term in Eq. (\ref{eq:lms-vbr}) represents the penalty on track change. Specifically, minimizing the second term aims at selecting  $\ell_t$ so that its difference from $\ell_{t-1}$, i.e., the track of the previous chunk,
is minimized.
%penalizes track level changes.
We consider two categories of chunks, Q4 and non-Q4 chunks when setting $\eta_t$, a weight factor. If the current and previous chunks fall in different categories, we set $\eta_t$ to 0; otherwise we set $\eta_t$ to 1 (so as to give equal weight to the two terms in Eq. (\ref{eq:lms-vbr}) since both are important for QoE).
%\footnote{The reason for setting $\eta_t$ to 1 is to give equal weight to the two terms in Equation~(\ref{eq:lms-vbr}) since both are related to important QoE metrics.}.
That is, we do not penalize quality change if two consecutive chunk positions are in different categories (and hence of different importance in viewing quality).
%for space
%\fengc{When $\eta_t=1$, the second term in (\ref{eq:lms-vbr}) is zero
%when the track  of the current chunk is chosen to be the same as that of the previous one, i.e., $\ell_t=\ell_{t-1}$, and is larger for larger track changes.}

Recall that $r(\ell)$ denotes the average bitrate of track $\ell$. 
%The  second term in Eq. (\ref{eq:lms-vbr}) uses
The penalty term on track change uses 
$r({\ell_t})- r({\ell_{t-1}})$, instead of $\ell_t-\ell_{t-1}$ or $R_t({\ell_t})- R_t({\ell_{t-1}})$. This is because $r({\ell_t})- r({\ell_{t-1}})$ reflects  track change, and yet is in the unit of bitrate,  which is the same as that of the first term in  Eq. (\ref{eq:lms-vbr}).
The  expression $\ell_t-\ell_{t-1}$
%We do not use $\ell_t-\ell_{t-1}$ because it
represents track change directly, whose unit is however different from that of the first term in  Eq. (\ref{eq:lms-vbr});
%which is a quantity incompatible with the first term in Eq. (\ref{eq:lms-vbr});
the expression $R_t({\ell_t})- R_t({\ell_{t-1}})$ represents bitrate change on a per chunk level, which is not meaningful for VBR videos since even chunks in the same track can have highly dynamic bitrate.

\iffalse
Recall that $r(\ell)$ denotes the average bitrate of track $\ell$. The  second term in Eq. (\ref{eq:lms-vbr}) uses
%Note that we use
$r({\ell_t})- r({\ell_{t-1}})$, instead of $\ell_t-\ell_{t-1}$,  to be of the same unit (in bitrate) as the first term in Eq. (\ref{eq:lms-vbr}).
In addition, it does not use $R_t({\ell_t})- R_t({\ell_{t-1}})$ since it is more meaningful to penalize track changes instead of bitrate changes for VBR videos, which can have highly dynamic bitrate even in the same track.
\fi

\textbf{Deflating/inflating network bandwidth estimate.} In Eq. (\ref{eq:lms-vbr}), $\alpha_t \widehat C_{k}$ represents the assumed network bandwidth for chunk $k$. As mentioned earlier, according to differential treatment principle (\textbf{P2}), we set $\alpha_t>1$ for complex scenes to inflate the bandwidth, and set $\alpha_t<1$ for simple scenes to deflate the bandwidth.  Determining the value of $\alpha_t$ presents a tradeoff: a large value for complex scenes can lead to higher quality level, at the cost of potential stalls; similarly, a small $\alpha_t$ value for simple scenes, while can save bandwidth and eventually benefit complex scenes, may degrade the quality of simple scenes too much. We have varied $\alpha_t$ for complex scenes from $1.1$ to 1.5, and varied $\alpha_t$ for simple scenes from 0.6 to 0.9. In the rest of the chapter, we set $\alpha_t=1.1$ for Q4 chunks and $\alpha_t=0.8$ for Q1-Q3 chunks, which lead to a good tradeoff observed empirically.
%

%In addition, if the current chunk is a non-Q4 chunk and the selected level is very low (i.e., level 1 or 2) under the above strategy, while the buffer level is above a threshold (chosen as 10 seconds in \S\ref{sec:perf}) indicating low risk of stalls, we do not deflate network bandwidth, and instead use $\alpha_t=1$. This heuristic avoids choosing unnecessarily low levels for Q1-Q3 chunks.

In addition, we use the following heuristic for Q1-Q3 chunks: if the current chunk is a Q1-Q3 chunk and the selected level is very low (i.e., level 1 or 2) under the above strategy, while the buffer level is above a threshold (chosen as 10 seconds in \S\ref{sec:cava_perf}) indicating low risk of stalls, we do not deflate network bandwidth, and instead use $\alpha_t=1$. This heuristic avoids choosing unnecessarily low levels for Q1-Q3 chunks.
Similarly, we can use a heuristic for Q4 chunks: if the current chunk is a Q4 chunk and the buffer level is below a threshold indicating high risk of stalls, we can also set $\alpha_t$ to 1 (i.e., do not inflate network bandwidth). This heuristic can lead to lower stalls at the cost of potentially reducing the quality of Q4 chunks. Our evaluations show that in the settings that we explore, the number of stalls in CAVA is already very low even without using this heuristic.
Therefore, in~\S\ref{sec:cava_perf}, we only report the results when this heuristic is disabled.

%To summarize, the formulation in (\ref{eq:lms-vbr}) aims to maximize the quality while minimizing quality changes, two important QoE metrics for video streaming. It uses track level to approximate quality since it is currently the only available quality information in predominant ABR standards; using such information only allows \piv to be easily deployable. In our performance evaluation (\S\ref{sec:perf}), we use a perceptual quality metric (VMAF) to evaluate the performance of \piv, and show that it achieves good perceptual quality although the formulation does not use a perceptual quality metric.

%\textbf{Computational overhead.}
\textbf{Time complexity.} Let $\mathcal{L}$ be the set of all track levels. The optimal track is
\vspace{-.1in}
\begin{eqnarray}
\label{eq:lms-vbr-solu}
\ell^*_t = \arg \min \limits_{\ell_t \in \mathcal{L}}  Q(\ell_t).
\end{eqnarray}
We can find $\ell^*_t$ easily by evaluating (\ref{eq:lms-vbr}) using all possible values of
$\ell_t \in \mathcal{L}$, and selecting the value that minimizes the objective function.
For every  $\ell_t \in \mathcal{L}$,
obtaining $Q(\ell_t)$ requires computation of $N$ steps. Therefore,
the total computational overhead of solving (\ref{eq:lms-vbr-solu}) is $O(N|\mathcal{L}|)$.
%for space
%\fengc{In \S\ref{sec:dash}, we evaluate the computation overhead in \texttt{dash.js} and show that it is negligible.}

%%wb, 6/19/2018, move to earlier to highlight the design
%\noindent{\textbf{Quality Representation.}}
%The formulation in (\ref{eq:lms-vbr}) uses track level to approximate quality since it is currently the only available quality information in predominant ABR standards; using such information only allows \piv to be easily deployable. In our performance evaluation (\S\ref{sec:perf}), we use a perceptual quality metric (VMAF) to evaluate \piv, and show that it achieves good perceptual quality although the formulation does not use a perceptual quality metric explicitly.

\iffalse
\noindent{\textbf{Choice of $W$.}} We refer to $W$ as \emph{inner controller window size}. Note that while $W \ge N$, setting $W$ too large is not desirable since in that case $\bar{R_t}(\ell_t)$ is averaged over many future chunks, and hence neglects the variability in bitrate.
%In the extreme case when $W$ is close to the length of the video, it essentially treats the VBR video as a CBR video with the bitrate of each chunk in a track as the average bitrate of the track, which can lead to a large amount of rebuffering (see \S\ref{subsec:cbr-schemes-for-vbr}).
We explore the choice of $W$ in \S\ref{sec:perf-parameter}.
\fi

\mysubsection{The Outer Controller} \label{sec:outer}
When using the inner controller alone, we observe that rebuffering sometimes happens when the network bandwidth is very low and the buffer level is also low (due to relatively large chunks downloaded earlier). While the inner controller reacts by selecting the track with the lowest bitrate, the reaction appears to be too late. To deal with such cases,
% (which may cause seconds of stalls),
we incorporate the {proactive principle} (\textbf{P3}) by adopting the concept of \emph{preview control} from control theory~\cite{takaba2003tutorial} to adjust the target buffer level proactively.
%, taking account of future chunk size variability.

%%wb, use \tilde{\ell} to represent reference track to avoid confusion with x_r where r represents target
%We next describe how we adjust the target buffer level.
Fig.~\ref{fig:design-illus}(b) illustrates our approach. The target buffer level is
set to the base target buffer level, added by a term that is determined by
%set to account for
future chunk size variability. Specifically, when there are large chunks in the future, and hence
%set to a base target buffer level, unless there are large chunks in the future. In that case, since there is
a higher chance of longer downloading time and slower buffer filling rate
that can lead to buffer underrun,
% in the future,
the target buffer level is adjusted to a higher level to react proactively (handled by the ``Long-term Statistical Filter'' in Fig.~\ref{fig:PIB-diagram}), as illustrated in the region around $t$ in Fig.~\ref{fig:design-illus}(b). This adjustment will reduce rebuffering because the controller aims to reach a higher target buffer level.
%for space
%\fengc{which acts as a cushion for the future larger chunks.}

Specifically, let $x_r$ denote a base target buffer level (we set it to 60 seconds in \S\ref{sec:cava_perf}; setting it to 40 seconds leads to similar results).
%\feng{[How do you set $x_r$?]}\yanyuan{It is set as the desired buffer at the steady state if there are no chunk size variations.}
The target buffer level at time $t$, $x_r(t)$, is set to be $x_r$ added by a term that accounts for the future chunk size variability. We use a track $\tilde{\ell}$ (e.g., a middle track)
%, Track 3, without loss of generality)
as the reference track.
If the average size of the future chunks in that track is above the mean value of that track, we proactively increase the target buffer level as: %Specifically, $x_r(t)$ is set as
\vspace{-.05in}
\begin{eqnarray}
\label{eq:targetbuffer}
x_r(t) = x_r + \max\left(\frac{\Sigma_{k=t}^{t+W'} {R_k(\tilde{\ell})}\Delta - r(\tilde{\ell}) W' \Delta}{r(\tilde{\ell})},0\right),
\end{eqnarray}
%\vspace{-.05in}
where $W'$ is referred to as \emph{outer controller window size} that represents how much to look ahead,
$\Delta$ is the chunk duration,
$r(\tilde{\ell})$ is the average bitrate of the reference track $\tilde{\ell}$, and $R_t(\tilde{\ell})$ represents the bitrate of the chunk at time $t$ (again we slightly abuse the notation to use $t$ to represent chunk index).
In the second term in Eq. (\ref{eq:targetbuffer}),
%\feng{which is computed by the ``Long-term Statistical Filter'' in Figure~\ref{fig:PIB-diagram},}
%In (\ref{eq:targetbuffer}), the first term in
%the numerator represents the sum of the next $W'$ chunk sizes, and the second term is the sum when assuming each chunk is of the average bitrate.
%Hence
the numerator represents how much (in bits) the sum of the next $W'$ chunks deviates from the {average} value; dividing the numerator by $r(\tilde{\ell})$ converts the unit of the deviation to seconds, compatible with how we represent buffer level.
To avoid pathological scenarios where $x_r(t)$ becomes too large, we limit its value to $2x_r$, which is chosen empirically.
The choice of $W'$ is explored in \S\ref{sec:perf-parameter}.

\mysubsection{Implementation} \label{sec:implementation}

We have implemented CAVA in \texttt{dash.js}, a production quality open-source web-based DASH video player~\cite{dash-js} (version v2.7.0).
Under the \texttt{dash.js} framework, we implemented a new
ABR streaming rule \texttt{CAVARule.js} (consisting of 520 LoC) to realize CAVA.
%
%We further added a bandwidth estimation module (by leveraging progress events) to estimate past throughput per second, which is not available through APIs in \texttt{dash.js}.
%
	%Exposing the segment size information to ABR logic.
In addition,
the interface of \texttt{dash.js} only exposes limited information including the track format, buffer occupancy level, bandwidth estimation, and declared bitrate. The segment size information, however, is not accessible by default.
%(Although according to the DASH standard, the manifest file could have such information in \texttt{meidaRange} filed).
We therefore add a new \texttt{LoadSegmentSize} class to expose such information to the ABR logic.
% It reads the segment size information from files in server before starting download the video.}
We further develop a bandwidth estimation module that responds to playback progress events to estimate throughput using a harmonic mean of the past 5 chunks.
%, which is not available through APIs in \texttt{dash.js}.

	\section{Performance Evaluation} \label{sec:cava_perf}

We explore  CAVA and its parameter settings across a  broad spectrum of real-world scenarios (covering both LTE and broadband), using  $16$ video clips (\S\ref{sec:cava_video-dataset}).
Each video is around 10 minutes long. The length is much longer than network RTT and the timescale of ABR rate adaptation.
	In addition, it allows the player to go beyond the start-up phase, allowing us to understand both start-up and longer-term behaviors. 
%\yanyuan{\footnote{The video length (10 mins) is much longer compared to the RTT of the network, timescales of TCP adaptation and ABR rate adaptation. The player is also able to go beyond the start-up phase, allowing understand both start-up behavior and longer-term behavior. We do not expect longer videos to influence the results.}}, 
These videos cover different content genres (i.e.,  animation, science fiction, sports, animal, nature and action), track encodings~(Youtube and \texttt{FFmpeg}~\cite{FFmpeg}),  codecs~(H.264~\cite{h264} and H.265~\cite{h265}), and chunk durations (2 and 5 seconds).
%, and a range of different real network conditions covering both LTE and broadband.

%talk about what is a trace: average throughput every second
\vspace{-0.05in}
\mysubsection{Evaluation Setup}\label{sec:cava_setup}

To ensure repeatable experimentations and evaluate different schemes under identical settings,  we use real-world network trace-driven replay experiments.
To make the large state space exploration scalable and tractable, we use both trace-driven simulations,  and  experiments with  our prototype implementation in \texttt{dash.js}.

%%wb, 6/13/2018, use two sets of traces
We use two sets of network traces.
%that cover both broadband and LTE 
%a wide range
%of real-world 
%scenarios to emulate real-world network conditions.
%
The first set, {\em LTE},
%contains 100  cellular network
%traces that we captured in commercial LTE networks, covering
%diverse scenarios of location, signal strength,  and user motion (stationary, walking and driving). The third,  {\em Cross Country Cellular (C3)},
contains 200 cellular network traces we captured in commercial LTE networks
with a collaborator driving coast-to-coast across the US.
%sets of cellular network traces
The LTE traces were represented
as per-second network throughput, recorded when
%using active
%measurements,
downloading a large file from a well-provisioned
server to a mobile phone.
%, represented as per-second network throughput.
The second set, {\em FCC},  contains 200 broadband traces, randomly chosen from the set of traces collected by the FCC~\cite{fcc-traces}, represented
as per-5 second network throughput. Each trace contains at least
18 minutes of bandwidth measurements.
%\sen{Check: is every  trace really around 18 min., incl. FCC?}

ABR logic operates at the application level, using application-level estimation of the network bandwidth~(based on the throughput for recently downloaded chunks) as part of the decision process to determine the quality of the next chunk to download~\cite{huang2014buffer, jiang2012improving, Xu17:Dissecting}.  The impact of lower level network characteristics such as signal strength,  loss, and RTT on ABR adaptation behavior is reflected
%affected solely
through their impact on the network throughput.
%Therefore, our network traces, by capturing the  bandwidth variability over time, accurately reflect the impact of these important factors.
{Therefore, our evaluation is through replaying the network traces that capture the network bandwidth variability over time.}
% {The impact of RTT is at  a timescale that is significantly smaller than the timescale of rate adaptation decisions, and is hence not evaluated explicitly.}

%wb, 6/13/2018, only compare with BBA-1 and RBA earlier on; here we only consider more recent/sophisticated schemes: MPC and PANDA-CQ variants
%\noindent
{\textbf{ABR Schemes.}} We compare CAVA with the following state-of-the-art rate adaptation schemes:

\BULLET	 MPC and RobustMPC~\cite{yin2015control}, which are based on model predictive control. RobustMPC is shown to outperform MPC under more dynamic network bandwidth settings~\cite{Mao17:pensieve}.

\BULLET	 PANDA/CQ~\cite{li2014streaming}, which incorporates video quality information to optimize the performance of ABR streaming using dynamic programming. Specifically, it considers a window of $N$ future chunks while making decisions. We investigate two variants, (i)  {\em max-sum} that maximizes the sum of the quality of the next $N$ chunks, and (ii)  {\em max-min} that maximizes the minimum quality of the next $N$ chunks.

\BULLET	 BOLA-E~\cite{Spiteri2018bolae}, which selects the bitrate to maximize a utility function
considering both rebuffering and delivered bitrate. It is an improved version of BOLA~\cite{spiteri2016bola}.

\BULLET	 BBA-1~\cite{huang2014buffer} and RBA~\cite{tong2017rbavbr}. Both are myopic schemes (\S\ref{sec:cava_principles}). Through extensive evaluation, we find that, consistent with the example in \S\ref{sec:cava_principles}, CAVA significantly outperforms these two schemes in all cases. In the interest of space, we omit the results for these two schemes.

%\yanyuan{Do we need to describe all the parameters in all schemes?}

%\end{itemize}
Since our focus is VBR-encoded videos, following the recommendation of each scheme described above, we use the actual size of a video chunk in making rate adaptation decisions.
Note that, of all the schemes, only
PANDA/CQ  relies on using video quality information. Such information, however, is not available in the dominant ABR protocols used today  (i.e., HLS and DASH), limiting the practical applicability and deployability of  such a scheme.
Still, using video quality information as part of the adaptation decision process can be potentially valuable, and for that reason, we include PANDA/CQ in our evaluations.
As we shall see later in \S\ref{sec:perf-comp}, while CAVA does not use such video quality information and relies only on information available in ABR streaming protocols,  it still outperforms PANDA/CQ.
%
%\shuai{Shall we move BBA and RBA up to the bullet lists? Reviewers may miss it in the text.}
%Finally, we have also evaluated BBA-1~\cite{huang2014buffer} and RBA~\cite{tong2017rbavbr}. We find that, consistent with the example in \S\ref{sec:principles}, \piv significantly outperforms BBA-1 and RBA  in all the settings. In the interest of space, we omit the results for these two schemes.
%
Learning based approaches such as~\cite{Mao17:pensieve} require extensive training on a large corpus of data, and are not included for comparison since our focus is on pure client-side based schemes that are more easily deployable.
%  in today's commercial systems.
%does rate adaptation at the server and hence is not included for comparison):

%, which relate to the magnitude of the level and level changes, respectively.
%
%\noindent
{\textbf{ABR Configurations.}}
An important parameter in all schemes is the playback startup latency, i.e.,
%%defined as
%The playback startup latency
the minimum number of seconds worth of video data that needs to be downloaded before the client begins playback.  We explore  a range of values  for this parameter, guided by commercial production players~\cite{Xu17:Dissecting},
%best practice recommendations and values typically used in commercial production players~\cite{Xu17:Dissecting},
and report results for one setting -- 10 seconds  (results for other practical settings  were similar).
A startup delay of 10 seconds corresponds to having two chunks in the buffer for chunk length of 5 seconds (as the case for YouTube videos), consistent with the recommendation of having at least two to three chunks before starting playback~\cite{Xu17:Dissecting}.
%
%
%Both \piv and RBA require estimating
%network bandwidth.
Another important factor is the
%client
network bandwidth estimation technique used by the ABR logic.
% (needed by all the above schemes except for BBA-1). 
For a fair comparison,
for all the schemes that need bandwidth estimation,
we use the harmonic mean of the past 5 chunks as the bandwidth prediction approach, as the technique has been shown to be robust to measurement outliers~\cite{jiang2012improving,yin2015control}.
We evaluate the impact of inaccurate bandwidth estimation on performance in \S\ref{subsec:bw-error}.
%$20$~sec., which has been shown to be robust to measurement outliers~\cite{jiang2012improving}.
%
Unless otherwise stated, the maximum buffer size is set to 100 seconds for all the schemes for apple-to-apple comparison. The client does not download the next chunk when the maximum buffer size is reached. 

For CAVA, the base target buffer level, $x_r$, is set to 60 seconds. The controller parameters, $K_p$ and $K_i$, are selected adopting the methodology outlined in~\cite{qin2017control}.
Specifically, we varied $K_p$ and $K_i$, and confirmed that, similar to~\cite{qin2017control},  a wide range of $K_p$ and $K_i$ values lead to good performance.
%\sen{SS: add a short intuitive description of the selection methodology}
%For a fair comparison with BBA-1, we set
%the base target buffer level, $x_r$, to 60 seconds.
%the weight factor $\eta$ is set to $1$,  i.e., giving equal weight to the two terms in Equation~(\ref{eq:lms-vbr}) since both are related to important QoE metrics.

%\yanyuan{We use four QoE metrics related to quality, quality variation and rebuffering/stalls, and another metric on data usage.}
%We use five QoE metrics related to quality, quality variation and rebuffering/stalls, and data usage.
% related to encoding quality, quality variation and rebuffering/stalls~\cite{dobrian2011understanding, krishnan2013video,Liu15:Deriving,ni2011flicker}.
%quality, quality changes and rebuffering~\cite{dobrian2011understanding, krishnan2013video,Liu15:Deriving,ni2011flicker}.
%that are related to the QoE of VBR videos.

%\noindent
{\textbf{Performance Metrics.}}
We use five metrics: four for measuring different aspects of user QoE and one for measuring resource efficiency, i.e., network data usage.
All metrics are computed with respect to the delivered video, i.e., considering the chunks that have been downloaded and played back.
We measure the video quality using VMAF (a recently proposed perceptual quality metric from Netflix, see \S\ref{sec:VBR-characteristics}).
%; as we show in \S\ref{sec:characteristics}, the quality measurement using VMAF is consistent with two other quality metrics, PSNR and SSIM.\sen{??? VMAF will not always be consistent with things like PSNR.  Else PSNR etc would be good and no need for VMAF. Yanyuan ???}.\yanyuan{VMAF is better to represent the perception quality of videos.}
Specifically, we use the VMAF phone model~(covers typical smartphone viewing settings) for the evaluations with the cellular traces, and the VMAF TV model~(covers typical larger screen TV viewing settings) for the evaluations with the FCC traces, based on the typical device types and viewing conditions prevalent in cellular and home networks, respectively.

The performance metrics are (i) {\em quality of Q4 chunks}: captures the video quality for the most complex scenes. (ii) {\em low-quality chunk percentage}: among all the chunks played back in a given streaming session, this  measures the proportion of chunks that were selected with low video quality.  For the experiments, we identify  VMAF values below 40 (representing poor or unacceptable quality~\cite{vmaf}) as low quality.  (iii) {\em rebuffering duration}:  measures the total rebuffering/stall time in a video session. (iv) {\em average quality change per chunk}: is defined as the average quality difference of two consecutive chunks in playback order (since human eyes are more sensitive to level changes in adjacent chunks) for a session, i.e., $\sum_{i} |q_{i+1}-q_i|/n$, where $q_i$ denotes the quality of  $i^{th}$ chunk~(in terms of playback order), and $n$ is the number of chunks. (v) {\em data usage}: captures the total amount of data downloaded (i.e., the  total network resource usage) for a session.
% measuring . This is  a measure of the  total network resource usage  for a given scheme.
While comparing two schemes, for each of the last 4 metrics,  a lower  value is preferable;
%  from the relevant perspective~(QoE or resource efficiency).
the only exception is the first metric (quality of Q4 chunks), where a higher value is preferable.

In the above performance metrics, we use perceptual quality metrics since they quantify the viewing quality more accurately compared to  bitrate that is commonly used in the literature.
	% of CBR streaming,
	%This is an important distinction between VBR and CBR videos.
As an example, the average bitrate of the delivered video, which is commonly used to quantify the  quality of a streaming session, is an especially poor metric for VBR videos. To see this, consider two cases with the same average bitrate. In the first case, the lowest track is selected for Q4 chunks, while higher tracks are selected for Q1-Q3 chunks; in the second case, a medium track is selected for Q4 chunks, while low or medium tracks are selected for Q1-Q3 chunks. These two cases clearly lead to different viewing quality, while will be classified as being the same quality based on the average bitrate. In addition, we use metrics to capture both the average and tail behaviors. Specifically, the percentage of low quality chunks is an important metric on tail behavior since it has been reported that human eyes are particularly sensitive to bad quality chunks~\cite{VMAF-aggr}. 
\vspace{0.1in}
%VMAF thresholds
%0-20 Unacceptable
%20-40 poor
%40-60 fair
%60-80 good
%80-100  Excellent

%%two small subfigures to show the impact of W
\begin{figure}[t]
	\centering
%	\vspace{-.1in}
	%\subfloat[Q4 chunk quality. \label{subfig:Q4}]{%
		\includegraphics[width=0.4\textwidth]{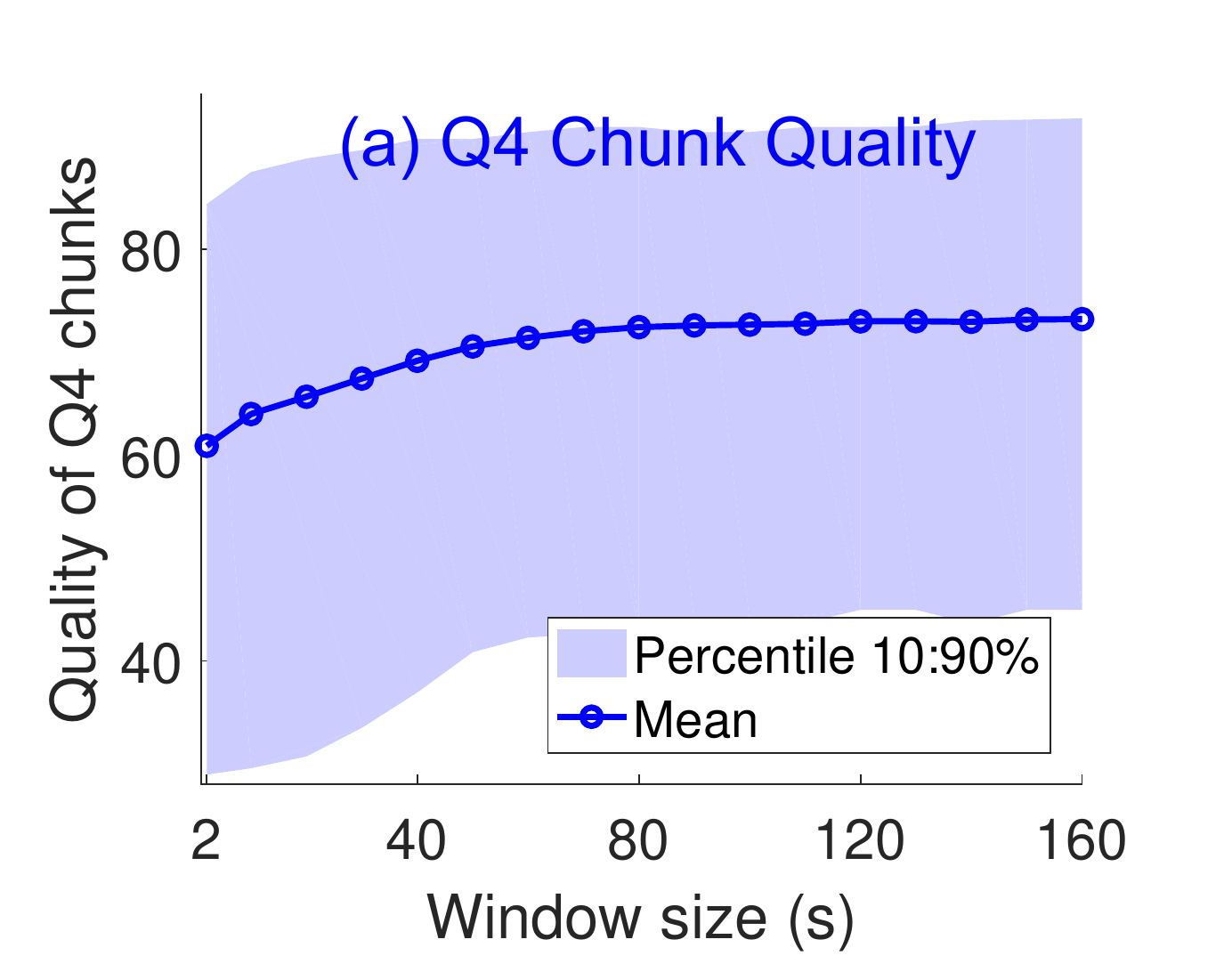}
	%}
	%\subfloat[Rebuffering\label{subfig:rebuffering}]{%
		\includegraphics[width=0.4\textwidth]{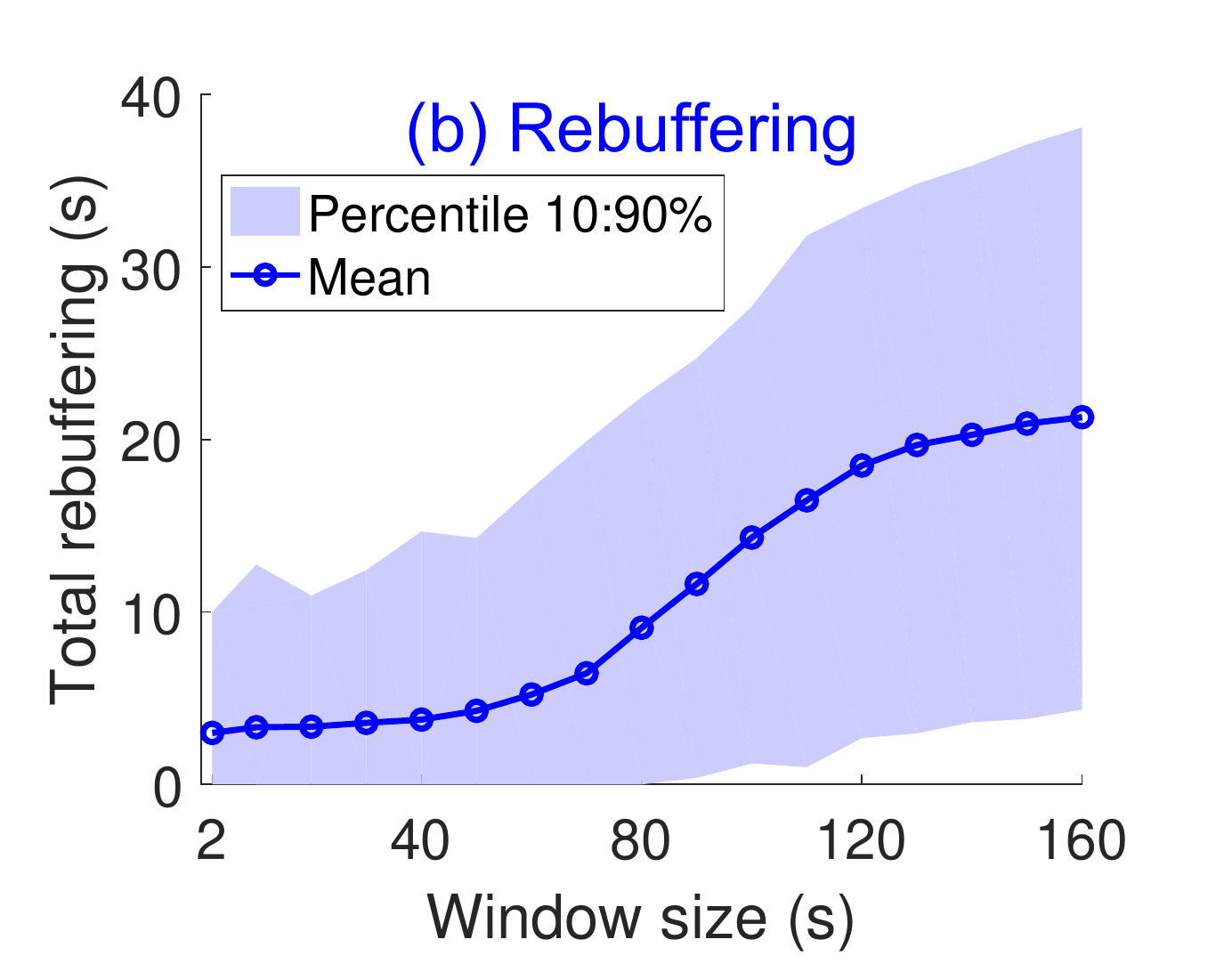}
	%}
	\caption{Impact of inner controller window size.
		% of \piv (Elephant Dream, \texttt{FFmpeg} encoded, H.264, LTE traces).
	}
	\label{fig:impact-W}
\end{figure}

%%only show one video in the interests of space
\begin{figure*}[ht]
	\centering
	%\subfloat[\label{subfig:CAVA-Q4}]{%
		\includegraphics[width=0.29\textwidth]{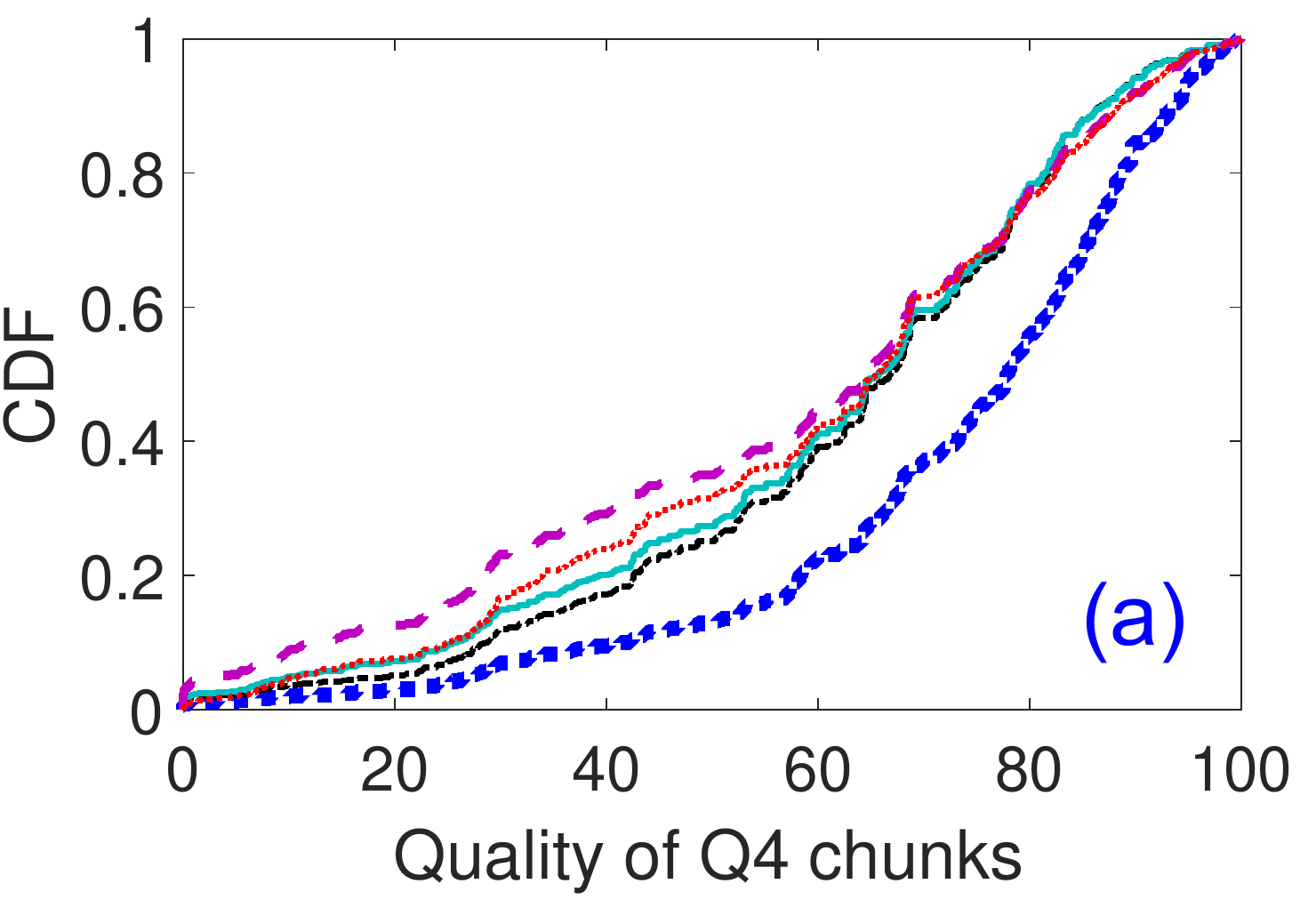}
	%}
	%\subfloat[\label{subfig:CAVA-low-quality}]{%
		\includegraphics[width=0.29\textwidth]{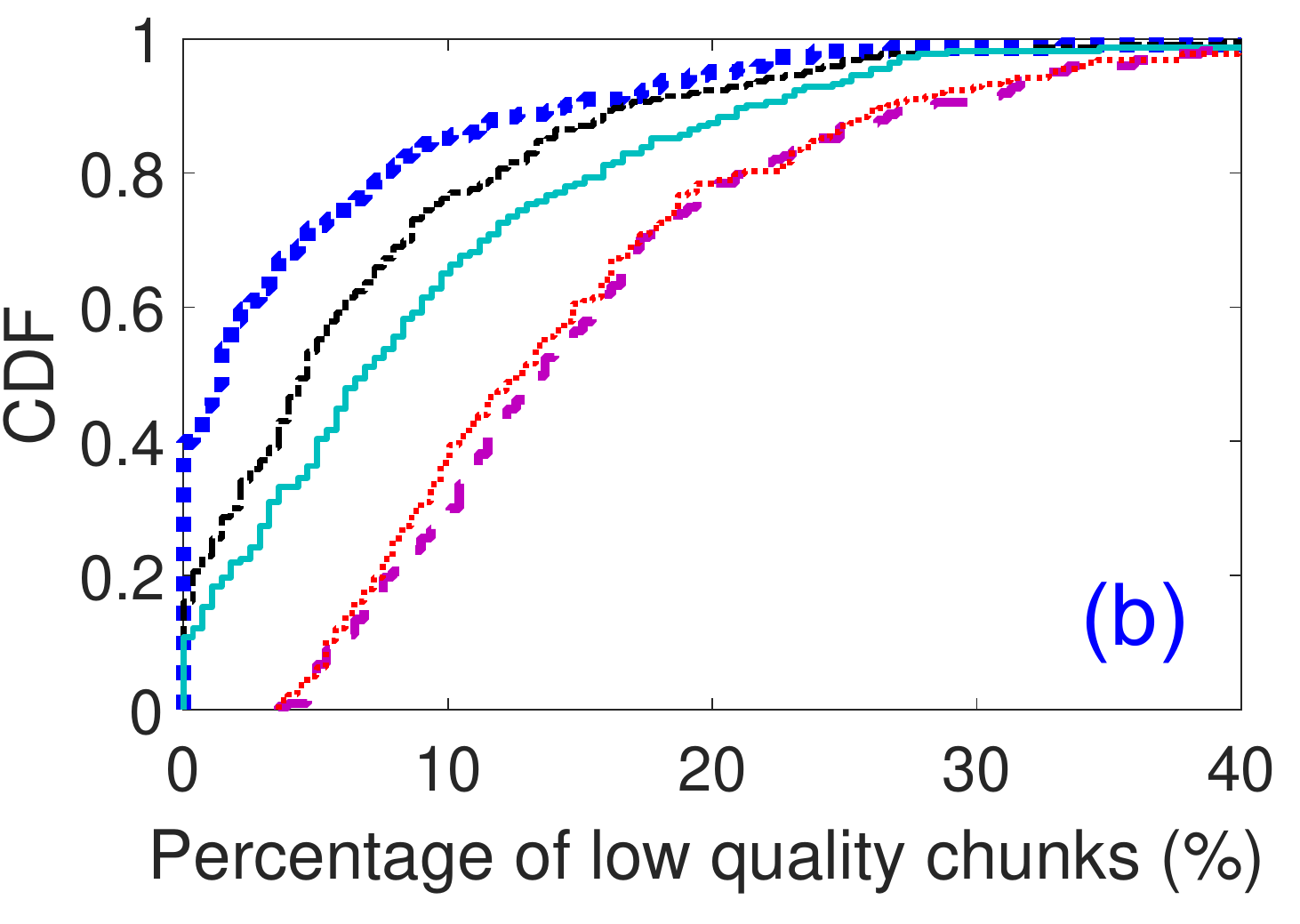}
	%}
	%\subfloat[\label{subfig:CAVA-rebuffering}]{%
		\includegraphics[width=0.29\textwidth]{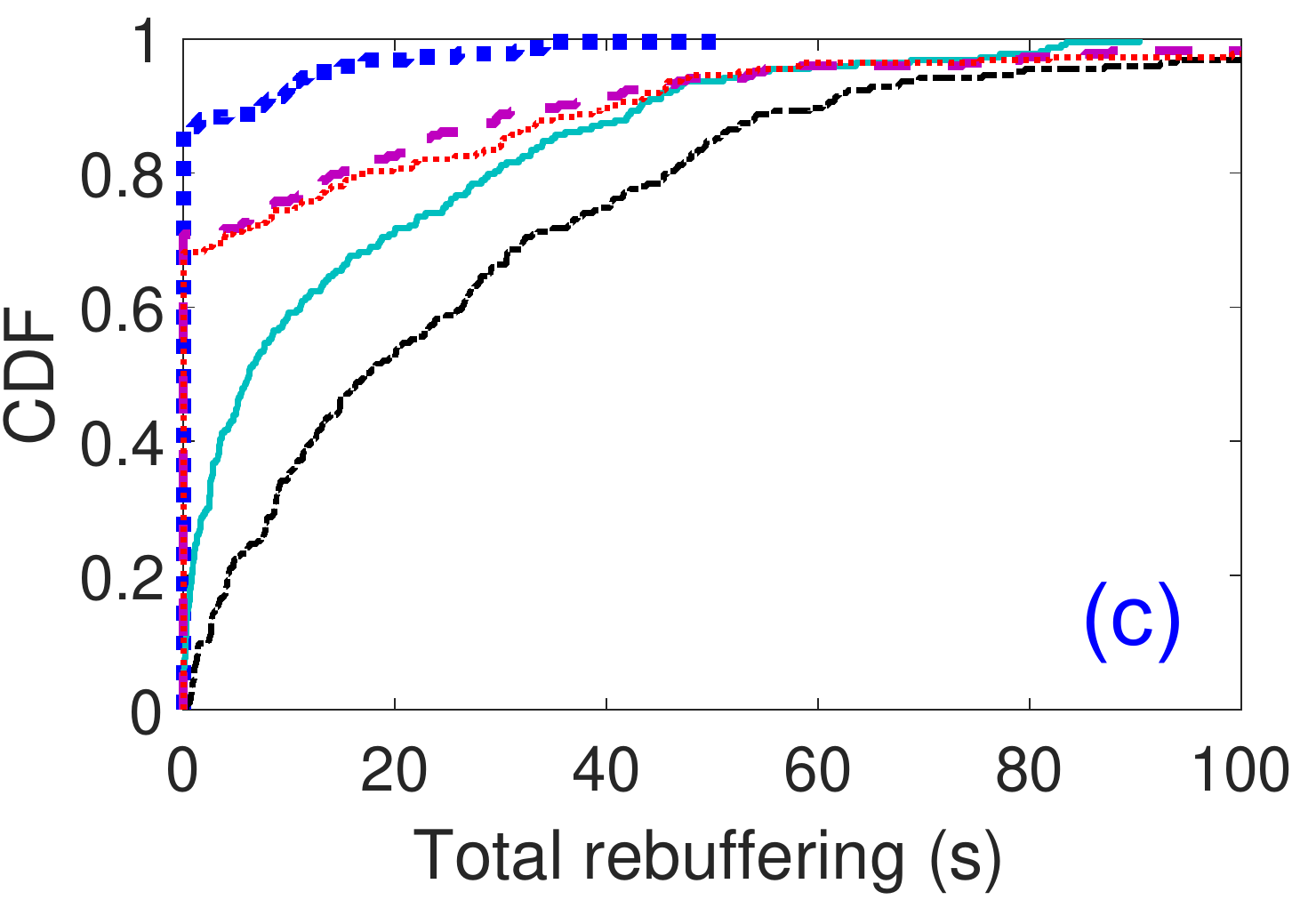}
	%}
	%\subfloat[\label{subfig:CAVA-change}]{%
		\includegraphics[width=0.29\textwidth]{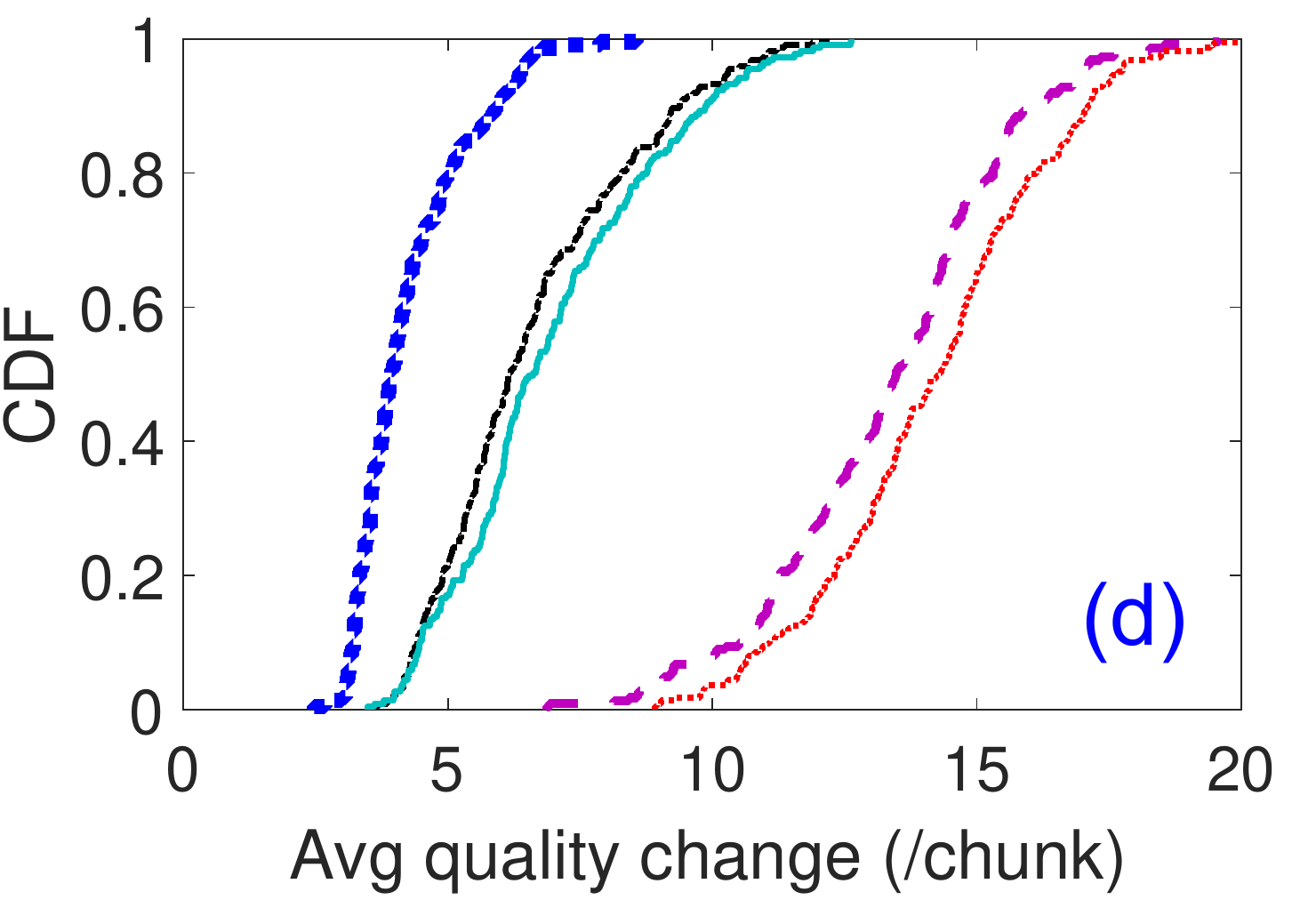}
	%}
	%\subfloat[\label{subfig:CAVA-data-usage}]{%
		\includegraphics[width=0.29\textwidth]{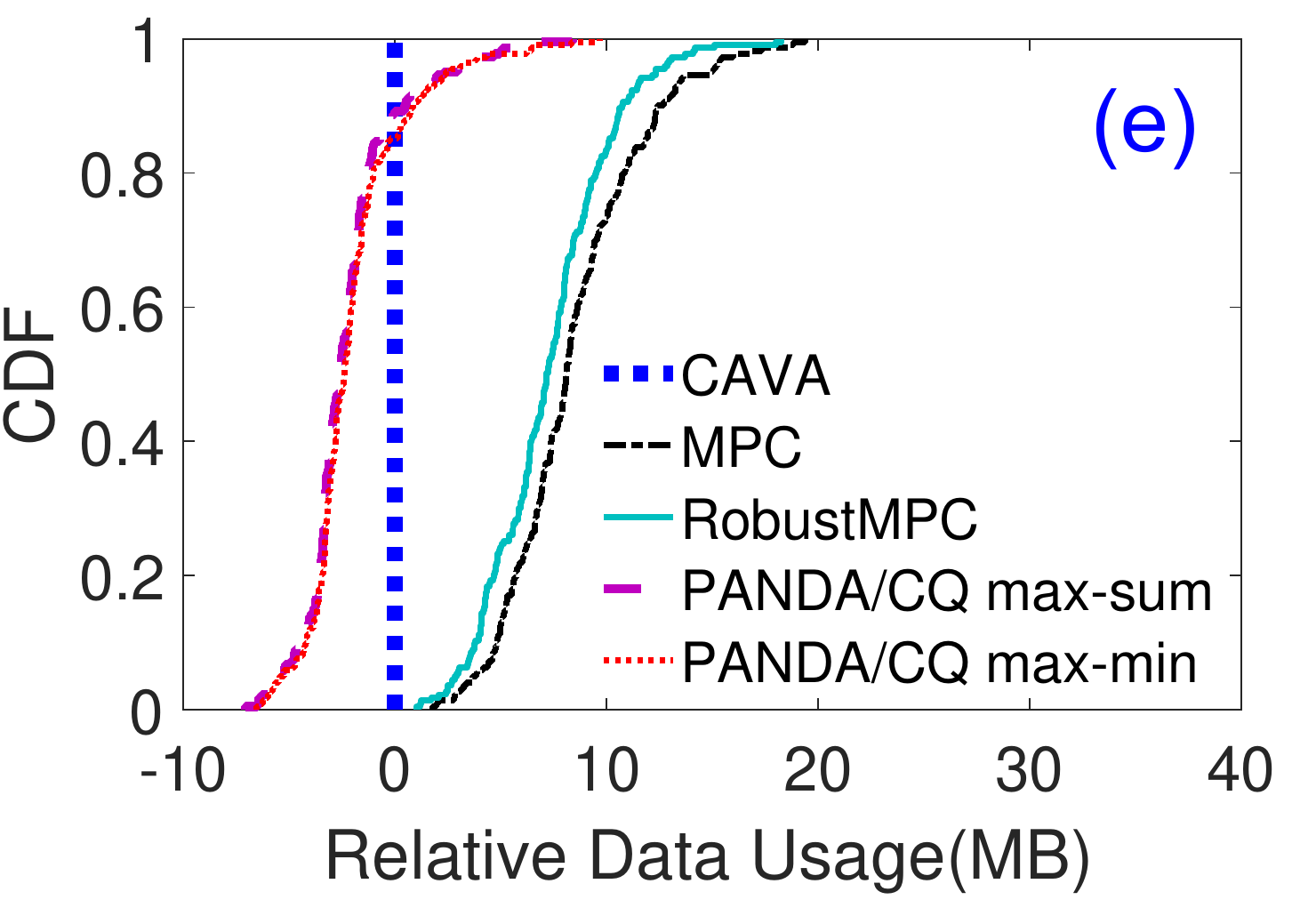}
	%}
	\caption{Performance comparison for one video (Elephant Dream, \texttt{FFmpeg} encoded, H.264) under LTE traces. }
	\label{fig:comparison-FCC-LTE-cc}
\end{figure*}

\mysubsection{Choice of Parameters for CAVA} \label{sec:perf-parameter}
%CAVA has two important parameters, inner controller window size $W$, and outer controller size $W'$. We explore the impact of these two parameters to choose their values. In the following, we first explore the impact using our encoded videos (chunk duration 2 s); at the end, we further consider YouTube encoded videos, and summarize our choice for these two parameters.

%\smallskip
\noindent{\textbf{Inner controller window size.}} Recall that the  inner controller in CAVA uses the average bitrate over a window of $W$ future chunks to represent the bandwidth requirement of the current chunk.
%We vary $W$ from $1$ to $80$ chunks (\shuai{with chunk size of 2s}), and investigate the impact of each choice on the performance.
We investigate the impact of $W$ on the performance of CAVA. Fig.~\ref{fig:impact-W} shows an example (Elephant Dream, \texttt{FFmpeg} encoded, H.264, LTE traces); the solid lines represent the average values, and the top and bottom of the shaded region represent the 90th and 10th percentile across the network traces, respectively.
%which plots the average Q4 chunk quality and the refuffering duration (along with the 5th and 95th percentiles in the shaded regions).
When increasing $W$, (i) the quality of the Q4 chunks first improves significantly and then flattens out, and (ii) the amount of rebuffering  increases slightly and then sharply. Increasing $W$ at the beginning is beneficial because  even averaging over a few chunks can smooth out the bitrate of the chunks, which allows higher levels to be chosen for Q4 chunks (since the smoothed bitrate for a Q4 chunk tends to be lower than its actual bitrate).
The impact on smoothing the bitrate diminishes after $W$ becomes sufficiently large, leading to diminishing gains for Q4 chunks. The increase in rebuffering when $W$ is large is because CAVA becomes less sensitive to the bitrate changes in that case.
We set $W$ to 40 seconds (i.e., 20 and 8 chunks for 2 and 5 seconds chunk durations, respectively) as  this leads to a good tradeoff between quality and rebuffering.
%We observe setting $W$ to 40 seconds (i.e., 20 and 8 chunks for 2 and 5 seconds chunk durations, respectively) leads to
%a good tradeoff between quality and rebuffering, which is used for the rest of the paper.
%We observe that for \texttt{FFmpeg} encoded videos (chunk duration of 2 second), setting $W$ to 20 chunks (i.e., 40 seconds) leads to a good tradeoff between quality and rebuffering. Setting $W$ to 40 seconds for YouTube encoded  videos (i.e., 8 chunks since chunk duration is 5 seconds) also leads to good performance.

%Based on the above results, we set the window size to 40 seconds (corresponding to 20 and 8 chunks for chunk duration of 2 and 5 seconds, respectively), which leads to a good tradeoff between quality and rebuffering for both the FFMpeg and YouTube encoded  videos~(with chunk duration 2 and 5 s, respectively).
% \sen{SS: confusing. So we selected different W  ( in unit of chunk) for FFMPEG and Youtube: ie 20 chunks and 8 chunks, but the same N = 5 chunks.  Whats the intuition?}

\noindent{\textbf{Outer controller window size.}} The outer controller in CAVA uses 
%the chunk size variability in 
a window of $W'$ future chunks to adjust the target buffer level.
%We next investigate the impact of $W'$ on the performance of
We vary $W'$
%from 0 to 150 chunks
to explore its impact.
In general, the amount of rebuffering decreases as $W'$ increases since the controller reacts more proactively with larger $W'$.
For some videos, however, the amount of rebuffering may start to increase as $W'$ increases further. This is because using very large $W'$ smoothes out the effect of the variability of the video (i.e., the average bitrate of the future $W'$ chunks becomes closer to the average bitrate of the track, causing little increment in the target buffer level as shown in Eq. (\ref{eq:targetbuffer})).
We set $W'$ to 200 seconds (i.e., 100 and 40 chunks for 2 and 5 seconds chunk durations, respectively), which leads to good results in reducing rebuffering.

\mysubsection{Comparison with Other Schemes}\label{sec:perf-comp}

We compare CAVA with other ABR schemes using all the videos under both LTE and FCC traces (the comparison with BOLA-E is deferred to~\S\ref{sec:cava_dash}).
Fig.~\ref{fig:comparison-FCC-LTE-cc}(a)-(e) compare the performance of CAVA with other schemes for one \texttt{FFmpeg} encoded video under the LTE traces.
Our results below focus on CAVA versus RobustMPC and PANDA/CQ max-min; MPC can have significantly more rebuffering than RobustMPC, and PANDA/CQ max-sum can have significantly lower quality for Q4 chunks than PANDA/CQ max-min.
We observe that CAVA  significantly outperforms RobustMPC and PANDA/CQ max-min.
(i)  With CAVA,  a much larger  proportion ($79$\%)  of the Q4 chunks are delivered at a quality higher than 60 (considered as good quality~\cite{vmaf}), compared to only $59$\% and $57$\% for RobustMPC and PANDA/CQ max-min, respectively.
The median VMAF value across Q4 chunks is $78$ under CAVA, $11$ and $12$ larger than the corresponding values for RobustMPC and PANDA/CQ max-min, respectively, indicating noticeable perceptual improvements (It has been reported that a difference of $6$ or more in VMAF would be noticeable to a viewer~\cite{Ozer:vmaf-jnd}).
(ii)  For the percentage of low-quality chunks, under CAVA, 40\% of the traces have no low-quality chunks (VMAF score < 40); and 82\% of the traces have less than 10\% low-quality chunks, significantly larger than the corresponding values 65\% and 40\% for RobustMPC and PANDA/CQ max-min, respectively.
(iii) CAVA has no rebuffering for $85\%$ of the traces, compared to only 20\% and 68\% of traces under RobustMPC and PANDA/CQ max-min, respectively.
(iv) For average quality change per chunk, CAVA realizes substantially lower changes than the other schemes~(on average, 67\% and 29\% of
%%1.5$\times$ and 3.5$\times$ lower than
the values of RobustMPC and PANDA/CQ max-min, respectively), resulting in less variability in playback quality, and therefore more consistent and smoother playback experience.
(v) CAVA has 5\% to 40\% lower data usage than RobustMPC. PANDA/CQ max-min has slightly lower data usage than CAVA (10\% at most), at the cost of lower Q4 quality and overall quality (see below).
Figures~\ref{fig:deep-dive}(a) and (b) show the quality of Q1-Q3 chunks and all the chunks, respectively. We observe that while CAVA does not lead to very high quality for Q1-Q3 chunks, it does not choose low quality for these chunks either. Overall, CAVA leads to a desired tradeoff in choosing less low-quality chunks compared to other schemes.

\begin{figure}[ht]
	\centering
	%\subfloat[ \label{subfig:Q13}]{%
		\includegraphics[width=0.43\textwidth]{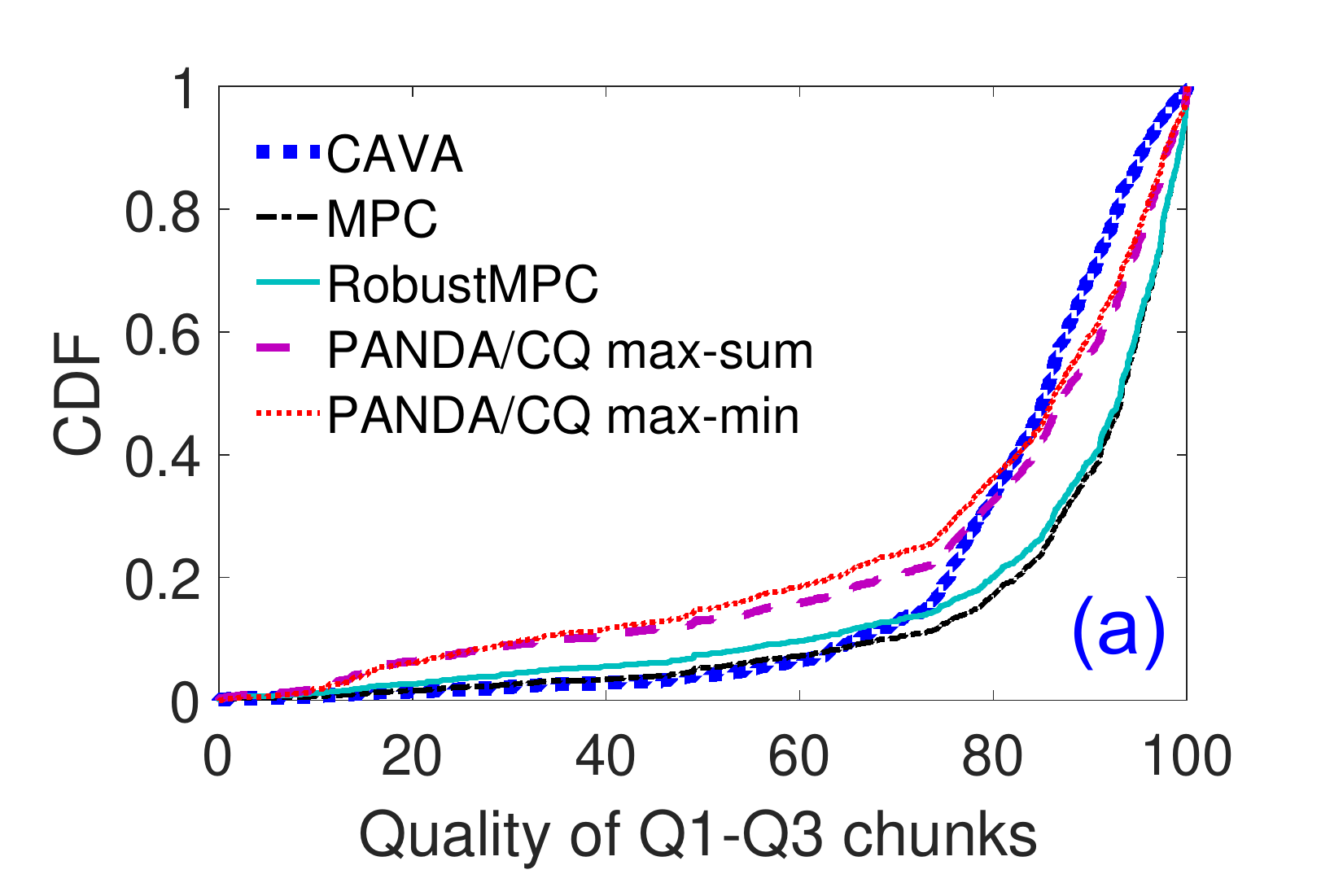}
	%}
	%\subfloat[\label{subfig:allchunks}]{%
		\includegraphics[width=0.43\textwidth]{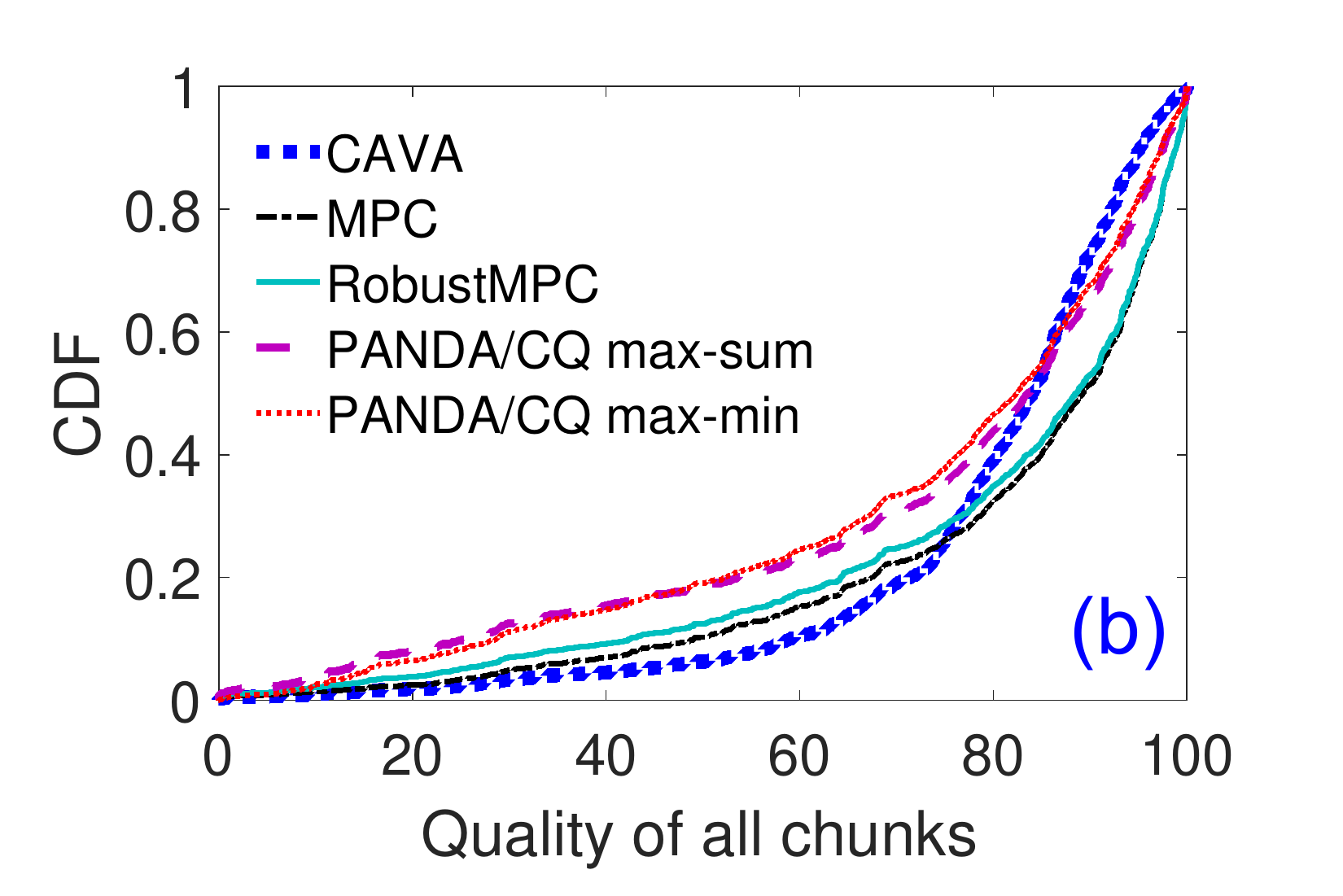}
	%}
	\caption{Quality of Q1-Q3 chunks and all the chunks.
		% and overall chunks \piv (Elephant Dream, \texttt{FFmpeg} encoded, H.264, LTE traces).
	}
	\label{fig:deep-dive}
\end{figure}

%%wb, 6/13/2018, show four columns, Q4, rebuffering, quality changes and percentage of low-quality chunks; the results for
The above results are for one \texttt{FFmpeg} video under LTE traces.
Table~\ref{perf-cmp-all-videos} shows the results for several YouTube videos under both LTE and FCC traces.
For Q4 chunk quality, we show two values: the metric value (averaged over all the runs) obtained by CAVA subtracted by that of RobustMPC and PANDA/CQ max-min, respectively.
For each of the other four performance metrics, we show two percentage values, (i)  the difference in the metric value (averaged over all runs) for CAVA subtracted by that for RobustMPC, as a percentage of the latter value, and
(ii) the corresponding value for CAVA vs. PANDA/CQ max-min.
%% and then divided by the latter, the other representing the corresponding value for PANDA/CQ max-min.
In Table~\ref{perf-cmp-all-videos}, $\uparrow$  indicates that CAVA has a higher value; while $\downarrow$ indicates that CAVA has a lower value.
%In Table~\ref{perf-cmp-all-videos}, $\uparrow$  represents that \piv increases the value; while $\downarrow$ represents that \piv decreases the value.
%
%several other settings.
The first part of Table~\ref{perf-cmp-all-videos} shows the results under LTE traces.
We observe that CAVA leads to average quality improvement for Q4 chunks in the range of 8-18 VMAF points compared to RobustMPC, and 3-9 VMAF points compared to PANDA/CQ max-min.
Also CAVA substantially reduces the average stall/rebuffering duration by 62\%-95\%,
reduces the quality change per chunk by 25\%-48\%,  and reduces the percentage of low-quality chunks by 4\%-61\%.
 Last, the data usage of CAVA is 7\%-11\% lower than RobustMPC, and 2\%-10\% lower than PANDA/CQ max-min.

The second part of Table~\ref{perf-cmp-all-videos} shows the results for FCC traces. Compared to the LTE trace results, the rebuffering for all the schemes becomes lower due to smoother network bandwidth profiles;  CAVA still leads to  lower rebuffering and better performance in all the performance metrics.

\begin{table}[]
\centering
\small{
	\caption{Performance comparison (YouTube videos).}
	\label{perf-cmp-all-videos}
		\begin{tabular}{|p{0.1cm}|p{1cm}|c|c|c|c|c|}
	\cline{2-7}
	\multicolumn{1}{c|}{} & Video & {\begin{tabular}[c]{@{}c@{}}Q4\\chunk\\quality\end{tabular}} &
	{\begin{tabular}[c]{@{}c@{}}Low-qual.\\chunks\\(\%)\end{tabular}} & {\begin{tabular}[c]{@{}c@{}}Stall\\duration\\(\%)\end{tabular}} &
	{\begin{tabular}[c]{@{}c@{}}Quality\\changes\\(\%)\end{tabular}} &
	{\begin{tabular}[c]{@{}c@{}}Data\\usage\\(\%)\end{tabular}}
	\\ \hline
	\multirow{8}{*}{\begin{sideways} LTE  \end{sideways}} &
%Youtube results
 BBB   & \textcolor{green}{$\uparrow$}13, \textcolor{green}{$\uparrow$}4  & \textcolor{green}{$\downarrow$}61, \textcolor{green}{$\downarrow$}31 & \textcolor{green}{$\downarrow$}62, \textcolor{green}{$\downarrow$}64 & \textcolor{green}{$\downarrow$}48, \textcolor{green}{$\downarrow$}37 & \textcolor{green}{$\downarrow$}11, \textcolor{green}{$\downarrow$}5 \\ \cline{2-7} %\cline{2-6}
& ED  & \textcolor{green}{$\uparrow$}10, \textcolor{green}{$\uparrow$}4 & \textcolor{green}{$\downarrow$}54, \textcolor{green}{$\downarrow$}37 & \textcolor{green}{$\downarrow$}73, \textcolor{green}{$\downarrow$}77 & \textcolor{green}{$\downarrow$}38, \textcolor{green}{$\downarrow$}36 & \textcolor{green}{$\downarrow$}9, \textcolor{green}{$\downarrow$}4 \\ \cline{2-7}%\cline{2-6}
& Sintel  & \textcolor{green}{$\uparrow$}8, \textcolor{green}{$\uparrow$}4 & \textcolor{green}{$\downarrow$}75, \textcolor{green}{$\downarrow$}42 & \textcolor{green}{$\downarrow$}83, \textcolor{green}{$\downarrow$}77 & \textcolor{green}{$\downarrow$}37, \textcolor{green}{$\downarrow$}32 & \textcolor{green}{$\downarrow$}8, \textcolor{green}{$\downarrow$}3\\ \cline{2-7}%\cline{2-6}
& ToS   &  \textcolor{green}{$\uparrow$}9, \textcolor{green}{$\uparrow$}6 & \textcolor{green}{$\downarrow$}72, \textcolor{green}{$\downarrow$}55 & \textcolor{green}{$\downarrow$}84, \textcolor{green}{$\downarrow$}81 & \textcolor{green}{$\downarrow$}39, \textcolor{green}{$\downarrow$}40 & \textcolor{green}{$\downarrow$}9, \textcolor{green}{$\downarrow$}3  \\ \cline{2-7}  %\cline{2-6}
& Animal & \textcolor{green}{$\uparrow$}15, \textcolor{green}{$\uparrow$}3 & \textcolor{green}{$\downarrow$}4, \textcolor{green}{$\downarrow$}7 & \textcolor{green}{$\downarrow$}83, \textcolor{green}{$\downarrow$}95 & \textcolor{green}{$\downarrow$}43, \textcolor{green}{$\downarrow$}38 & \textcolor{green}{$\downarrow$}11, \textcolor{green}{$\downarrow$}10  \\ \cline{2-7}
& Nature & \textcolor{green}{$\uparrow$}10, \textcolor{green}{$\uparrow$}3 & \textcolor{green}{$\downarrow$}17, \textcolor{green}{$\downarrow$}11 & \textcolor{green}{$\downarrow$}80, \textcolor{green}{$\downarrow$}97 & \textcolor{green}{$\downarrow$}40, \textcolor{green}{$\downarrow$}41 & \textcolor{green}{$\downarrow$}7, \textcolor{green}{$\downarrow$}10 \\ \cline{2-7}
& Sports & \textcolor{green}{$\uparrow$}18, \textcolor{green}{$\uparrow$}9 & \textcolor{green}{$\downarrow$}30, \textcolor{green}{$\downarrow$}2 & \textcolor{green}{$\downarrow$}83, \textcolor{green}{$\downarrow$}83 & \textcolor{green}{$\downarrow$}35, \textcolor{green}{$\downarrow$}31 & \textcolor{green}{$\downarrow$}8, \textcolor{green}{$\downarrow$}2 \\ \cline{2-7}
& Action & \textcolor{green}{$\uparrow$}14, \textcolor{green}{$\uparrow$}6 & \textcolor{green}{$\downarrow$}49, \textcolor{green}{$\downarrow$}27 & \textcolor{green}{$\downarrow$}79, \textcolor{green}{$\downarrow$}74 & \textcolor{green}{$\downarrow$}38, \textcolor{green}{$\downarrow$}25 & \textcolor{green}{$\downarrow$}8, \textcolor{green}{$\downarrow$}3 \\ \hline \hline
\multirow{4}{*}{\begin{sideways}  FCC  \end{sideways}} &
%%results under FCC traces (TV model); we use YouTube encoded;
 BBB   & \textcolor{green}{$\uparrow$}11, \textcolor{green}{$\uparrow$}4  & \textcolor{green}{$\downarrow$}70, \textcolor{green}{$\downarrow$}28 & \textcolor{green}{$\downarrow$}26, \textcolor{green}{$\downarrow$}51 & \textcolor{green}{$\downarrow$}48, \textcolor{green}{$\downarrow$}32 & \textcolor{green}{$\downarrow$}8, \textcolor{green}{$\downarrow$}4\\ \cline{2-7}
& ED  & \textcolor{green}{$\uparrow$}8, \textcolor{green}{$\uparrow$}3 & \textcolor{green}{$\downarrow$}33, \textcolor{green}{$\downarrow$}11 & \textcolor{green}{$\downarrow$}79, \textcolor{green}{$\downarrow$}34 & \textcolor{green}{$\downarrow$}36, \textcolor{green}{$\downarrow$}31 & \textcolor{green}{$\downarrow$}7, \textcolor{green}{$\downarrow$}2\\ \cline{2-7}
& Sintel  & \textcolor{green}{$\uparrow$}7, \textcolor{green}{$\uparrow$}4 & \textcolor{green}{$\downarrow$}87, \textcolor{green}{$\downarrow$}70 & \textcolor{green}{$\downarrow$}94, \textcolor{green}{$\downarrow$}85 & \textcolor{green}{$\downarrow$}35, \textcolor{green}{$\downarrow$}26 & \textcolor{green}{$\downarrow$}6, \textcolor{green}{$\downarrow$}2 \\ \cline{2-7}
& ToS   &  \textcolor{green}{$\uparrow$}6, \textcolor{green}{$\uparrow$}6 & \textcolor{green}{$\downarrow$}69, \textcolor{green}{$\downarrow$}62 & \textcolor{green}{$\downarrow$}90, \textcolor{green}{$\downarrow$}6 & \textcolor{green}{$\downarrow$}40, \textcolor{green}{$\downarrow$}38 & \textcolor{green}{$\downarrow$}7, \textcolor{green}{$\downarrow$}1 \\ \hline \hline
\multicolumn{2}{|c|}{} &  \textcolor{green}{$\uparrow$} better & \textcolor{green}{$\downarrow$} better & \textcolor{green}{$\downarrow$} better & \textcolor{green}{$\downarrow$} better & \textcolor{green}{$\downarrow$} better \\ \hline
%\cline{3-7}
\multicolumn{7}{l}{* The two numbers in each cell represents the changes by CAVA} \\
\multicolumn{7}{l}{ relative to RobustMPC and PANDA/CQ max-min, respectively.}
%\multicolumn{5}{l}{PANDA/CQ max-min, respectively.}
\end{tabular}
}
\vspace{-.2in}
\end{table}

\mysubsection{Impact of  CAVA Design Principles}
Recall CAVA incorporates three key design  principles~(\S\ref{sec:cava_principles}). In the following, we investigate the contribution of each principle to the performance of CAVA.
As detailed in \S\ref{sec:cava_principles}, the first (i.e., non-myopic) principle is used in the inner controller of CAVA: it uses the average bitrate of multiple future  chunks to represent the bandwidth requirement of the current chunk when deciding the track level for the current chunk.  The second (i.e., differential treatment) principle favors Q4 chunks over Q1-Q3 chunks.
% to account for the characteristics.
% that current encoding techniques lead to lower quality for Q4 chunks compared to other chunks. 
The third (i.e., proactive) principle is used in the outer controller: it considers long-term future (within tens of chunks) and increases the target buffer level to be prepared for incoming large chunks in the future.
Of the three principles, the first principle is the basic principle, while the other two principles are used on top of the basic principle.
%Specifically,
In the following, we consider three variants of CAVA, referred to as CAVA-p1, CAVA-p12 and CAVA-p123, which includes the first principle, the first two principles, and all three principles, respectively.

\begin{figure}[ht]
	\centering
	\vspace{-.16in}
	%\subfloat[Q4 chunk quality. \label{subfig:Q4}]{%
		\includegraphics[width=0.43\textwidth]{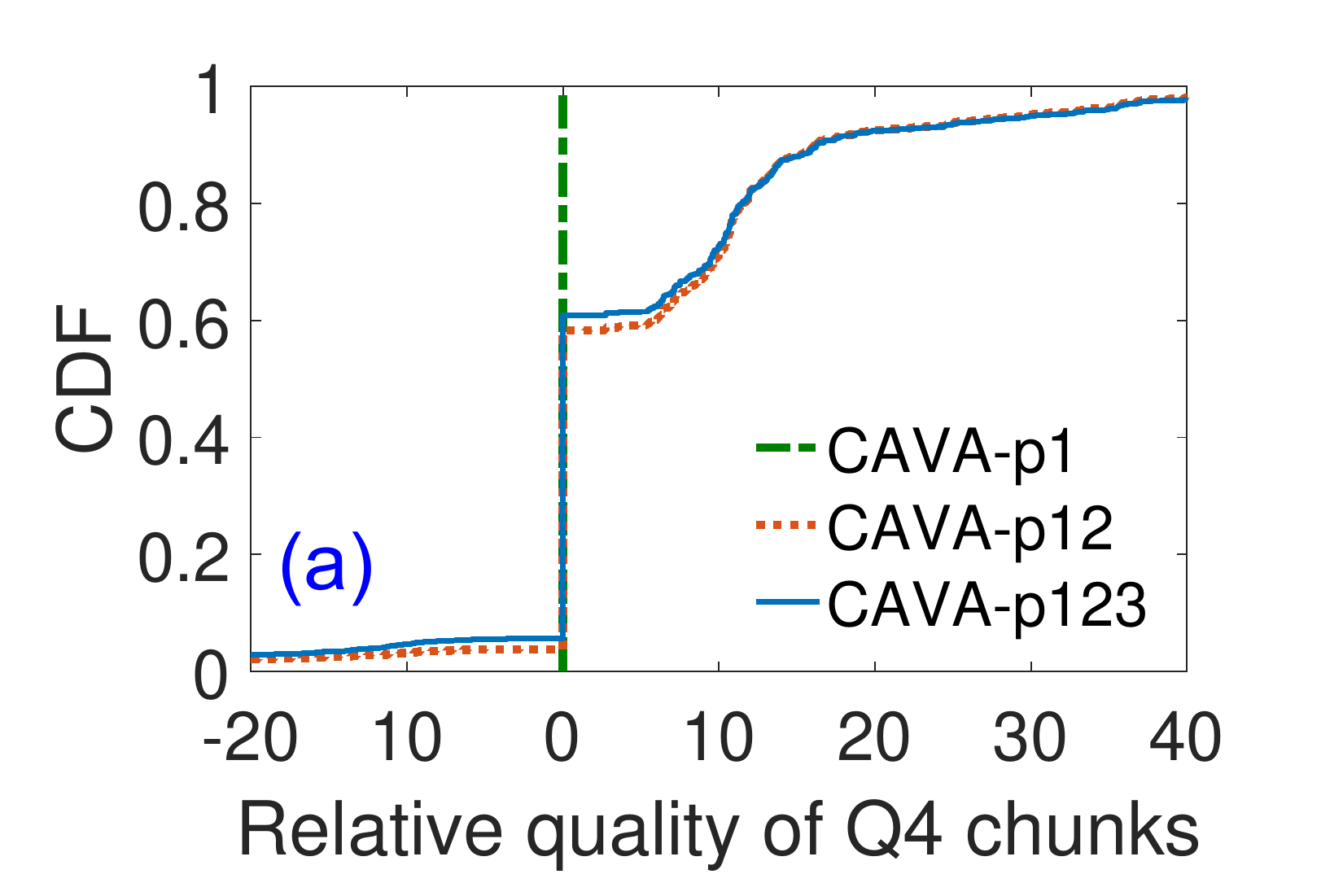}
	%}
	%\subfloat[Rebuffering\label{subfig:rebuffering}]{%
		\includegraphics[width=0.43\textwidth]{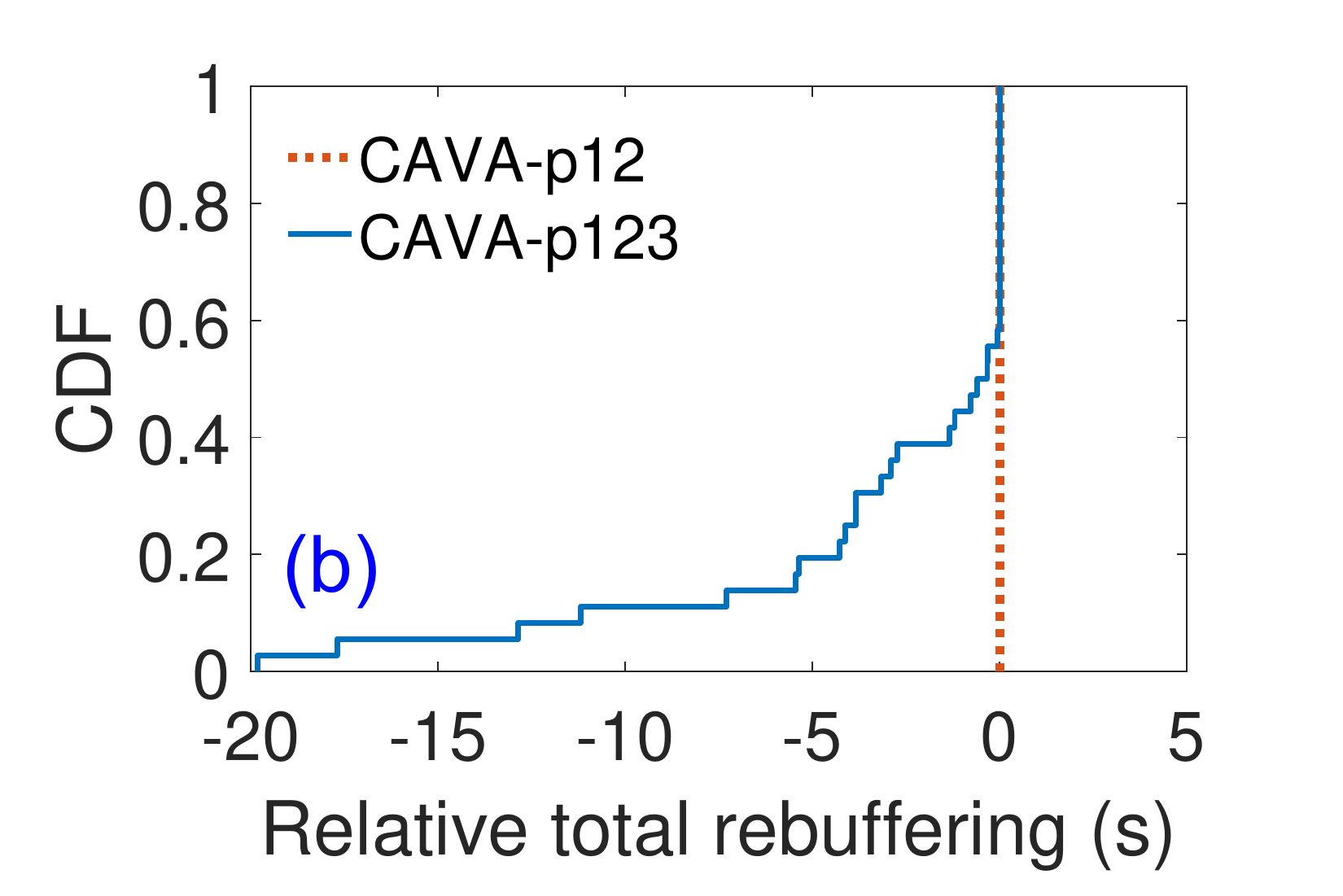}
	%}
	\vspace{-.12in}
	\caption{Impact of the design principles of CAVA.
		% (Elephant Dream, \texttt{FFmpeg} encoded, H.264, LTE traces).
%
	}
	\label{fig:impact-principles}
	\vspace{-.2in}
\end{figure}

Our evaluation confirms that the three principles indeed successfully play their individual roles. Specifically, CAVA-p1 already leads to fast responses to dynamic network bandwidths, consistent playback quality and low rebuffering; CAVA-p12 leads to higher Q4 chunk quality, while CAVA-p123 leads to lower rebuffering. Fig.~\ref{fig:impact-principles} shows an example for one video (ED, \texttt{FFmpeg} encoded, H.264) under LTE traces.
Fig.~\ref{fig:impact-principles}(a) plots the Q4 chunk quality under the two variants with the differential treatment principle (i.e., CAVA-p12 and CAVA-p123) relative to CAVA-p1. Specifically, it shows the
	distribution of the metric value under CAVA-p12 and CAVA-p123 minus the value under CAVA-p1 across all runs. We observe that for around 40\% of the Q4 chunks,  CAVA-p12 and CAVA-p123 lead to higher Q4 chunk quality.
	For only around 5\% of the Q4 chunks, the quality under CAVA-p12 and CAVA-p123 is lower than CAVA-p1.
%We observed from Fig.~\ref{fig:impact-principles}(a) that CAVA-p12 leads to higher Q4 chunk quality compared to CAVA-p1, confirming the benefits of incorporating the differential treatment principle.
Fig.~\ref{fig:impact-principles}(b) shows the total rebuffering of CAVA-p123 relative to CAVA-p12 for traces that lead to rebuffering under either of these two variants (35 out of 200 traces fall into this category).
Specifically, it shows the metric value under each variant minus the value under CAVA-p12. We see that CAVA-p123 leads to lower  rebuffering than CAVA-p12 in 55\% of these traces and the reduction is up to 20 seconds.

%wb, 6/16/2018, move to earlier part

%%dash results
%wb, 10/8/2018, use BBB instead of ED 
\begin{figure*}[ht]
	\centering
	%\subfloat[\label{subfig:CAVA-Q4-dash}]{%
	\includegraphics[width=0.3\textwidth]{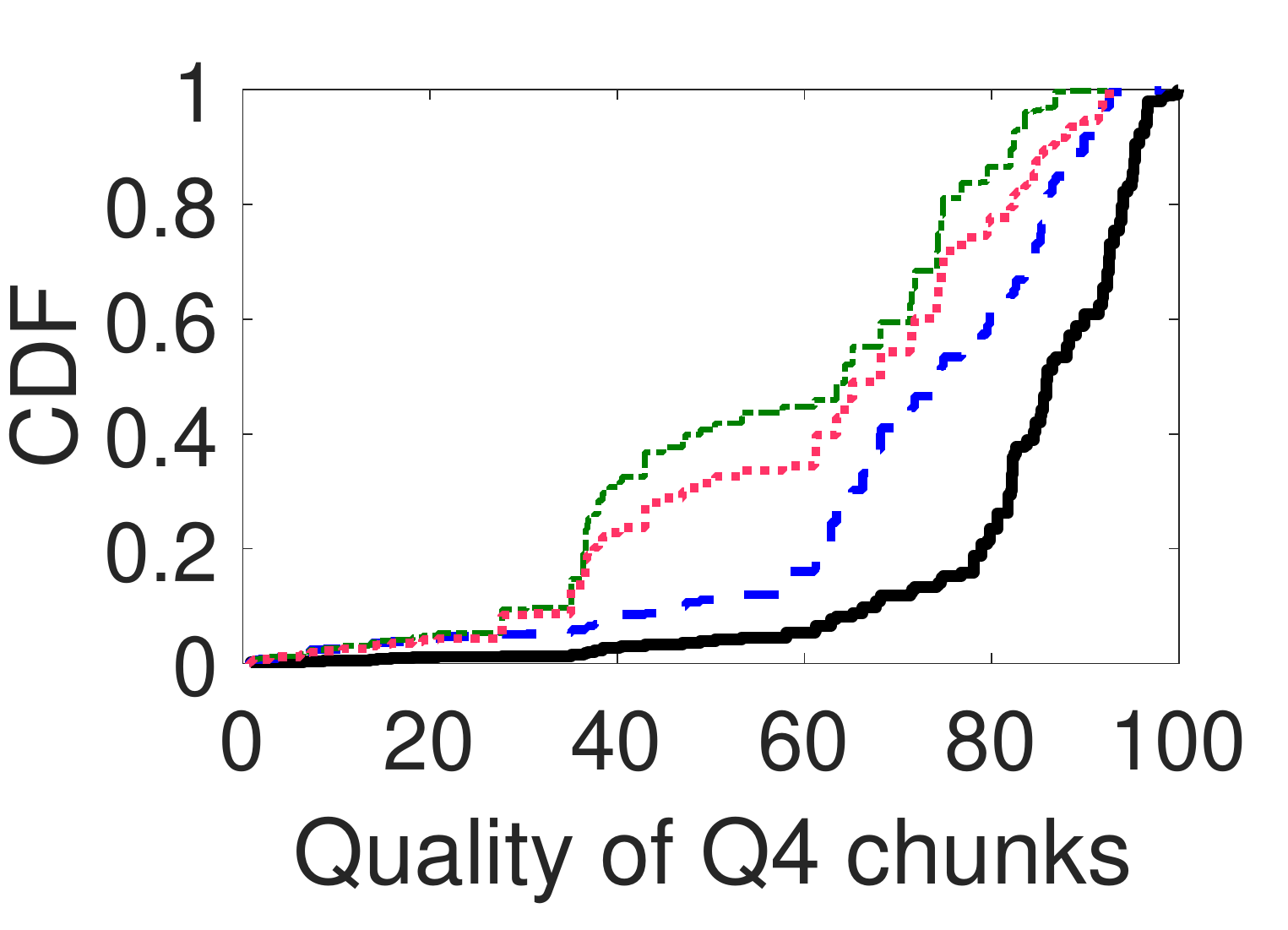}
	%}
	%\subfloat[\label{subfig:CAVA-low-quality-dash}]{%
	\includegraphics[width=0.3\textwidth]{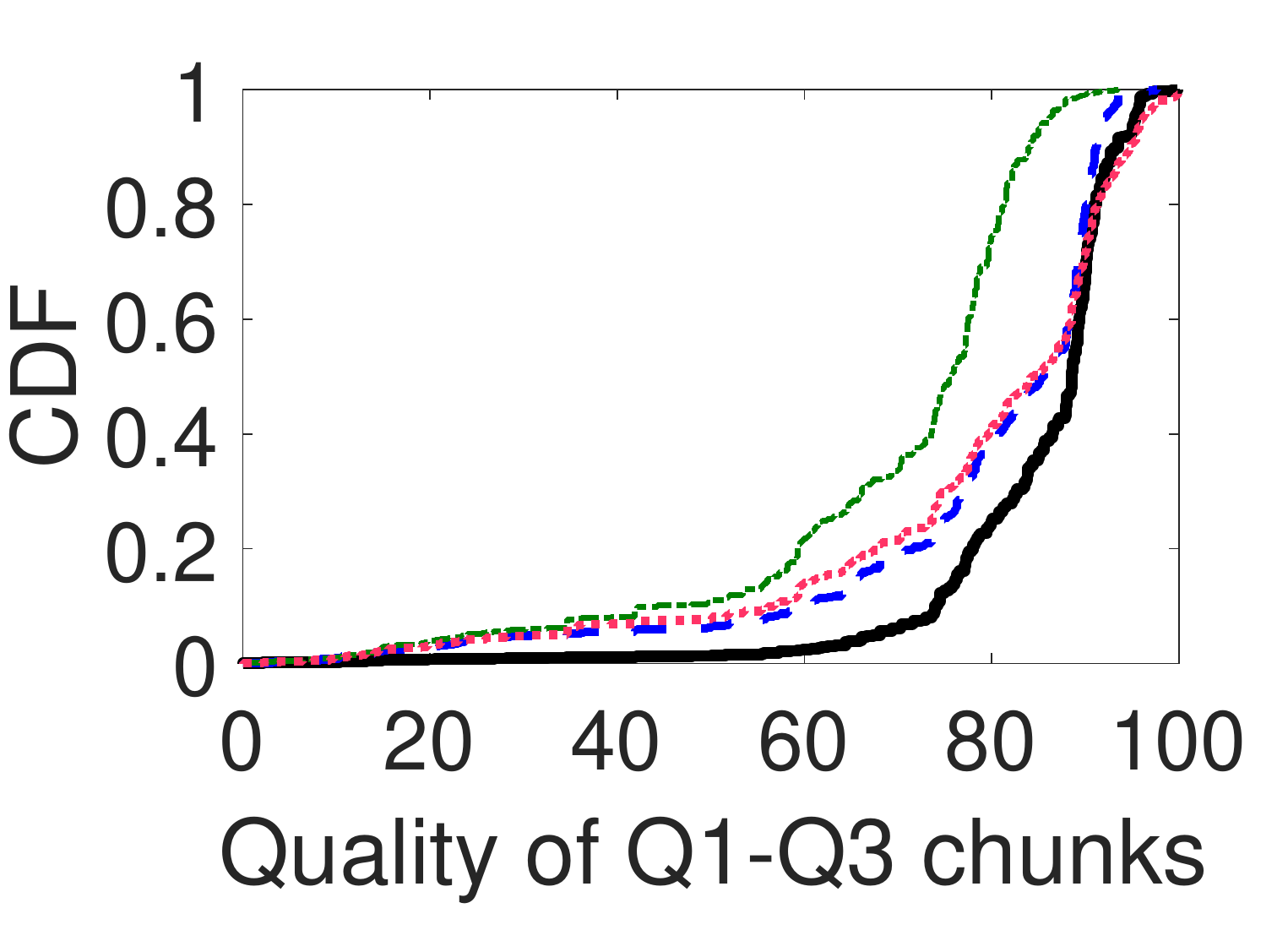}
	%\subfloat[\label{subfig:CAVA-low-quality-dash}]{%
	\includegraphics[width=0.3\textwidth]{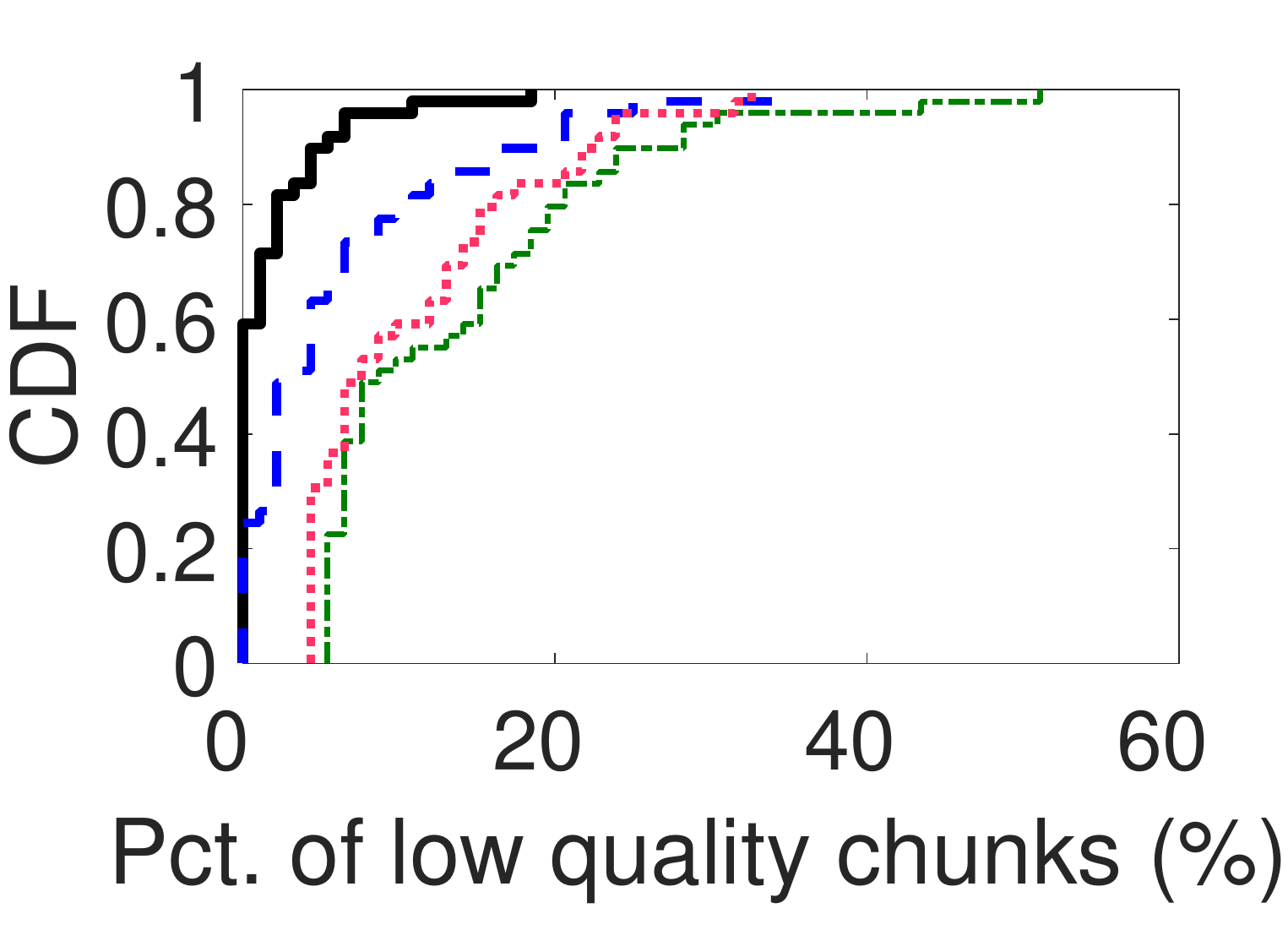}
	%}
	%\subfloat[\label{subfig:CAVA-rebuffering-dash}]{%
	\includegraphics[width=0.3\textwidth]{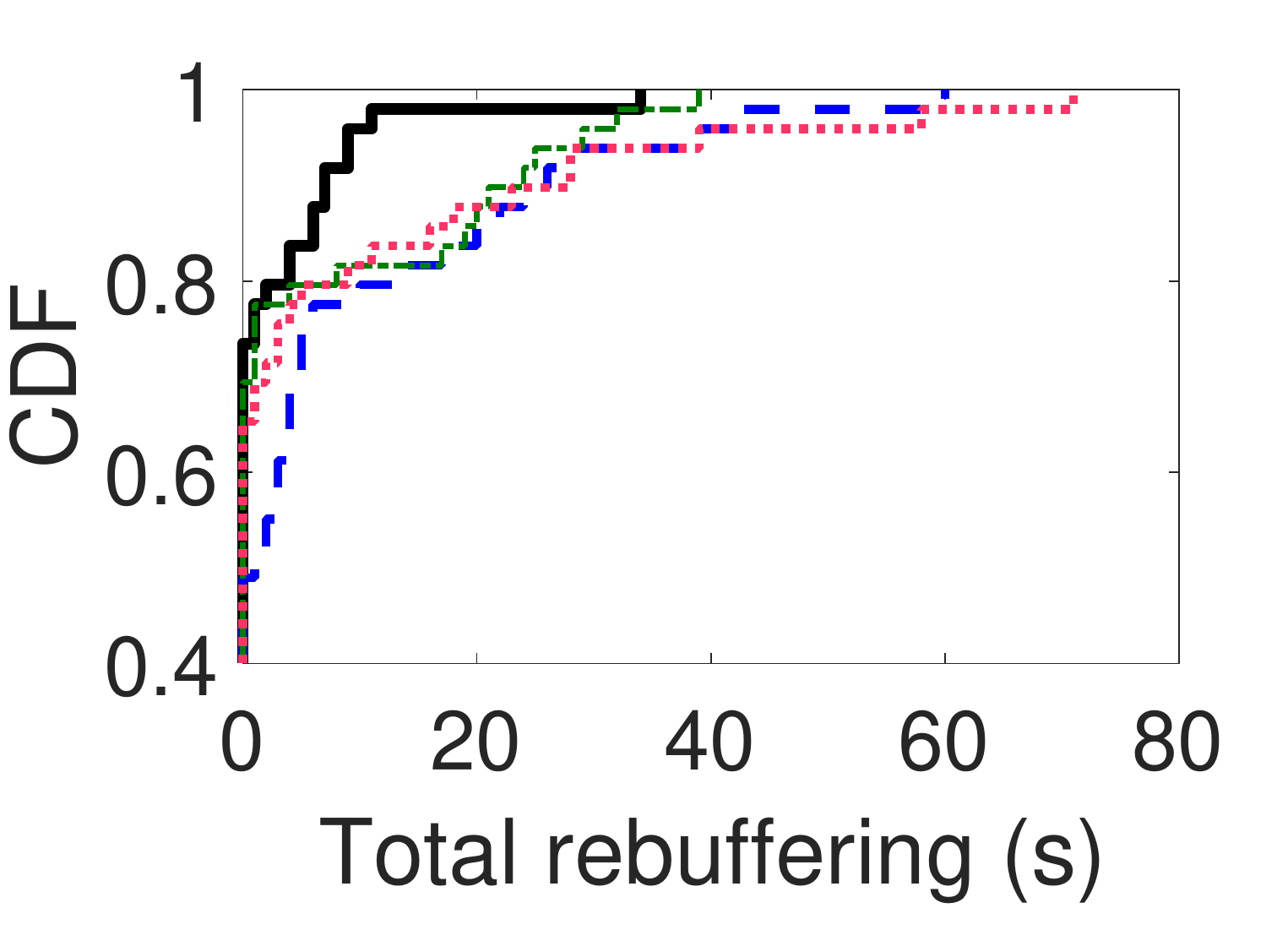}
	%}
	%\subfloat[\label{subfig:CAVA-change-dash}]{%
	\includegraphics[width=0.3\textwidth]{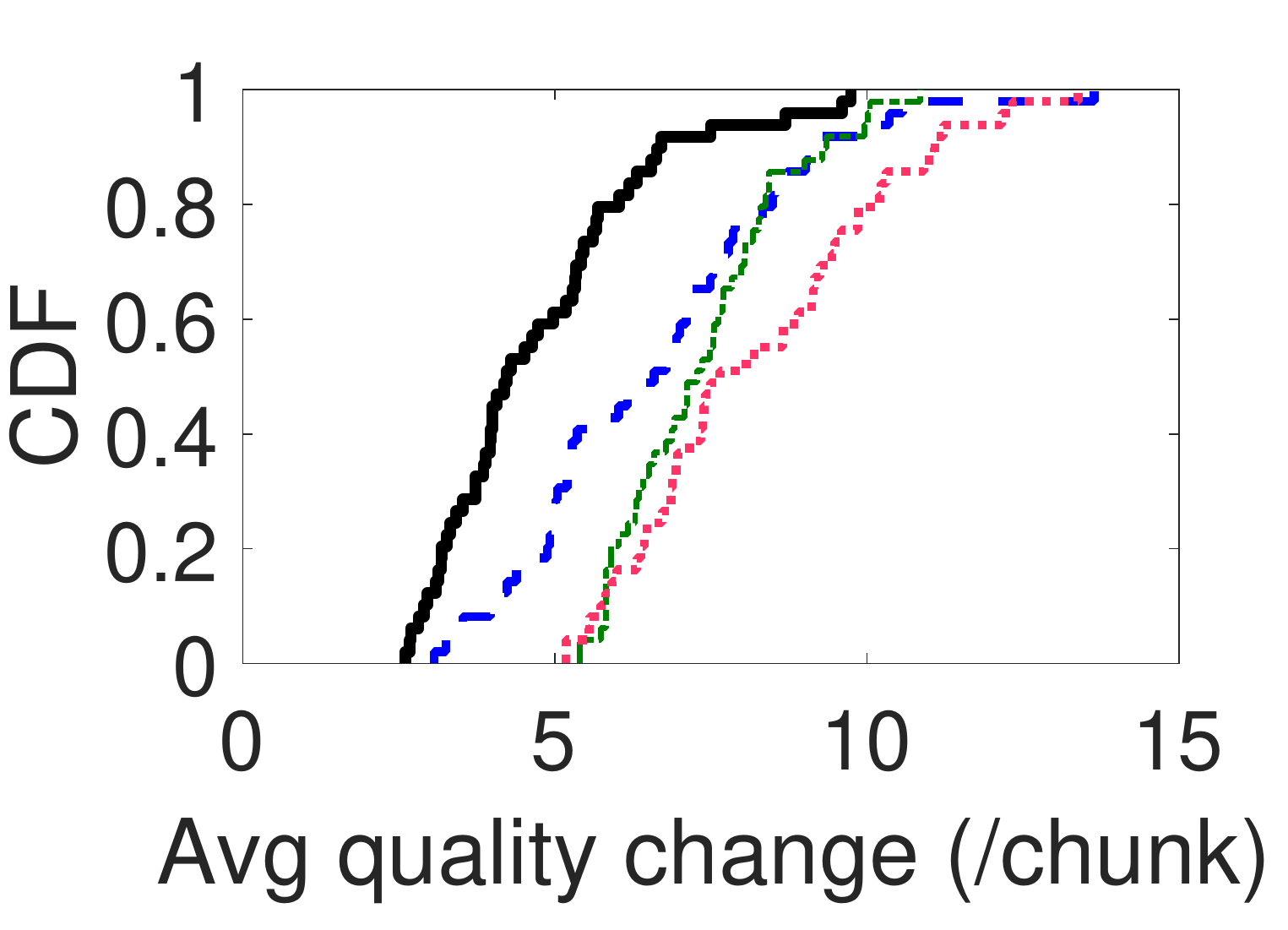}
	%}
	%\subfloat[\label{subfig:CAVA-data-usage-dash}]{%
	\includegraphics[width=0.3\textwidth]{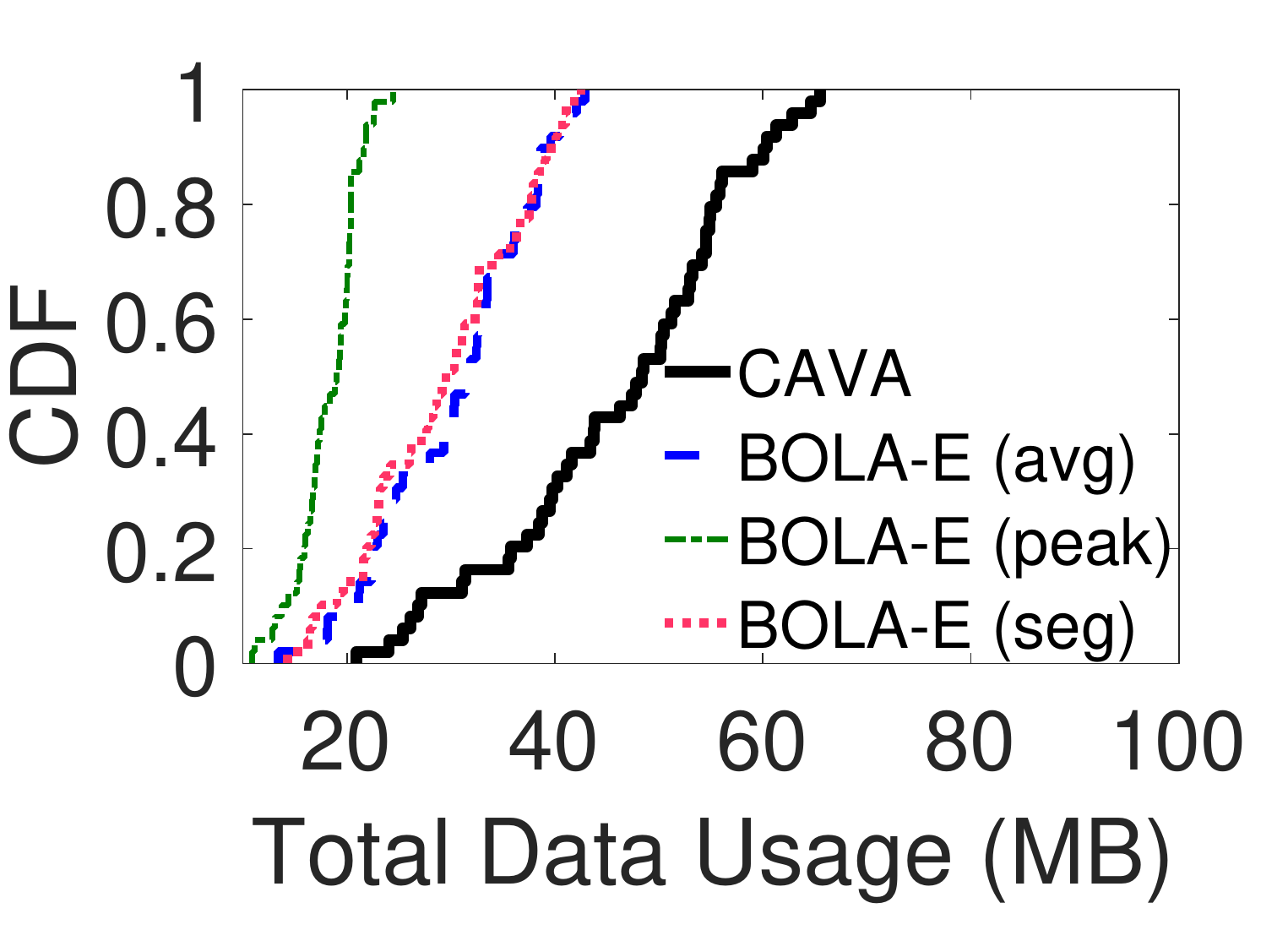}
	%}
	\vspace{-.1in}
	\caption{Performance comparison in \texttt{dash.js} (Big Buck Bunny, YouTube encoded, H.264, LTE traces).}
	\label{fig:cava_comparison-dash-js}
	\vspace{-.1in}
\end{figure*}

\mysubsection{Codec Impact} \label{sec:perf-h265}

%%only H.265

For each video, the performance of all the schemes under H.265 encoding is better than that under H.264 encoding (figures omitted). This is due to the significantly lower bitrate requirement of H.265 encoding.
We observe that CAVA also significantly outperforms all the other schemes under H.265 encoding. Compared to RobustMPC and PANDA/CQ max-min, on average, Q4 chunk quality under CAVA is 7-12 higher, the percentage of low-quality chunks is 51\%-82\% lower, rebuffering is 52\%-91\% lower, and  average quality change is 27\%-72\% lower. Also CAVA has similar data usage compared to other schemes.

\mysubsection{Impact of Higher Bitrate Variability} \label{sec:perf-large-cap} 
%\bing{Do we still want to keep this section?}
We further use the 4$\times$ capped VBR-encoded video (Elephant Dream \texttt{FFmpeg} encoded, in H.264) to explore the impact of higher bitrate variability. In this case, we observe similar trends as the earlier results.
%CAVA still outperforms  MPC and PANDA/CQ based schemes. 
For CAVA, the average Q4 chunk quality is 65 across the LTE traces, 8 and 7 larger than RobustMPC and PANDA/CQ max-min, respectively. The average quality change is 42\% and 68\% lower, the rebuffering is 90\% and 89\% lower,
and the average percentage of low-quality chunks is 39\% and 57\% lower.

%scenes when a VBR video has larger variability. Specifically, we use a video with bitrate cap of 4$\times$. Large cap will bring more challenges to streaming VBR videos. 

\mysubsection{Impact of Bandwidth Prediction Error} \label{subsec:bw-error}

%% yanyuan, show CAVA is resilent to bandwidth prediction
%%wb, 6/21/2018, for space
\iffalse
\begin{figure*}[ht]
	\centering
	\subfloat[\label{subfig:CAVA-Q4-bw-error}]{%
		\includegraphics[width=0.20\textwidth]{newfigs/NF-264-ED/BW-quality-q4-LTECC-_phone}
	}
	\subfloat[\label{subfig:CAVA-low-quality-bw-error}]{%
	\includegraphics[width=0.20\textwidth]{newfigs/NF-264-ED/BW-quality-low-LTECC-_phone}
	}
	\subfloat[\label{subfig:CAVA-all-bw-error}]{%
		\includegraphics[width=0.20\textwidth]{newfigs/NF-264-ED/BW-rebufferingLTECC-_phone}
	}
	\subfloat[\label{subfig:CAVA-change-bw-error}]{%
		\includegraphics[width=0.20\textwidth]{newfigs/NF-264-ED/BW-avg-quality-change-LTECC-_phone}
	}
	\subfloat[\label{subfig:CAVA-data-usage-bw-error}]{%
		\includegraphics[width=0.20\textwidth]{newfigs/NF-264-ED/relativeBW-realR-LTECC-_phone}
	}

	\vspace{-.15in}
	\caption{Impact of bandwidth prediction error (Elephant Dream, \texttt{FFmpeg} encoded, H.264, LTE traces).}
	\label{fig:comparison-bw-error}
	\vspace{-.15in}
\end{figure*}
\fi

So far, we have used a particular bandwidth estimation algorithm (harmonic mean of past 5 chunks) to predict available network bandwidth. We next investigate the impact of bandwidth prediction errors on different schemes in a controlled manner as follows.
%  and compare  the performance of \piv with MPC and PANDA/CQ max-min. Specifically,
We assume the bandwidth prediction error is $err$. That is, if the actual network bandwidth at time $t$ is $C_t$, we assume that the predicted network bandwidth is uniformly randomly distributed in $C_t  (1 \pm err)$. We vary $err$ from 0 to 50\% with a step of 25\%.
%Fig.~\ref{fig:comparison-bw-error} plots the results for \piv, RobustMPC and PANDA/CQ max-min, where $err$ is 0 or 50\% (the results for $err$ as 25\% show similar trend).
We observe that CAVA is insensitive to bandwidth prediction errors: the quality of Q4 chunks, the amount of rebuffering and the percentage of low quality chunks when $err$=50\% are similar to those  when $err$=0 (i.e., perfect prediction), demonstrating that CAVA is adaptive and tolerant to relatively significant bandwidth prediction errors (due to the control-theoretical underpinning).
% the average quality change increases with $err$, since this performance metric is concerned with two consecutive chunks, and bandwidth prediction errors;
% the amount of data usage when $err$=50\%  is similar to that when $err$=0  across the traces. \sen{ Did not fully follow the reasoning for average quality change argument. What is it ? }
 In comparison, MPC leads to significantly more rebuffering and data usage when $err$=50\%, compared to those when $err$=0.  PANDA/CQ max-min leads to  noticeably more rebuffering when $err$ is increased from 0 to 50\%.
 %Add explanation by \piv is resilent to bandwidth estimation errors.
The reason why CAVA is resilent to bandwidth estimation errors is because it is a principled approach based on control theory -- it continuously monitors the  ``errors" in the system, namely  the difference between the target and current buffer levels of the player, and adjusts the control signal to correct ``errors" caused by  inaccurate bandwidth estimation.

%The increase of rebuffering under PANDA/CQ max-min when $err$ is increased from 0 to 50\% is leads to  noticeably more rebuffering when $err$=50\%

%the only noticeable dfference when $err$=0 (i.e., perfect prediction) or 50\% is the average quality change .

%\iffalse
%\yanyuan{ CAVA is resilient to bandwidth prediction error.
%Fig.~\ref{fig:comparison-bw-error} shows the sensitivity analysis of different schemes on bandwidth prediction error.
%
%\BULLET  As the bandwidth prediction error increases, the avg quality changes increases. Percentage of low quality chunks increase rebuffering has almost no change.
%
%The reason is as follows:
%Large prediction error will make the bandwidth prediction value more extreme, (either much higher or much lower), leading to more higher track chunks and more lower track chunks. Therefore, the CDF of chunk quality become more horizontal. As the diminishing return of the BD curve (quality to bitrate), the influence on the low quality part is more visible than the high quality part. As the same reason, the quality change will be larger as the bandwidth prediction error increase.
%
%\BULLET  If the prediction error within 50\%, CAVA still get good performance close to the case with perfect prediction.
%}
%\fi

%\input{YouTube-results}

%%wb, 1/23/2018, present this earlier on to motivate the design principles of CAVA
%%\input{case-study}

	\mysubsection{DASH Implementation Results} \label{sec:cava_dash}
%We  implemented  \piv,  BBA-1 and RBA in \texttt{dash.js}, an open-source web-based DASH video player~\cite{dash-js} (version v2.4.1).
%The evaluation of the various schemes is through an emulation environment that we create.

%%BBB, Elevant dream, animals, sports
We evaluate CAVA using the \texttt{dash.js} implementation (\S\ref{sec:implementation}), and compare its performance with that of BOLA-E~\cite{Spiteri2018bolae} implemented in \texttt{dash.js} version 2.7.0, which contains a stable release of BOLA-E.
%The original BOLA implementation takes a single bitrate value from each track; we modified it to use the actual chunk sizes as suggested by the paper~\cite{spiteri2016bola} for VBR encodings.
%
Our evaluations below use  two computers (Ubuntu 12.04 LTS, CPU Intel Core 2 Duo, memory 4GB) with a 100 Mbps direct network connection to emulate the video server (Apache httpd) and client. At the client, we use Selenium~\cite{selenium}
%(an automated software testing tool for web applications )
to programmatically run a Google Chrome web browser and use \texttt{dash.js} APIs to  collect the streaming performance data.
We use \texttt{tc}~\cite{tc} to emulate  real-world variable network conditions by programmatically ``replaying'' the network traces (\S\ref{sec:cava_setup}).

We find  that CAVA is very light-weight -- even with the current prototype implementation, the runtime overhead (from profiling) is only around 56ms for a 10-minute video.

We compare the performance of CAVA with three versions of BOLA-E. Two versions are based on the
original BOLA-E implementation, which takes a single bitrate value from each track (i.e., the declared bitrate from the manifest file). Specifically, we consider BOLA-E (peak), where the declared bitrate for a track is the peak bitrate of the track, and BOLA-E (avg), where the declared bitrate for a track is the average bitrate of the track. In the third version, referred to as BOLA-E (seg), we modified the original BOLA-E implementation to use the actual chunk size for each chunk, as suggested by the paper~\cite{spiteri2016bola} for VBR encodings.
All the three versions use the default parameters in BOLA-E.
%\bingc{Do we need to mention explicitly the parameter values, e.g., the maximum buffer size is 30 seconds?}
Fig.~\ref{fig:cava_comparison-dash-js} shows the performance of CAVA and the three versions of BOLA-E for one Youtube video.
We observe that CAVA outperforms the three versions of BOLA-E in having higher Q4 chunk quality, lower percentage of low-quality chunks, lower rebuffering, and lower quality changes. While the data usage of BOLA-E (all three variants) is lower than that of CAVA, it comes at the cost of worse performance in other performance metrics. The lower data usage of BOLA-E could  be because it sometimes pauses before fetching the next chunk under various scenarios. In addition, the implementation uses bitrate capping when switching to a higher bitrate to avoid bitrate oscillations.
 %(i.e., the BOLA-O variant in \cite{spiteri2016bola}).
 
 Of the three BOLA-E variants, not surprisingly, BOLA-E (peak) is the most conservative variant in choosing bitrate, since it uses the peak bitrate of a track as the bitrate of every chunk in the track,  and hence overestimates the bandwidth requirements of the individual chunks. In comparison, BOLA-E (avg) is the most aggressive,
% since it underestimates the
and BOLA-E (seg) is between BOLA-E (peak) and BOLA-E (avg). We see that for Q1-Q3 chunks, BOLA-E (seg) can choose even higher tracks than  BOLA-E (avg), since BOLA-E (seg) uses the actual chunk size and can choose higher quality for some small chunks (that fall into Q1-Q3). We further see that using the actual chunk size in BOLA-E (seg) leads to more significant quality changes compared to  BOLA-E (peak) and BOLA-E (avg).
The above behaviors demonstrate once again that  while it is important to consider individual chunk sizes to handle ABR streaming of VBR videos, 
%it is not sufficient for existing schemes (that were designed for CBR videos)
%make their adaptation decisions using individual chunk size information rather than the peak or average bitrate.  
the actual decision logic needs to be developed with VBR characteristics in mind; simply plugging in the individual chunk sizes is insufficient.

Table~\ref{perf-cmp-dash} compares the performance of CAVA and BOLA-E (seg)
%, i.e., the suggested BOLA-E variant for VBR videos,
 for four YouTube videos, all  under LTE traces. We observe that while the data usage of BOLA-E (seg) is lower,  CAVA outperforms BOLA-E (seg) in all the other metrics: on average, Q4 chunk quality under CAVA is 10-21 higher, the percentage of low-quality chunks is 73\%-87\% lower,  rebuffering is 15\%-65\% lower, and quality changes is 24\%-45\% lower.

%the importance of designing ABR streaming schemes to explicitly account for the characteristics of VBR

\begin{table}[]
	\centering
	\small{
		\caption{CAVA versus BOLA-E in \texttt{dash.js}.}
		\label{perf-cmp-dash}
		\begin{tabular}{|c|c|c|c|c|c|}
			\hline
			%\cline{1-6}
			 Video & {\begin{tabular}[c]{@{}c@{}}Q4\\chunk\\quality\end{tabular}} &
			{\begin{tabular}[c]{@{}c@{}}Low-qual.\\chunks\\(\%)\end{tabular}} & {\begin{tabular}[c]{@{}c@{}}Stall\\duration\\(\%)\end{tabular}} &
			{\begin{tabular}[c]{@{}c@{}}Quality\\changes\\(\%)\end{tabular}} &
			{\begin{tabular}[c]{@{}c@{}}Data\\usage\\(\%)\end{tabular}}
			\\ \hline
			%%H.264 results
			%\multirow{4}{*}{\begin{sideways} LTE, phone \end{sideways}} &
			BBB   & \textcolor{green}{$\uparrow$}21 & \textcolor{green}{$\downarrow$}87 & \textcolor{green}{$\downarrow$}65 & \textcolor{green}{$\downarrow$}45  & \textcolor{red}{$\uparrow$}56\\\hline  %\cline{2-6}
			ED  & \textcolor{green}{$\uparrow$}20 & \textcolor{green}{$\downarrow$}73 & \textcolor{green}{$\downarrow$}15 & \textcolor{green}{$\downarrow$}35  & \textcolor{red}{$\uparrow$}50\\\hline %\cline{2-6}
			Sports   &  \textcolor{green}{$\uparrow$}10& \textcolor{green}{$\downarrow$}75 & \textcolor{green}{$\downarrow$}29 & \textcolor{green}{$\downarrow$}40  & \textcolor{red}{$\uparrow$}25\\ \hline
			
			ToS  & \textcolor{green}{$\uparrow$}12 & \textcolor{green}{$\downarrow$}76& \textcolor{green}{$\downarrow$}19& \textcolor{green}{$\downarrow$}24  & \textcolor{red}{$\uparrow$}45 \\\hline %\cline{2-6}
			%\cline{2-6}
%			\multicolumn{2}{|c|}{} &  \textcolor{green}{$\uparrow$} better & \textcolor{green}{$\downarrow$} better & \textcolor{green}{$\downarrow$} better & \textcolor{green}{$\downarrow$} better & \textcolor{green}{$\downarrow$} better \\ \hline
		\end{tabular}
	}
\end{table}

\section{Related Work}\label{sec:cava_related}

There is a large body of existing work on ABR streaming~\cite{Cicco11:feedback,tian2012towards,yin2015control,huang2014buffer,li2014streaming,spiteri2016bola,jiang2012improving,Xie15:piStream,Chen13:Scheduling,Zou15:video,sun2016cs2p,qin2017control,tong2017rbavbr,Mao17:pensieve,akhtar2018oboe}.
%We briefly describe a few representative schemes below. 
Rate-based~(e.g., \cite{Kuschnig10:evaluation,jiang2012improving,tong2017rbavbr}) ABR schemes select bitrate simply based on network bandwidth estimates. Buffer-based schemes,  e.g., BBA~\cite{huang2014buffer}, BOLA~\cite{spiteri2016bola}, and BOLA-E~\cite{Spiteri2018bolae}, select bitrate only based on buffer occupancy.
Most recent schemes combine both rate-based and buffer-based techniques.
%consider both network bandwidth and buffer status to determine the bitrate for the next chunk.
For instance, MPC~\cite{yin2015control} and PIA~\cite{qin2017control} use control-theoretic approaches to design rate adaptation  based on both network bandwidth estimates and buffer status. Oboe~\cite{akhtar2018oboe} dynamically adjusts the parameters of an ABR logic to adapt to various network conditions.
Another group of related studies~\cite{Chiariotti2016OLA,Mao17:pensieve}  uses machine learning based approaches, which ``learns" ABR schemes from data.

While content providers have recently begun adopting VBR encoding, their treatment to VBR videos is simplistic (e.g., by simply assuming that the bandwidth requirement of a chunk equals to the peak bitrate of the track)~\cite{Xu17:Dissecting}. As we have seen in \S\ref{sec:cava_dash}, such simplistic treatment can lead to poor performance.
Of the large number of existing studies, only a few targeted VBR videos~\cite{li2014streaming,huang2014buffer,tong2017rbavbr};
most others were designed for and evaluated in the context of CBR encodings.
While several schemes  in the latter category can be applied for VBR videos by explicitly incorporating the actual bitrate of each chunk, they were not designed ground up to account for VBR characteristics, nor had their performance been evaluated for that case.
We evaluate the performance of schemes from both categories in this chapter
(see~\S\ref{sec:cava_principles} and \S\ref{sec:cava_perf}), and show that
they can suffer from a large amount of rebuffering and/or significant viewing quality impairments when used for VBR videos.

Our study differs from all the above works in several important aspects, including developing insights based on analyzing content complexity, encoding quality, and encoding bitrates, identifying general design principles for VBR streaming, and developing a concrete, practical ABR scheme for VBR streaming that significantly outperforms the state-of-the-art schemes.

	%\section{Conclusions} \label{sec:concl}
\mysection{Conclusions and Future Work} \label{sec:concl}

Through analyzing a number of VBR-encoded videos from a variety of genres,
we identified a number of characteristics related to scene complexities, bandwidth variability, and quality levels, that are relevant to ABR streaming and QoE.
Based on these insights, we proposed 3 key principles to guide rate adaptation for VBR streaming.
  We also designed CAVA, a practical control-theoretic rate adaptation scheme that explicitly leverages our proposed principles. Extensive evaluations demonstrate that CAVA substantially outperforms existing ABR schemes even in challenging network conditions common in cellular networks. The results demonstrate the importance of utilizing these design principles in developing ABR adaptation schemes for VBR encodings.
  In this chapter, we primarily focused on the VoD setting.
As future work, we will explore  extending CAVA and its concepts to ABR streaming of live VBR encoded videos.

 %}

	\newpage

\chapter{Quality-aware Strategies for Optimizing ABR Video Streaming QoE and Reducing Data Usage} \label{chapter:quad}

\mysection{Introduction}

This chapter  attempts to answer the following question: how can we effectively reduce the bandwidth consumption for mobile video streaming while minimizing the impact on users' quality of experience (QoE)?
	This is a highly important and practical research problem since cellular network bandwidth is a relatively scarce resource.
	The average data plan for a U.S. cellular customer is only 2.5 GB per month~\cite{ericsson-mobility-report}, while streaming just one-hour High Definition (HD) video on mobile Netflix can consume
	 %as much as
	  3 GB data.
	%the \emph{metered} nature of cellular data networks.
	%Studies have shown that the average data plan for a U.S. cellular customer is only 2.5 GB per month, with most of the data being used for video~\cite{xxx}. Even for the expensive ``unlimited'' data plans, the data rate can be throttled after the usage hitting a threshold. Meanwhile, streaming just one-hour HD video on mobile Netflix can consume as much as 3GB data.
	Therefore, the capability to make more efficient use of  data while still hitting quality targets is the key to enabling users to consume more content within their data budgets without adversely impacting QoE. In addition, downloading less data for a video session also translates to lower radio energy consumption~\cite{yuemmsys2020}, less thermal overhead on mobile devices, as well as
	potentially better QoE for other users sharing the same cellular RAN {(radio access network)} or base station.

Existing ABR adaptation schemes (\S\ref{sec:related work}) focus mainly on maximizing the video quality and user QoE.
{While some schemes do conservatively utilize the bandwidth, their decisions are primarily driven by QoE impairment concerns such as the possibility of stalls caused by the potentially inaccurate bandwidth estimation.}
%While the schemes do  differ in their data usage, such behaviors are driven primarily by QoE impairment concerns, e.g., that the network bandwidth estimation at the client may not be accurate, thus selecting very high tracks can increase the chance of stalls.
For example, when selecting the next chunk, the default adaptation scheme in ExoPlayer~\cite{exoplayer} (a popular open-source player that is used in more than 10000 apps) only considers tracks whose declared bitrates\fengc{(typically set around the peak bitrate of that track)} are at least $25\%$ lower than the estimated network bandwidth.
%{While such a design does reduce the data usage, existing ABR schemes do not explicitly consider bandwidth efficiency as a key factor in making the track selection decisions.}
Such a design saves data just by being conservative. As we show later, it can lead to significantly lower quality/QoE (\S\ref{sec:exoplayer}).
Existing ABR schemes do not explicitly consider
bandwidth efficiency together with quality in making the track selection decisions.
%One challenge that existing ABR schemes do not explicitly consider
%bandwidth efficiency and quality together  as a key factor in making the track selection decisions.

Some mobile network operators and commercial video services  provide users with
certain ``data saver'' options.
%~(\S\ref{sec:current-practice-save-data}).
%the option of certain ``data saver'' modes~(\S\ref{sec:current-practice-save-data}).
{These options take either a service-based approach or a network-based approach.
The former limits the highest level of quality/resolution/bitrate by, for example, streaming only  {standard definition} (SD) content over cellular networks.
A network-based approach instead limits the network bandwidth.
}
%When activated, these options either (i)  limit the highest quality/screen resolution/bitrate track (e.g., stream only Standard Definition content on cellular) that the ABR logic will ever download~(service-based) or (ii)  limit the network bandwidth~(network-based),  which indirectly has a similar effect.
%These approaches primarily focus on data savings. Their implications for the delivered video {QoE} is little understood.
We survey the ``quality throttling'' mechanisms used by today's commercial video content providers for saving bandwidth for mobile devices, and pinpoint their inefficiencies~(\S\ref{sec:quality-bitrate-tradeoffs}-\S\ref{sec:reduce-data}).
Despite their simplicity, we find such approaches often achieve tradeoffs between video quality and bandwidth usage that are far from what viewers desire. A key reason is that for  state-of-the-art encoders (e.g., H.264~\cite{h264}, H.265~\cite{h265} and VP9~\cite{vp9}), the actual perceived picture quality exhibits significant variability across different chunks within the same track, for both Constant Bitrate (CBR) and Variable Bitrate~(VBR) encodings.
Therefore, a data-saving approach that
%uses a track-level filter
simply removes high tracks
%is bound to
leads to quality variations.\fengc{(e.g., streaming simple scenes at an unnecessarily high quality and complex scenes at a much lower quality).}
%, when there is ample network bandwidth available.
Such quality variability impairs user QoE, and makes suboptimal use of the network bandwidth.

% Whats missing is approaches that save data while not impacting quality.

To address the drawbacks of the existing practices, we propose a \emph{quality-aware data saving} strategy~\cite{sen2021chunk}, which provides data savings while allowing users' direct control of the video quality.
Specifically, a user selects from multiple quality options (e.g., good, better, best), and the player will map the selection to a \emph{target quality} (\S\ref{sec:reduce-data}).
Within this strategy, we propose two new schemes that explicitly consider the bandwidth efficiency by matching the fetched content quality against a target quality. In this way, the player will avoid fetching chunks whose qualities are beyond the target quality, leading to bandwidth savings. Specifically, the two new schemes are:

%To better balance the tradeoffs among the video quality, rebuffering, quality changes, and bandwidth usage, we propose two new schemes that explicitly consider the bandwidth efficiency by matching the fetched content quality against a pre-defined target quality. In this way, the player will avoid fetching chunks whose qualities are too high, leading to bandwidth savings.

\BULLET \emph{Chunk-Based Filtering (CBF).}
We propose a novel, {quality-variability aware} scheme called Chunk-Based Filtering (CBF) (\S\ref{sec:need-for-efficiency}) that can be retro-fitted into existing ABR adaptation schemes~\cite{sen2021perceptual}. Its high-level idea is as follows.
For every chunk position in the video, CBF limits the choice of the highest quality chunk for ABR rate adaptation %is  limited
to the chunk whose quality is {closest to} the target quality.
CBF thereby steers an ABR adaptation scheme to the set of more desirable choices {(from
the perspective of balancing the tradeoff between the video quality and bandwidth usage)},
%both the quality and the bandwidth usage perspective),
and helps the ABR scheme achieve better streaming performance than it would by itself.
Using CBF in conjunction with existing ABR adaptation schemes is also attractive from a practical incremental deployment perspective.  CBF has no dependencies on and requires no changes to the complex ABR adaptation logic, and can be relatively easily inserted into the existing streaming  workflow~(\S\ref{sec:cbf-deployment}).

\BULLET \emph{Quality-aware ABR Adaptation.} From a performance and bandwidth-efficiency tradeoff perspective, it is possible to do even better, {if the ABR scheme itself can
explicitly integrate the goal of approaching the target quality into
its
rate adaptation logic.}
%directly make tradeoff decisions between the quality and data usage when selecting the next chunk.
To this end,
we develop \emph{QUality-Aware Data-efficient  streaming~(QUAD)}, a holistic rate adaptation algorithm that is designed ground up (\S\ref{sec:design}).
%that explicitly leverages CBF as a building block while also integrating other essential components such as bandwidth awareness and buffer control (\S\ref{sec:design}).
%
QUAD jointly considers three aspects when making rate adaptation decisions:
pacing {the selected chunks'
	%encoded
	 bitrate} to the estimated network bandwidth {to prevent stalls},
adapting to the target quality to reduce bandwidth consumption, and
minimizing the inter-chunk quality change to enhance the playback smoothness.
%Developed upon robust control theoretic foundations,
%the overall design of \qea is robust and incurs very low runtime overhead.
{QUAD is robust and lightweight as it is developed upon solid control theoretic foundations.}

We implemented CBF and QUAD (\S\ref{sec:setup}) in two popular state-of-the-art open source ABR video players: \texttt{dash.js}~\cite{dash-js} and ExoPlayer~\cite{exoplayer}.
% (which is used in more than 10000 apps~\cite{exoplayer_why}).
%We extensively evaluate both
Our evaluation of these two techniques uses  a  diverse set of real videos and different encoding schemes (VBR and CBR),
{under both emulated and real-world LTE networks~(\S\ref{sec:setup}).}
%by using a combination of real-world LTE trace-driven simulations, emulations over WiFi, and experiments in production cellular networks~(\S\ref{sec:setup}).
The key evaluation results include the following.

\BULLET CBF significantly improves the performance of existing state-of-the-art ABR schemes~(\S\ref{sec:perf-CBF}). Specifically, after employing CBF as a prefiltering step, the average deviation from the target quality is reduced by 37-67\%; the average quality variation is reduced by 7-31\%; and the data usage is reduced by 34-67\% even in
challenging network conditions
 (in easier conditions with ample bandwidth, the reduction is even more).
We further experimentally demonstrate that CBF is significantly more effective in steering existing ABR schemes to more desirable rate adaptation decisions than {traditional}  service-based or network-based data saving approaches.
%track-based filtering.

\BULLET Compared to existing schemes enhanced with CBF, QUAD
%outperforms existing schemes with CBF in
achieves even better performance in approaching the target quality with low quality variations {while still achieving good QoE (\S\ref{sec:qea-eval}).}
For instance, our evaluation on \texttt{dash.js} shows that, compared to a state-of-the-art ABR scheme, BOLA-E~\cite{Spiteri2018bolae} enhanced with CBF, QUAD leads to 37\% fewer low-quality chunks
%, 80\% reduction in rebuffering,
and 12\% reduction in quality variation.
Compared to the default rate adaptation of ExoPlayer,
QUAD reduces the deviation from the target quality by 64\%, reduces the number of low-quality chunks by 81\%, and reduces the quality variation by 43\%.
{Compared to an optimized version of ExoPlayer's algorithm 
%that is further 
enhanced with CBF, the corresponding reductions brought by QUAD are 40\%, 46\% and 22\%, respectively.}

\section{Quality and Bitrate Tradeoffs} \label{sec:quality-bitrate-tradeoffs}
In this section, we describe the  encoding characteristics of both VBR and CBR videos to motivate 
our proposed solutions.
\fengc{the quality-aware data saving strategy (\S\ref{sec:reduce-data}), the chunk-based filtering (CBF) approach (\S\ref{sec:need-for-efficiency}), and a grounds-up new ABR adaptation algorithm  design (\S\ref{sec:design}).}
%\fengc{We first describe the video dataset and then the encoding characteristics.}

\subsection{Video Dataset}\label{sec:video-dataset}

%Our chunk-based filtering approach  is motivated by the encoding characteristics of both VBR and CBR videos.
%In this section, we present the video dataset that is used in  our study.

Our video dataset includes 18 VBR and 4 CBR videos.
%Variable Bitrate (VBR) videos and 4 Constant Bitrate (CBR) videos.
We consider both VBR and CBR videos because both are widely used in practice~\cite{Xu17:Dissecting},
with the trend of wider adoption of VBR videos thanks to their many advantages over CBR videos~\cite{lakshman1998vbr}.
Each video is around 10 minutes long, and encoded at 6 tracks/levels\footnote{We use the terms \emph{track} and \emph{level} interchangeably in this chapter.  Other equivalent terms include representation and rendition, which are also widely used in the literature.} for ABR streaming,
with resolutions of 144p, 240p, 360p, 480p, 720p, and 1080p.
%, consistent with the recommendation in~\cite{youtube-bitrate-levels,netflix:per-title}.
%wb, this bitrate ranges are for YouTube encoded videos
%The average bitrate is 73 to 100 kbps for the lowest track, and 2.1 to 2.8 Mbps for the highest one.
%For the lowest track, the average bitrate is from 73 to 100 kbps; for the highest track, the average bitrate is from 2.1 to 2.8 Mbps.
%
%We next explain in detail how we obtain our dataset.

\textbf{VBR Videos.}
All the 18 VBR videos were encoded by YouTube (YouTube has adopted VBR encoding~\cite{lin2015multipass}). Four videos, Elephant Dream (ED), Big Buck Bunny (BBB), Sintel, and Tears of Steel (ToS), were encoded from publicly available \emph{raw} videos~\cite{xiph}.
%, which can be directly used to obtain perceptual quality (see \S\ref{sec:perceptual-quality}).
Specifically, we uploaded the raw videos to YouTube and downloaded the encoded videos using \texttt{youtube-dl}~\cite{youtube-dl}.
These four videos are in the categories of animation and science fiction.
%To make  video dataset more diverse,
We further downloaded 14 other videos, in a wide range of categories, including sports, animal, nature, action movies, family drama, comedy, and documentary, using \texttt{youtube-dl}.
%
%For these four videos, we use the track at 1080p as the reference track to measure the video quality of the lower tracks, as we do not have the raw source for these videos.
%As the raw video footage is of higher quality than the 1080p track, the VMAF values might be higher than the values when using the  raw video footage as the reference video. On the other hand, we expect that this compromise only leads to slight overestimation, particularly considering the small screen size of a phone.
%
All the above videos are encoded using the H.264 codec~\cite{h264}, with chunk duration of around 5 seconds, consistent with~\cite{lin2015multipass} (the encoding is multi-pass; readers can find the detailed encoding settings in~\cite{lin2015multipass}). Ten out of the aforementioned 14 videos were also available in another codec, VP9~\cite{vp9}, encoded by YouTube.
%\yanyuan{(The readers are refered to ~\cite{lin2015multipass} for the detailed encoding settings of YouTube service).}
In addition, we further use \texttt{FFmpeg} to encode the four publicly available videos using a more recent and efficient codec, H.265~\cite{h265},
following  the ``three-pass'' encoding
	%described
	in~\cite{de2016large}.\fengc{In \S\ref{quality-vs-bitrate}, we investigate the characteristics of H.264, H.265 and VP9 encoded YouTube videos.}
%\yanyuan{The H.265 encodings follow the ``three pass'' encoding described in ~\cite{de2016large} with same encoding parameters except the 5 seconds chunk duration.}

%{The average bitrate ranges $73-100$~kbps for the lowest track, and $2.1-2.8$ Mbps for the highest one (i.e., 1080p).}

\textbf{CBR Videos.}

We also created CBR encodings for the four publicly available raw videos  using  \texttt{FFmpeg}~\cite{FFmpeg}, a popular open source encoder.
Specifically, we used the one-pass CBR encoder, the default  CBR encoder in \texttt{FFmpeg}, which is often used instead of multi-pass encoders due to latency considerations, particularly in live streaming.
Each video is encoded using H.264 into six tracks (144p to 1080p), and each track is segmented into 5-sec chunks, consistent with the ABR track configurations
%and characteristics
of YouTube.\footnote{As an example, the command for encoding an input video into a 1080p track is  {\small \texttt{ffmpeg -i  inputvideo -b:v 2500k -minrate 2500k -maxrate 2500k -bufsize 2500k  -r 24 -profile:v high -x264-params nal-hrd=cbr:keyint=120 :min-keyint=120:scenecut=0 -vcodec libx264 -vf scale=1920:1080  -preset fast output.mp4.}}
}

%\vspace{-.1in}
\subsection{Video Quality Metrics} \label{sec:perceptual-quality}
%We use multiple qulity metrics to measure the video quality, and explore the tradeoffs of quality and bitrate.
%Specifically, we use three quality metrics,
We use two quality metrics, Peak Signal-to-Noise Ratio (PSNR)
%, Structural Similarity Index (SSIM)~\cite{Wang04:SSIM},
and Video Multimethod Assessment Fusion (VMAF)~\cite{vmaf}.
%The results below primarily focus on VMAF (the results for PSNR and SSIM are consistent).
PSNR is a traditional image quality metric.
%The PSNR of a chunk is represented as the median PSNR of all frames in that chunk. SSIM is a widely used perceptual quality metric; it utilizes local luminance,
%contrast and structure measures to predict quality.
%PSNR and SSIM are traditional quality metrics.
VMAF is a recently proposed perceptual quality metric that correlates quality strongly with subjective scores, and has been validated independently in~\cite{Rassool17:VMAF-rep}.
%\senc{There are two VMAF models: \emph{TV model} for larger screens (TV, laptop, tablet), and \emph{phone model} for smaller screens (phones).}
VMAF provides different models tailored to various screen sizes, such as phone and TV.
We focus on the VMAF phone model
%(for smaller screens)
in this chapter since phones are the dominant platform for viewing videos over cellular networks.
%In the following, we only present the results under VMAF phone model; the results under other quality metrics show similar trend.
A VMAF score is between 0 and 100: a score of 0-20 is considered as unacceptable,
20-40 as poor,
40-60 as fair,
60-80 as good,
and 80-100  as excellent~\cite{vmaf}.
Similarly, different ranges of PSNR values are used to categorize picture qualities~\cite{de2016large}.
%The aggregate VMAF of a chunk can be calculated in multiple ways~\cite{VMAF-aggr}.
The aggregate VMAF of a chunk is set as the median VMAF of all the frames in a chunk (using mean leads to similar values for videos in our dataset)~\cite{VMAF-aggr}.
The same approach is used for PSNR.
% and SSIM.
%of a chunk is represented as the median PSNR of all frames in that chunk.

To calculate PSNR and VMAF for a video, we need a \emph{reference video}, i.e., a pristine high quality copy of  the video against which to compute these metrics. For the four VBR and CBR videos that were encoded using the publicly available raw videos, the corresponding raw videos are used as the reference videos. For the other VBR videos
downloaded from YouTube, we do not have the raw video footage, and use the top track (1080p) as the reference track to measure the video quality of the lower tracks.
To understand the impact of this approximation, we calculate the quality values for the four videos that we have raw videos under two options, one with the raw video and the other with the 1080p track as the reference video.
%Not surprisingly, the latter leads to higher VMAF values. In addition, we also observe
We find empirically that the latter leads to lower variability across the chunks in the same track for both PSNR and VMAF,
% \yanyuan{(e.g. for VMAF, the median value is increased by 0.3 to 3.8 while the std recuced by 0.3 to 2.3)},
a point that we will return to in \S\ref{quality-vs-bitrate}.

\begin{figure}[t]
	\centering

	\subfigure[VBR, VMAF\label{subfig:pdf}]{%
		\includegraphics[width=0.43\textwidth]{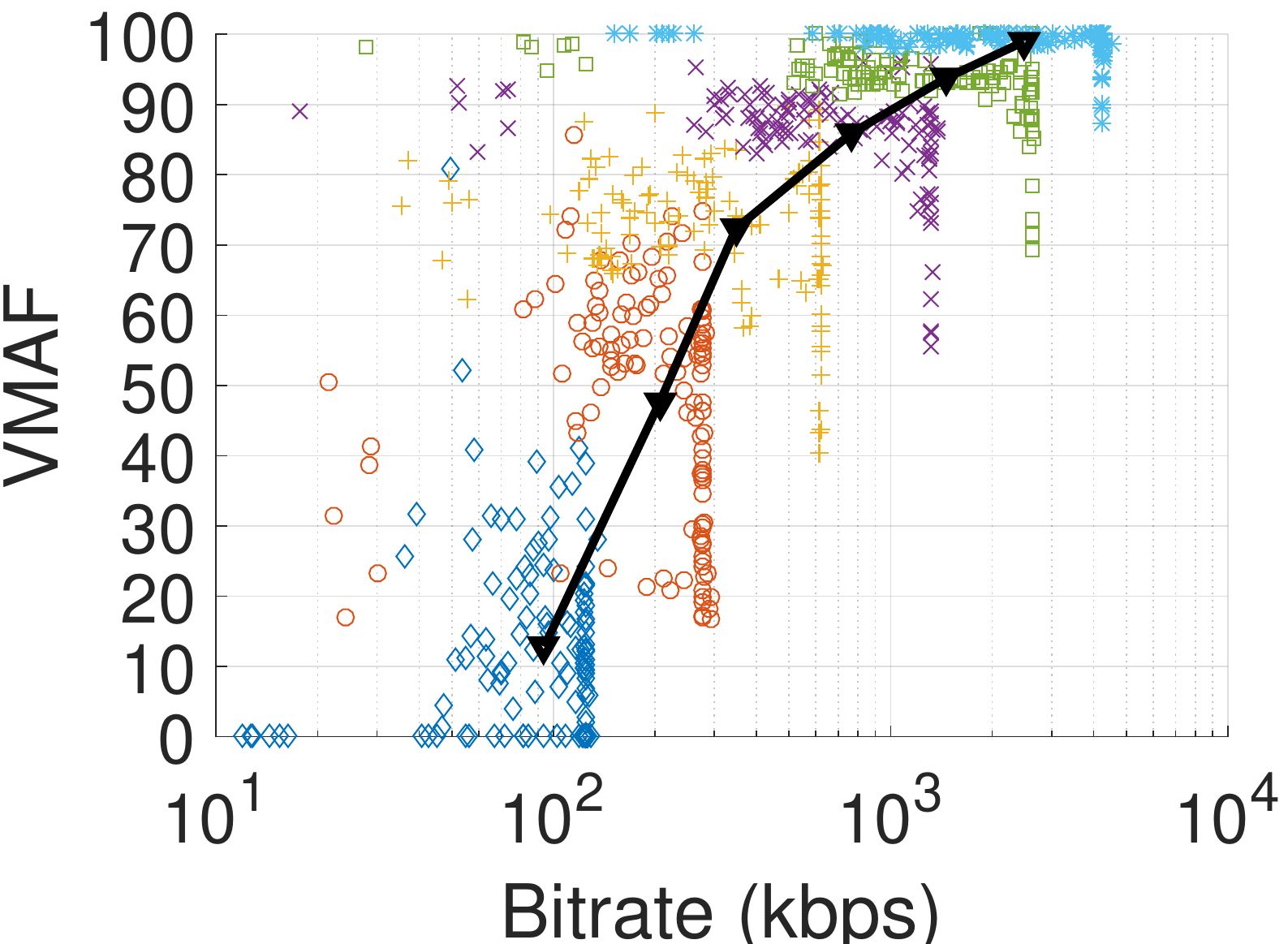}
	}
	\subfigure[VBR, PSNR\label{subfig:RDcurve}]{%
		\includegraphics[width=0.43\textwidth]{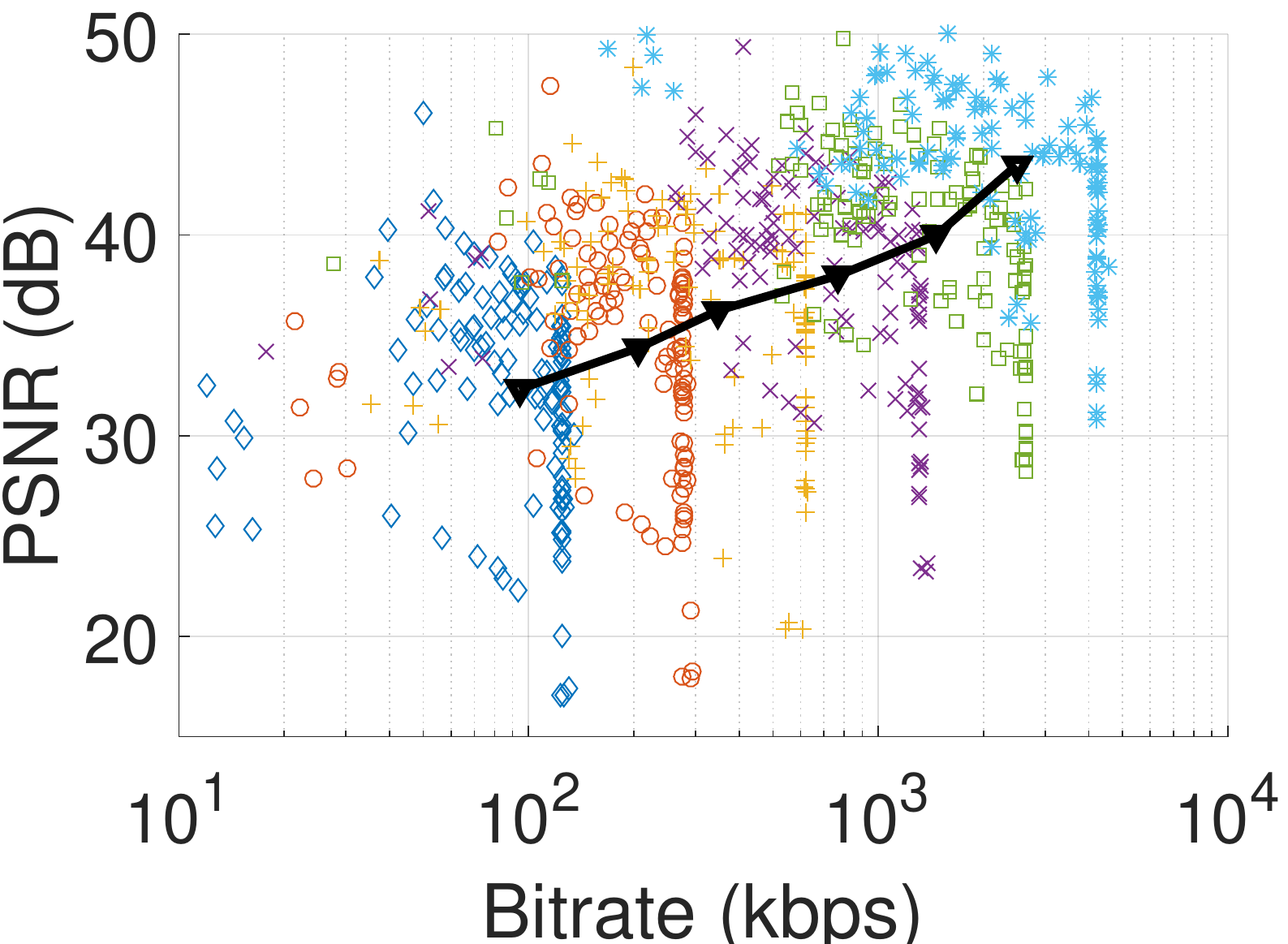}
	} \\

	\subfigure[CBR, VMAF\label{subfig:RDcurve}]{%
		\includegraphics[width=0.43\textwidth]{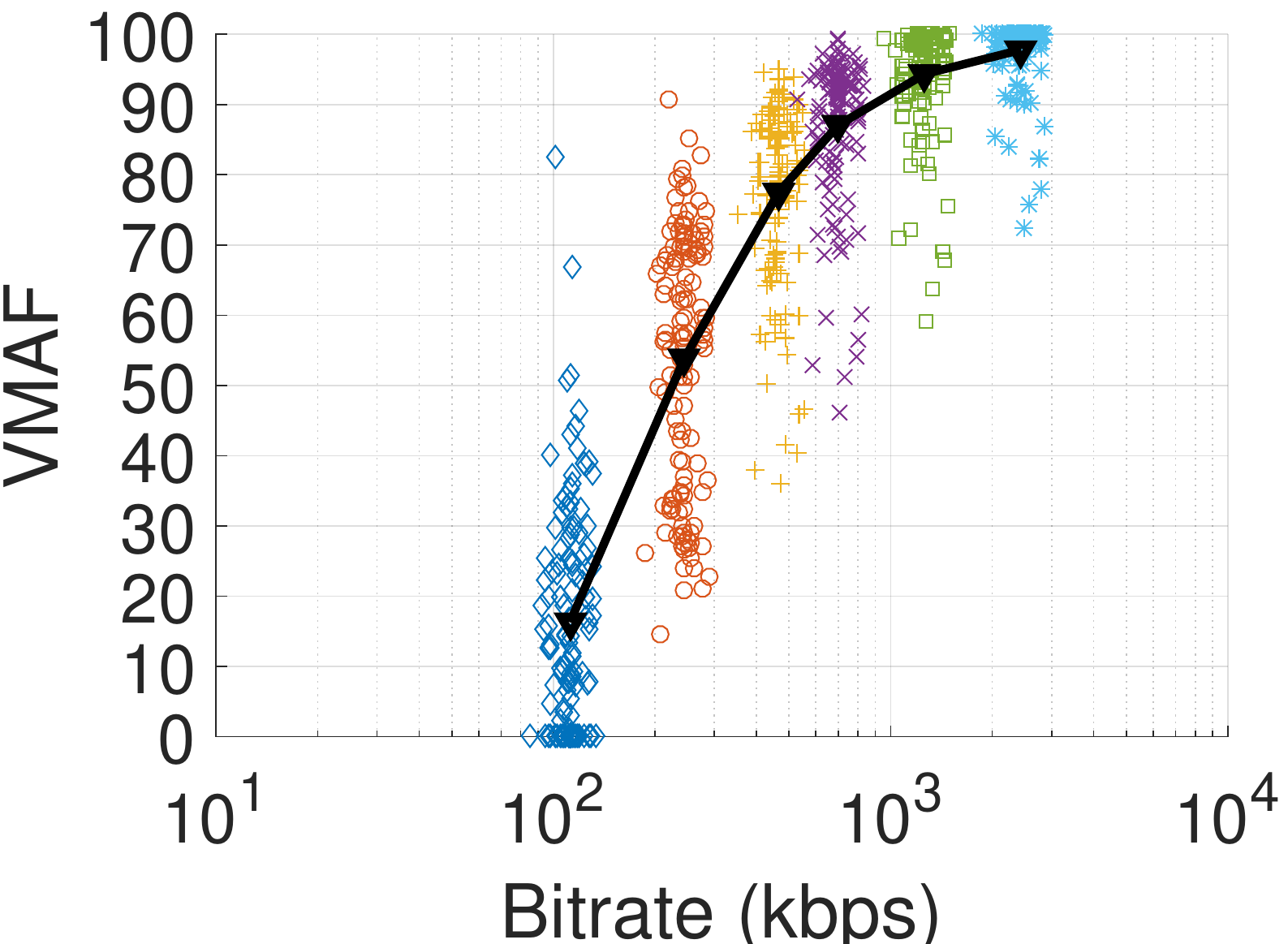}
	}
	\subfigure[CBR, PSNR\label{subfig:pdf}]{%
	\includegraphics[width=0.43\textwidth]{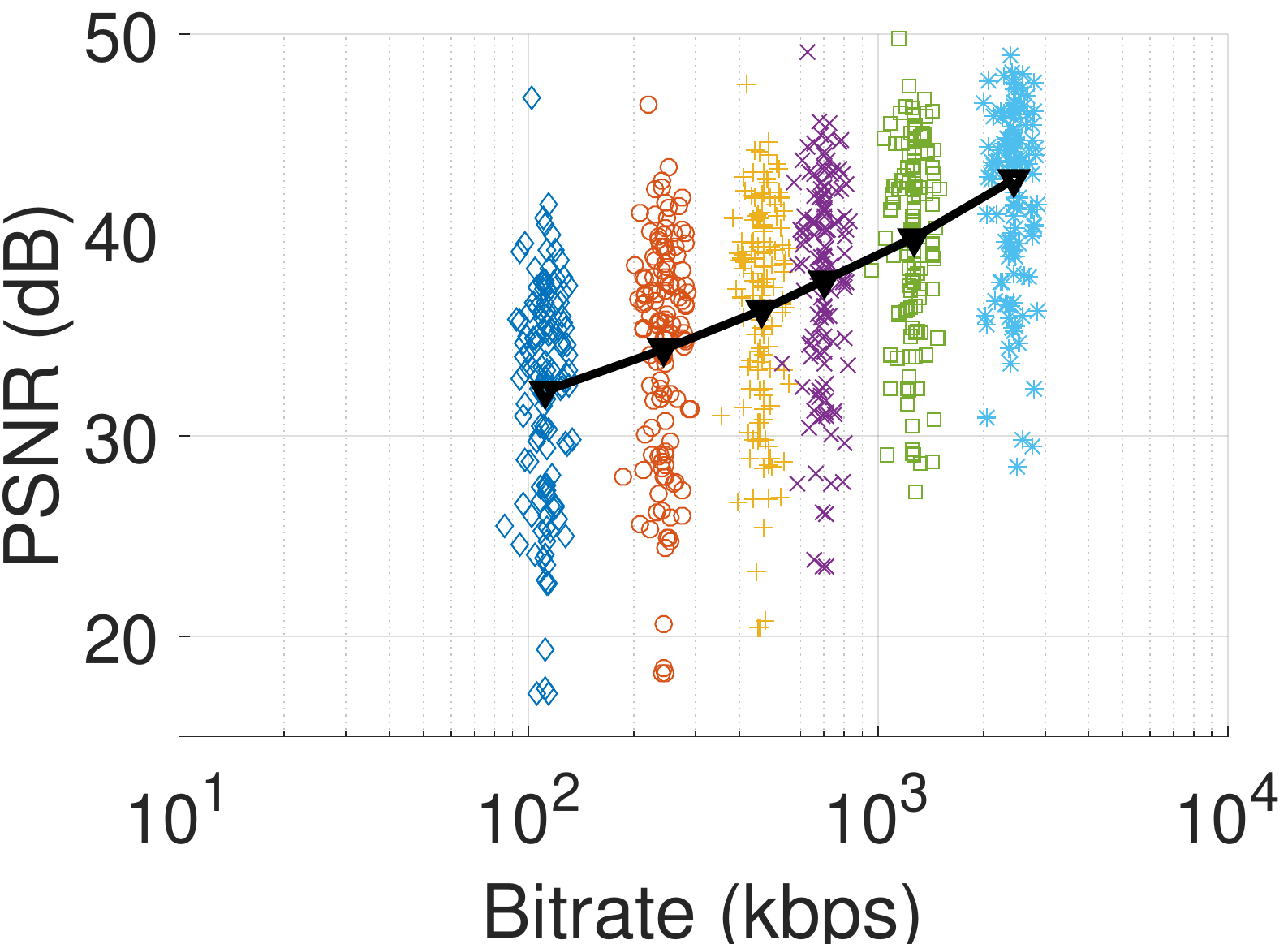}
	}

	\caption{Quality vs. bitrate for one video (ED).}
	\label{fig:quality-bitrate}
	%\vspace{-.1in}
\end{figure}

% for a set of VBR and CBR videos
\begin{figure}[t]
	\centering

	\subfigure[YouTube VBR (H264) \label{subfig:pdf}]{%
		\includegraphics[width=0.43\textwidth]{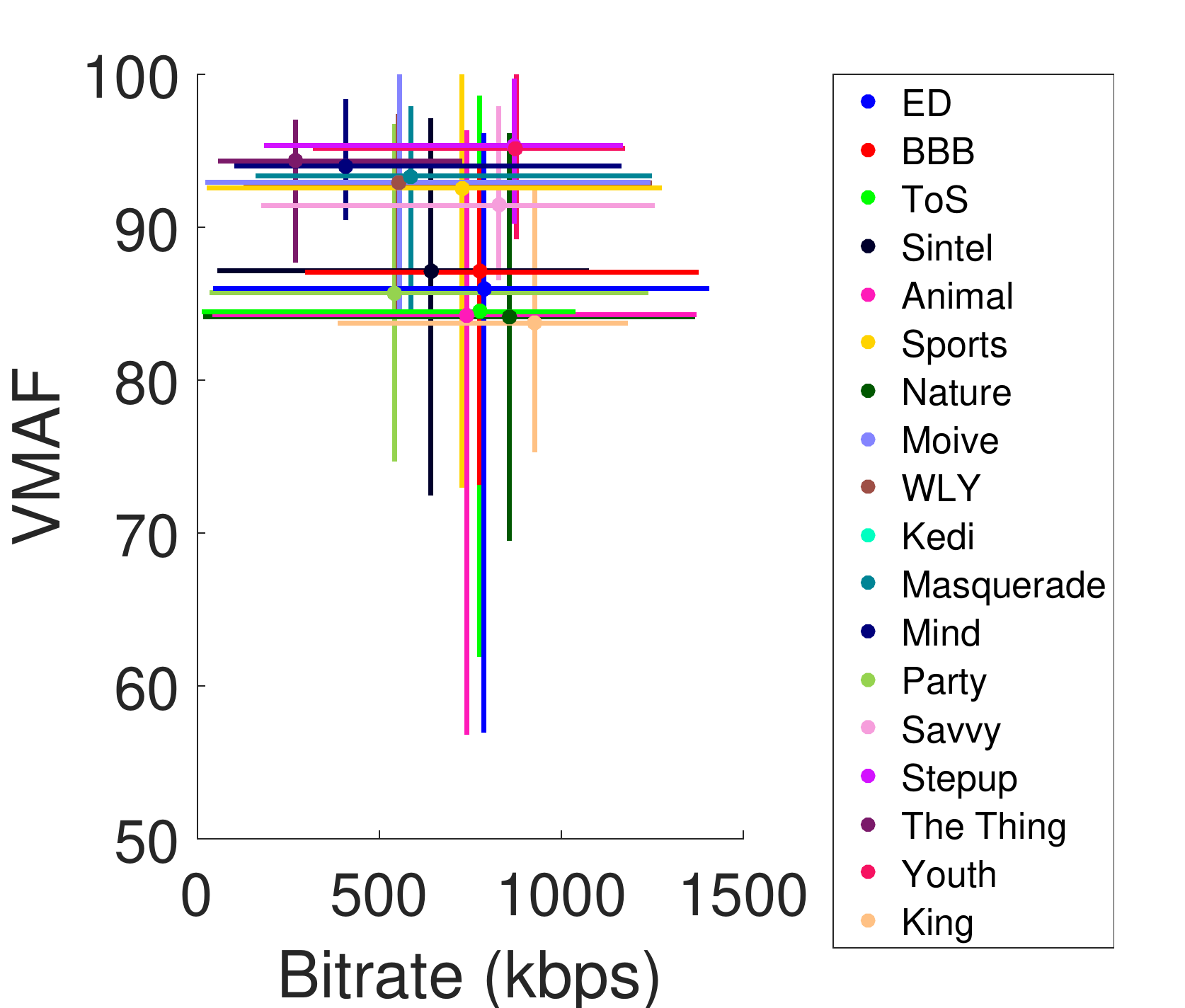}
	}
	\subfigure[FFmpeg CBR (H264)\label{subfig:RDcurve}]{%
		\includegraphics[width=0.43\textwidth]{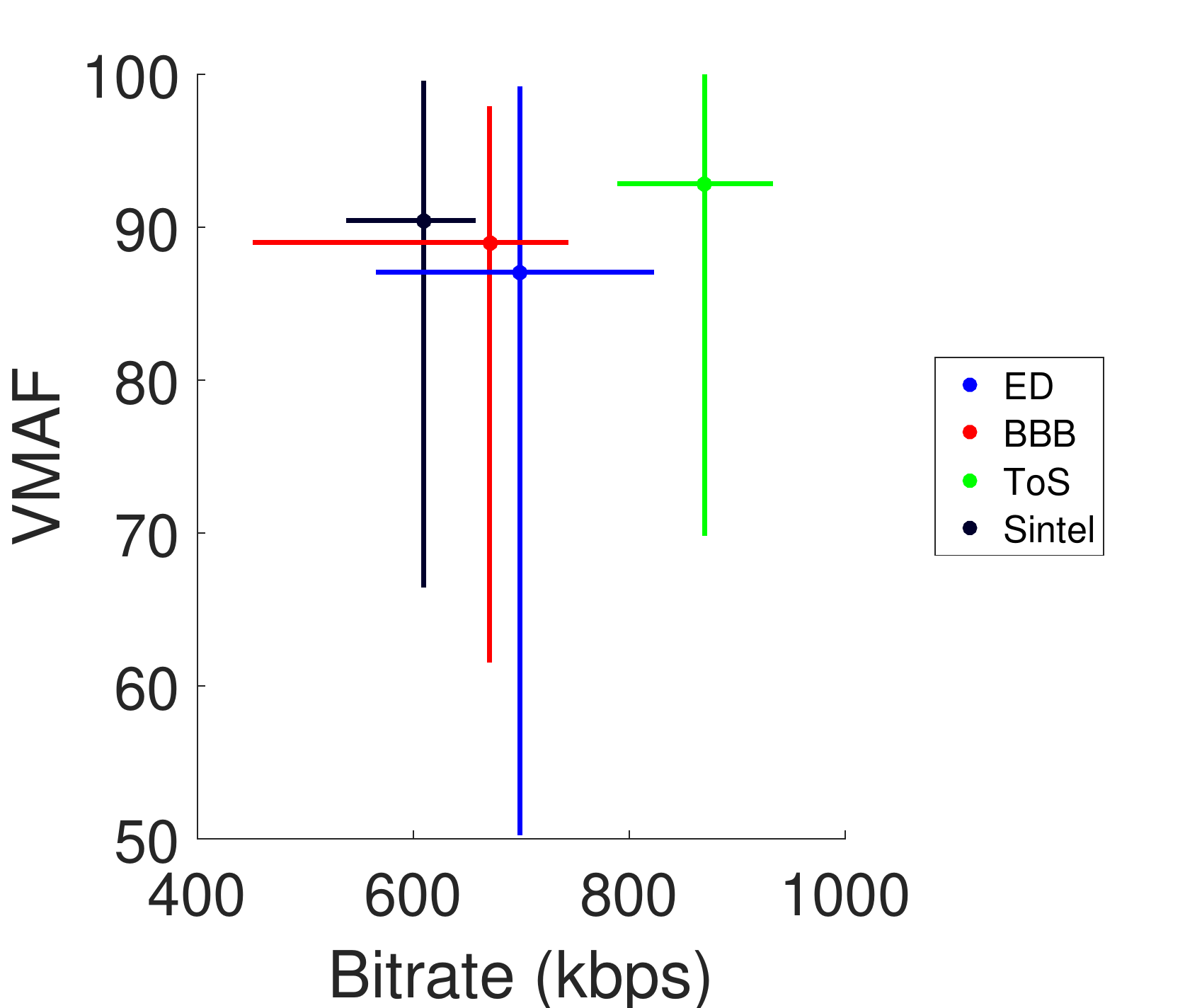}
	}

	\caption{Quality vs. bitrate for track 4 (480p).}
	\label{fig:RD-all-video}
\end{figure}

\subsection{Perceptual Quality vs. Encoding Bitrate} \label{quality-vs-bitrate}
%It is well known that increasing bitrate for an encoding leads to diminishing gains in viewing quality~\cite{}.
We first use an example video
%(of VBR and CBR encodings)
to illustrate the tradeoffs between quality and bitrate, and then describe the observations across the video dataset.
Figures~\ref{fig:quality-bitrate}(a) and (b) plot quality versus bitrate for a VBR video (with 123 chunks, each of 5-sec duration, and 6 tracks), using VMAF and PSNR as the quality metric, respectively. In these figures, different colors  represent the chunks in the different tracks; the black curve shows the average quality versus average bitrate for each track.
%This video has 123 chunks (each of 5.33 seconds) and 6 tracks.
%
%\shuai{We should remind reviewers that the x-axis is in log scale. The diminishing return (i.e. decreasing slope) is much more severe than linear scale.}
%
We see a diminishing gain in increasing bitrate on quality
%for all the chunk indices
(note the log scale in x-axis), consistent with the observations in~\cite{de2016complexity}. Furthermore, even under VBR encoding,  the chunks in the same track have significantly variable perceptual quality (a VMAF difference of $6$ or more would be noticeable to a viewer~\cite{Ozer:vmaf-jnd,de2016complexity}). Specifically, for a track, the standard deviation of quality across the chunks varies from 2 to 14 for VMAF, 4.1 to 5.4 in PSNR (note  PSNR values in log scale), %and 0.02 to 0.09 in SSIM (0-1), \bingc{double check the values,}
with the {middle} tracks exhibiting more variability. We make similar observations for CBR video. Figures~\ref{fig:quality-bitrate}(c) and (d) plot VMAF quality versus bitrate for a CBR video, where all the chunks in same track have similar bitrates.
%We observe the same two quality and bitrate relationships: (1) diminishing gain in increasing bitrate on quality improvement, and (2) significantly variable quality for the chunks in the same track, even more significant than that in VBR encoding.

Fig.~\ref{fig:RD-all-video} plots the quality (in VMAF) and bitrate variations for track 4 for all the videos we consider (18 VBR videos in Fig.~\ref{fig:RD-all-video}(a) and 4 CBR videos in Fig.~\ref{fig:RD-all-video}(b)); the results under PSNR show a similar trend. Each video is represented by two error bars, representing the 1st and 99th percentile in quality and bitrate, respectively.
%\yanyuan{10\% chunks with the largest variability is missing in the figures if we use 5\% to 95\%}
We see the same two observations hold for all the videos we consider. Note that Fig.~\ref{fig:RD-all-video}(a) includes the 14 VBR videos for which we use the 1080p tracks as the reference videos (due to unavailability of the raw videos),  which tend to underestimate the quality variability across the chunks in the same track (see empirical observations in \S\ref{sec:perceptual-quality}). We see significant variability even with the underestimation.
%\yanyuan{we also check another widely used codec -- VP9 from YouTube for the same set of videos, our above observations still hold. Moreover, we check another popular video streaming service--Twitch. We select five popular videos in different genres based on the number of viewers. The result shows that the quality various even within the same track.
%}
%This holds obviously  for Constant-Bit-Rate (CBR) encodings,  which encode the entire video at a relatively fixed bitrate, allocating the same bit budget to  both \emph{simple scenes} (i.e., low-motion or low-complexity scenes) and \emph{complex scenes} (i.e., high-motion or high-complexity scenes). More interestingly and somewhat counterintuitively, this holds  even for  real-world Variable-Bit-Rate~(VBR)  encodings which aim to encode \emph{simple scenes}  with fewer bits and \emph{complex scenes} with more bits to achieve a  relatively more consistent quality throughout the track.
%The above results use VMAF as the perceptual quality metric. We observe similar results when using PSNR and SSIM. In particular, the CoV varies from 0.01 to 0.17 in PSNR and 0.02 to 0.1 in SSIM across the tracks.

The above results are for H.264~\cite{h264} videos. To ensure that the above observations are general, we further investigate the quality and bitrate relationships for other encoders and content. Specifically,
we examine (i) H.265~\cite{h265}, a more recent and efficient codec than H.264, using the four publicly available videos,  (ii) VP9~\cite{vp9}, another widely used codec in YouTube, using ten VP9 videos downloaded from YouTube, and (iii) Twitch, another popular video streaming service, by selecting five popular videos in different genres based on the number of viewers. For all the three cases, our results confirm
%similar observations as described earlier:
the same two quality and bitrate relationships observed earlier: (1) increasing bitrate leads to diminishing gain in quality improvement, and (2) the chunks in the same track have significantly variable quality.
%, even more significant than that in VBR encoding.
%Last, the above two observations on YouTube, \texttt{FFmpeg} and Twitch encoded videos
These two observations are consistent with the results of Netflix encoded videos~\cite{de2016complexity,de2016constant}, indicating that they hold widely across encoding platforms and videos.

The property that chunks within the same track have highly variable quality holds obviously  for CBR encodings,  which encode the entire video at a relatively fixed bitrate, allocating the same bit budget to both \emph{simple scenes} (i.e., low-motion or low-complexity scenes) and \emph{complex scenes} (i.e., high-motion or high-complexity scenes). The fact that it also holds for VBR is somewhat counterintuitive since
VBR allocates bits according to scene complexity to achieve a more consistent quality throughout a track.
%VBR aims to encode \emph{simple scenes}  with fewer bits and \emph{complex scenes} with more bits to achieve a  relatively more consistent quality throughout the track, and is
Part of the reason is the inherent complexity of encoding and the difficulty of handling scenes of diverse complexities~\cite{de2016complexity,lin2015multipass}.

\section{Reducing Data Usage} \label{sec:reduce-data}
The diminishing gains in increasing bitrate on quality improvement demonstrated in the previous section indicate that an ABR logic that simply aims to maximize quality is not bandwidth efficient.
% \yanyuanc{In cellular networks, where data is a scarce resource, achieving high bandwidth efficiency is tremendously beneficial for both network providers and end users.}
  %\yanyuan{Based on current practice, the cost of data plan service is proportional to the amount of data budget. From the user perspective, reducing data rate but having optimized QoE will allow them watch more quality videos without exceeding data plan limit (i.e. pay extra money for the data).}
\fengc{Several mobile network operators and commercial video service providers have provided users with some data saving options.}
%This motivates the practices for limiting bandwidth for video streaming
 %(see \S\ref{sec:current-practice-save-data})
In the following, we first describe the current data saving practices and show that they are inefficient. We then propose a quality-aware strategy for reducing data usage.

\subsection{Current Data Saving Practices} \label{sec:current-practice-save-data}

Certain cellular network operators provide users with options to limit the network bandwidth for a streaming session (e.g., ~\cite{attstreamsaver, tmobilebingeon}).
%(e.g., AT\&T Stream Saver~\cite{attstreamsaver}, T-Mobile Binge On~\cite{tmobilebingeon}).
%AT\&T Stream Saver~\cite{attstreamsaver} (automatically turned on for some AT\&T users), the network speed for content that is recognized as video is capped at 1.5 Mbps, which generally is equivalent to Standard Definition (SD) quality, similar to DVD (about 480p).
The rationale is that the bandwidth cap may lead an ABR player to avoid bandwidth-consuming High-Definition (HD) tracks so as to save data.

\begin{table}[t]
	\centering
	\setlength{\arrayrulewidth}{1pt}	
	\begin{tabular}{lllll}
		Option & Top Track & Declared Bitrate & Resolution \\ \hline
		Data saver         & 3      & 120kbps    & 288p                   \\
		Good               & 6     & 450kbps     & 360p                 \\
		Better             & 7    &  650kbps     & 396p                 \\
		Best               & 8    &  1000kbps    & 480p
	\end{tabular}
	\caption{Data saving options in Amazon Prime video.}
	\label{table:Amazonphone}
	\vspace{-0.3in}
\end{table}

Various commercial video streaming services have also provided users with options
to save  data.
For instance, the YouTube phone app provides
%a data saving
an option called ``Play HD on Wi-Fi only'', i.e.,
%HD videos will only be streamed when a phone is connected to WiFi networks; when connected to cellular networks,
only Standard Definition (SD) videos will be streamed over cellular networks.
To understand the behavior of this option, we stream eight videos in different categories using the YouTube app over a commercial LTE network.
We observe that, when the option is on, even if the network bandwidth is very high (over tens of Mbps), the 480p track is selected throughout the video.\fengc{(consistent with the specification: tracks above 480p are considered HD).} The fact that the selected tracks  never exceed 480p despite significantly higher bandwidth indicates that the data saving is achieved by capping the top track to the 480p track. Henceforth, we refer to this practice as \emph{Track-based Filtering (TBF)}. We find that data saving options in the Amazon Prime Video app are also achieved by capping the top track.  Table~\ref{table:Amazonphone} summarizes the measurement results, showing the top track for the four options varies from track 3 to 8.

The above two current practices both have drawbacks. The network-based approach forces an ABR scheme to choose lower tracks due to the network bandwidth limit. It provides no explicit control on what quality will be chosen for a particular chunk position, thus leading to highly variable quality across the rendered chunks (\S\ref{sec:network-vs-prefiltering}).
The practice of TBF  does not account for the high quality variability across chunks within the same track. For example, the purple points in Fig.\ref{fig:quality-bitrate}(b) represent chunks encoded at track 4 (or 480p). When using TBF with 480p as top track\fengc{(used by YouTube and Amazon)}, the quality for some chunk positions is lower than 60  (i.e., the threshold for good quality in VMAF~\cite{vmaf}),
\emph{no matter what ABR scheme is being used and how much network bandwidth is available}.

%\subsection{Problem Formulation}

\subsection{Quality-aware Data Saving}
To address the drawbacks of the current data saving practices, we propose a quality-aware strategy for reducing data usage.
We assume that a user is provided with multiple viewing quality options (e.g., good, better, best),
with the understanding that the saving is higher under a lower viewing quality option and vice versa. For an option chosen by a user,  the player will map it to a particular quality value, referred to as \emph{target quality},
and the goal of the ABR logic is to maintain the quality to be close to the target quality, subject to the network bandwidth constraints.
In contrast to the current data saving practices, the above target quality based strategy provides data savings while directly controlling the quality level.

The target quality can be specified in terms of a wide range of perceptual quality metrics. While there is no {single} agreed-upon way of defining a good perceptual quality, existing literature has established certain metrics, e.g., through threshold values in VMAF and PSNR~(see~\S\ref{sec:perceptual-quality}).
As an example, VMAF values of 60 and 80 are the thresholds for ``good'' and ``very good'' quality, respectively~\cite{vmaf}.
The player can set the target quality in VMAF values based on the viewing quality option that a user chooses: 
when the ``good'', ``better'', ``best'' option is chosen, the target quality is set to VMAF 60, 70, 80, respectively.\fengc{when  ``good'' option is chosen, the target quality is set to VMAF 60; when ``best'' option is chosen, the target quality is set to 80; and when ``better'' option is chosen, the target quality is set to 70.}
Users do not need to numerically specify the target quality. Instead, they only need to select a desired viewing quality option such as good/better/best---similar to the current practice in commercial streaming systems.  For a given video, the quality metrics such as PSNR and VMAF can be calculated by the server after the video is encoded, and then shared with the client.\fengc{We will discuss possible mechanisms for a server to make the quality information available to a client in \S\ref{sec:cbf-deployment}.}
In addition,\fengc{In addition to what is described above, where a video  player determines a target quality based on a viewing quality option specified by a user,}
a video player can also automatically decide the target quality
based on the user's cellular data plan, the cellular data budget, the video content type, and the user's historical preferences. Furthermore, the target quality can be changed over time during the playback; our schemes in \S\ref{sec:need-for-efficiency} and \S\ref{sec:design} can be applied to dynamic target qualities.
%In contrast to the current data saving practices, the above target quality based strategy provides data savings while directly controlling the quality level.

\section{ Chunk-based Filtering (CBF)} \label{sec:need-for-efficiency}
We first describe the CBF approach, and then detail its deployment scenarios and how to leverage it in ABR streaming.

\subsection{CBF Approach}
CBF is motivated from the two video quality and bitrate tradeoffs  in \S\ref{sec:quality-bitrate-tradeoffs}, i.e., (i) increasing bitrate leads to diminishing gain in improving quality, and (ii) the chunks in the same track exhibit significantly variable quality.
Specifically, for a given target quality, $Q_r$, CBF filters the tracks that are undesirable on \emph{a per-chunk basis} as follows.
For the $i$-th chunk position, let $q_{i,\ell}$ denote the quality for track $\ell$, which can be obtained right after the encoding process at the server.
Then, for a given $Q_r$, CBF sets the top level for chunk position $i$ to $\bar{\ell}_i$ so that the corresponding quality, $q_{i,\bar{\ell}_i}$, is closest to the target quality, $Q_r$, among all the tracks for chunk position $i$ (i.e., $|q_{i,\bar{\ell}_i}-Q_r|$ is the smallest).
% (note that $q_{i,\bar{\ell}_i}$ can be larger or smaller than the target quality).
In other words, for chunk position $i$,  all the encodings (i.e., tracks) that are above  $\bar{\ell}_i$  will not be considered in ABR streaming.

\begin{figure}[t]
	\centering
	\vspace{-.1in}
	\subfigure[Top track when applying CBF.\label{subfig:CBFLevel}]{%
		\includegraphics[width=0.43\textwidth]{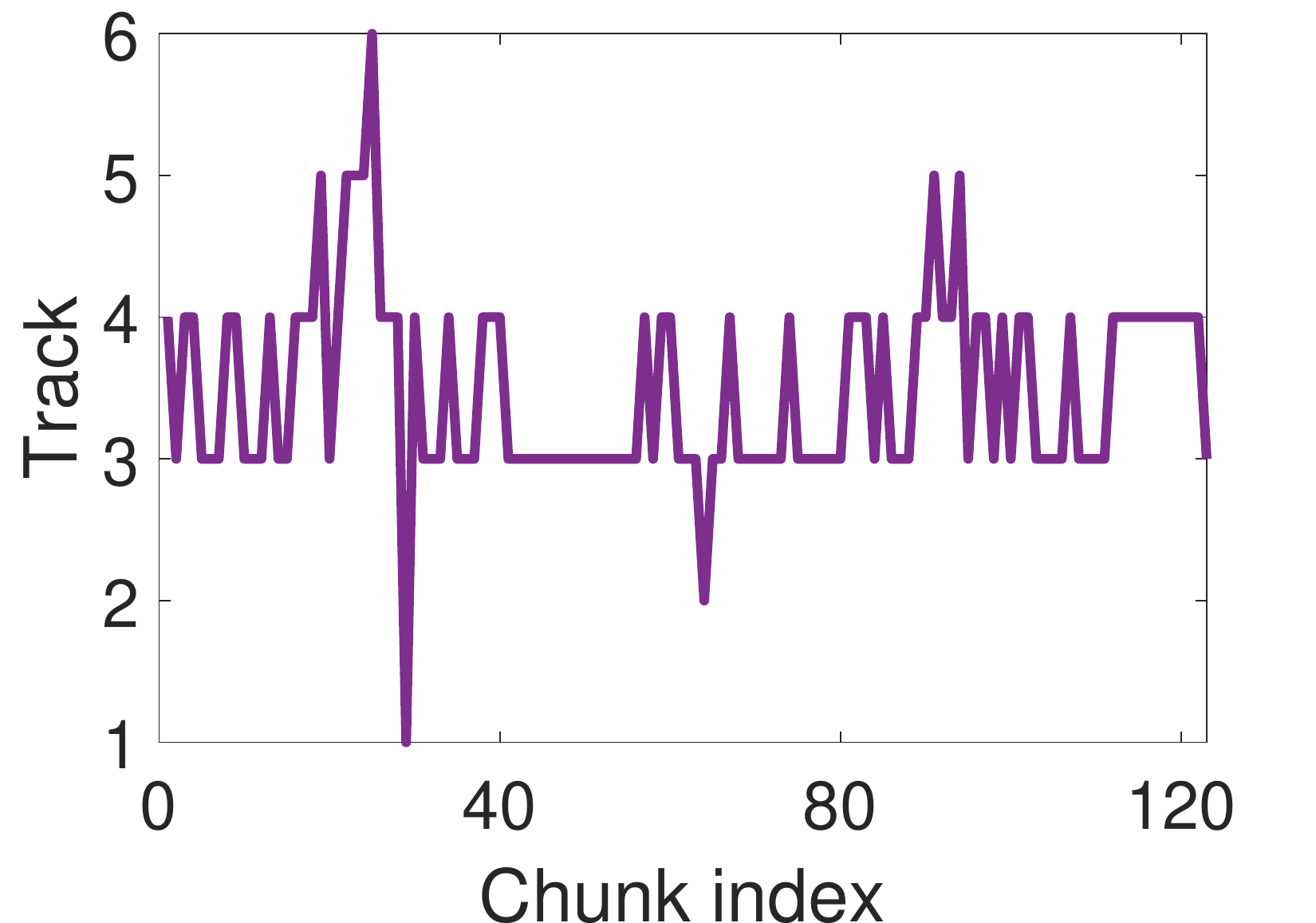}
	}
	\subfigure[Quality for top track with CBF.\label{subfig:CBFQuality}]{%
	\includegraphics[width=0.43\textwidth]{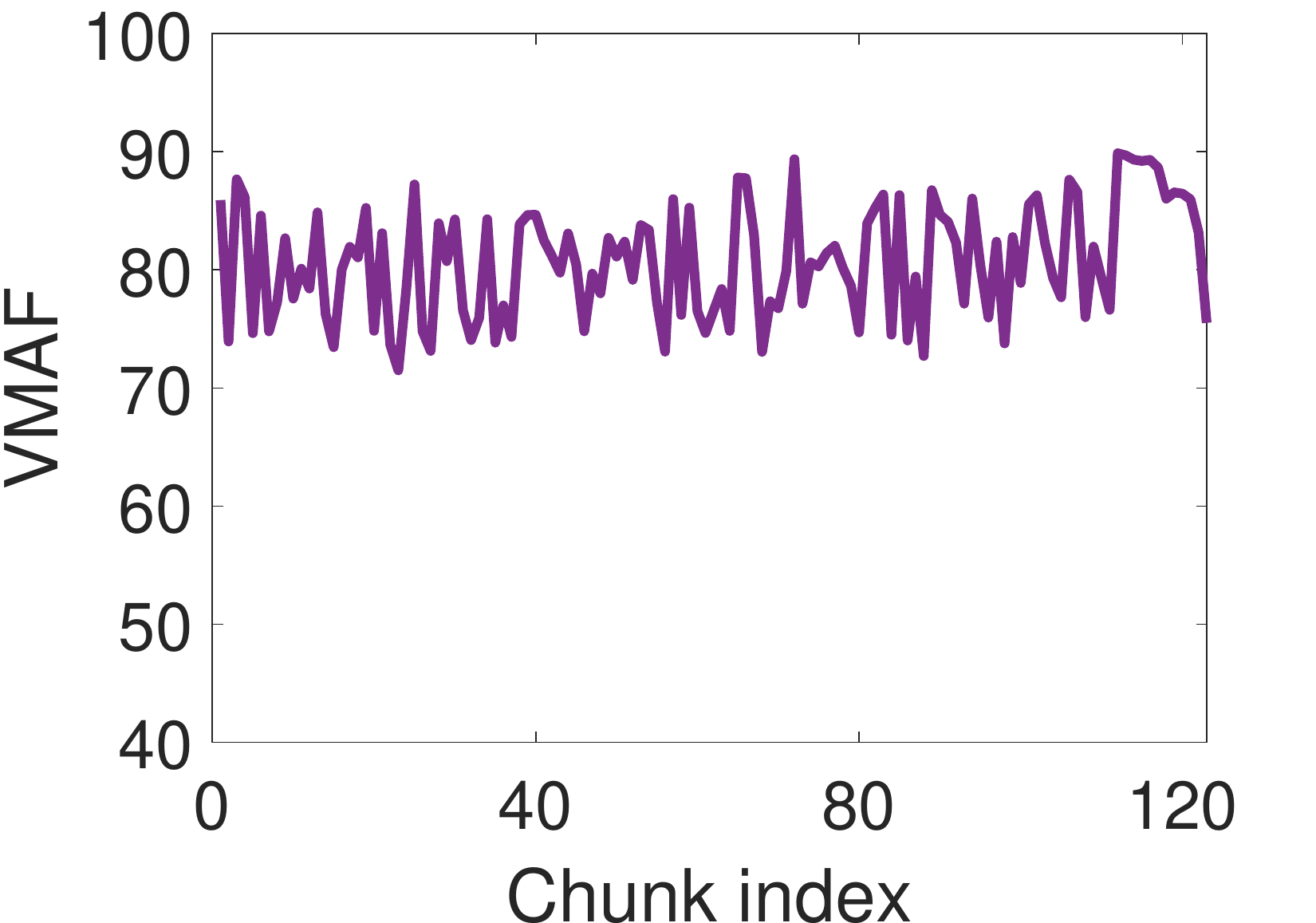}
	}
    \vspace{-.1in}
	%	\caption{Top quality/track when applying CBF (ED, target quality 80).}
	\caption{Illustration of CBF (ED, $Q_r=80$).}
	%YouTube encoded, target quality 80).}
	\label{fig:quality-CBF}
	\vspace{-.1in}
\end{figure}

%Essentially, CBF caps the top track on a per-chunk basis. One example is shown in
Fig.~\ref{fig:quality-CBF}(a) illustrates CBF using an example. It plots the top track after filtering by CBF for each chunk position when the target quality is 80 (VMAF). The video has 6 tracks originally. We see that the top track after CBF varies from $1$ to $6$ for the different chunk positions (with 53\% and 39\% as track 3 and 4, respectively, and 6\% above 4, and 2\% below 3).
As an example, for the 29th chunk, the lowest track (i.e., track 1) is sufficient to achieve the target quality 80. A manual inspection reveals that this chunk contains very simple scenes that require less bits to encode. 
Fig.~\ref{fig:quality-CBF}(b) plots the highest quality variant for each chunk position that CBF retains.
We see that  95\% of the chunk positions have top quality within 10\% of the target quality (i.e., between 72 and 88).

\begin{figure}[t]
	\centering
	\subfigure[Quality for top track]{%
		\includegraphics[width=0.43\textwidth]{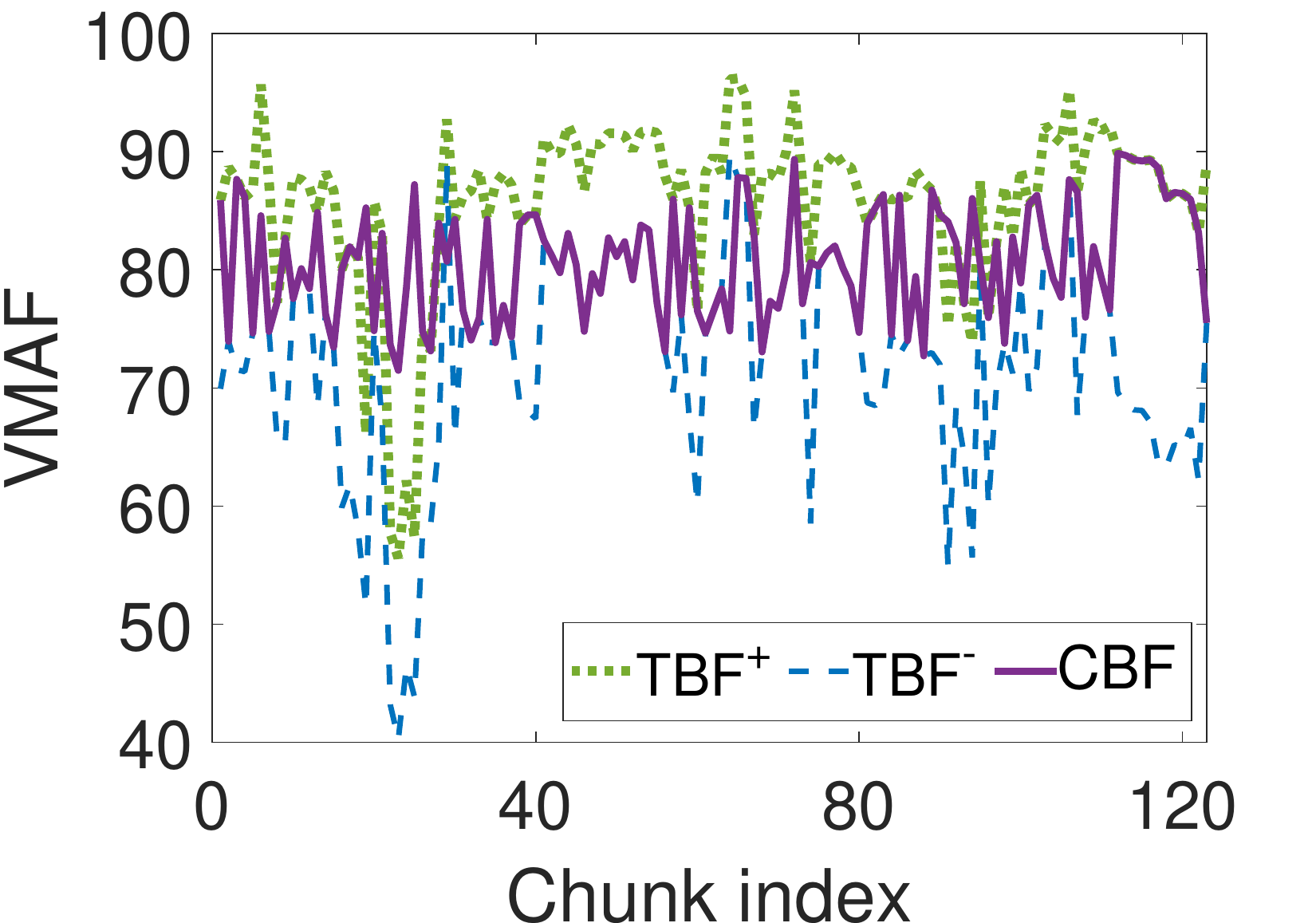}
	}
	\subfigure[CDF for quality of top track]{%
		\includegraphics[width=0.43\textwidth]{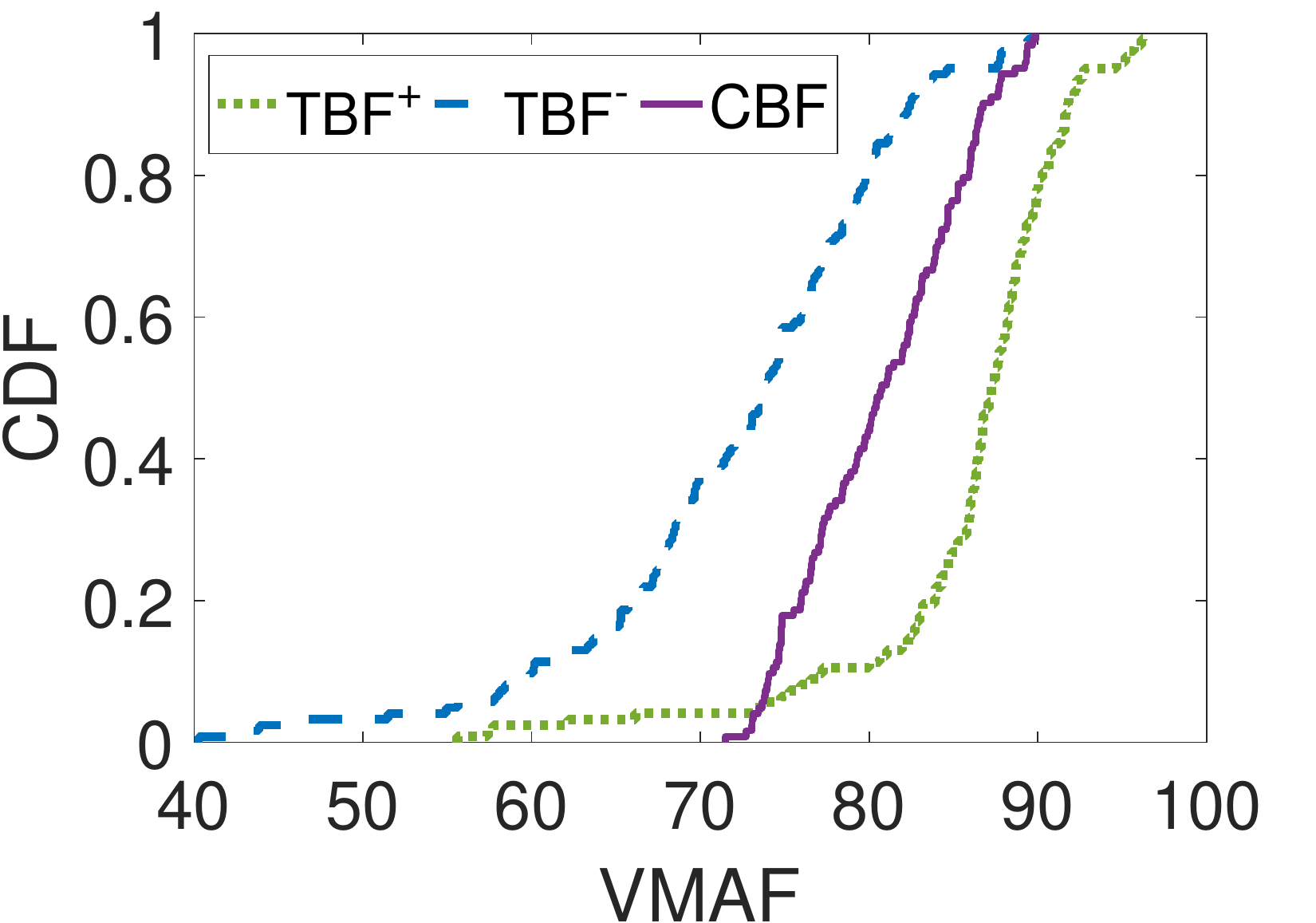}
	}
	\caption{CBF, $\mbox{TBF}^-$, and $\mbox{TBF}^+$ (ED,  $Q_r=80$).}
%	\caption{Quality of top track under CBF, $\mbox{TBF}^-$, and $\mbox{TBF}^+$ (ED, YouTube encoded, target quality 80).}
	\label{fig:quality-CBF-vs-rmTracks}
\end{figure}

\subsection{CBF vs. TBF} \label{sec:cbf-tbf-comp}
We next compare CBF and TBF.
%Existing practice of TBF does not factor in a target quality. \sen{ Meaning?? May need to rephrase.  NOTE: When they select  TBF, sometimes they may do some subjective testing (I hope) to decide the track.
%So one can argue some tenuous connection to quality, though not precise or accurate or even a single target quality.}
%To make CBF and TBF comparable,
Specifically, we consider two variants of TBF as follows.
For a video, let $\widetilde{Q}(\ell)$ represent the average quality of all the chunks in track $\ell$. Let $\ell^{-}$ and ${\ell}^{+}$ denote two adjacent tracks, $\ell^{-}={\ell}^{+}-1$, satisfying that $\widetilde{Q}({\ell^{-}}) \le Q_r$ and $\widetilde{Q}({\ell^{+}}) > Q_r$.
The first variant of TBF, denoted as  $\mbox{TBF}^-$, caps the top track to $\ell^{-}$, i.e., it removes all the tracks that are higher than ${\ell}^{-}$.
The second variant of TBF, denoted as  $\mbox{TBF}^+$, caps the top track to $\ell^{+}$.
%, i.e., it removes all the tracks that are higher than ${\ell}^{+}$.
Clearly, $\mbox{TBF}^-$ is more aggressive in filtering out tracks than $\mbox{TBF}^+$.

%We see from the above example that the track needs to be capped at different values for different chunk indices, which indicates that capping all the chunks at the same level as in track-based filter will lead to suboptimal results. We next show an example.
Fig.~\ref{fig:quality-CBF-vs-rmTracks}(a) plots the quality of the top track for each chunk position under CBF, $\mbox{TBF}^-$, and $\mbox{TBF}^+$ for one video when the target quality is 80.  For this setting, $\ell^{-} =3$ and $\ell^{+} =4$.
In $\mbox{TBF}^+$, all the tracks above track 4 (with resolution 480p) are removed, which coincides with YouTube's data saving option~(\S\ref{sec:current-practice-save-data}).
%what YouTube does in practice when its bandwidth-saving mode is activated~(\S\ref{sec:current-practice-save-data}).
%\senc{(the bandwidth saving option in YouTube removes all HD tracks, i.e., the tracks with resolution above 480p, in cellular networks).}
In Fig.~\ref{fig:quality-CBF-vs-rmTracks}(a), the quality for a given chunk position represents the maximum achievable quality when the network bandwidth is sufficiently large. We see that  the quality under CBF is overall much closer to the target quality than that under the two TBF variants. Fig.~\ref{fig:quality-CBF-vs-rmTracks}(b) shows the cumulative distribution function (CDF) corresponding to the quality values in Fig.~\ref{fig:quality-CBF-vs-rmTracks}(a). For $\mbox{TBF}^-$ and $\mbox{TBF}^+$, only 56\% and 53\% of the chunk positions have quality within 10\% of the target quality, compared to 95\% under CBF.
We observe similar results as above for other videos and target quality investigated.\fengc{This is not surprising. In fact,}
It is easy to prove that, for any chunk position, the top quality under CBF is no farther away from the target quality than that under the two TBF variants.
%\fengc{In \S\ref{sec:track-vs-chunk-perf}, we show that the fine-grain capping of top track in CBF leads existing ABR schemes to better performance than the coarse-grain TBF.}

\subsection{Deployment Scenarios}\label{sec:cbf-deployment}
From a practical perspective, a key advantage of CBF is that it can be  \emph{incrementally} deployed in the existing DASH and HLS streaming pipelines at either the server or client side. In both deployment scenarios, the server does not remove any chunk from its storage. Rather, it modifies or extends the manifest file that it transmits to the client. We next describe the two deployment scenarios, and end the section with a brief description of using CBF in ABR streaming.

\begin{figure}[t]
	\centering
	\vspace{-.1in}
	\subfigure[Server side deployment\label{subfig:RDcurve}]{%
		\includegraphics[width=0.43\textwidth]{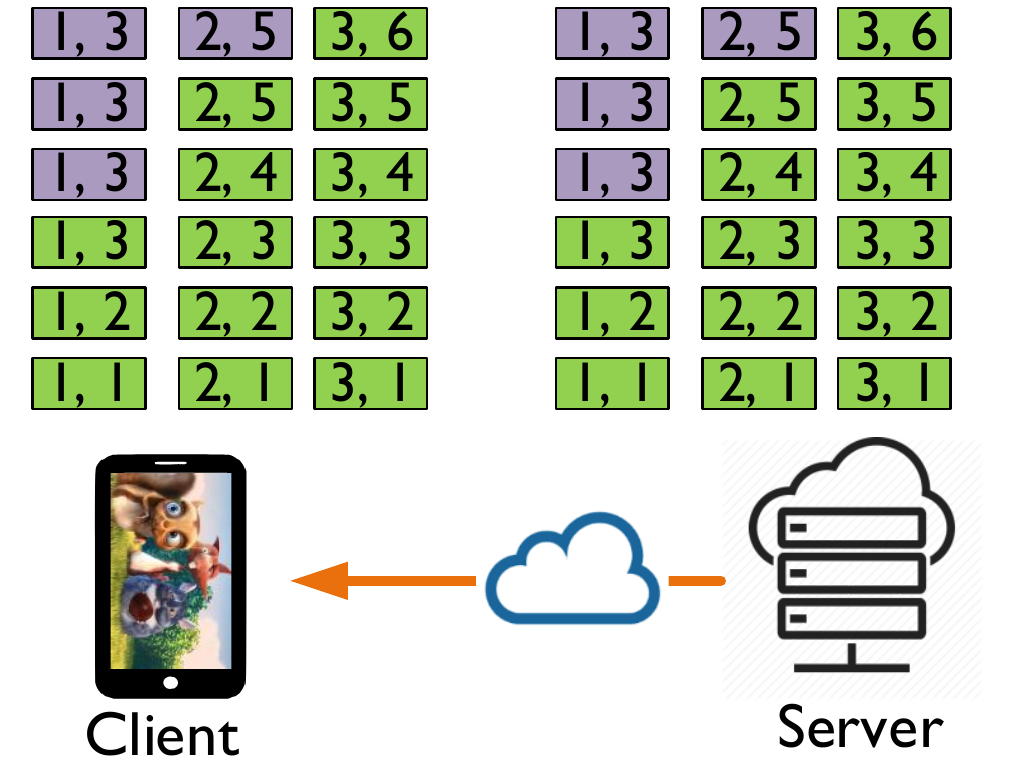}
	}
	\subfigure[Client side deployment\label{subfig:pdf}]{%
		\includegraphics[width=0.43\textwidth]{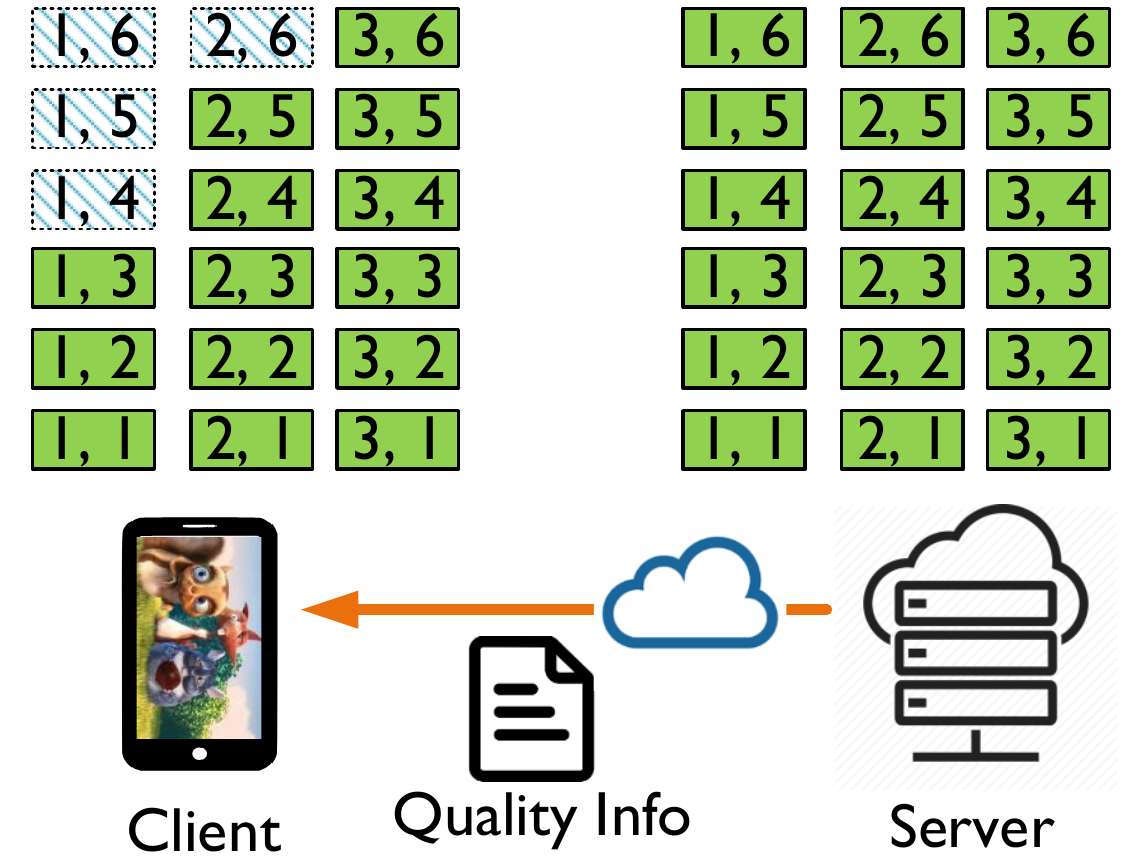}
	}	\vspace{-.15in}
	\caption{Illustration of CBF deployment.}
	% at the server and client, respectively.}
	\label{fig:chunk-level-prefiltering}
	\vspace{-.2in}
\end{figure}

%\bingn
{\textbf{Server Side Deployment.} Two approaches, called \emph{chunk variant trimming} and \emph{chunk variant substitution}, can be used for server side deployment. In {chunk variant trimming}, the server simply modifies the manifest file so that,  for each chunk position, it only lists the chunk variants that remain after CBF filtering. As an example, in Fig.~\ref{fig:chunk-level-prefiltering}(a), for chunk 1, only the three lowest track variants will be listed in the manifest file. In {chunk variant substitution}, the server makes the filtering {completely} transparent to the client {operation} by  substituting the information for certain chunk variants
	%(e.g.,  the URLs or the byte ranges of these chunks,
%	their declared bitrate)
	as follows. Consider the chunks at position $i$. Let $i.\ell$ represent the chunk at level $\ell$. Let $\bar{l}$ denote the top track after the filtering by CBF. Then the server modifies the manifest file so that the information for chunk $i.\ell$, $\ell > \bar{\ell}$, is replaced with the information of chunk $i.\bar{\ell}$.
	% (see an example in Fig.~\ref{fig:chunk-level-prefiltering}(a)).
	%
	In this way, each playback position still has the same number of levels, and the changes through CBF is transparent to the client.
	For example, in Fig.~\ref{fig:chunk-level-prefiltering}(a),
	%for the chunks with index 1,
	for chunk 1,
	levels 4, 5 and 6  are filtered according to CBF;
	the server modifies the manifest file to replace the information for chunks 1.4, 1.5 and 1.6 with that of 1.3.

We have verified that the above two approaches work in the context of both DASH and HLS protocols and common packaging formats such as Fragmented MP4 and MPEG-2 TS~\cite{dash-standard,apple-hls}. %The verification is through experiments using ExoPlayer communicating with DASH and HLS servers, respectively. 
The  {chunk variant substitution} approach clearly works when the media format does not include separate initialization segments (i.e., each chunk is self-initializing). It can also be realized for media formats where 
%Even in the case when 
separate initialization segments (each containing information required to initialize the video decoder to decode a particular chunk) are included, as long as a proper initialization segment is specified for each chunk in the manifest file.  We have confirmed  through experiments that both DASH and HLS  have ways to specify such associations, and that, when presented with the appropriately modified manifest file,
the player was able to  correctly decode and play the associated video. In DASH, this can be achieved using ``Period" construct~\cite{dash-standard}. 
In HLS, this can be achieved through extra ``EXT-X-MAP" tags~\cite{apple-hls}.
}

%When CBF is deployed at the server, chunk quality information does not need to be transmitted from the server to the client.
\textbf{Client Side Deployment.} When CBF is deployed at the client, the server further needs to transmit the quality information for each chunk to the client, e.g., by
%\yanyuan{carrying quality metric in ISO base media file~\cite{ISO23001}  or}
including the information in the manifest file, as our prototype implementation of CBF in two ABR streaming platforms (\S\ref{sec:setup}).
%Fig.~\ref{fig:chunk-level-prefiltering}(b) illustrates the deployment at the client.
The client, when making rate adaptation decision for a chunk position,
%after receiving chunk quality information,
will exclude the levels that are above the top track for that chunk position (illustrate as shaded chunks in  Fig.~\ref{fig:chunk-level-prefiltering}(b)).
%In the example, for the chunks with index 1,  the client will exclude levels 4, 5 and 6 (marked by chunks 1.4, 1.5 and 1.6 in the figure) when making rate adaptation decisions.
Note that while quality metrics can be carried in the media file, e.g., following the ISO standard~\cite{ISO23001}, for rate adaptation purposes, the quality information needs to be known to the client beforehand to assist the decision making, instead of extracting after a chunk is downloaded. Therefore, our implementation embeds the quality information in the manifest file instead of the media file.

\textbf{Leveraging CBF in ABR Streaming.} CBF can be  retrofitted to, and improve the performance and data efficiency of  existing ABR schemes  that may not be quality-aware themselves.
This  can be achieved either through server-side or client-side deployment of CBF.
Specifically, under server-side deployment of CBF, an existing ABR scheme simply selects from the remaining levels for each chunk position.
The scheme does not need to leverage or even be aware of any quality information. It can simply aim at maximizing the bitrate (as an indirect way of maximizing the quality) with other QoE considerations, as in most existing schemes.
Under client-side deployment of CBF, the client can add a function that applies CBF, and then pass the information of the remaining tracks for a chunk position to an existing ABR algorithm.

\section{Grounds-up Design: QUAD} \label{sec:design}

{Besides integrating CBF into existing ABR schemes, another approach to design target quality aware ABR adaptation schemes is to develop them from the ground up. Such schemes, since explicitly designed with the target quality in mind, have the potential to outperform existing schemes enhanced with CBF.
%
%As mentioned earlier, a family of quality-aware schemes can be designed from the ground up when chunk quality information is available at the client.
%
%Such schemes are  designed groud-up to achieve the target quality, while optimizing other QoE metrics such as minimizing the quality changes and rebuffering. They can use CBF to reduce the problem space, and can be used to stream both VBR and CBR videos.
To demonstrate this approach, we propose one design, called QUAD (QUality Aware Data-efficient streaming),  based on control theory. QUAD explicitly integrates the goal of approaching the target quality into its online optimization framework. As a result, it is more capable of maintaining the target quality, and more adaptive to the fluctuating network conditions compared to using CBF with existing schemes.\fengc{as will be shown  in \S\ref{sec:cbf-qea}.}
%; other possible approaches are discussed in \S\ref{sec:conclusion}.
%

As shown in Fig.~\ref{fig:diagram},
QUAD takes a
%user-specified
target quality as input, and leverages an optimization formulation and feedback control to optimize the QoE metrics while approaching the target quality.\fengc{as much as possible.}

%\subsection{\qea based on PID Control}

\begin{figure}[t]
	%\vspace{-.1in}
	\centering
	\includegraphics[width=0.85\textwidth]{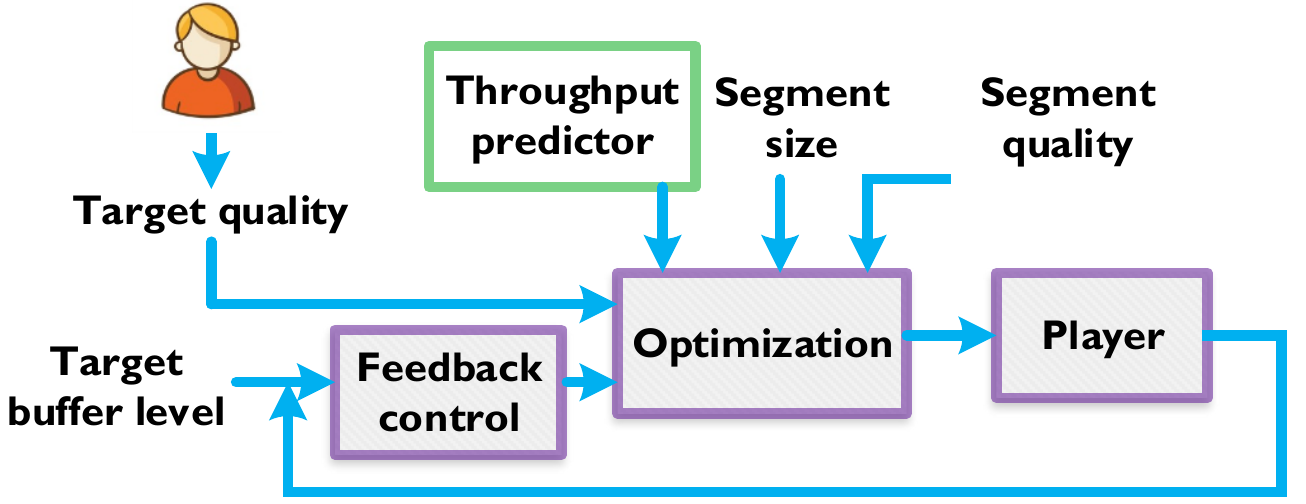}
	%\vspace{-.1in}
	\caption{Design diagram  of QUAD.}
	\label{fig:diagram}
	\vspace{-.1in}
\end{figure}

%\smallskip
\textbf{Target Quality based Optimization.}
%The feedback control block in the above maintains the buffer at the client close to the target level, which helps to avoid stalls.
The optimization formulation below aims to make the chosen chunks' quality approach the target quality while
minimizing
rebuffering and quality changes.
%Specifically, suppose that at time $t$, we need to determine the track for chunk $t$.
Let $\ell_t$ denote the track number selected at time $t$. Let $R_t(\ell_t)$ denote the corresponding bitrate of the selected chunk. We use  $R_t(\ell_t)$ instead of $R(\ell_t)$ to accommodate VBR encoding whose bitrate is both a function of $\ell_t$ and time $t$ since the bitrate can vary significantly even within a track.
The client selects $\ell_t$ from the set of levels that remains after CBF.
%Since it is undesirable to choose a chunk with higher than the target quality, the client uses chunk-based prefiltering first, and only selects from the set of levels that remain after chunk-based prefiltering.
The optimization problem is to determine the track, $\ell_t$,  so that the following objective function is minimized:
\begin{equation}
\label{eq:qea_obj}
J(\ell_t) =  {\max\left(0, u_{t}R_t(\ell_t) - \widehat C_{t}\right)}^2 +  \alpha {Q_{r} - Q_t(\ell_t)}^2 \nonumber + \eta {Q_t({\ell_t}) - Q_{t-1}({\ell_{t-1}})}^2,
\end{equation}
where $u_t$ is the controller output and $\widehat C_{t}$ is  the estimated link bandwidth at time $t$,  $\alpha>0$  and $\eta>0$ are parameters for the second and third terms, respectively,
%$R_t(\ell_t)$ is the bitrate of the  chunk at track $\ell_t$,
$Q_{r}$ is the target quality specified by the user, $Q_t(\ell_t)$ denotes the quality of the chunk at track $\ell_t$ for time $t$, and $Q_{t-1}({\ell_{t-1}})$ represents the quality of the previous chunk (here we slightly abuse the notation by using $t$ to represent the index of the chunk for time $t$ and use $t-1$ to represent the index of the previous chunk).
%The double vertical bars repersent normalization (see below).

The formulation in Eq. (\ref{eq:qea_obj}) is a least-square optimization problem.
In the first term,  $u_{t}R_t(\ell_t)$ represents the bandwidth requirement of the selected track, derived from the feedback control that we will explain shortly. For now readers can regard it as a black box.
%(we use $u_{t}R_t(\ell_t)$ due to our adoption of feedback control; see details later).
The first term is zero if the bandwidth requirement of the selected track is no more than the estimated network bandwidth; otherwise, a stall may potentially occur, so it incurs a penalty that equals to the amount of bandwidth that is exceeded by the bandwidth requirement.
The second term depends on how much the chosen quality for the chunk deviates from the target quality. The sum of the first and second terms allows the chosen track to be as close to the target quality as possible, while does not overly exceed the network bandwidth (to avoid stalls). The last term penalizes quality changes between two adjacent chunks, in order to maintain a more consistent quality and smooth playback.

%\textbf{Normalization:}
We apply normalization in Eq. (\ref{eq:qea_obj}) since the first term is of a different unit from the second and third terms. Specifically, we normalize all the three terms to be unitless as follows.
% Each term is normalized to be unitless.
 The first term is normalized by $\widehat C_{t}$, the estimated bandwidth, and the other two terms are normalized by
 %the maximum quality, which is
 $Q_r$ (since QUAD selects from the tracks that remain after CBF, the maximum quality is approximately $Q_r$).
In Eq. (\ref{eq:qea_obj}), $\alpha$ and $\eta$ represent the weights for the second and third terms, respectively. We set both of them to 1 since all the three terms in  Eq. (\ref{eq:qea_obj}) are important QoE metrics.

{We see from Eq. (\ref{eq:qea_obj}) that choosing $\ell_t$ above the target quality $Q_r$ is not beneficial in minimizing the objective function. Therefore, we may apply CBF before solving the optimization problem to reduce the problem space and improve the runtime efficiency.}
Let $\mathcal{L}_t$ be the set of track levels for chunk $t$ (retained after CBF). We can find the optimal solution to (\ref{eq:qea_obj}) by evaluating Eq. (\ref{eq:qea_obj}) using all possible values of $\mathcal{L}_t$, leading to computational overhead $O({|\mathcal{L}_t|})$. In \S\ref{sec:qea-eval}, we show that QUAD is very lightweight using our implementation.

%\smallskip
\textbf{Feedback Control Block.} %\label{sec:basic-control}
%PID control is a widely used feedback control technique.
In the first term in (\ref{eq:qea_obj}), $u_{t}R_t(\ell_t)$, is derived from the feedback control block shown in Fig.~\ref{fig:diagram}.
%The feedback control block maintains the buffer at the client close to the target buffer level, which helps to avoid stalls.
We use PID control~\cite{astrom08:feedback} as the underlying control framework since it is simple and robust for ABR streaming~\cite{qin2017control}.
Specifically, the PID control block works by continuously monitoring  the difference between the target and current buffer levels  of the video player, and adjusting the control signal to maintain the target buffer level, which helps to avoid stalls.
%
%Specifically, let $C_t$ denote the network bandwidth at time $t$.
%Let $\ell_t$ denote the track number selected at time $t$, with the corresponding bitrate of the chunk denoted as $R_t(\ell_t)$
We define the controller output, $u_t$, as:
\begin{equation}
\label{eq:pid_ut}
u_t = \frac{C_t}{R_t(\ell_t)},
\end{equation}
where $C_t$ denotes the network bandwidth at time $t$, and $R_t(\ell_t)$ denotes the bitrate of the chunk selected for time $t$.
%The controller output defined above is a unitless quantity representing
%the relative buffer filling rate.
The control policy is defined as:
\begin{eqnarray}
u_t = {K_p}({x_r} - x_t) + {K_i}\int_0^t {(x_r - x_\tau)d\tau} + \textbf{1}(x_t - \Delta)
\label{eq:PIA-control}
\end{eqnarray}
where $K_p$ and $K_i$ denote respectively the parameters for proportional and integral control (two key parameters in PID control), $x_r$ is the target buffer level, $x_t$  is the current buffer level  (in seconds) at time $t$, $\Delta$ denotes the playback duration of a chunk, and the last term, $\mathbf{1}(x_t - \Delta)$, is an indicator function (1 when  $x_t\geq \Delta$ and 0 otherwise), which makes the feedback control system linear, and hence easier to control and analyze. %controller a linear controller, which is easier to analyze and set parameters~\cite{qin2017control}.
From (\ref{eq:pid_ut}), we derive $C_t = u_{t}R_t(\ell_t)$ and plug it into (\ref{eq:qea_obj}).

%\smallskip
\textbf{Further Reducing Rebuffering.}
To avoid rebuffering when the current buffer level is low, we further use a heuristic. Specifically, if $x_t < 4 \Delta$, i.e.,  there are less than four chunks in the buffer,
% ($\Delta$ denotes the playback duration of a chunk),
then the track is selected as $\min(\ell_{f},\widehat C_{t}/u_t )$, where $\ell_{f}$ is the lowest track with fair quality for that chunk position. In other words, when the current buffer level is low\fengc{(and hence there is a risk of rebuffering)}, we ignore the goals of achieving the target quality and reducing quality changes (i.e., the last two terms in (\ref{eq:qea_obj})), and only consider the first term (to reduce the risk of stalls). In that case, we first select the track based on $\widehat C_{t}/u_t $. Since $\widehat C_{t}$ can be an overestimate of the  actual network bandwidth, we further bound the selected track to be no more than level $\ell_{f}$. For the videos that we use in our evaluation (see \S\ref{sec:setup}), we set $\ell_{f}$ to 2, which has
%higher bandwidth requirement and
significantly higher quality than track 1.

%talk about what is a trace: average throughput every second
%\vspace{-0.05in}

\section{Implementation \& Evaluation Setup}\label{sec:setup}

%\subsection{Implementation of CBF and QUAD} \label{sec:imple}
\noindent{\textbf{Implementation.}}
We implemented client-based CBF and QUAD in two popular players, \texttt{dash.js}~\cite{dash-js} and ExoPlayer~\cite{exoplayer}.
We made a set of non-trivial changes. First, we included the quality information of each chunk in the manifest file. We then modified \texttt{dash.js} and ExoPlayer so that the chunk quality information can be passed to the ABR logic. We also implemented a new module \texttt{cbf.js} to realize CBF, and a new rate adaptation module \texttt{quad.js} to realize QUAD in \texttt{dash.js}.
{In ExoPlayer, we created two new classes, \texttt{cbf.java} and \texttt{quad.java}, for CBF and QUAD.}
The total number of LoC that is involved in the above changes is 336 in  \texttt{dash.js} and 396 in ExoPlayer.
In  \texttt{dash.js}, we developed a bandwidth estimation module that responds to playback progress events and estimates network throughput using the harmonic mean of the last 5 chunks.
%In  Exoplayer, we use the default module that uses sliding percentile to estimate  bandwidth.

\textbf{Evaluation Setup.} Our evaluations use a combination of controlled lab experiments with our implementations in \texttt{dash.js} and ExoPlayer, as well as simulations, all driven by real-world network bandwidth traces collected from commercial cellular networks.
% and real-world trace-driven
This methodology allows
%To ensure
repeatable experiments and
to evaluate different schemes under identical settings.
%, most of our evaluations use real-world network traces collected from commercial LTE networks.
In \S\ref{sec:in-the-wild}, we also run in-the-wild tests using ExoPlayer on a phone over an LTE network.
%\qea in ExoPlayer on a phone over a LTE network for in-the-wild tests.

\textbf{Network Traces and Videos.} We collected a total of 42 hours of network traces over two large commercial LTE networks in the U.S.
Our traces consist of per-second network bandwidth measurements, which are collected on a phone, by recording the throughput of a large file downloading from a well-provisioned server. They cover a diverse set of scenarios, including different time of day, different locations,
 % (in three states in the US),
 %CT, NJ and NY),
 and different movement speeds (stationary, walking, local driving, and highway driving).
 We selected 50 \emph{challenging} traces, each of 700 seconds, with the average network bandwidth below 1 Mbps.
 %These low-bandwidth settings represent challenging conditions that do occur in  real networks, e.g., when the load is high or when the signal is poor.
 The reason for choosing such traces is because even with target quality of 80 (in VMAF, regarded as very good quality~\cite{vmaf}, the highest quality we evaluate), the average bandwidth requirement is  below 1 Mbps (varies from 500 to 800 kbps) for the videos we use.
 Using higher bandwidth traces will diminish the differences among different schemes;
 %On the other hand,
 these low-bandwidth traces, measured from commercial cellular networks, represent challenging conditions that do occur in  real networks, e.g., when the network is congested or the signal is poor.
% Fig.~\ref{fig:traces}(a) shows a boxplot that shows the minimum, first quartile, median, third quartile, and maximum bandwidth of each trace, where the traces are sorted by the median bandwidth.
%\textbf{Videos.}
We use 4 VBR and 4 CBR videos in our evaluation. They are encoded using the four raw videos (ED, BBB, Sintel, ToS) that we have access to (see \S\ref{sec:video-dataset}), and hence we can calculate the perceptual quality using raw videos as the reference. The results in \S\ref{sec:perf-CBF} and \S\ref{sec:qea-eval} focus on VBR videos; the results for CBR videos are consistent and omitted due to space.

%Our evaluation setup does not consider lower level network characteristics (e.g., signal strength,  loss,  RTT); their impact is solely reflected through network throughput, which is used by the ABR logic that operates at application level.

%. This is because ABR adaptation operates at application level, using estimation of the network bandwidth (based on the throughput for recently downloaded chunks); the impact of lower level network characteristics is solely reflected through network throughput.

%ABR logic operates at the application level, using application-level estimation of the network bandwidth~(based on the throughput for recently downloaded chunks) as part of the decision process to determine the quality of the next chunk to download~\cite{huang2014buffer, jiang2012improving, Xu17:Dissecting}.  The impact of lower level network characteristics such as signal strength,  loss, and RTT on ABR adaptation behavior is affected solely through their impact on the network throughput. Therefore, our network traces, by capturing the  bandwidth variability over time, accurately reflect the impact of these important factors.

%wb, 6/13/2018, only compare with BBA-1 and RBA earlier on; here we only consider more recent/sophisticated schemes: MPC and PANDA-CQ variants
%\noindent
{\textbf{CBF with Existing ABR Schemes.}}
%We compare consider the following state-of-the-art rate adaptation schemes:
We use CBF as a prefilter for the following state-of-the-art rate adaptation schemes:
(i) \textbf{RobustMPC}~\cite{yin2015control}, which is a well-known scheme 
based on model predictive control.\fengc{and has been shown to outperform its more aggressive variants, MPC and FastMPC, under more dynamic network bandwidth settings~\cite{Mao17:pensieve}.}
%RobustMPC is shown to outperform MPC under more dynamic network bandwidth settings~\cite{Mao17:pensieve}.
(ii) \textbf{PANDA/CQ}~\cite{li2014streaming}, which directly incorporates video quality information in ABR streaming. It maximizes the minimum quality for the next $N$ chunks, which achieves better fairness (in terms of quality) among multiple chunks.
%We investigate one variant that maximizes the minimum quality of the next $N$ chunks while making decisions, which
% (e.g., the variant that maximizes the sum of the quality for next $N$ chunks),
%achieves better fairness (in terms of quality) among multiple chunks than other variants.
(iii) \textbf{BOLA-E}~\cite{Spiteri2018bolae,spiteri2016bola}, which selects the bitrate to maximize a utility function
considering both rebuffering and delivered bitrate.
%It is based upon and improves an earlier version, BOLA~\cite{spiteri2016bola}.
(iv) \textbf{ExoPlayer}'s ABR adaptation~\cite{exoplayer}, referred to as \emph{Exo} henceforth, which is essentially a rate based logic, i.e., selecting the track based on bandwidth estimation.
%We compare \qea and BOLA in \texttt{dash.js}, leveraging the implementation of BOLA in version xx; and compare \qea and the default Exoplayer algorithm in Exoplayer; for the rest of the algorithms, the comparison is through trace-driven simulation.
%Learning based approaches such as~\cite{Mao17:pensieve} require extensive training on a large corpus of data, and are not included for comparison since our focus is on pure client-side based schemes that are more easily deployable.
%The BOLA-E based comparison  is done in \texttt{dash.js} (version 2.4.1),  the Exoplayer
%, leveraging the implementation of BOLA-E in version 2.4.1; the results for the default Exoplayer algorithm are obtained in Exoplayer;
The results for RobustMPC and PANDA/CQ are obtained through trace-driven simulation; the evaluation involving BOLA-E and ExoPlayer ABR logic is done using open source implementation in \texttt{dash.js} and ExoPlayer, respectively.

\textbf{Offline Optimal Scheme.}
We further use an offline optimal solution as a baseline to evaluate the performance of various schemes. This offline scheme
assumes that the entire network bandwidth is known beforehand.
It considers three QoE metrics related to target quality, quality changes, and stalls.
% and the quality information is available at the client.
Specifically, for a video with $n$ chunks, it selects tracks $\ell_1,\ldots,\ell_n$  to minimize
%so that the following objective function is minimized:
%
\begin{eqnarray}
\label{eq:qoe-opt}
%\vspace{-0.4in}
J(\ell_1,\ldots,\ell_n) &=& \sum_{t = 1}^{n} \left( Q_{r} - Q_{t}(\ell_t)\right)^2
+  \sum_{t = 1}^{n-1} \left( Q_{t+1}({\ell_{t+1}}) - Q_{t}({\ell_{t}}) \right)^2 \nonumber \\
&+&  \gamma T_{r}(\ell_1,\ldots,\ell_n) \nonumber
%\vspace{-0.2in}
\end{eqnarray}
where $Q_r$ is the target quality, $Q_{t}(\ell_t)$ is the quality for $\ell_t$, and $T_r(\ell_1,\ldots,\ell_n)$ is the rebuffering duration,  and $\gamma$ is the weight for rebuffering. The results below use $\gamma=100^2$, i.e.,
we penalize each second of rebuffering with{
	%two orders
square of the maximum VMAF quality.}

%\noindent
{\textbf{ABR Configurations.}}
%An important parameter in all schemes is the playback startup latency.
Following practices in commercial production players~\cite{Xu17:Dissecting},
we set the player to start
%the player starts
the playback when two chunks are downloaded into the buffer.
%we set the startup latency to be $2\Delta$, where $\Delta$ denotes chunk duration.
The track for the first chunk is selected to be the middle track (i.e.,  track 3).
%\fengc{, i.e., the client downloads two chunks before beginning playback.}
%
%
%Another key factor is the
%network bandwidth estimation technique used by the ABR logic.
Unless otherwise stated, we use the harmonic mean of the  average download throughout for the past 5 chunks as the bandwidth prediction, as it  has been shown to be robust to measurement outliers~\cite{jiang2012improving,yin2015control}. For the results using ExoPlayer, the bandwidth prediction uses the built-in sliding percentile technique~\cite{exoplayer}.
% \bingc{say a bit about it.}\yanyuan{-sliding percentile, which calculate 50 percentile over a sliding window of weighted values of past transfer rate observations.}
%\fengc{For all the schemes, video chunks that are downloaded before the playback time are stored in the client buffer.}
Unless otherwise stated, for all schemes, we set the maximum client-side buffer size to 120 seconds.
%\fengc{(i.e., it can store 120 seconds worth of video content).}
This is reasonable for Video on Demand (VOD) streaming, and is consistent with existing practices that set the maximum buffer limit to hundreds of seconds~\cite{dash-js,huang2014buffer,Xu17:Dissecting}. An ABR scheme may use thresholds to control when to stop and resume downloading based on the buffer level. When comparing with another scheme, we ensure that our schemes use the same thresholds as used in that scheme.
%The client stops downloading video when the buffer is full, and resumes downloading when there is space in the buffer. For \qea, we set the target buffer level to 60 seconds.
For QUAD, the controller parameters, $K_p$ and $K_i$, are selected by adopting the methodology outlined in~\cite{qin2017control}.
Specifically, we varied $K_p$ and $K_i$, and confirmed that
%, similar to~\cite{qin2017control},
a wide range of $K_p$ and $K_i$ values lead to good performance.

\textbf{Perceptual Quality Metric.} We measure video quality using VMAF (\S\ref{sec:quality-bitrate-tradeoffs}). Specifically, we use VMAF phone model (instead of TV model) given our focus on cellular networks. In the testing, we assume that a user selects from three quality options, good/better/best.
%, with the understanding that higher quality leads to more data consumption.
Correspondingly, the player maps these quality options to VMAF values of 60 (for ``good''), 70 (for ``better''), and 80 (for ``best'').

{\textbf{Performance Metrics.}}
We use five metrics: four for measuring different aspects of user QoE and one for measuring data usage.
All metrics are computed with respect to the delivered video, i.e., considering the chunks that have been downloaded and played back.
The metrics are listed as follows. (i) {\em Quality of all the chunks}: measures how the quality of each chunk differs from the target quality. (ii) {\em Low-quality chunk percentage}: measures the proportion of the chunks that were selected with low quality during a streaming session.  The reason for using this metric is because human eyes are sensitive to bad quality chunks~\cite{VMAF-aggr}. We identify VMAF values below 40  as low-quality based on~\cite{vmaf}.  (iii) {\em Rebuffering duration}:  measures the total rebuffering/stall time in a streaming session. (iv) {\em Average quality change per chunk}: defined as the average quality difference of two consecutive chunks in playback order for a streaming session (since human eyes are more sensitive to level changes in adjacent chunks). 
%\fengc{, i.e., $\sum_{i} |q_{i+1}-q_i|/(n-1)$, where $q_i$ denotes the quality of the $i${th} chunk, and $n$ is the number of chunks.} 
(v) {\em Data usage}: measures the total amount of data downloaded for a streaming session.
For metrics (ii)-(v),  a lower  value is preferable;
for (i), we measure how close it is to the target quality. 
In addition to the above metrics, we further explored the number of stalls during a session in various cases. Our results show that using CBF in existing schemes reduces the number of stalls, and QUAD leads to fewer stalls compared to existing scheme enhanced with CBF in almost all the cases.
	%comment on the number of stalls during a session in various cases.}

%%RobustMPC and PANDA/CQ with and without CBF
\begin{figure*}[ht]
	\vspace{-.1in}
	\centering
	%\subfloat[\label{subfig:CAVA-Q4}]{%
	\includegraphics[width=0.296\textwidth]{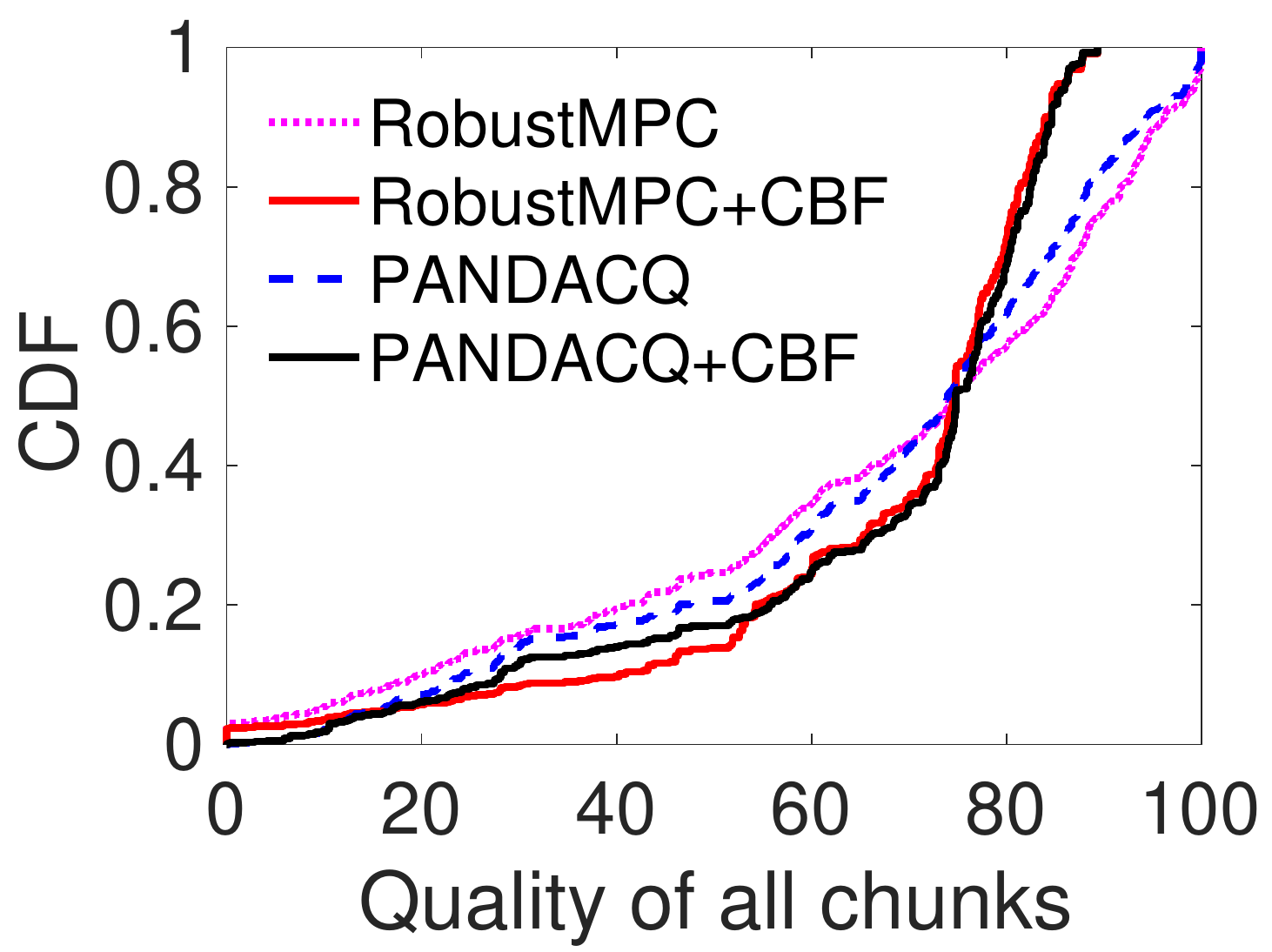}
	%}
	%\subfloat[\label{subfig:CAVA-low-quality}]{%
	\includegraphics[width=0.296\textwidth]{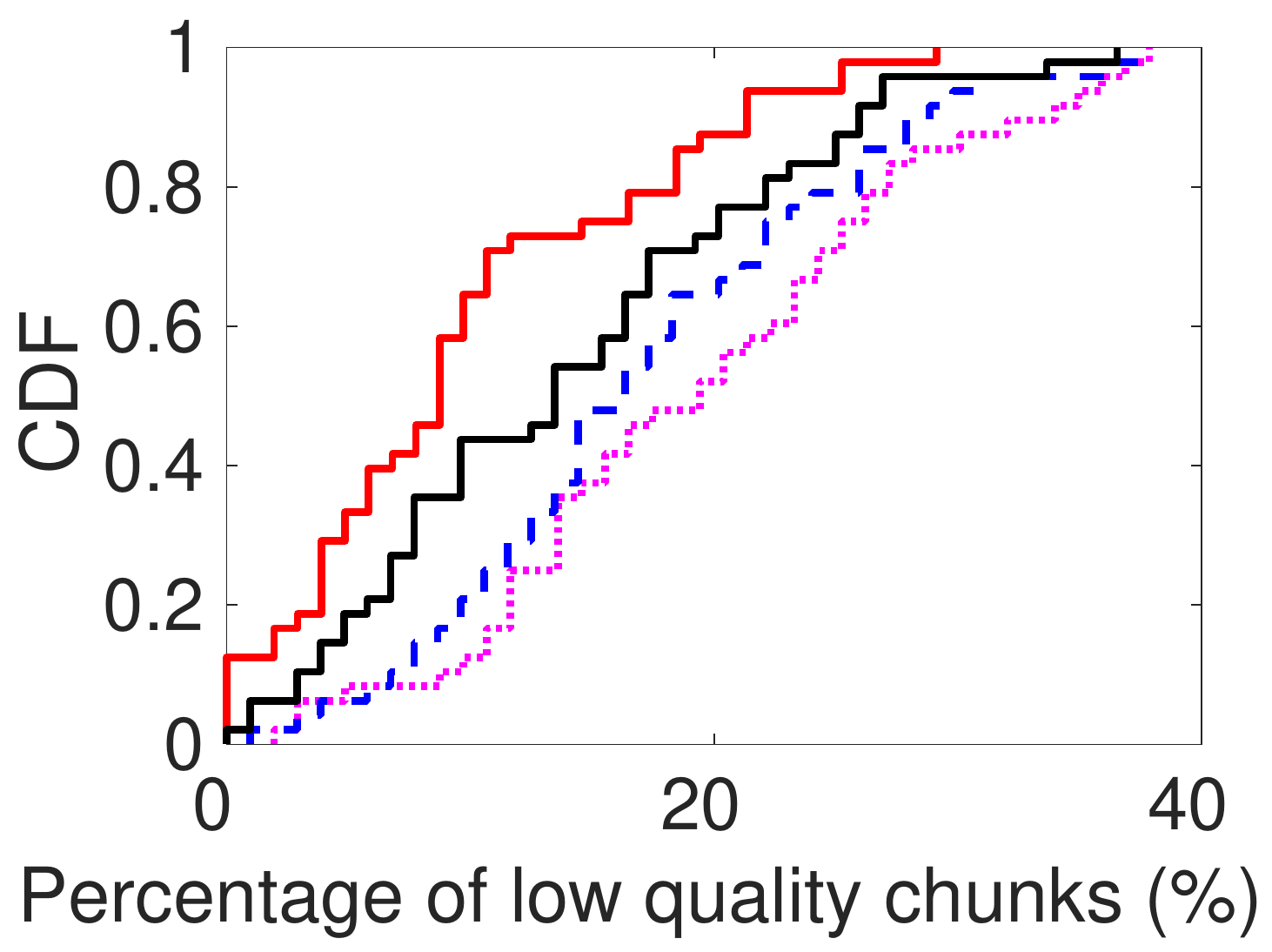}
	%}
	%\subfloat[\label{subfig:CAVA-change}]{%
	\includegraphics[width=0.296\textwidth]{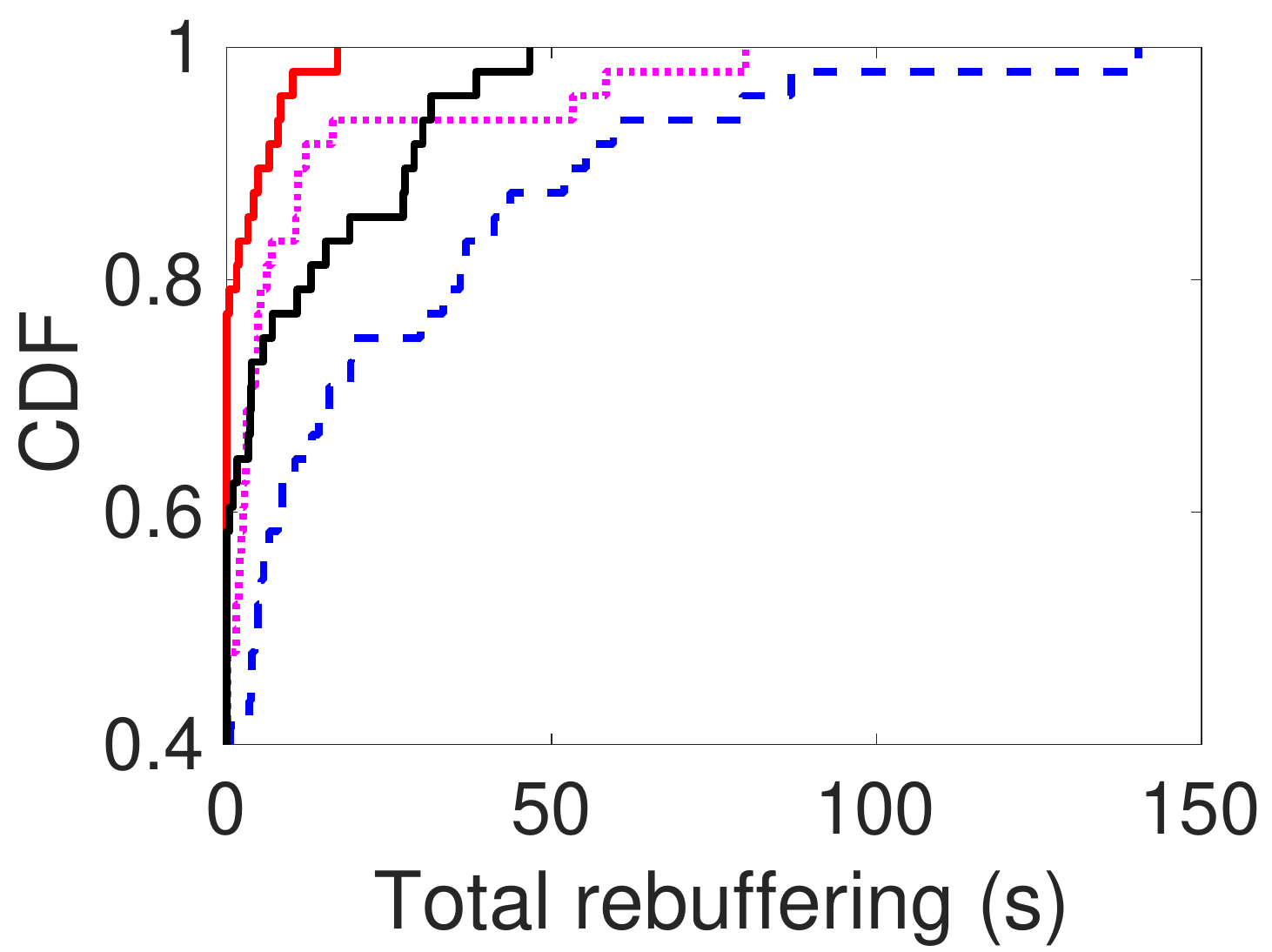}
	%}	
	%\subfloat[\label{subfig:CAVA-rebuffering}]{%
	\includegraphics[width=0.296\textwidth]{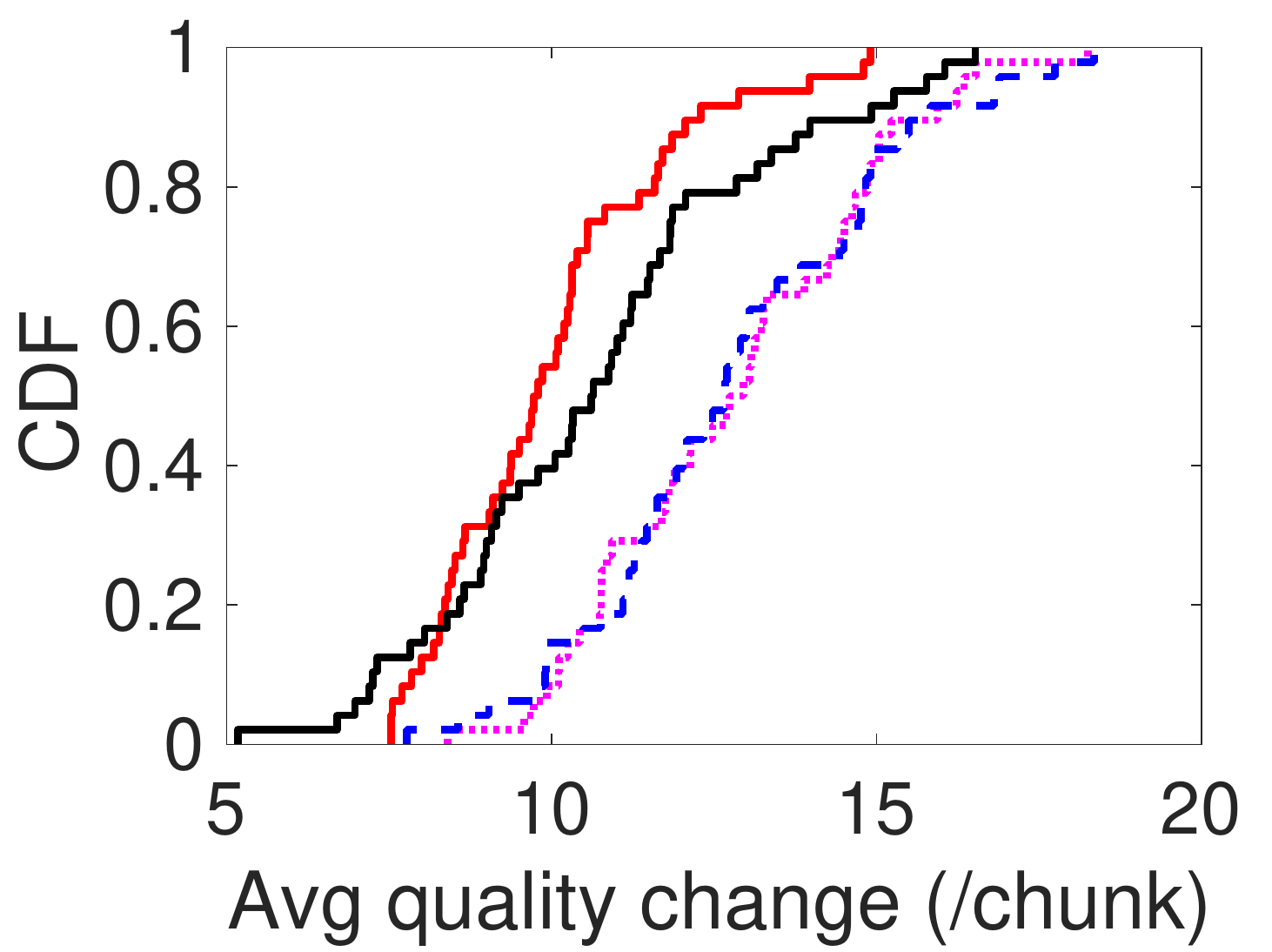}
	%}
	%\subfloat[\label{subfig:CAVA-data-usage}]{%
	\includegraphics[width=0.296\textwidth]{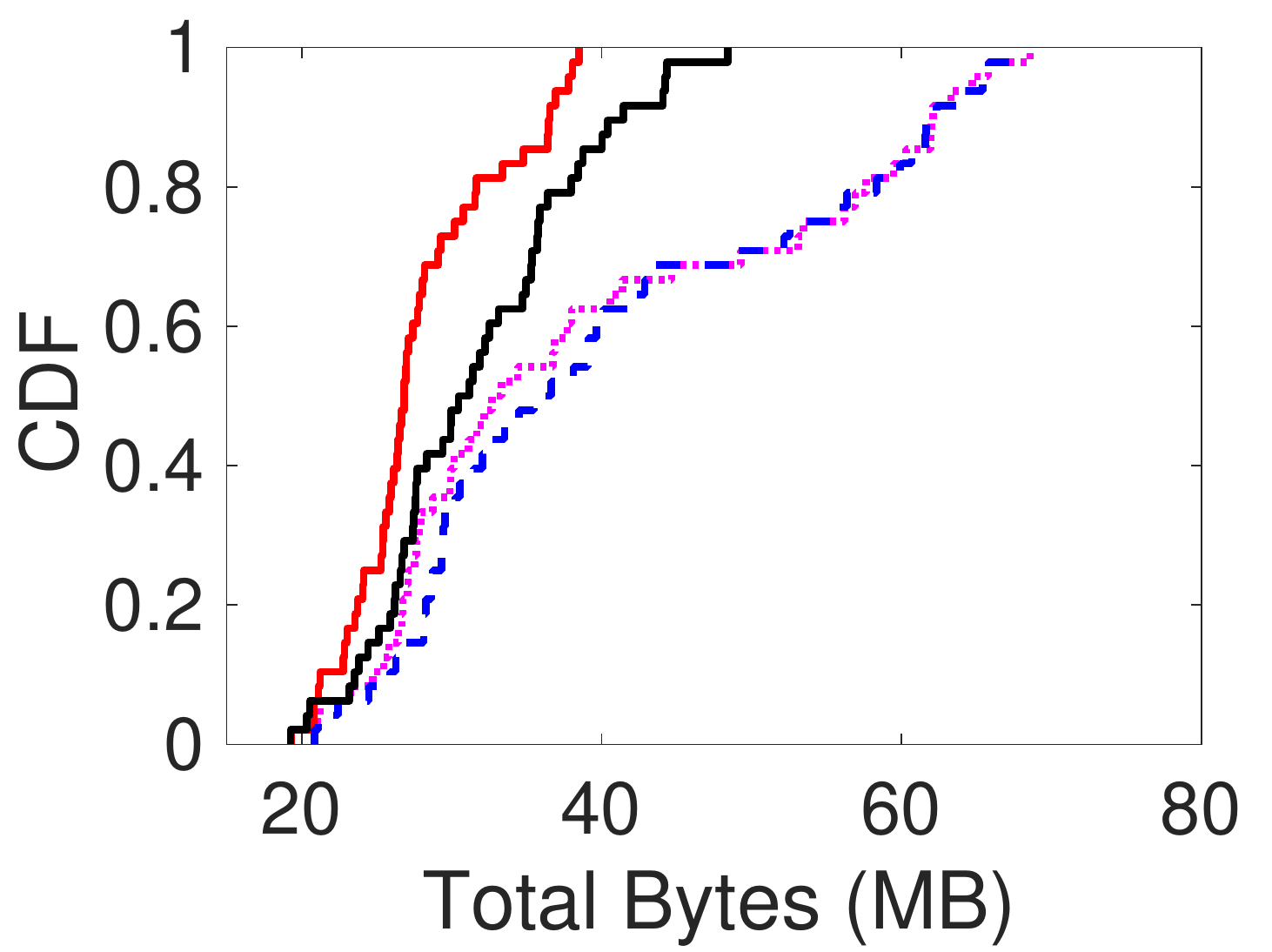}
	%}
	\vspace{-.1in}
	\caption{Two existing ABR schemes with and without CBF (ED, YouTube encoded, target quality 80). }
	\label{fig:CBF-benefits-80}
	%\vspace{-.1in}
\end{figure*}

%%with bandwidth constraints
 \begin{figure*}[ht]
 	\centering
 	%\subfloat[\label{subfig:CAVA-Q4}]{%
 	\includegraphics[width=0.296\textwidth]{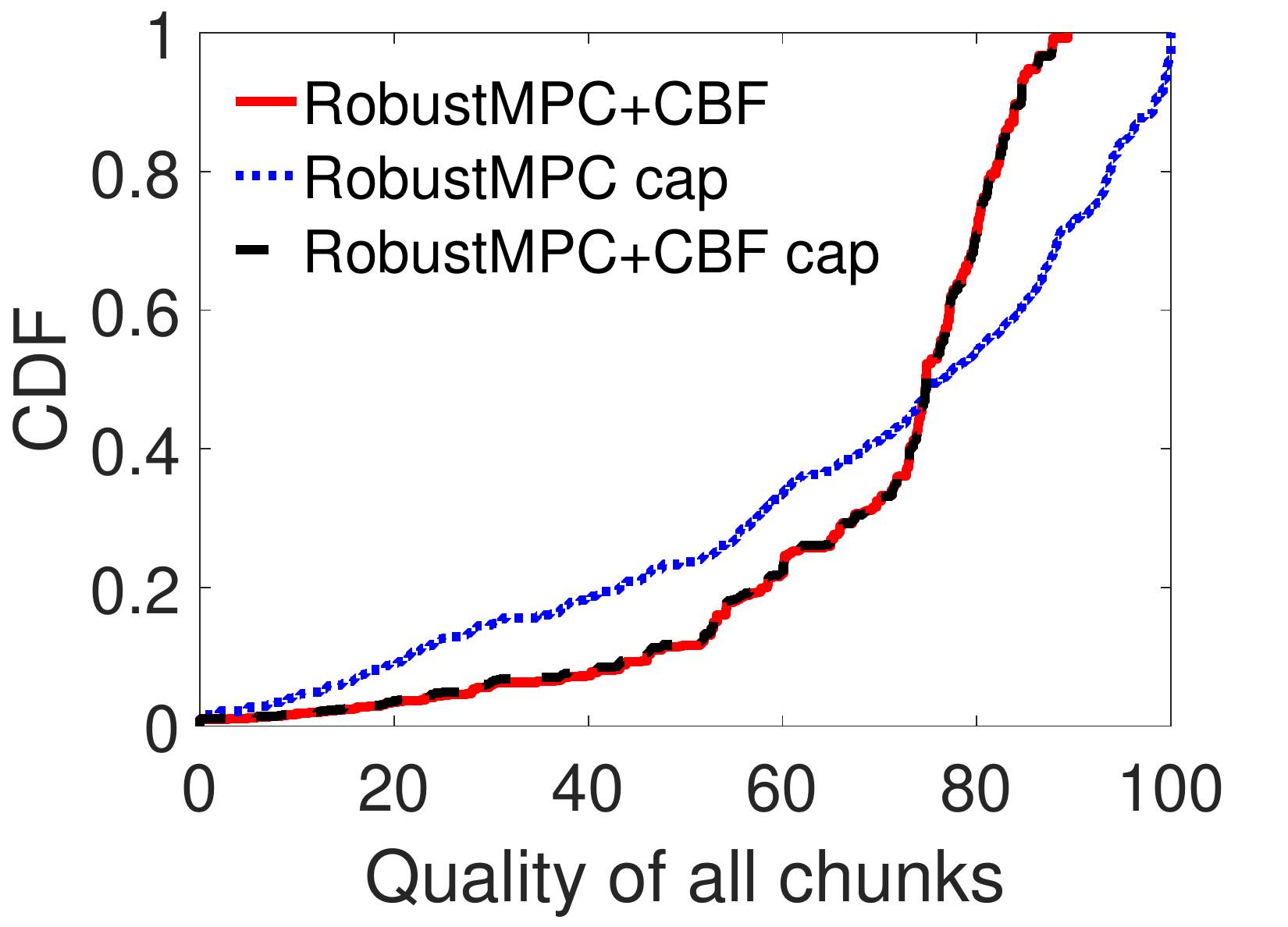}
 	%}
 	%\subfloat[\label{subfig:CAVA-low-quality}]{%
 	\includegraphics[width=0.296\textwidth]{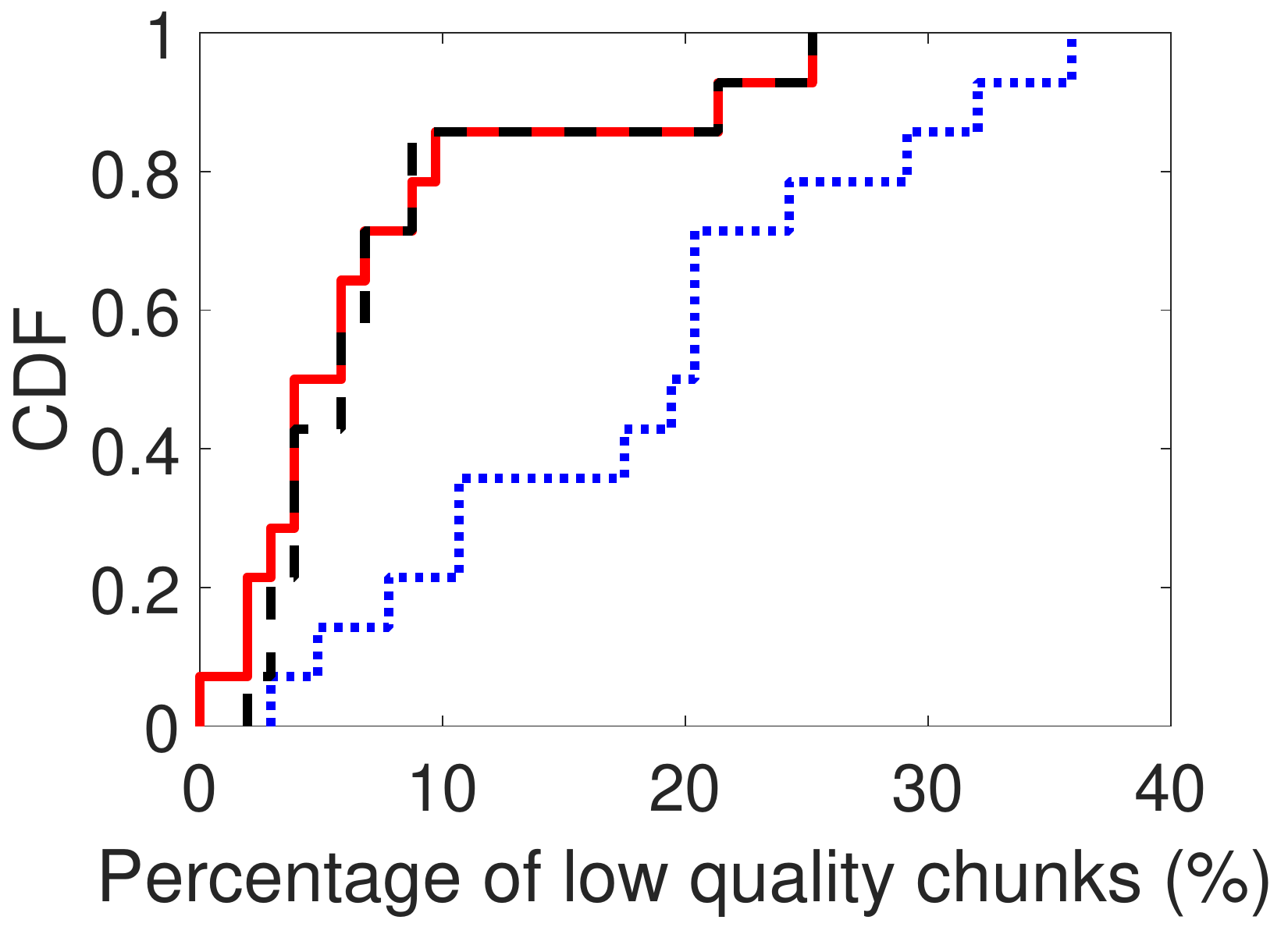}
 	%}
 	%\subfloat[\label{subfig:CAVA-change}]{%
 	\includegraphics[width=0.296\textwidth]{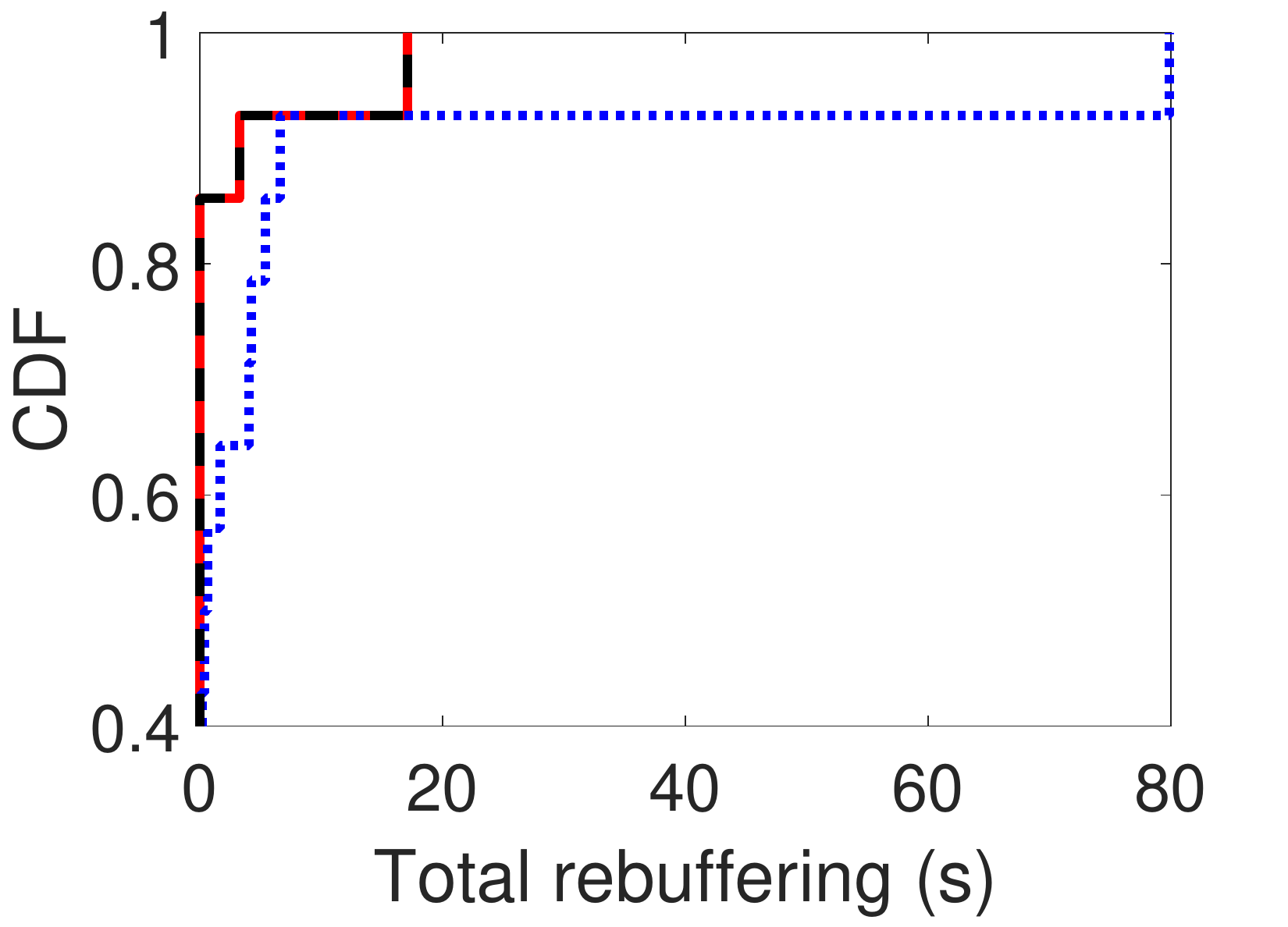}
 	%}
 	%\subfloat[\label{subfig:CAVA-rebuffering}]{%
	 \includegraphics[width=0.296\textwidth]{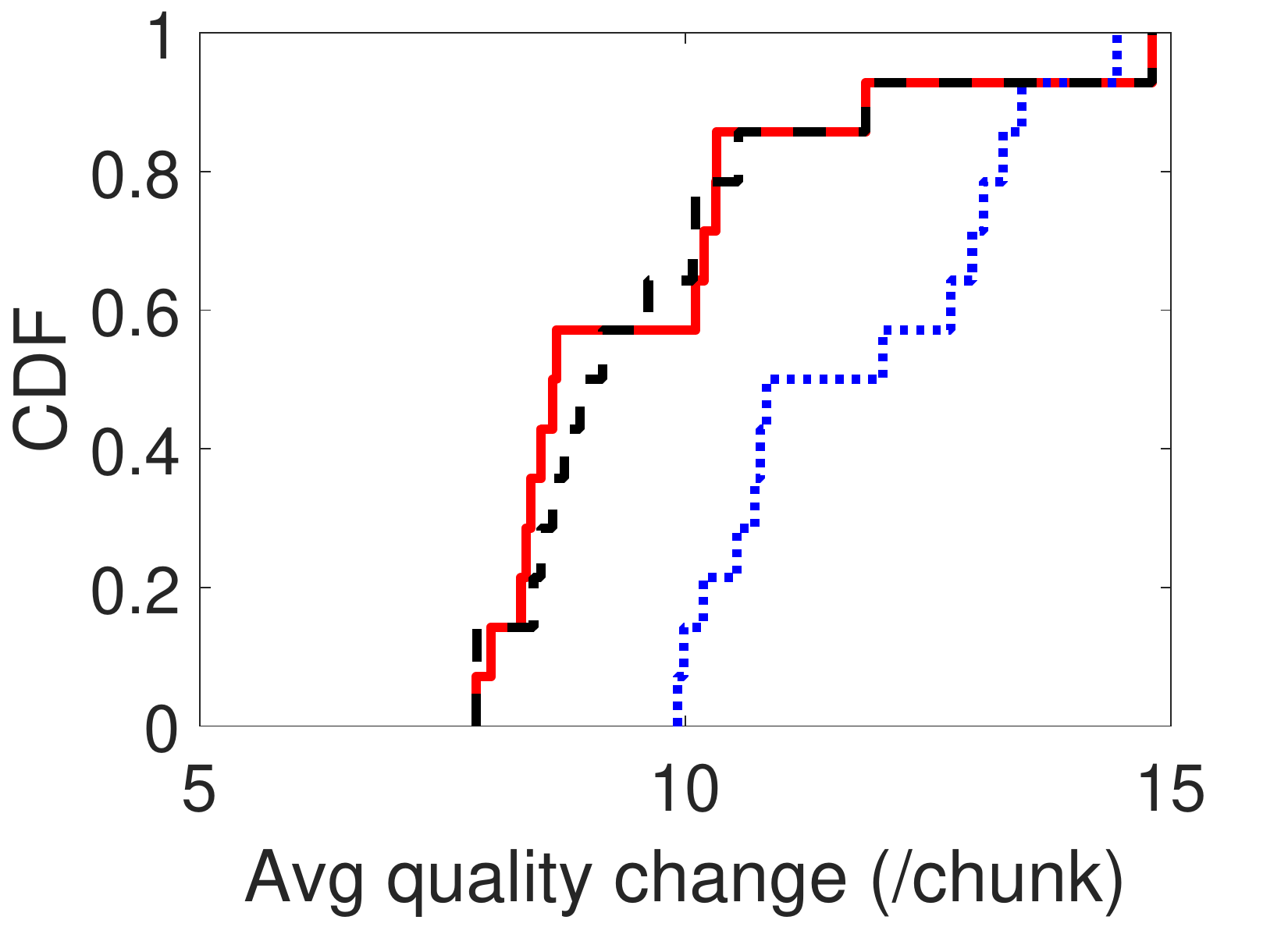}
 	%}	
 	%\subfloat[\label{subfig:CAVA-data-usage}]{%
 	\includegraphics[width=0.296\textwidth]{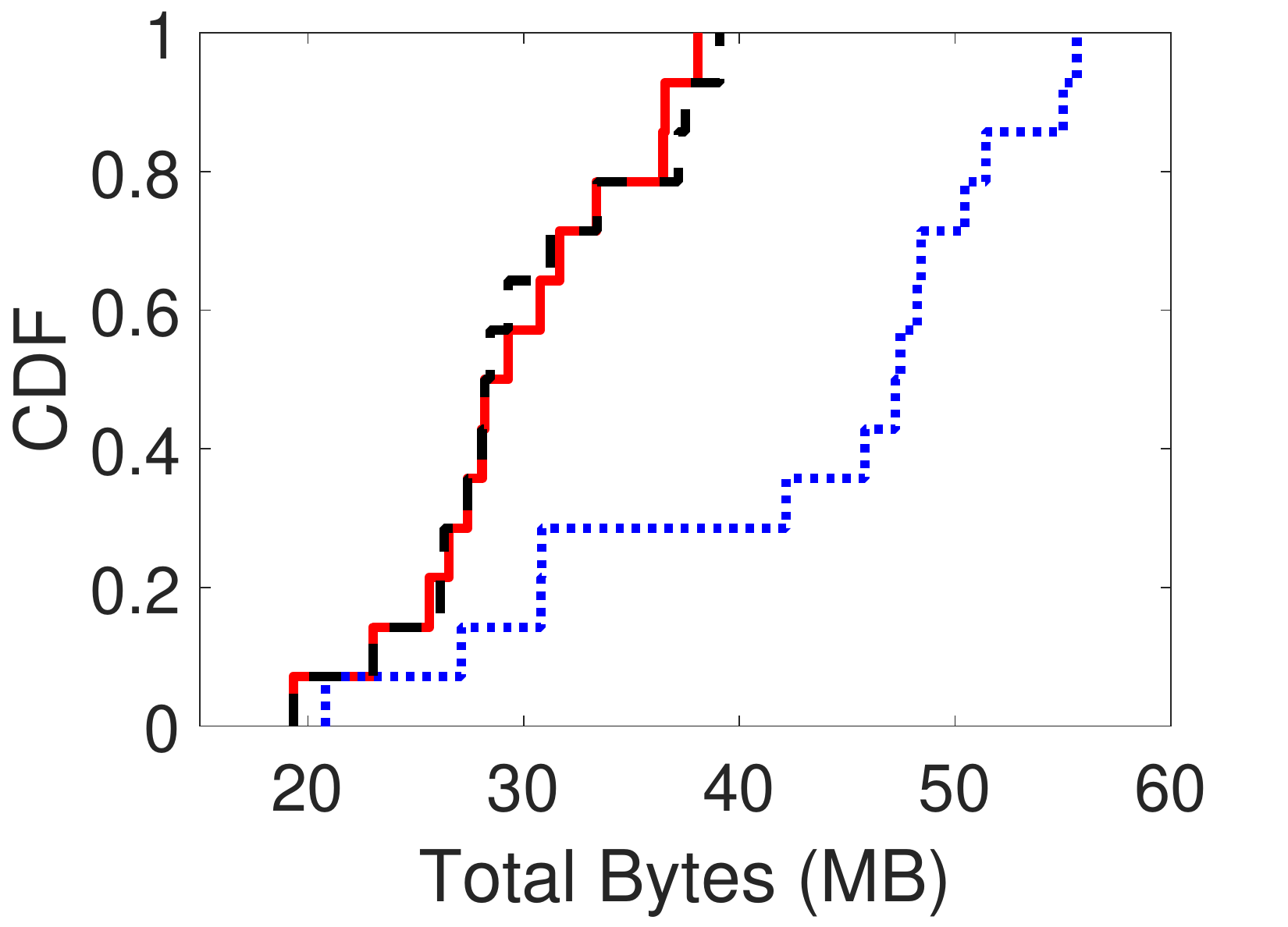}
 	%}
 	\vspace{-.1in}
 	\caption{CBF versus capping network bandwidth (ED, YouTube encoded, target quality 80). }
 	\label{fig:network-vs-chunk-prefiltering-80}
 	%\vspace{-.1in}
 \end{figure*}

\section{Evaluation Existing ABR Schemes Enhanced with CBF} \label{sec:perf-CBF}
We evaluate the performance of adding CBF to existing ABR schemes.
%\bing{Our goal is to demonstrate that CBF can be directly retrofitted to existing schemes and improve the performance of these schemes.}
%
%\fengc{
%We first show that CBF significantly benefits existing schemes  in reducing the data usage while approaching the target quality. We further show that
%CBF leads an existing scheme to significantly better performance compared to  the current two practices of saving data in cellular networks (see \S\ref{sec:current-practice-save-data}).}
%
We focus on two existing schemes, RobustMPC and PANDA/CQ. The performance of other schemes with CBF is deferred to \S\ref{sec:qea-eval}.

%compare a scheme with CBF and without CBF
\mysubsection{Benefits of CBF} \label{sec:cbf-benefits}
Fig.~\ref{fig:CBF-benefits-80} shows the performance of
%two existing schemes,
RobustMPC and PANDA/CQ with and without CBF for one video across the network traces when the target quality is 80. The results for the five performance metrics are shown in the five subplots in the figure: the first subplot is the CDF across individual chunks across all runs; the rest four are CDFs across runs, with each run corresponding to a different network trace. We observe that
%for both schemes, %adding CBF improves the performance of the schemes across all the performance metrics.
CBF improves performance for both ABR schemes, across all metrics.
Specifically, compared to the original schemes, adding CBF leads the quality to be closer to the target quality, and reduces the  percentage of low-quality chunks, the quality changes, rebuffering, and  data usage.

\begin{table}[]
	\centering
	\small{
		\caption{RobustMPC vs. RobustMPC+CBF.}
		\label{tb:perf-all-videos-CBF-RoMPC}
		\begin{tabular}{|p{0.1cm}|p{0.8cm}|c|c|c|c|c|}
			\cline{2-7}
			\multicolumn{1}{c|}{} & Video & {\begin{tabular}[c]{@{}c@{}}Avg. dev.\\from\\target\\quality\end{tabular}} &
			{\begin{tabular}[c]{@{}c@{}}\% of traces\\w/ $>20$\%\\low-qual.\\chunks \end{tabular}}
			& {\begin{tabular}[c]{@{}c@{}}Avg.\\stall\\dura.\\(s)\end{tabular}} &
			{\begin{tabular}[c]{@{}c@{}}Avg.\\quality\\change\end{tabular}} &
			{\begin{tabular}[c]{@{}c@{}}Data\\usage\\(MB)\end{tabular}}
			\\ \hline
			\multirow{4}{*}{\begin{sideways} VBR, $60$   \end{sideways}} &
			%Youtube results
			ED   &25, 9&48\%, 10\%&8, 0 &13, 9&39, 19\\ \cline{2-7} %\cline{2-7}
			& BBB  & 27, 9&37\%, 8\%&6, 0& 14, 11&36, 14 \\ \cline{2-7} %\cline{2-7}
			& Sintel  &28, 11&6\%, 0\%&8, 0&10, 12&55, 18\\ \cline{2-7}%\cline{2-7}
			& ToS   &26, 14& 56\%, 40\% &5, 0& 13, 12&46, 22  \\ \cline{2-7}  %\cline{2-7}
			\hline \hline
			\multirow{4}{*}{\begin{sideways} VBR, $80$   \end{sideways}} &
			%Youtube results
			ED        &23, 15& 48\%, 12\% &8, 2&13, 10 &39, 28\\ \cline{2-7} %\cline{2-7}
			& BBB     &22, 14& 37\%, 8\% &6, 1&14, 13&36, 21 \\ \cline{2-7}%\cline{2-7}
			& Sintel  & 19, 9& 6\%, 0\% &8, 0&10, 9&55, 29\\ \cline{2-7}%\cline{2-7}
			& ToS    &24, 16& 56\%, 21\% &5, 2&13, 10&46, 35 \\ \cline{2-7}  %\cline{2-7}
		 \hline %\hline						
			\multicolumn{7}{l}{* The two numbers in each cell are the results for RobustMPC} \\
			\multicolumn{7}{l}{and RobustMPC+CBF, respectively.}
			%			\multicolumn{6}{l}{	and RobustMPC with CBF , respectively.}
		\end{tabular}
	}
	\vspace{-.1in}
\end{table}

Table~\ref{tb:perf-all-videos-CBF-RoMPC} summarizes the results for four VBR videos for RobustMPC with and without CBF for target quality of 60 and 80, respectively. The first column is the average deviation from the target quality across all the network traces; the value for one trace is $(\sum |q_i - Q_r|)/n$, where $q_i$ is the quality of chunk $i$ in the rendered video, $Q_r$ is the target quality, and $n$ is the number of chunks in the video.
The second column shows the percentage of the traces that have  more than 20\% of low-quality chunks. The third column shows the average rebuffering duration for the traces where either case (i.e., with or without CBF) has rebuffering. The last two columns show the average quality changes and the data usage, averaged across the network traces.
%We see that CBF leads the scheme 8-16 points (in VMAF) closer to the target quality,
We see that CBF reduces the deviation from  the target quality by 37-67\%,
%reduces the tail of low-quality chunks by  6-42\%,
reduces the number of traces with frequent low-quality chunks by 6-42\%,
and reduces the average quality change by 7-31\%.
% and reduces the duration of stalls by 3-8 seconds on average.
For many videos, across all network traces, using CBF leads to no stalls compared to substantial amount of stalls without CBF.
The data usage with CBF is 34-67\% lower than that without CBF.
%\bing{The wide range of data usage saving is due to the variations of the network bandwidth traces: since CBF limits the highest quality chunk to the one with the quality closest to the target quality, when the network bandwidth is high,
%	the saving provided by CBF is high; when the network bandwidth is lower, the saving is lower, thus leading to a range of bandwidth savings.}
%
In the extreme case when the bandwidth is sufficiently high, RobustMPC without CBF will choose the highest track (i.e., track 6).  The data usage for the four videos will be 203, 179, 240, and 209 MB respectively, 5.6-7.5 times higher than the corresponding values with CBF under target quality 80 (the ratios are even higher for target quality 60).
%
%\fengc{Last, adding CBF slightly reduces quality changes; we will see in \S\ref{sec:cbf-qea} that our newly designed scheme, \qea, leads to significantly lower quality changes  compared to existing schemes.}
%

%We see similar results for PANDA/CQ with and without CBF. Table~\ref{tb:perf-all-videos-CBF-PANDACQ} shows the results for PANDA/CQ with and without CBF. We see that adding CBF significantly improves all the metrics. Compared to the case when adding CBF to RobustMPC, the benefits of reducing the tail of low-qulity chunks is slightly lower. On the other hand, the benefits of reducing rebuffering is more significant in this case.

We see similar results for PANDA/CQ with and without CBF.\fengc{(table omitted in the interest of space).}
The above results demonstrate that CBF can significantly improve the performance of existing ABR schemes by prefiltering the chunks whose qualities are higher than the target quality, and hence steering the schemes to the set of more desirable choices.
%eliminating candidate chunks with low bandwidth efficiency based on the target quality.
%The above results show that adding CBF to an existing scheme significantly improves the performance of the scheme. This is because CBF eliminates the undesirable choices for an ABR scheme, helping the scheme to make more desirable decisions.

%\vspace{-.12in}
\mysubsection{CBF vs. Network Bandwidth Cap} \label{sec:network-vs-prefiltering}
 We now compare CBF with an existing practice for saving data by capping network bandwidth (\S\ref{sec:current-practice-save-data}).
 % (used by AT\&T and T-mobile).
 Specifically, for an existing ABR scheme, we consider three cases: (i)
 %The first case uses the original scheme and the network bandwidth profile, serving as a baseline.
 %In the first case,
 we assume that the cellular network provider caps the network bandwidth to 1.5 Mbps;
 % (as illustrated in Fig.~\ref{fig:traces}(b)).
 %In the second case,
 (ii) there is no cap on the network bandwidth, while the scheme is used together with CBF;
 % In the third case,
 and (iii) there is a bandwidth cap and the scheme is used with CBF.
 Since the network bandwidth of some traces is low and imposing the constraint of 1.5 Mbps leads to little impact,
 %to compare the performance of the above three cases
 we choose a subset of network traces where the cap meaningfully changes the available bandwidth.
 Specifically, a network trace is chosen if the
 bandwidth estimated using a window of 1 second
 %average network bandwidth per second
 is larger than 1.5 Mbps for at least 10\% of the time. %(the maximum percentage is 15\% among all the traces).
 The results below are obtained from 20 traces selected as above.

 Fig.~\ref{fig:network-vs-chunk-prefiltering-80} shows the results for RobustMPC with target quality 80; the results for PANDA/CQ show a similar trend.
 We see that the two variants with CBF\fengc{(i.e., with and without network bandwidth cap)} achieve similar performance. Both of them
 significantly outperform the other variant (i.e., with bandwidth cap but no CBF) in all performance metrics.
The results demonstrate the effectiveness of CBF compared to the network bandwidth capping approach.\fengc{The results demonstrate that CBF is more effective in helping an ABR scheme achieve good performance compared to the network bandwidth capping approach.}
They also indicate that CBF\fengc{does not interfere with and} can co-exist with the network bandwidth capping approach.
 The above results are for one video with target quality 80. We observe similar results for other videos and target qualities.
 For lower target qualities (i.e., 60 and 70), we observe that
 CBF achieves even better performance.\fengc{an even larger performance gap between the variant with network bandwidth cap only and the two variants with CBF.}
 %

%RobustMPC, Q_r=80, for ToS, RobustMPC- uses track 3;  RobustMPC+ uses track 4 (which has resolution 480p, which corresponds to what YouTube does)
\begin{figure*}[ht]
	\centering
	\vspace{-.1in}
	%\subfloat[\label{subfig:CAVA-Q4}]{%
	\includegraphics[width=0.296\textwidth]{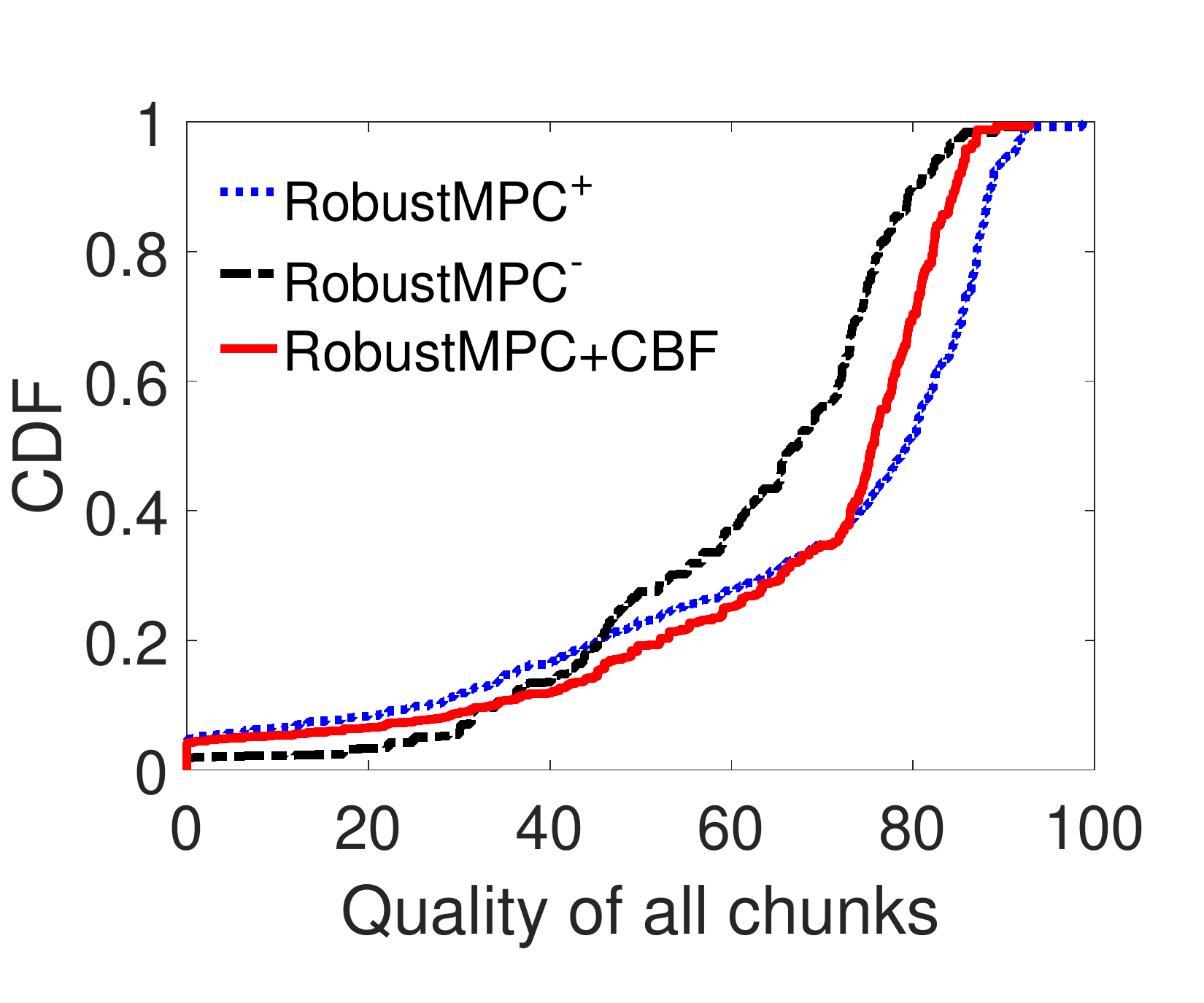}
	%}
	%\subfloat[\label{subfig:CAVA-low-quality}]{%
	\includegraphics[width=0.296\textwidth]{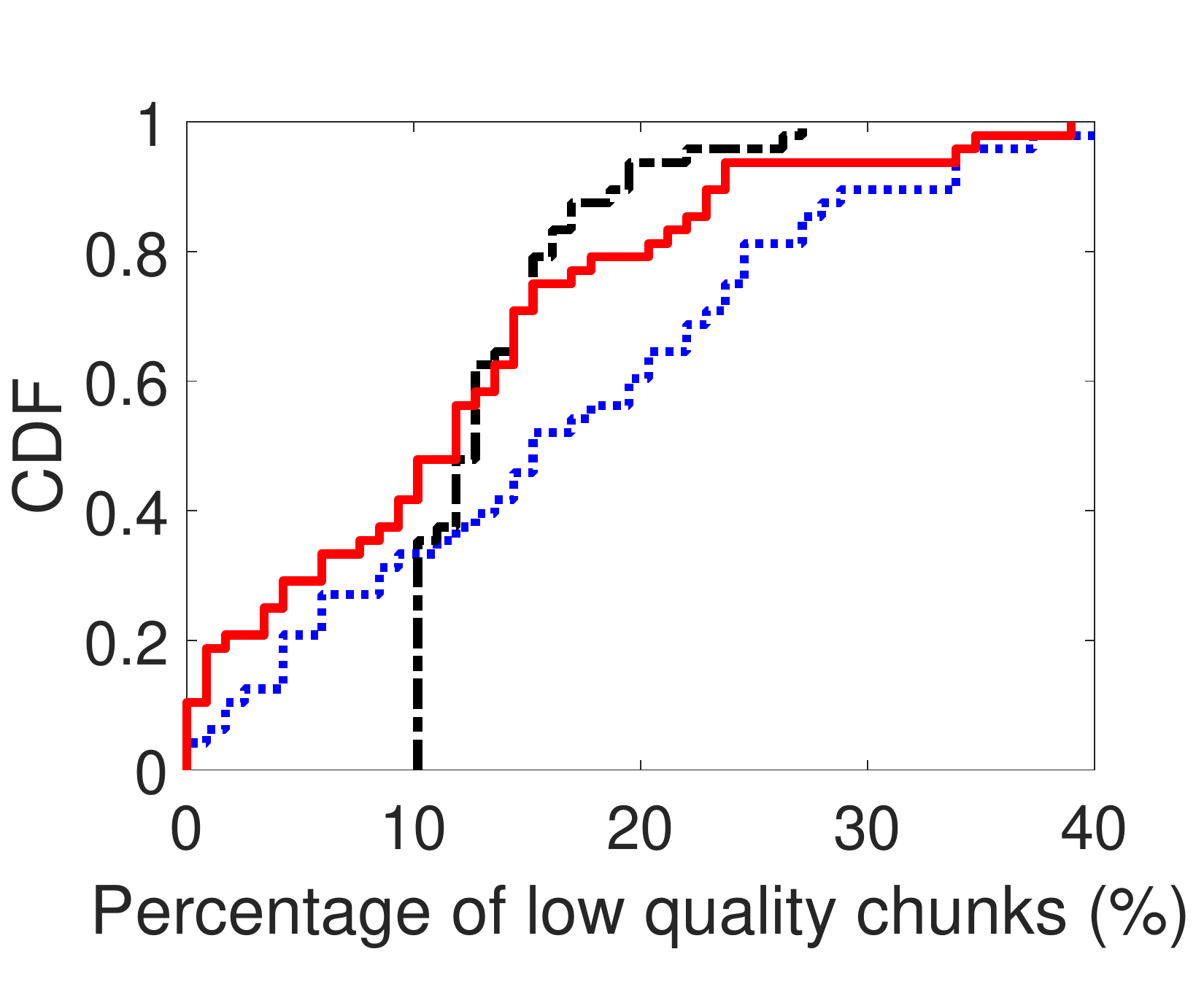}
	%}
	%\subfloat[\label{subfig:CAVA-rebuffering}]{%
	\includegraphics[width=0.296\textwidth]{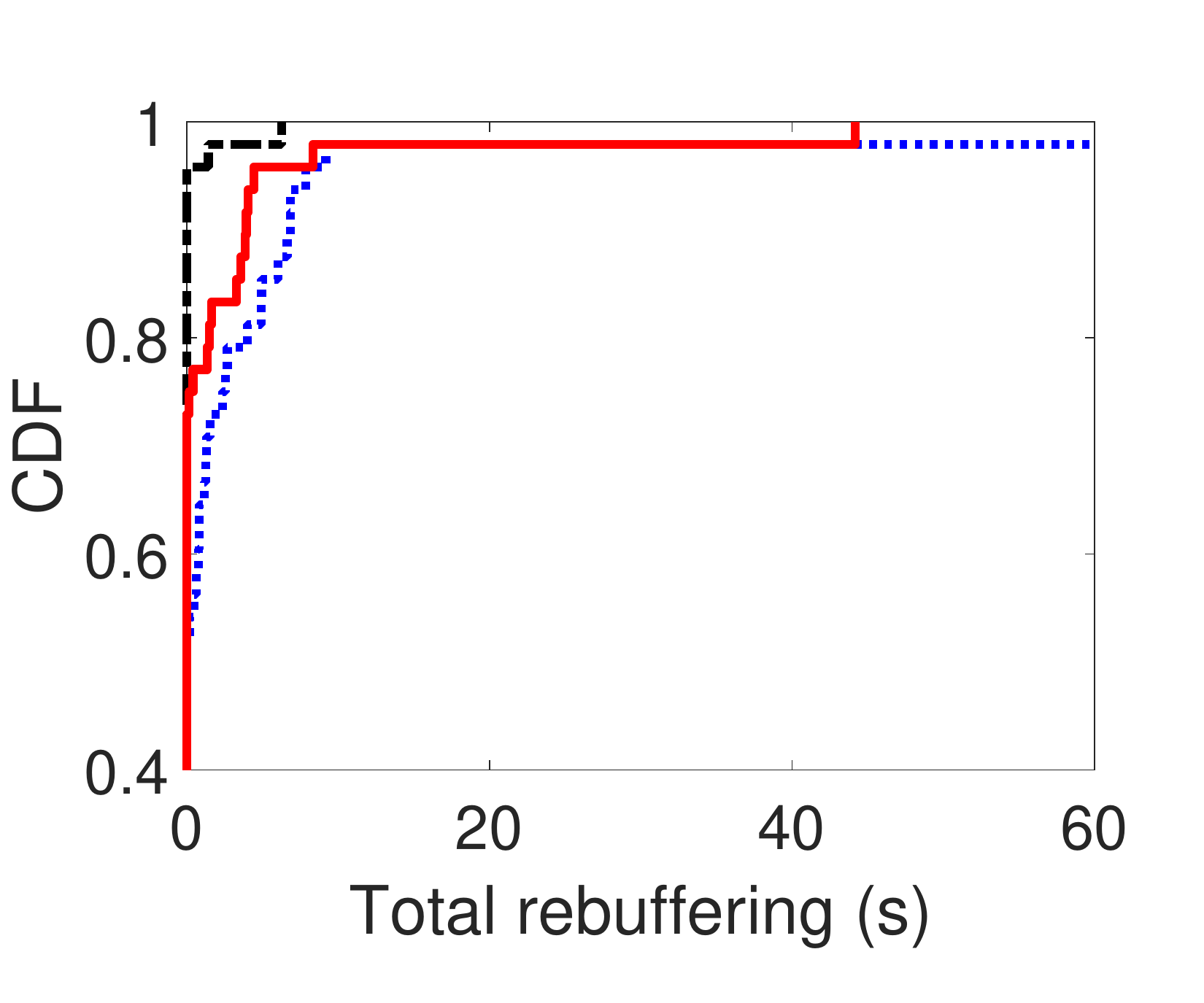}
	%}
	%\subfloat[\label{subfig:CAVA-change}]{%
	\includegraphics[width=0.296\textwidth]{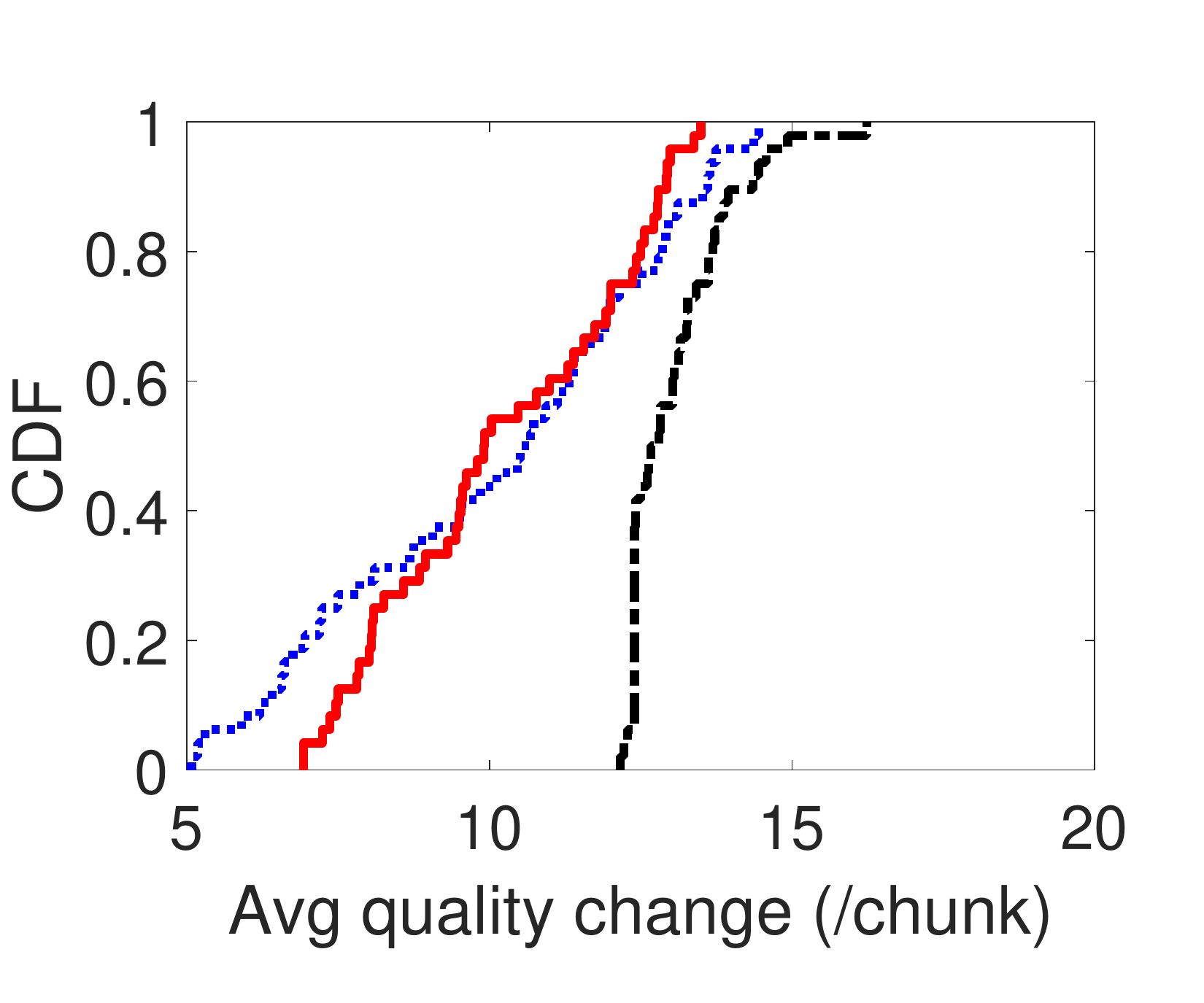}
	%}
	%\subfloat[\label{subfig:CAVA-data-usage}]{%
	\includegraphics[width=0.296\textwidth]{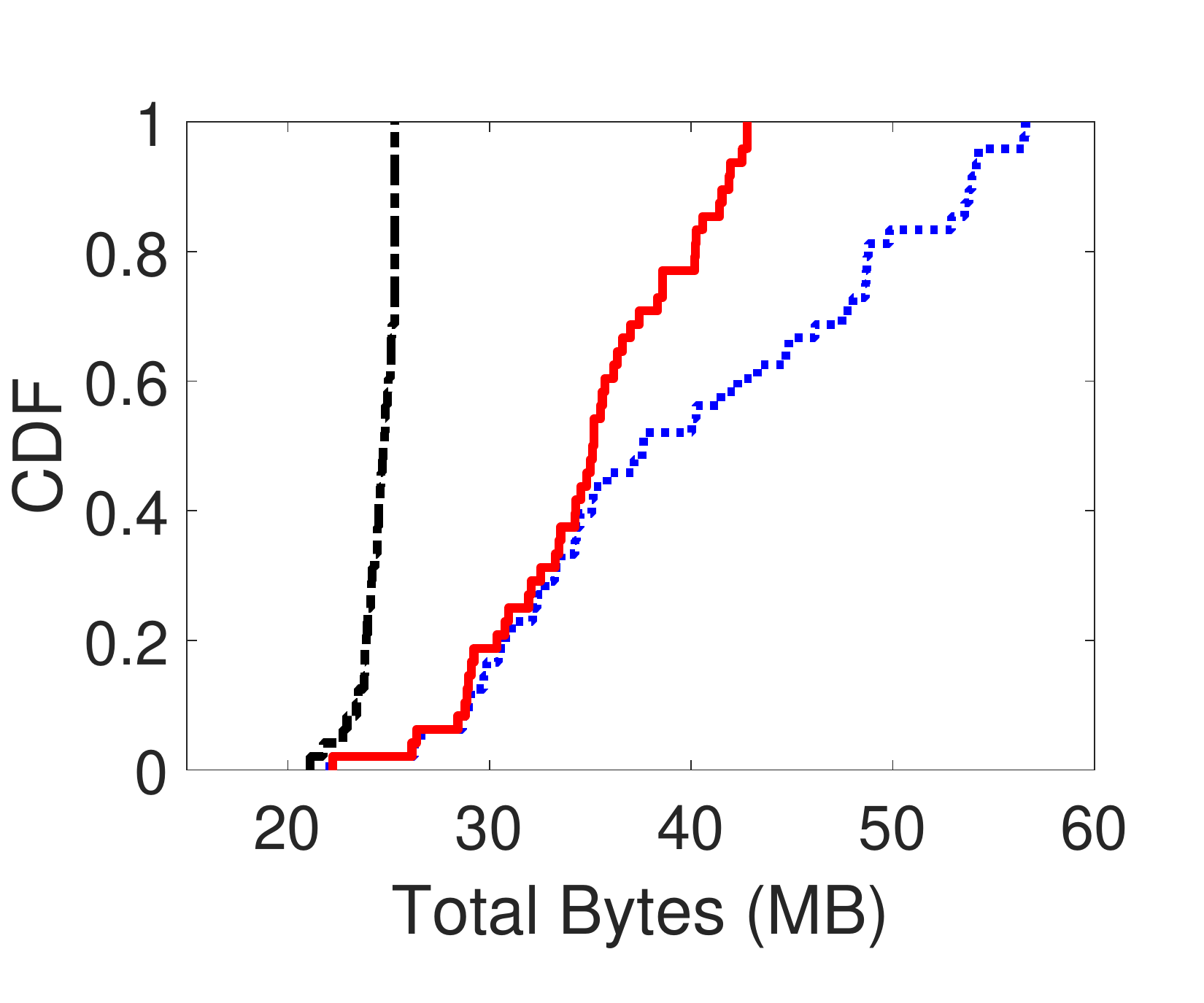}
	%}
	\vspace{-.1in}
	\caption{RobustMPC with CBF vs TBF (ToS, YouTube encoded, target quality 80). }
	\label{fig:comparison-ED-MPC-80}
	%\vspace{-.15in}
\end{figure*}

%the performance of RobustMPC- under this video seems to be pretty good, use ToS instead
\iffalse
%RobustMPC, Q_r=80, for ED, RobustMPC- uses track 3;  RobustMPC+ uses track 4 (which has resolution 480p, which corresponds to what YouTube does)
 \begin{figure*}[ht]
 	\centering
 	%\subfloat[\label{subfig:CAVA-Q4}]{%
 	\includegraphics[width=0.19\textwidth]{figs/YouTube-ED/MPC-quality-ChallengeLTE-80-_phone}
 	%}
 	%\subfloat[\label{subfig:CAVA-low-quality}]{%
 	\includegraphics[width=0.19\textwidth]{figs/YouTube-ED/MPC-quality-low40-ChallengeLTE-80-_phone}
 	%}
 	%\subfloat[\label{subfig:CAVA-rebuffering}]{%
 	\includegraphics[width=0.19\textwidth]{figs/YouTube-ED/MPC-rebuffering-ChallengeLTE-80-_phone}
 	%}
 	%\subfloat[\label{subfig:CAVA-change}]{%
 	\includegraphics[width=0.19\textwidth]{figs/YouTube-ED/MPC-qualitychange-ChallengeLTE-80-_phone}
 	%}
 	%\subfloat[\label{subfig:CAVA-data-usage}]{%
 	\includegraphics[width=0.19\textwidth]{figs/YouTube-ED/MPC-data-ChallengeLTE-80-_phone}
 	%}
 	\vspace{-.1in}
 	\caption{RobustMPC with CBF versus track-based filtering (ED, YouTube encoded, target quality 80). }
 	\label{fig:comparison-ED-MPC-80}
 	\vspace{-.1in}
 \end{figure*}
\fi
%\vspace{-0.1in}
\mysubsection{CBF vs. TBF} \label{sec:track-vs-chunk-perf}
We now compare CBF with TBF, which is used by commercial streaming services such as YouTube and Amazon for reducing data usage
%in cellular networks
(see \S\ref{sec:current-practice-save-data}). Specifically, we consider two variants of TBF, i.e., $\mbox{TBF}^-$ and $\mbox{TBF}^+$ (see \S\ref{sec:cbf-tbf-comp}).\fengc{where $\mbox{TBF}^-$ is  filtering out more tracks
 (and hence has fewer remaining tracks)
than $\mbox{TBF}^+$.}

Fig.~\ref{fig:comparison-ED-MPC-80} plots the performance of RobustMPC with CBF, and with the two TBF variants, referred to as $\mbox{RobustMPC}^-$ and $\mbox{RobustMPC}^+$. The results are for one VBR video when the target quality is 80. For this setting,
the top track in $\mbox{RobustMPC}^-$ and $\mbox{RobustMPC}^+$ is track 3 and 4, respectively.
%\fengc{
%The latter coincides with the practice of the bandwidth saving option in YouTube in cellular networks (which removes all HD tracks, i.e., the tracks with resolution above 480p).}
%
We make the following observations. (i) $\mbox{RobustMPC}^-$ is quite conservative. It provides low rebuffering time. However, the negative side is that it undershoots the target quality (its average deviation from the target quality is 20 VMAF points across the chunks, compared to 15 under RobustMPC+CBF), and leads to a higher percentage of low-quality chunks (it has at least 10\% low-quality chunks for 100\% of the traces, compared to 58\% and 68\% of the traces under RobustMPC+CBF and $\mbox{RobustMPC}^+$, respectively). (ii) $\mbox{RobustMPC}^+$, on the other hand, is too aggressive. It overshoots the target quality (it exceeds the target quality in 50\% of the chunks, compared to 30\% under RobustMPC+CBF), incurs the highest rebuffering time, and consumes the highest network bandwidth among the three schemes. Also, it results in highly variable quality by choosing high quality for some chunks while leaving a higher percentage of chunks with low quality. (iii) RobustMPC+CBF strikes a better balance among the aforementioned factors. It also achieves a quality that is closest to the target quality.
%
%%wb, show a table of the average quality of all the chunks and the average data usage of three schemes, with two target qualities, 60 and 80, for two representive VBR and CBR videos (i.e., four videos in total)
%The above results are for RobustMPC using one video with target quality 80.
%For other videos and target qualities, the relative performance of the three variants of RubostMPC varies, depending on the video and the target quality. Overall, RubostMPC+CBF achieves the best quality in all the cases we explore.
We observe similar results for other videos and target qualities.
% we also observe that RubostMPC+CBF achieves the best quality among the three variants of RubostMPC.
We further compare the three variants of PANDA/CQ and observe that PANDA/CQ+CBF achieves better performance than the others.
%achieves the best quality among the three variants of PANDA/CQ.
%Figures~\ref{fig:comparison-ED-PANDACQ-60} and \ref{fig:comparison-ED-PANDACQ-80} compare the performance of PANDA/CQ with track-based and chunk-level prefiltering.
The reason, as explained in~\S\ref{sec:cbf-tbf-comp}, is that compared to TBF, CBF filters out chunks at a finer granularity (on the basis of chunks instead of tracks), thus allowing it to make better choices.\fengc{The reason why a scheme with CBF outperforms the scheme with TBF is, as explained in \S\ref{sec:cbf-tbf-comp}, CBF filters out chunks at a finer granularity (on the basis of chunks instead of tracks), leading the scheme to make better choices.}

%\vspace{-.1in}
\section{Evaluation of QUAD} \label{sec:qea-eval}
%\shuaic{Having demonstrated the benefits of adding CBF to existing schemes, in this section, }
%
In this section, we compare QUAD
%, the scheme that we designed group-up based on the target quality,
and existing schemes enhanced with CBF. We also evaluate them\fengc{explore the practical viability of CBF and \qea by evaluating
them} in
%showing the results in
%\shuaic{two widely used ABR streaming platforms,}
\texttt{dash.js} and ExoPlayer.
\mysubsection{QUAD vs. CBF} \label{sec:cbf-qea}

%target quality = 80, ED
 \begin{figure*}[ht]
	\centering
	\vspace{-.1in}
	\subfigure[\label{subfig:CAVA-Q4}]{%
	\includegraphics[width=0.296\textwidth]{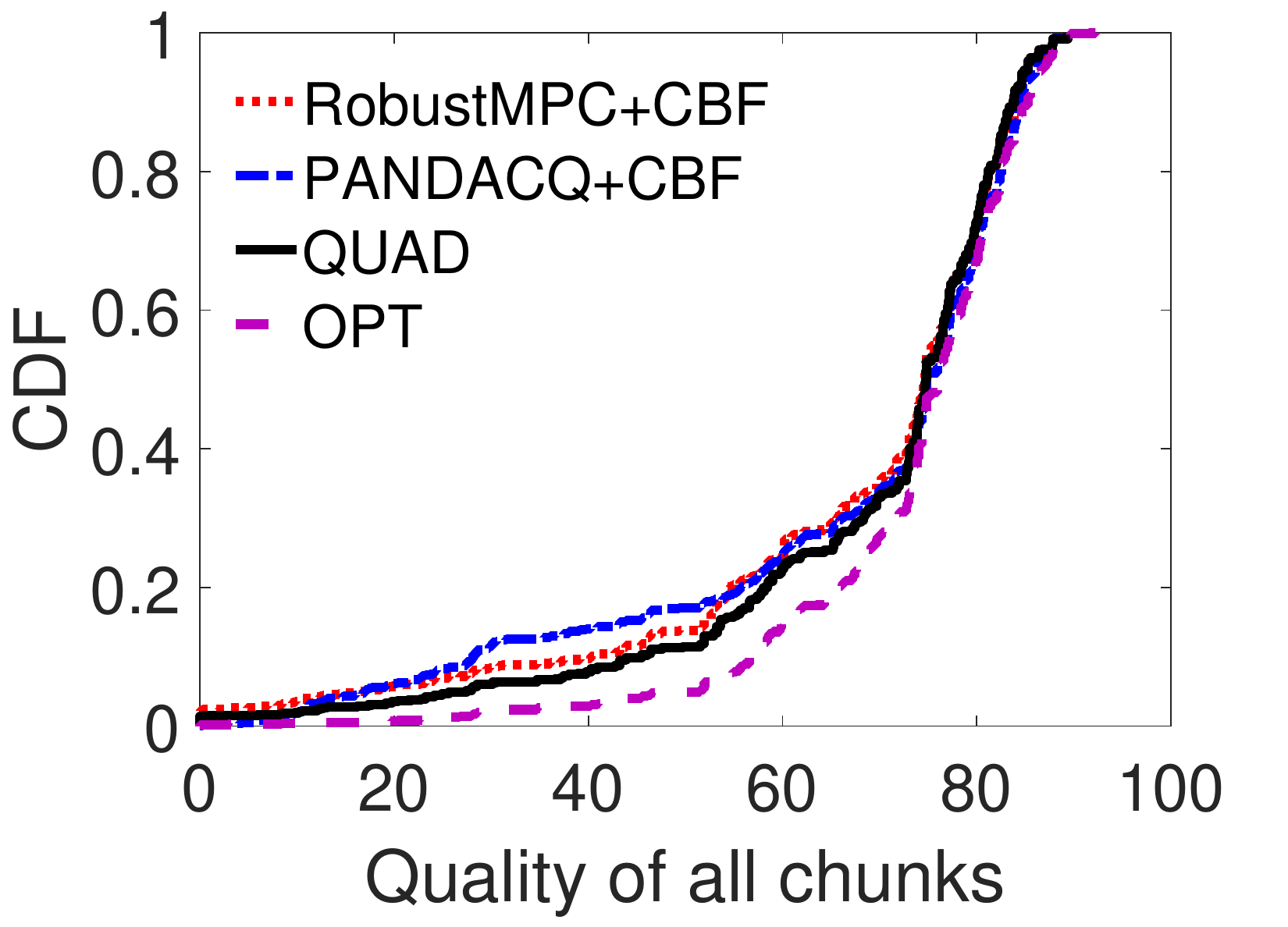}
	}
	\subfigure[\label{subfig:CAVA-low-quality}]{%
	\includegraphics[width=0.296\textwidth]{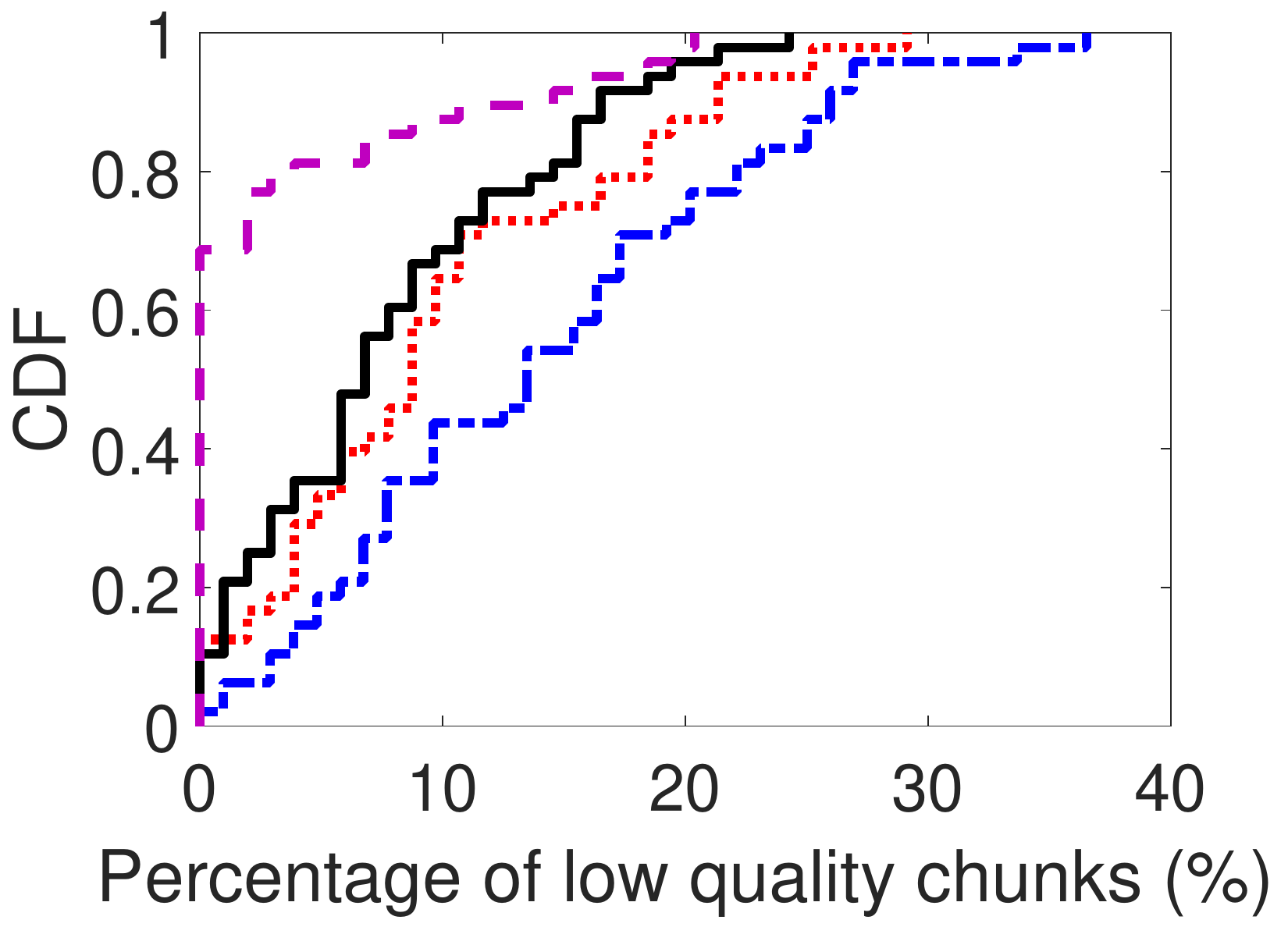}
	}
	\subfigure[\label{subfig:CAVA-rebuffering}]{%
	\includegraphics[width=0.296\textwidth]{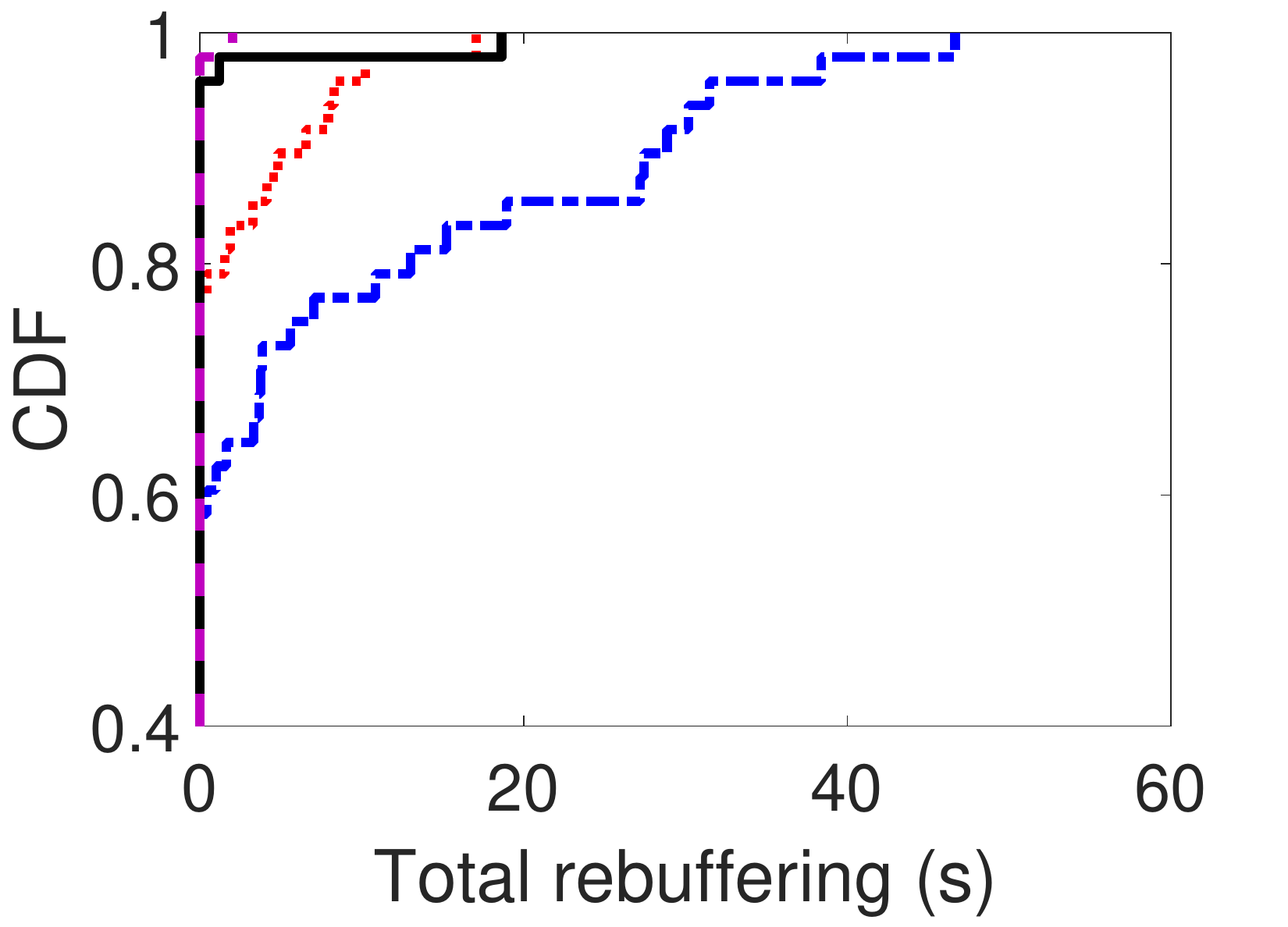}
	}
	\subfigure[\label{subfig:CAVA-change}]{%
	\includegraphics[width=0.296\textwidth]{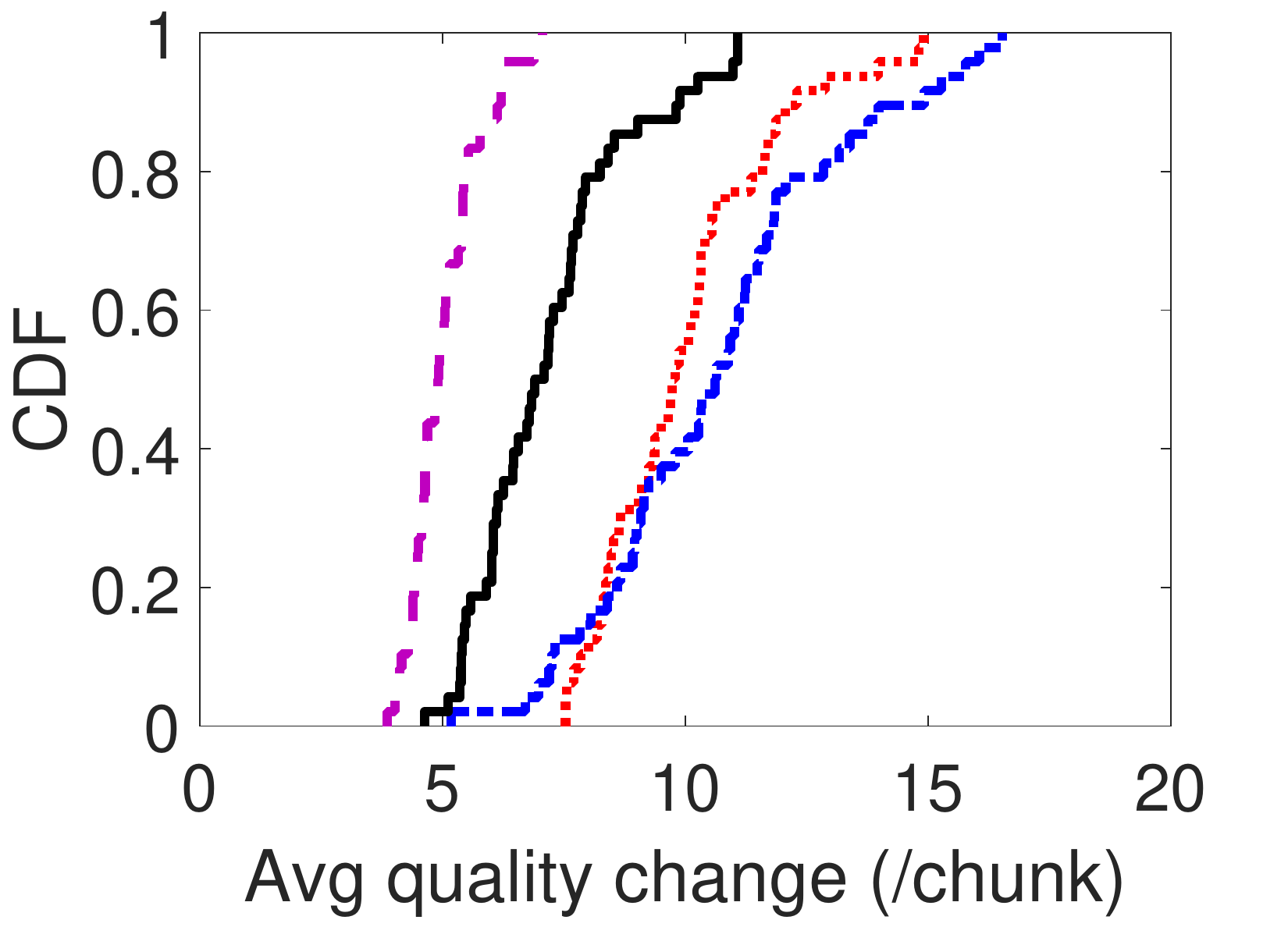}
	}
	\subfigure[\label{subfig:CAVA-data-usage}]{%
	\includegraphics[width=0.296\textwidth]{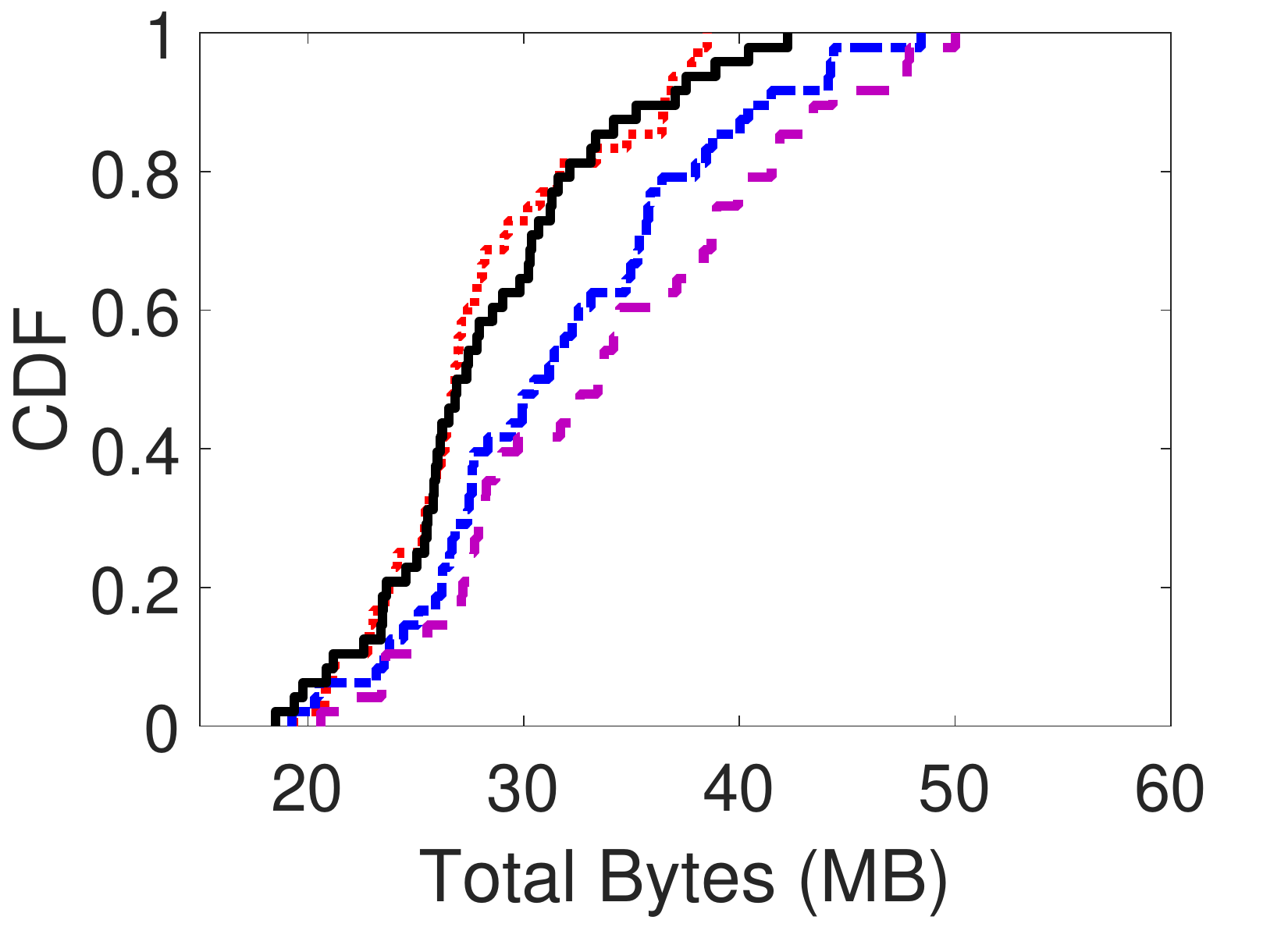}
	}
	%\vspace{-.2in}
	\caption{QUAD vs. two existing schemes with CBF (ED, YouTube encoded, target quality 80). }
	\label{fig:comparison-ED-80-qea}
%	\vspace{-.15in}
\end{figure*}
%
%%target quality = 80, BBB
\begin{figure*}[ht]
	\centering
	\subfigure[\label{subfig:CAVA-Q4}]{%
	\includegraphics[width=0.296\textwidth]{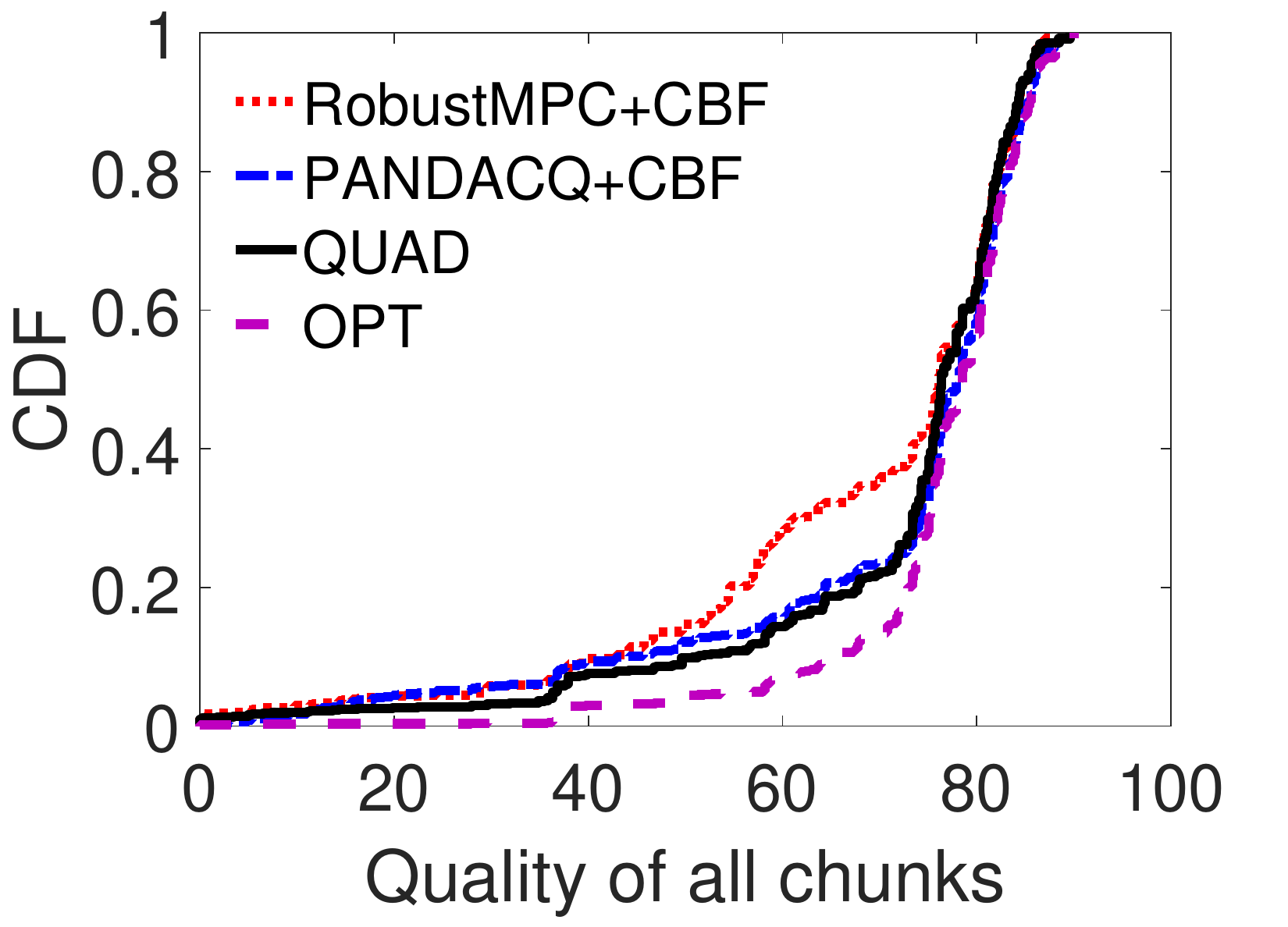}
	}
	\subfigure[\label{subfig:CAVA-low-quality}]{%
	\includegraphics[width=0.296\textwidth]{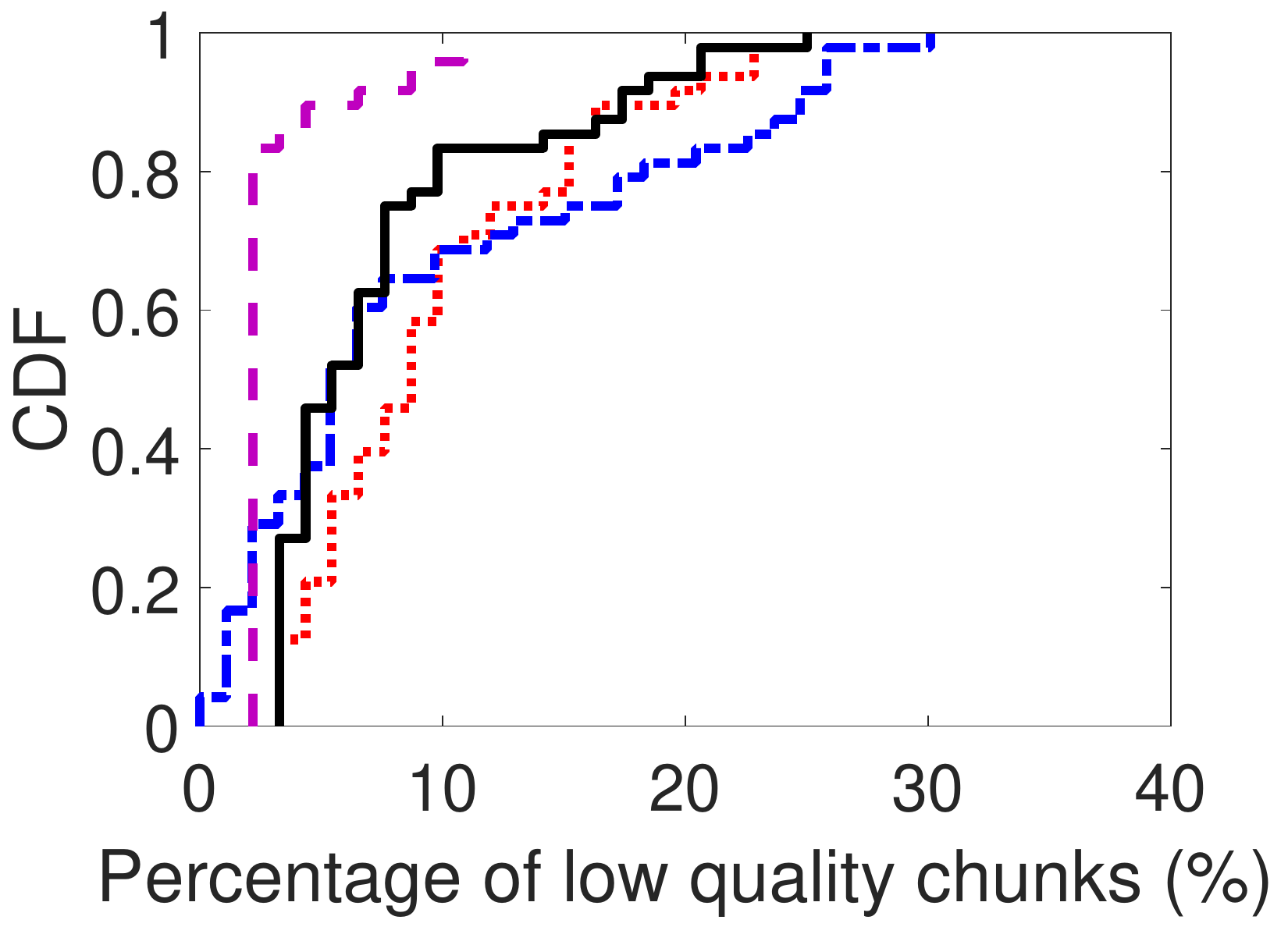}
	}
	\subfigure[\label{subfig:CAVA-rebuffering}]{%
	\includegraphics[width=0.296\textwidth]{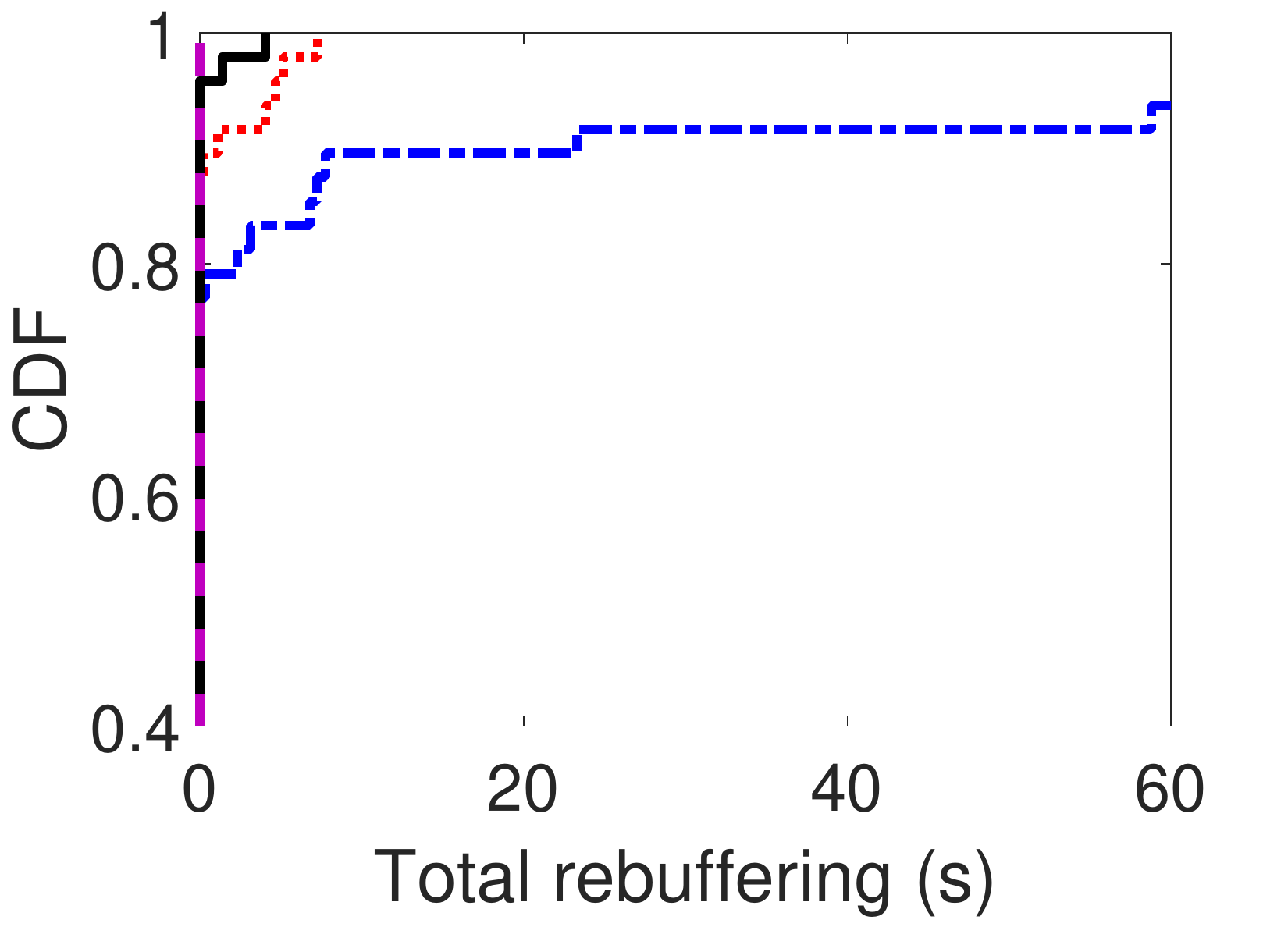}
	}
	\subfigure[\label{subfig:CAVA-change}]{%
	\includegraphics[width=0.296\textwidth]{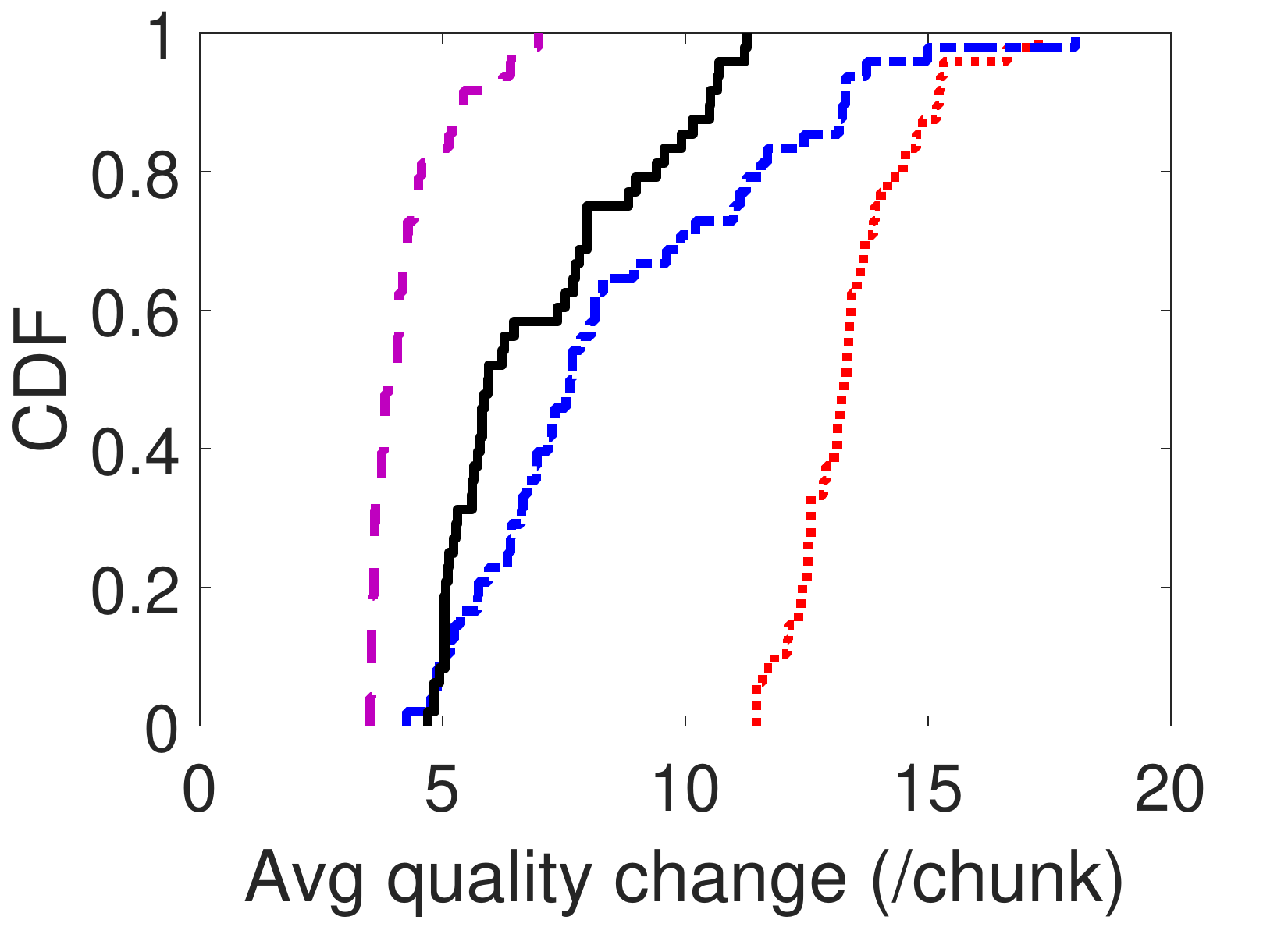}
	}
	\subfigure[\label{subfig:CAVA-data-usage}]{%
	\includegraphics[width=0.296\textwidth]{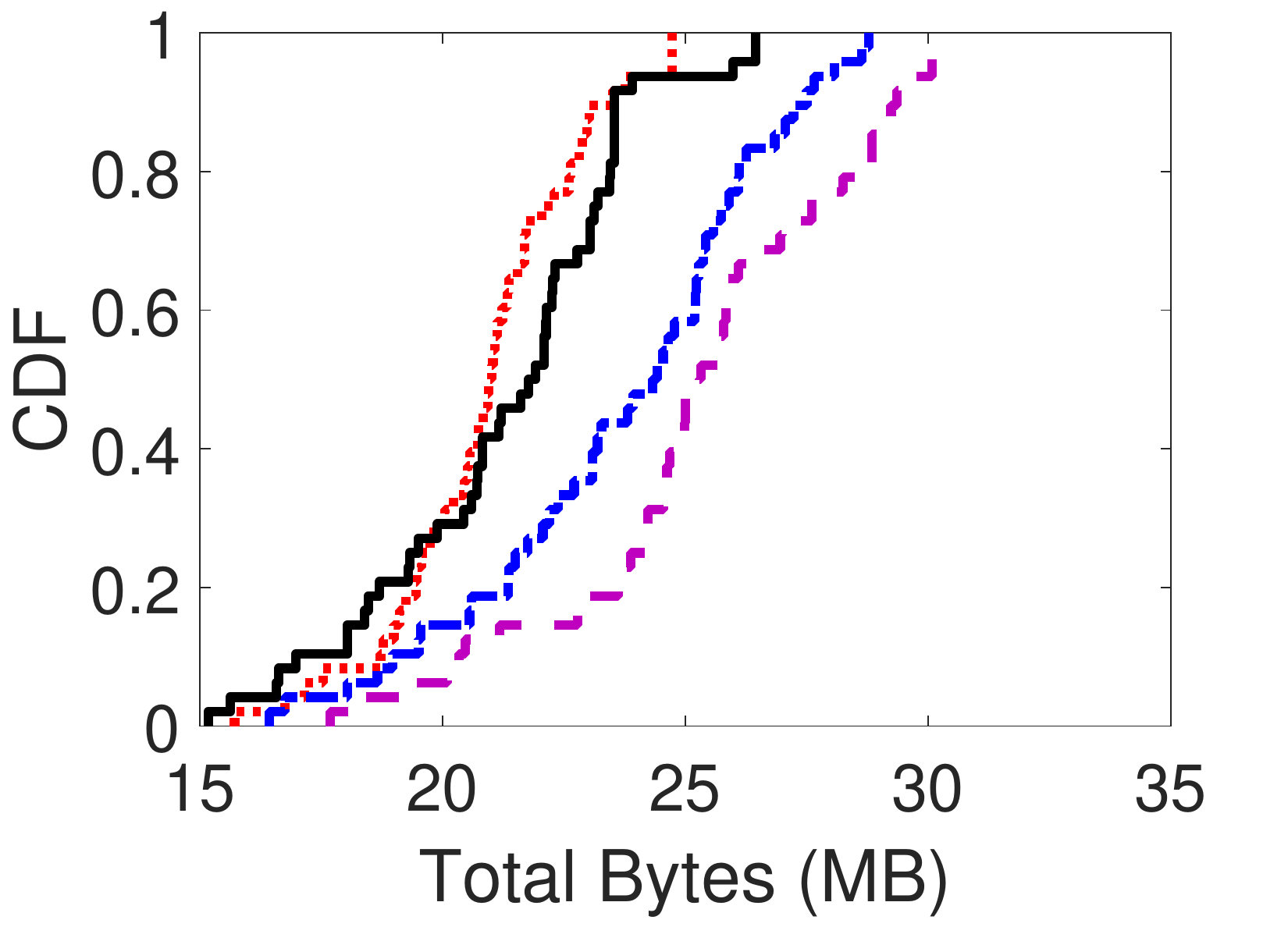}
	%\fi
	}
	\caption{QUAD vs. two existing schemes with CBF (ED, BBB, YouTube encoded, target quality 80). }
	\label{fig:comparison-80-qea-cbf}
	%\vspace{-.1in}
\end{figure*}

%We compare the performance of \qea and two existing schemes (RobustMPC and PANDA/CQ) with CBF.
%Fig.~\ref{fig:comparison-80-qea-cbf} shows the results for two videos.
Fig.~\ref{fig:comparison-80-qea-cbf}  plots the performance of QUAD and two existing schemes (RobustMPC and PANDA/CQ) with CBF for two videos, based on trace-driven simulations.
%
%\shuaic{(ED on the top row and BBB on the bottom row),} with target quality 80.
%
It also plots the results of the offline optimal scheme (see \S\ref{sec:setup}).\fengc{(designed to optimize the quality, quality changes and rebuffering jointly assuming the network bandwidth for the entire playback is known beforehand, see \S\ref{sec:setup}).}
%We see that the performance of \qea is closest to the offline optimal in \emph{all} the performance metrics.
We observe that for all five metrics except the data usage, QUAD achieves performance closest to that of the offline optimal;
% in all the cases we explore. The
for data usage, the offline optimal uses more data than other schemes, consistent with the best quality that it achieves.\fengc{We observe that for four out of five metrics (chunk quality, the percentage of low-quality chunks, quality changes, and rebuffering), \qea achieves performance closest to that of the offline optimal;
% in all the cases we explore. The
for the last metric, data usage, the offline optimal uses more data than other schemes, consistent with the best quality that it achieves.}
QUAD outperforms RobustMPC+CBF and PANDA/CQ+CBF for both videos in Fig.~\ref{fig:comparison-80-qea-cbf}. For video BBB (Fig.~\ref{fig:comparison-80-qea-cbf} bottom row), while RobustMPC+CBF has similar rebuffering as QUAD, it leads to a noticeably worse quality (the average deviation from the target quality is 14 VMAF points across the runs under RobustMPC+CBF, and 10 under QUAD). For video ED (Fig.~\ref{fig:comparison-80-qea-cbf} top row), the overall quality of the two schemes is similar, while RobustMPC+CBF has more rebuffering than QUAD (it has rebuffering in 23\% of the runs, compared to only 4\% in QUAD).
For both videos, the average quality change per chunk under QUAD is much lower compared to RobustMPC+CBF:
%(\shuaic{the average quality change across all the runs is }
7 under QUAD for both videos, compared to 10 and 13 under RobustMPC+CBF.
The overall quality of  PANDA/CQ+CBF is comparable to that of QUAD  for both videos, but has significantly more rebuffering, higher quality changes, and uses more data.

%Overall, while maintaining a similar target quality compared to RobustMPC and  PANDA/CQ with CBF, \qea outperforms the other two schemes
% \shuaic{in several metrics such as the rebuffering time, percentage of low-quality chunks, and quality changes,}
%due to its grounds-up design based on a control-theoretical framework.

%\mysection{DASH Implementation Results} \label{sec:dash}

\mysubsection{\texttt{dash.js} based Evaluation} \label{sec:dash}
%from Real Implementation

%Compare \qea with BOLA+CBF, Q_r=80,
\begin{figure*}[ht]
	\vspace{-.1in}
	\centering
	%\subfloat[\label{subfig:CAVA-Q4}]{%
	\includegraphics[width=0.296\textwidth]{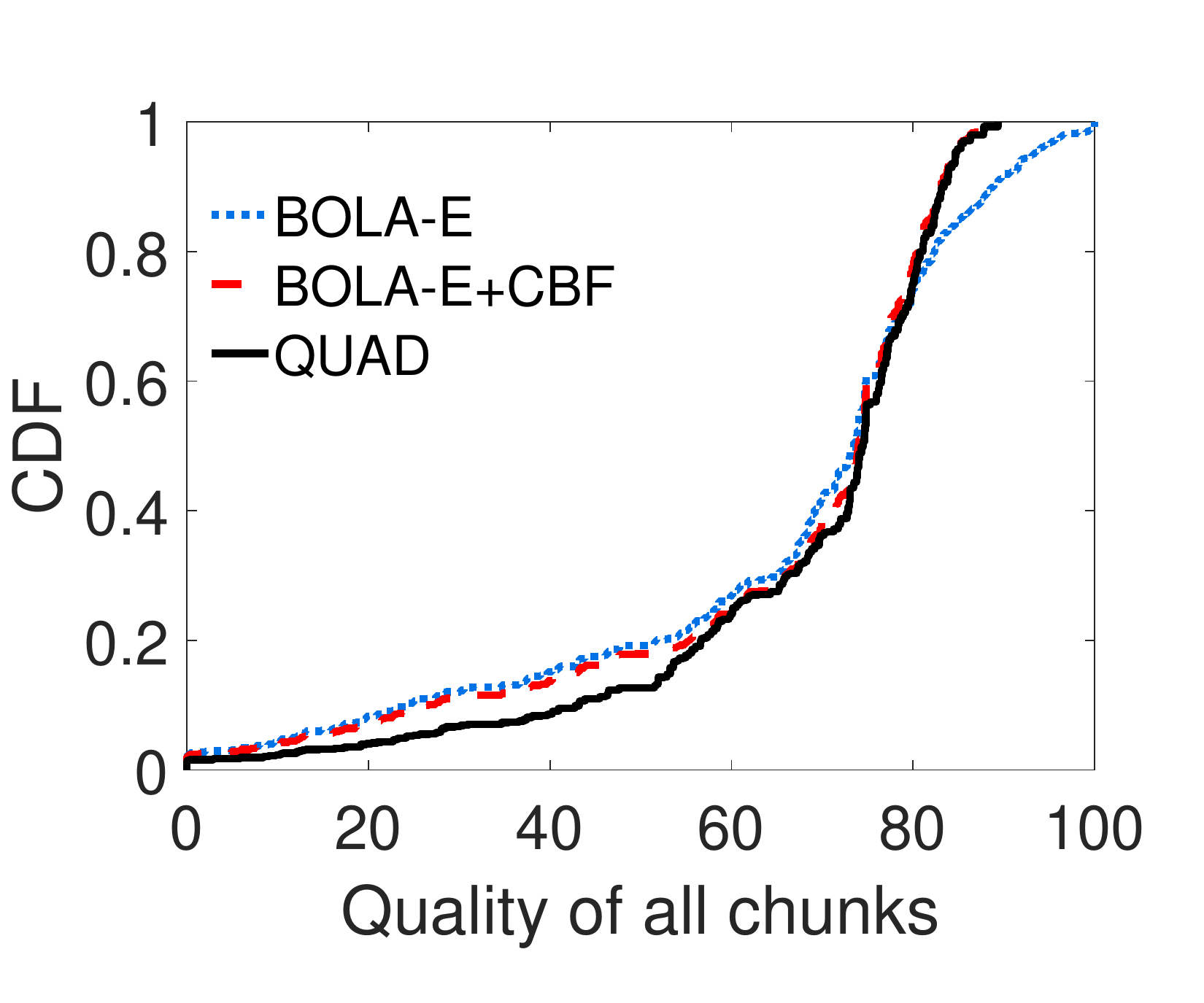}
	%}
	%\subfloat[\label{subfig:CAVA-low-quality}]{%
	\includegraphics[width=0.296\textwidth]{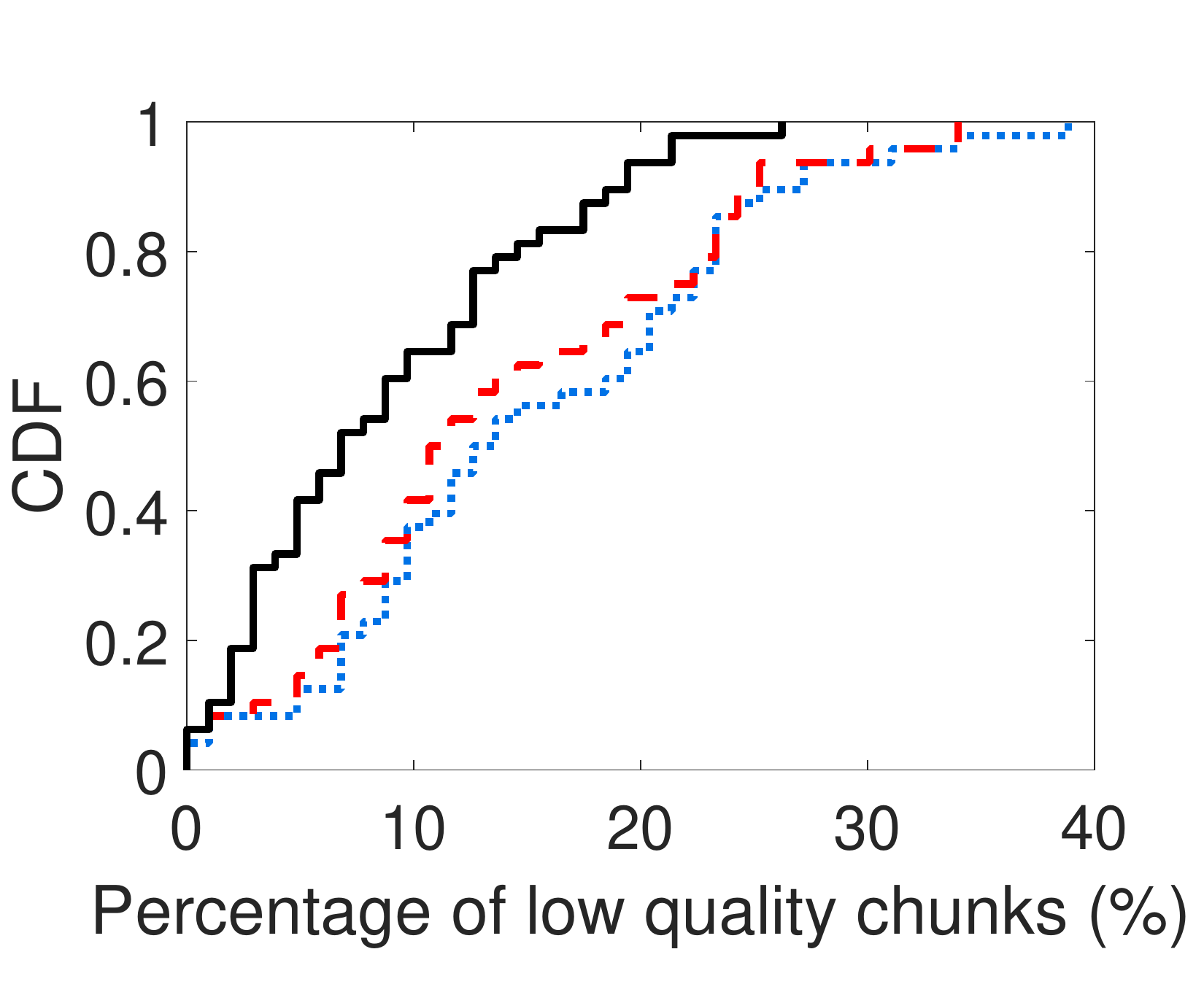}
	%}
	%\subfloat[\label{subfig:CAVA-rebuffering}]{%
	\includegraphics[width=0.296\textwidth]{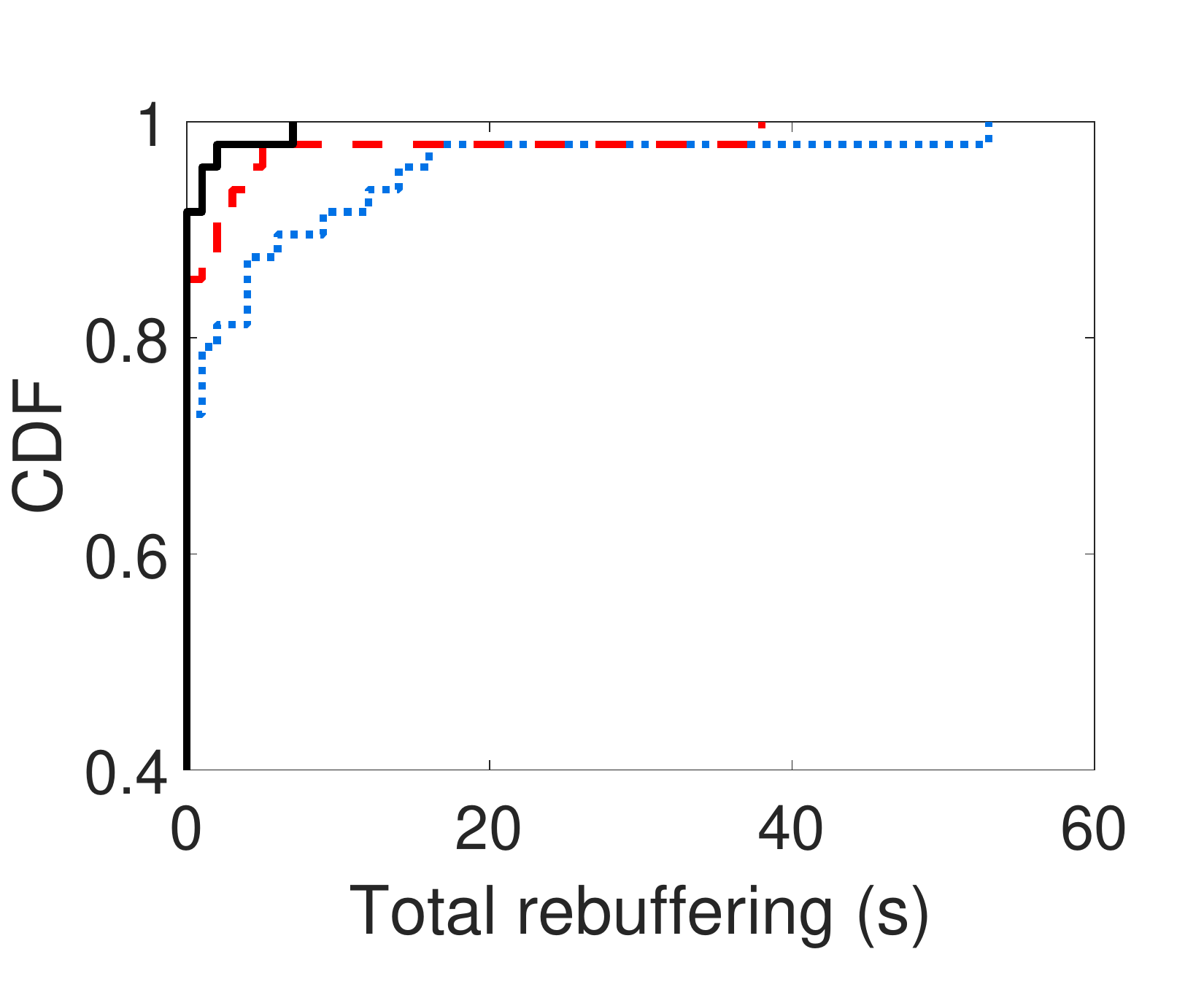}
	%}
	%\subfloat[\label{subfig:CAVA-change}]{%
	\includegraphics[width=0.296\textwidth]{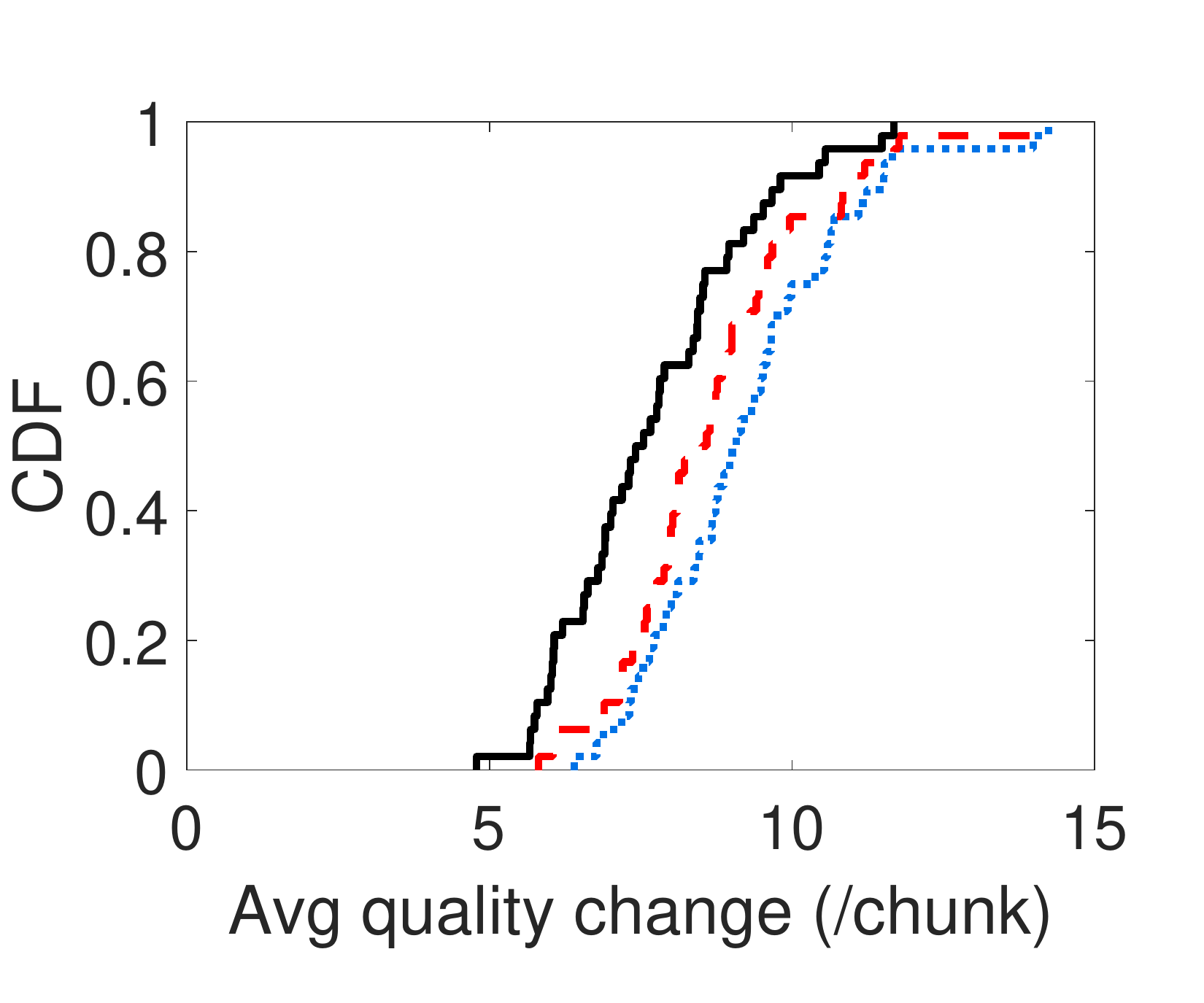}
	%}
	%\subfloat[\label{subfig:CAVA-data-usage}]{%
	\includegraphics[width=0.296\textwidth]{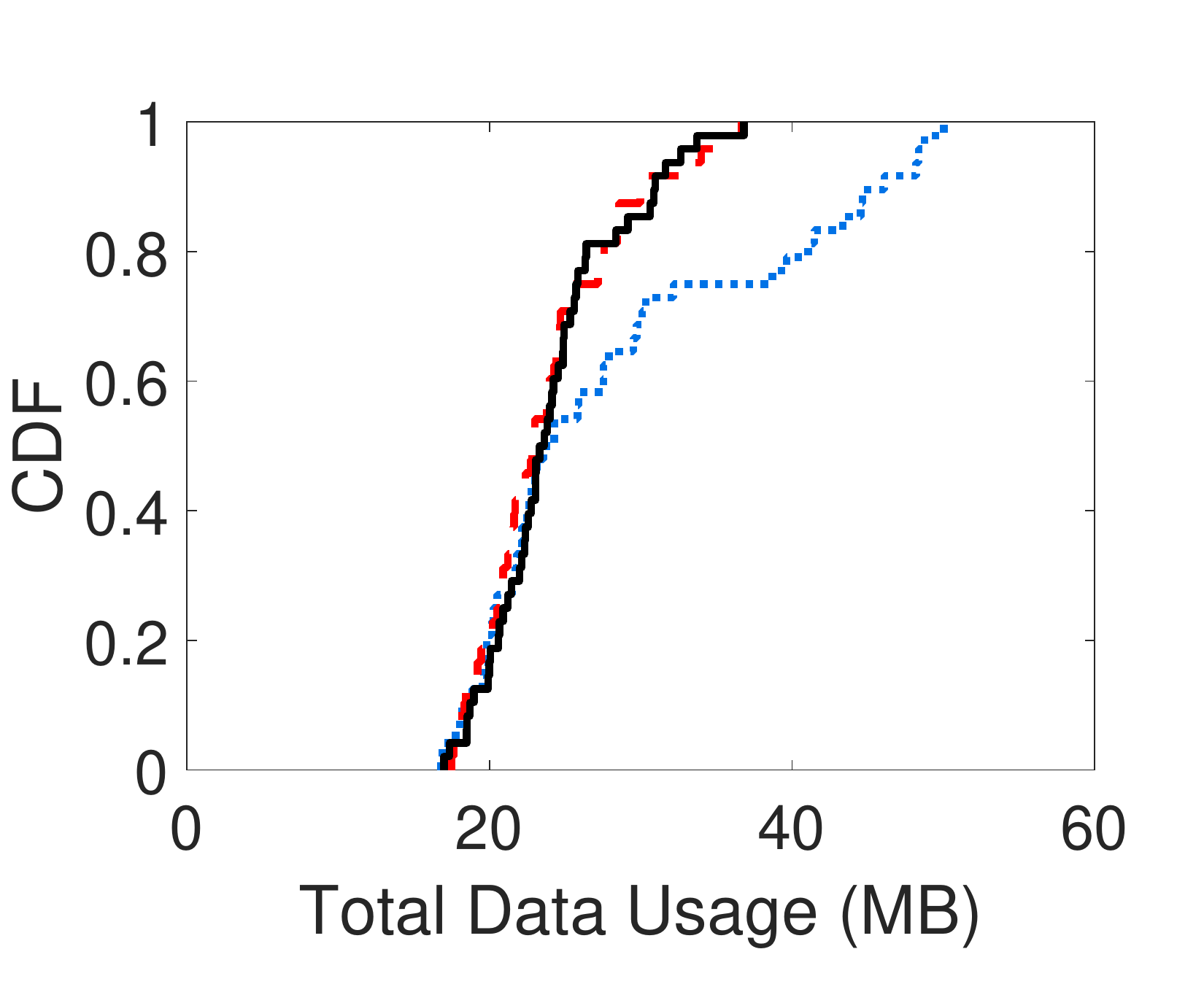}
	%}
	\vspace{-.1in}
	%\caption{\qea and BOLA-E+CBF in \texttt{dash.js} (ED, YouTube encoded, target quality 80). }
	\caption{QUAD vs.  BOLA-E with and without CBF in \texttt{dash.js} (ED, YouTube encoded, target quality 80). }
	\label{fig:comparison-dash-js}
	%\vspace{-.1in}
\end{figure*}

%%\subsection{\texttt{dash.js} Implementation} \label{sec:qea-implementation}

Using our developed QUAD in \texttt{dash.js} (version 2.4.1), we compare its performance with BOLA-E~\cite{spiteri2016bola}
%\shuaic{which has been implemented in \texttt{dash.js},}
and BOLA-E+CBF.
%\shuaic{, which is BOLA-E combined with the CBF module that we implemented}.
%We compare  the performance of \qea with that of BOLA-E+CBF in \texttt{dash.js}.
The original BOLA-E design takes a single bitrate rate for each track, which is suitable for CBR videos; for VBR videos, we improved the design by using the actual chunk sizes in the ABR logic based on~\cite{spiteri2016bola}.
% as suggested by~\cite{spiteri2016bola}.
Our evaluation below uses two computers (Ubuntu 12.04 LTS, CPU Intel Core 2 Duo, 4GB memory) with a 100 Mbps direct network connection to emulate the video server (Apache \texttt{httpd}) and client. At the client, we use \texttt{selenium}~\cite{selenium}
%(an automated software testing tool for web applications )
to
%programmatically
 run a Google Chrome web browser and use \texttt{dash.js} APIs to  collect the streaming performance data.
We use \texttt{tc} to emulate  real-world variable mobile network conditions by ``replaying'' the network traces (\S\ref{sec:setup}).
We find QUAD is very light-weight. With the current prototype,
the total execution time of QUAD
%the runtime overhead (from profiling)
is only about 10ms for a 10mins video.

%We find that \qea is very light-weight -- even with the current prototype implementation,
%the total execution time of QUAD
%the runtime overhead (from profiling)
%is only around 10 ms for a 10-minute video.

Fig.~\ref{fig:comparison-dash-js} shows the performance of QUAD, BOLA-E, and BOLA-E+CBF for one video with target quality 80. We observe that BOLA-E+CBF  significantly outperforms BOLA-E in all performance metrics. QUAD further outperforms BOLA-E+CBF in achieving  37\% reduction in average percentage of low-quality chunks and 12\% reduction in average quality changes  across the runs.
QUAD has rebuffering in 8\% of the runs, compared to 15\% under BOLA-E+CBF.
The chunk quality of BOLA-E+CBF is close to that of QUAD,  with data usage close to
%slightly lower
than that of QUAD. The results for other videos and target qualities show a similar trend.

%Compare \qea with default ExoPlayer rate adaptation algorithm with prefiltering.
\begin{figure*}[ht]
	\centering
	%\subfloat[\label{subfig:CAVA-Q4}]{%
	\includegraphics[width=0.396\textwidth]{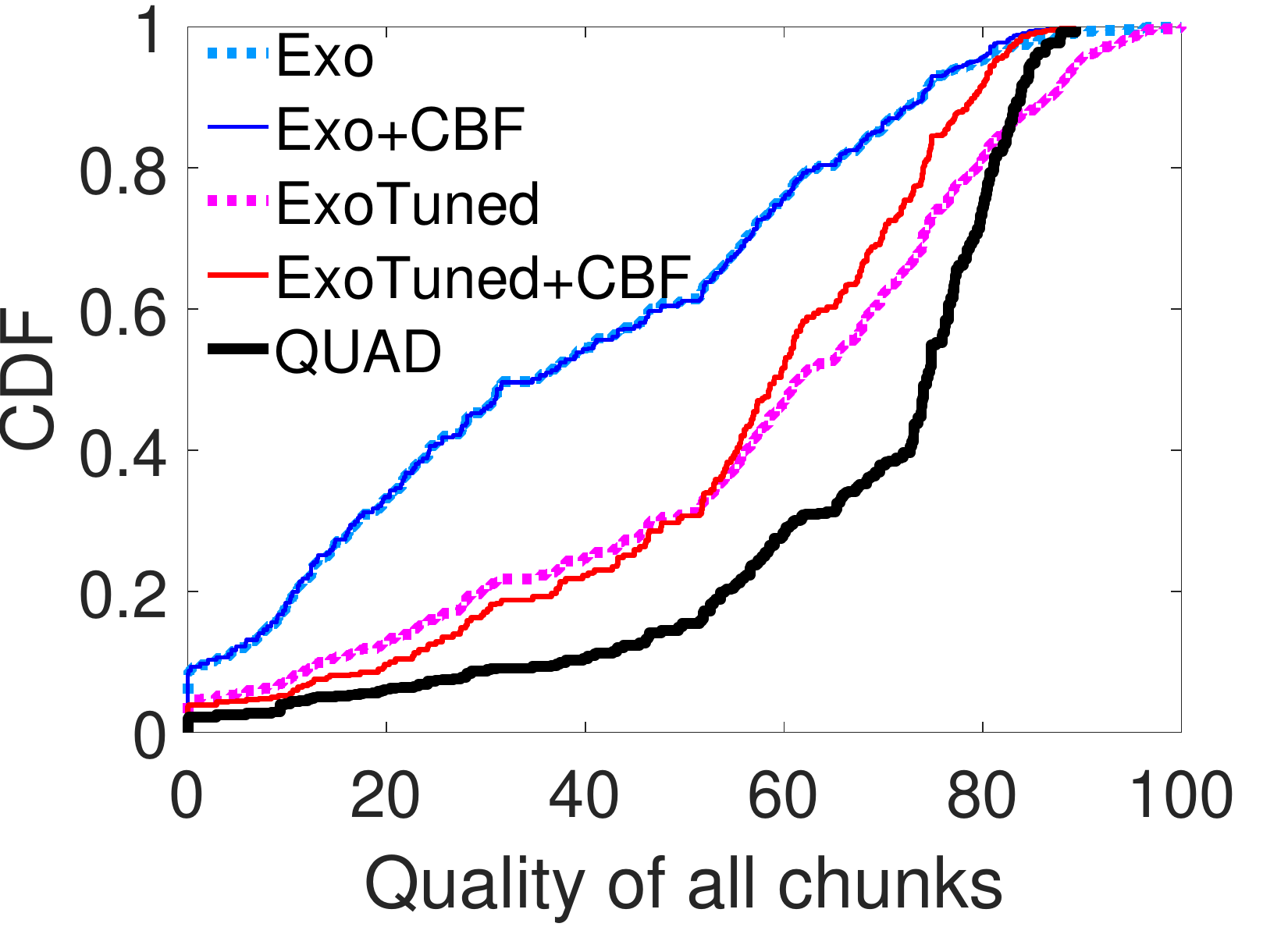}
	%}
	%\subfloat[\label{subfig:CAVA-low-quality}]{%
	\includegraphics[width=0.396\textwidth]{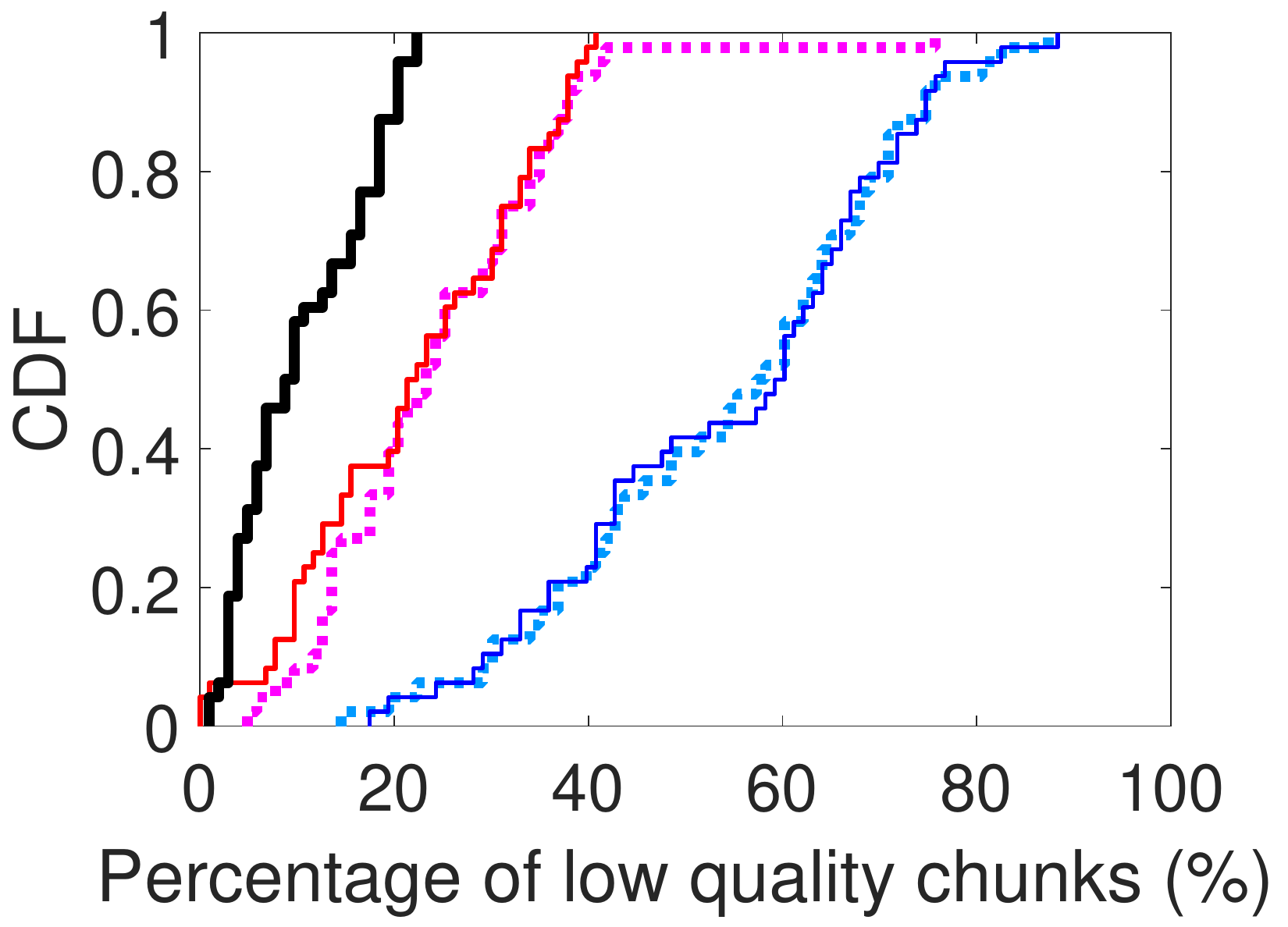}
	%}
	%\subfloat[\label{subfig:CAVA-rebuffering}]{%
	\includegraphics[width=0.396\textwidth]{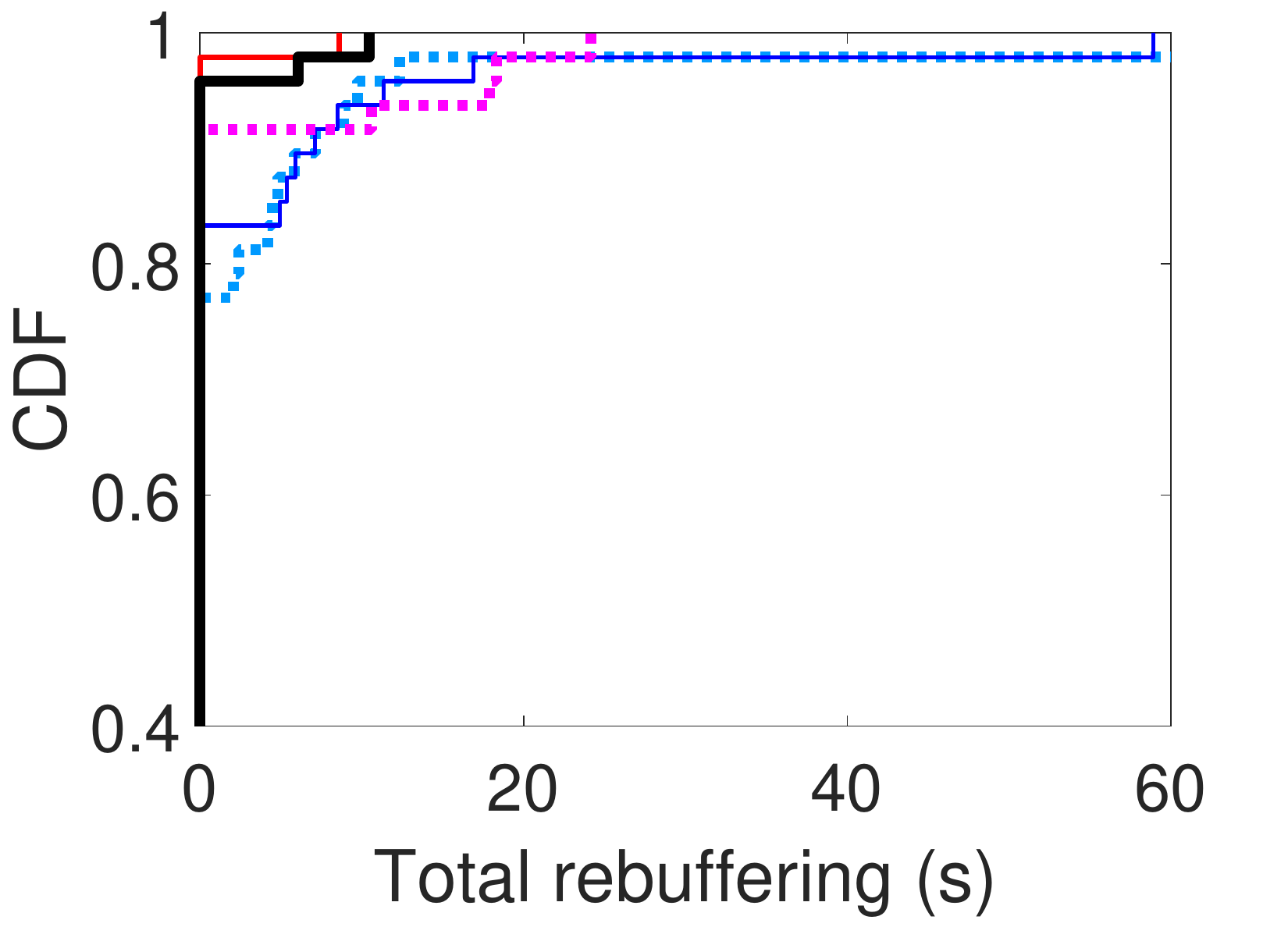}
	%}
	%\subfloat[\label{subfig:CAVA-change}]{%
	\includegraphics[width=0.396\textwidth]{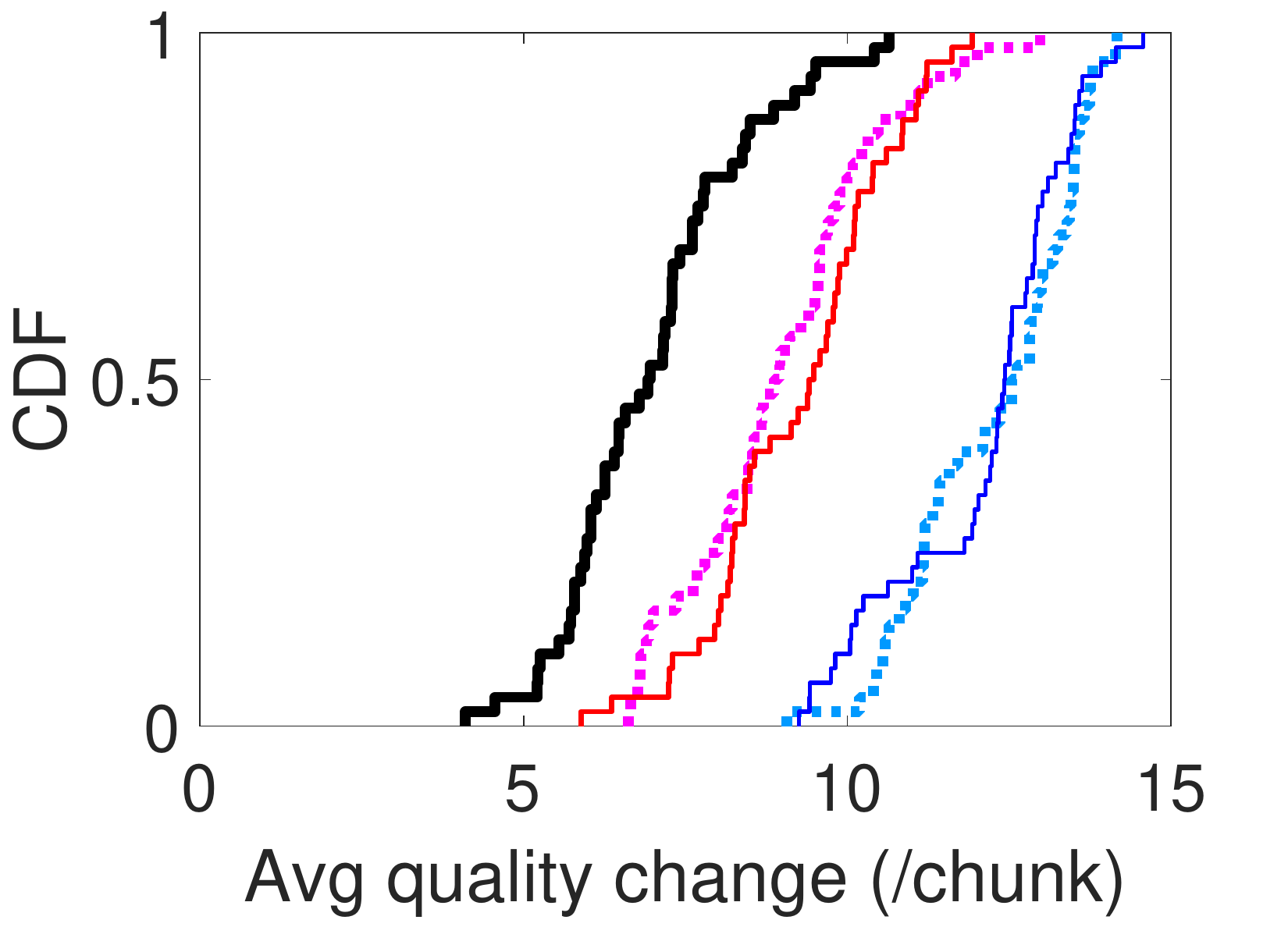}
	%}
	%\subfloat[\label{subfig:CAVA-data-usage}]{%
	\includegraphics[width=0.396\textwidth]{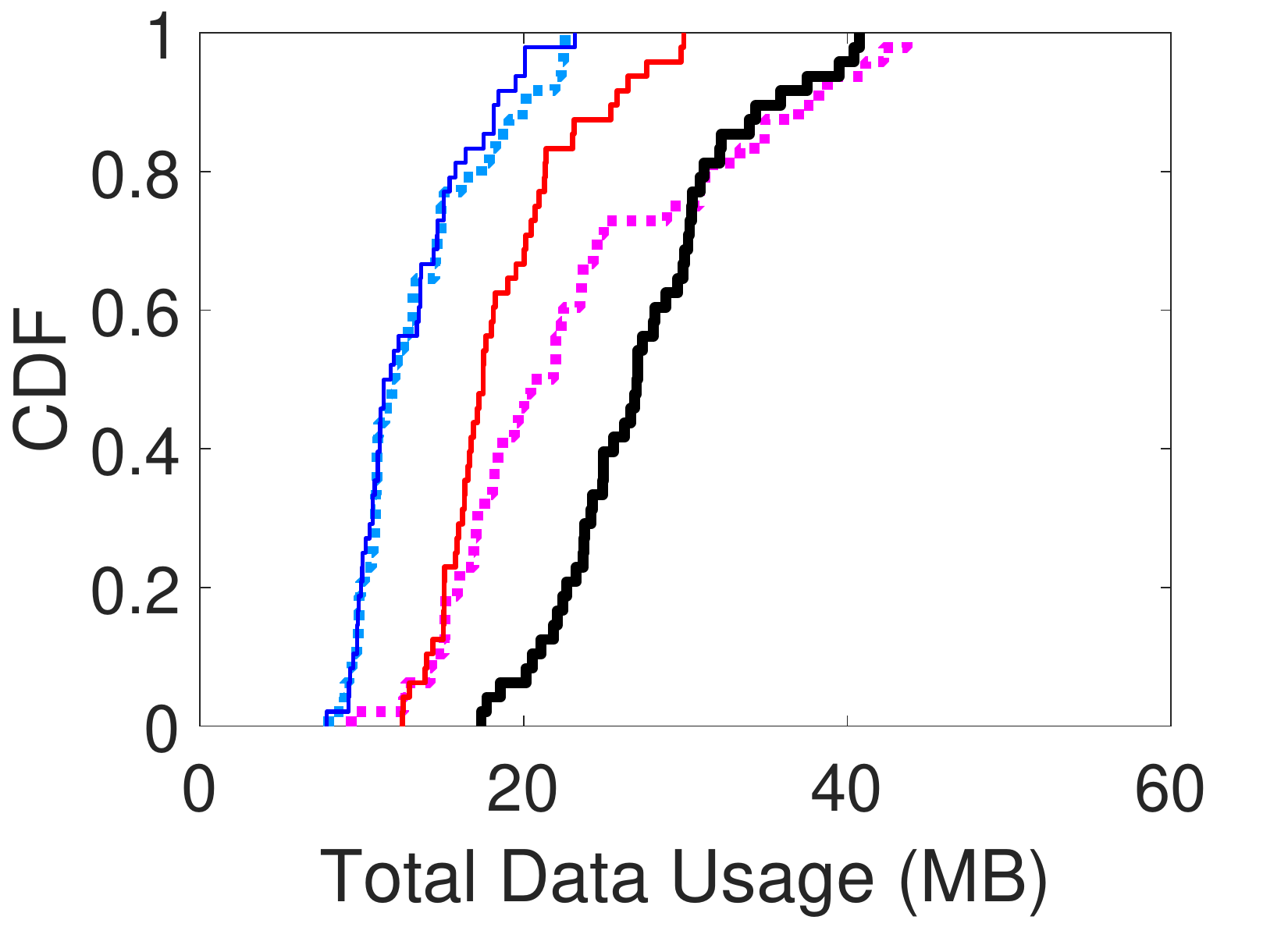}
	%}
	\vspace{-.1in}
	\caption{QUAD vs. Exo   with and without CBF in ExoPlayer (ED, YouTube encoded, target quality 80). }
	\label{fig:comparison-exoplayer}
	%\vspace{-.1in}
\end{figure*}

%\section{Test in the wild} \label{sec:exoplayer}
\mysubsection{ExoPlayer based Evaluation} \label{sec:exoplayer}

We compare the performance of QUAD (using our implementation)
%the default ExoPlayer algorithm
with Exo (the default
%ExoPlayer
ABR algorithm, see \S\ref{sec:setup}, with and without CBF)
in ExoPlayer (version 2.4.4). The client is an LG V20 phone (Qualcomm Snapdragon 820, 64GB storage and 4 GB RAM) with Android 7.0. The experiments are conducted over a WiFi network (with consistent tens of Mbps bandwidth). We again emulate the 50 challenging cellular network traces (\S\ref{sec:setup}) by ``replaying'' the traces.\fengc{in-the-wild results over cellular networks are in
%deferred to
\S\ref{sec:in-the-wild}.}

Fig.~\ref{fig:comparison-exoplayer} shows the results for two variants of Exo:
the first using the default parameters, and the second (referred to as \emph{ExoTuned}) using tuned parameters to improve its quality.
In the first variant (i.e., the default ABR logic),
%The default ABR logic in Exoplayer selects tracks based on historical bandwidth estimation.
	% It allows to increase the track only if buffer is larger than 10 seconds; decrease the track only when buffer level is less than 25 seconds.
	the downloading stops when the buffer level reaches 30 seconds and resumes when the buffer is less than 15 seconds. In ExoTuned, the parameters are set so that the player stops downloading  when the buffer reaches 120 seconds (the value we have used for other schemes, see \S\ref{sec:setup}), and resumes downloading when the buffer level drops to 105 seconds.
%(the default values for these two parameters are 30 and 15 seconds respectively).
Correspondingly, the two parameters for QUAD are set to 120 and 105 seconds as well. We see that ExoTuned indeed achieves better performance compared to Exo since the
 larger buffer allows the player to download and store more content in the buffer.
For both Exo and ExoTuned, adding CBF avoids downloading chunks with excessively high quality, which reduces rebuffering and the data usage (for these two schemes, adding CBF does not lead to much improvement in quality since their quality selections are already quite conservative). We further see that QUAD significantly outperforms all Exo variants.
%based algorithms.
%variants of Exo. logic.
%the ExoPlayer algorithm.
Compared to Exo,
%(i.e., no parameter tuning and no CBF),
QUAD reduces the average deviation from the target quality by 64\%, reduces the average percentage of low-quality chunks across all runs by 81\%, reduces the average rebuffering (over of the subset of runs where either algorithm has rebuffering) by 86\%, and reduces the average quality change across the runs by 43\%.
Compared to %the best Exo variant,
%ExoPlayer algorithm with tuned parameters+CBF,
ExoTuned+CBF, the corresponding reductions in the deviation from the target quality, the average percentage of low-quality chunks, and the quality changes are 40\%, 46\%, and 22\%, respectively, albeit QUAD uses more data.
%higher data usage.

%\qea achieves closer to the target quality (reduces the average deviation from the target quality by 40\%) and lower percentage of low-quality chunks (reduces the average percentage of low-quality chunks across the runs by 46\%),  and reduces the average percentage of quality change across the runs by 22\%, albeit at slightly higher data usage.

%\setlength{\belowdisplayskip}{1pt}
\begin{table}[t]
	\centering
	\setlength{\arrayrulewidth}{1pt}	
\small
	\caption{In-the-wild tests using ExoPlayer.}
	\label{tb:wildTest}
%\begin{tabular}{|c|c|c|c|c|c|}	
\begin{tabular}{l c c c c c }
	\hline
	& \begin{tabular}[c]{@{}c@{}}Avg. dev.\\ from\\ target\\ quality\end{tabular} & \begin{tabular}[c]{@{}c@{}}Avg. \% of \\ low\\ quality\\ chunks\end{tabular} & \begin{tabular}[c]{@{}c@{}}Avg.\\ stall\\ dura.\\ (s)\end{tabular} & \begin{tabular}[c]{@{}c@{}}Avg.\\ quality\\ change\end{tabular} & \begin{tabular}[c]{@{}c@{}}Data\\ usage\\ (MB)\end{tabular} \\ \hline
	\begin{tabular}[c]{@{}c@{}}Low\\bw.\end{tabular}    & \begin{tabular}[c]{@{}c@{}}  19.5$\pm$6.1 \\20.0$\pm$6.2 \\ 11.9$\pm$2.9  \end{tabular}                                                                      & \begin{tabular}[c]{@{}c@{}}  17.3$\pm$10.2
	\\ 14.2$\pm$8.5 \\6.5$\pm$3.7          \end{tabular}                                                                                         & \begin{tabular}[c]{@{}c@{}} 2.4$\pm$4.6 \\ 0 \\0                   \end{tabular}                                              & \begin{tabular}[c]{@{}c@{}}  7.8$\pm$0.9\\ 8.5$\pm$1.1 \\ 5.6$\pm$1.2        \end{tabular}                                                         & \begin{tabular}[c]{@{}c@{}}  26.1$\pm$6.5 \\   21.0$\pm$4.3\\ 28.8$\pm$ 3.7   \end{tabular}                                                   \\ \hline
	\begin{tabular}[c]{@{}c@{}}High\\bw. \end{tabular}   & \begin{tabular}[c]{@{}c@{}}17.9$\pm$0.6\\ 3.9$\pm$0.1\\ 4.2$\pm$0.1\end{tabular}                         & \begin{tabular}[c]{@{}c@{}}0\\ 0\\ 0\end{tabular}                                                  & \begin{tabular}[c]{@{}c@{}}0\\ 0\\ 0\end{tabular}                  & \begin{tabular}[c]{@{}c@{}}1.5$\pm$0.3\\ 5.0$\pm$0.1\\ 4.6$\pm$0.1\end{tabular}                & \begin{tabular}[c]{@{}c@{}}181.0$\pm$9.9\\ 47.7$\pm$0.6\\ 45.2$\pm$0.1\end{tabular}       \\ \hline
	\multicolumn{6}{l}{* The three rows in each cell are the results for ExoTuned,} \\
	\multicolumn{6}{l}{  ExoTuned+CBF, and QUAD, respectively.} 	
%	\multicolumn{6}{l}{* The three rows in each cell are the results for Exoplayer} \\
%	\multicolumn{6}{l}{ tuned, Exoplayer tuned+CBF and \qea, respectively.} 
\end{tabular}
\end{table} 

\mysubsection{In-the-wild Tests}\label{sec:in-the-wild}
So far, our evaluation has been through large-scale trace-driven simulation and experimentation using real systems. We next present in-the-wild test results, by running QUAD and ExoTuned (which outperforms Exo)
%\shuaic{(which achieves better overall performance than that using the default parameters)}
over a commercial LTE network.
%We use the ED video (YouTube encoded).
The video we use is ED (YouTube encoded).
We consider two settings, one with poor signal conditions and hence low bandwidth (consistently less than 1Mbps and unstable), and the other with good signal conditions and hence high bandwidth.  The first setting is in a residential home, and the second one is  in an office building. For each setting, we make 10 runs, each consisting of three schemes (QUAD, ExoTuned and ExoTuned+CBF, in a random order).
The results are shown in Table~\ref{tb:wildTest}, which lists the mean and standard deviation across the runs for each case.
%The standard deviation under low bandwidth is significantly larger than that under high bandwidth, consistent with the much dynamic network conditions ()
Under high bandwidth conditions, compared to ExoTuned alone, we see significant benefits of CBF and QUAD in reducing data usage and achieving quality close to the target quality.  Under low bandwidth conditions,
%due to a higher amount of network dynamics, the standard deviation for each case is larger than that under high bandwidth conditions.
the results exhibit more variations due to the fluctuating network bandwidth caused by the poor signal strength. Despite that, compared to ExoTuned,
we still observe that CBF significantly reduces the percentage of low-quality chunks, and QUAD achieves the best QoE overall.
 %(despite slightly higher data usage).
 ExoTuned+CBF and QUAD have no rebuffering,
while ExoTuned shows non-negligible rebuffering duration for the 10-minute video.
%We consider four settings, two with poor signal conditions (measured in a residence home), with and without capping network bandwidth to 1.5 Mbps;  and the other with good signal conditions (measured in an office setting), again with and without bandwidth cap. The settings with bandwidth cap (achieved by capping the bandwidth at the server using \texttt{tc}~\cite{tc}) emulate the practice of some cellular network providers (see \S\ref{sec:current-practice-save-data}). For each setting, we make ten runs, with three schemes (\qea, ExoPlayer algorithm with and without CBF, in random order).

%	\vspace{-0.1in}
\mysection{Related work} \label{sec:related work}

\textbf{Improving Video QoE.}
In addition to the schemes already described in \S\ref{sec:setup},
QDASH~\cite{mok2012qdash} tries to reduce quality switches during adaptation.
%
%PANDA/CQ~\cite{li2014streaming} jointly considers network bandwidth and video bitrate variability.
%
BBA~\cite{huang2014buffer} proposes adaptation schemes based on client-side buffer information.
%
%MPC~\cite{yin2015control} proposes a model predictive control algorithm that optimally combines throughput and buffer occupancy information.
%
PIA~\cite{qin2017control} designs a PID-based framework to account for various requirements of ABR streaming.
Pensieve~\cite{Mao17:pensieve} proposes a system that generates ABR algorithms using reinforcement learning.
%
%BOLA-E~\cite{Spiteri2018bolae} enhances BOLA to be responsive to user events such as startup and seeking.
%
Oboe~\cite{Akhtar18:oboe} pre-computes the best possible ABR parameters for different network conditions and dynamically adapts the parameters at run-time.
	CAVA~\cite{qinvbr}\fengc{characterizes the VBR encoding characteristics,} proposes design principles for VBR-based ABR streaming and a concrete scheme that instantiates these design principles.
	 %In contrast to this work, it does not assume that the video quality information is available to the ABR logic.
	{PANDA/CQ}~\cite{li2014streaming} directly incorporates video quality information in ABR streaming and maximizing QoE by dynamic programming.
The study in~\cite{Ali2016} outlines the challenges and open issues in consistent-quality streaming such as the scheme in~\cite{li2014streaming}.
None of the above studies explicitly considers data efficiency.\fengc{which is a key concern of our study.}
%\yanyuan{Ali~\cite{Ali2016} presents how the impact of quality fluctuations can be reduced if video content's natural variability is taken into consideration.}	
%\yanyuan{However, these works do not consider data efficiency, which is the focus of this paper.}

\textbf{Reducing Data Usage.}
%
%Streaming data usage can be reduced both during the content encoding and client adaptation process.
%
%For the former, the general idea is to improve encoding efficiency by searching the parameter space.
%
A number of studies have proposed techniques for reducing data in the content encoding process.
Chen et al.~\cite{chen2017encoding} propose an optimization framework to identify the optimal encoding bitrates that minimize the average streaming bitrate, subject to a given lower bound on delivered quality.
De Cock et al.~\cite{de2016constant} present a constant-slope rate allocation approach to improve the Bitrate-Distortion rate.
Aaron et al.~\cite{netflix:per-title} propose per-title encoding, i.e., each title should receive a bitrate ladder, tailored to its
%specific
complexity characteristics.
Katsavounidis et al.~\cite{netflix:dynamic-optimizer} develop a dynamic optimizer framework that searches for optimized encoding parameters.
% by multiple encoding at different resolutions and quality parameters.
Toni et al.~\cite{Toni15:representation-selection} determine the optimal selection of tracks for encoding.
% and show that it can provide network bandwidth savings.
Our work differs from the above by jointly improving video QoE and reducing data usage in the streaming process.\fengc{after the encoding is done, i.e., for given a set of encoded tracks.} Therefore, our work is orthogonal to those on reducing data in the encoding process, and can complement those efforts.
%
%For the latter, the common practice is to make a tradeoff decision between content quality and perceived QoE.
%

The study in~\cite{chen2012qava} manages the tradeoff between monthly data usage and video quality by leveraging the compressibility of videos and predicting consumer usage behavior throughout a billing cycle. Our study differs from it in that we consider the data usage and quality tradeoffs when streaming a video.
%The studies closest to ours are QBR~\cite{cooper2017qbr} and~\cite{rainer2017statistically}.
%QBR~\cite{cooper2017qbr} and the study in~\cite{rainer2017statistically} have similar goals as our study.
%It can be retrofitted to existing schemes, which is similar to CBF.
The study in~\cite{rainer2017statistically} observes that, for some chunks, lower bitrate tracks may be of similar perceptual quality as higher bitrate tracks.
%Given a  set of encoded ABR tracks, it proposes a server-side approach to further modify them by determining the statistically indifferent quality variation (SIQV) of adjacent video tracks (per resolution) so that a higher bitrate chunk can be replaced with a lower bitrate chunk and statistically indifferent quality.
%It proposes an approach to
%is based on the observation that, for some chunks, lower bitrate tracks may of similar perceptual quality as higher bitrate tracks. Therefore, the authors
%propose an approach to
%determine
Given a set of encoded ABR tracks, it proposes to perform a server-side chunk replacement (within the same resolution) so that a higher bitrate chunk can be replaced by a lower bitrate chunk with a perceptually similar quality.
%so lower bitrate tracks to replace higher bitrate tracks.
%and use lower bitrate tracks to replace higher bitrate tracks.
Unlike our work, this study does not perform an in-depth exploration of how their approach interacts with existing ABR rate adaptation algorithms.\fengc{The study does not explore the interaction between statistically indifferent quality variation and ABR rate adaption.}
%While the approach can be used with existing ABR schemes, the SIQV information is not used in the ABR logic. Rather, after track selection, it simply downgrades a track to one with lower bitrate and statistically indifferent quality.
%Their approach is not used with existing ABR schemes.
QBR~\cite{cooper2017qbr} aims to improve the efficiency of existing ABR schemes by reducing the data usage while potentially increasing QoE. Specifically, a QBR server provides additional metadata hints to a client, allowing the client to
request a reduced bitrate for chunks of low complexity.
%and a higher bitrate for those of greater complexity.
%As a proprietary technology, the technical details of QBR are not available to the public.
QBR is not designed to achieve a target quality based on a user-specified option. In addition, it does not provide a grounds-up design as QUAD.
%Neither QBR nor~\cite{rainer2017statistically} is designed to achieve a target quality based on a user-specified option. In addition, neither provides a grounds-up design such as QUAD.
%The approach can be used with existing ABR schemes (similar as CBF) and the authors use case studies to show that the proposed approach leads to bandwidth savings. Our study presents a detailed evaluation of CBF combined with existing schemes, and develop \qea, a grounds-up design that outperforms existing schemes with CBF.

%\vspace{-0.1in}
\section{Summary} \label{sec:quad-conclusion}
 Existing data saving practices for ABR videos often incur undesired and highly variable video quality, without making the most effective use of the available network bandwidth.
 In this chapter, we identify underlying causes for this behavior and design two novel approaches, CBF and QUAD, to achieve better tradeoffs among video quality, rebuffering, quality variations, and cellular data usage. Evaluations demonstrate that compared to the state of the art,  these two schemes achieve  quality closer to desired levels, lower stalls, and more efficient data usage.
Specifically, using CBF with existing schemes leads to significant benefits in all performance metrics, and
 QUAD achieves even better QoE compared to existing schemes enhanced with CBF. 
 
 Aligned with this work, we further propose a novel framework for quality-aware Adaptive Bitrate (ABR) streaming involving a per-session data budget constraint. Under the framework, we develop two planning based strategies, one for the case where fine-grained perceptual quality information is known to the planning scheme, and another for the case where such information is not available. More details can be found at ~\cite{qin_mmsys2021}.
%We  are currently exploring a number of future directions, including extending  \qea design concepts to other video types (e.g., 360 videos) and adaptation schemes (such as learning-based~\cite{Mao17:pensieve}).
\fengc{We are currently exploring a number of future directions, including extending  \qea design conceptsto other video types such as 360-degree videos and other rate adaptation schemes such as those based on machine learning~\cite{Mao17:pensieve}.}

%\bibliographystyle{acm}
%\bibliographystyle{ACM-Reference-Format}
%\bibliographystyle{unsrt}
%\bibliography{ref-bing}

%\end{document}

%%% Local Variables:
%%% mode: latex
%%% TeX-master: t
%%% End:

\chapter{ABR Streaming with Separate Audio and Video Tracks: Measurements and Best Practices} \label{chapter:audio}

	% \input{Audio/abstract}
	% \maketitle

\vspace{-0.15in}
\section{Introduction}

Little is known on how best to mesh together audio and video rate adaptation in ABR streaming.
In this chapter, we first examine the current practice
in the predominant DASH~\cite{dash-standard} and HLS~\cite{apple-hls} protocols and three popular players, ExoPlayer~\cite{exoplayer}, Shaka Player~\cite{shakaplayer}, and \texttt{dash.js} player~\cite{dash-js} (see \S\ref{sec:method}). Using a combination of code inspection and experiments, we identify a number of limitations in existing practices both in the protocols and the player implementations (see \S\ref{sec:behaviors}).
We identify the root causes of these limitations, and present a set of best practices and guidelines (see \S\ref{sec:suggestions}).
Our work makes the following contributions:

\BULLET We find that all three players have various issues in handling demuxed audio and video tracks (\S\ref{sec:behaviors}). These issues are not simple {implementation} bugs,
but rather arise  due to a confluence of {sometimes subtle} interacting factors, {ranging from  a {lack of sufficient insights}, design/engineering {inadequacies}  to architectural deficiencies}. Specifically, these issues include (i) limitations in  how DASH/HLS handles demuxed content,
(ii) player designs to circumvent the limitations in DASH/HLS,
(iii) little understood differences between HLS and DASH with important implications,
(iv) attempts in applying the same player logic designed for one protocol (DASH/HLS) to the other, and
(v)  inadequate attention to synchronizing a player's video and audio downloading.

\BULLET Our findings demonstrate that meshing together audio and video rate adaptation is a very challenging problem -- it involves many pieces (in ABR protocols, server and player designs)
%(in both ABR protocols and player designs)
and needs a holistic solution. As a starting point in addressing the challenges, we provide a set of best practices and guidelines (\S\ref{sec:suggestions}) for the ABR protocol level (manifest file specification enhancements), the server side (appropriately using manifest to convey needed information to the client), and the player side (audio/video prefetching, and joint adaptation for audio and video).

{Note that in this work we consider coordinated ABR rate adaptation for streaming  demuxed video and audio to the  \emph{same user} to maximize QoE. The demuxed video and audio tracks may be located at different servers and hence may not necessarily share  the same bottleneck link. This problem is very different from the multi-user rate adaptation problem where videos are streamed to \emph{multiple users} while sharing the same bottleneck link~\cite{akhshabi2012happens,jiang2012improving,huang2012confused,li2014probe}.

We hope that our study brings the problem of ABR streaming with demuxed audio and video tracks to the community's attention
{With coordinated efforts, we hope that }
the community can come up with effective solutions soon.
%\yanyuan{in a short amount of time. }
		
	\vspace{-0.15in}
\mysection{Motivation and Methodology} \label{sec:method}

\mysubsection{Motivation}
One way of treating demuxed audio and video tracks is  determining their rate adaptation independently, and combining the
audio and video chunks at playback time.
While this approach is easy to implement, it has two major drawbacks.
First, it can lead to some audio and video combinations that are clearly undesirable (e.g., lowest quality audio with highest quality video, or vice versa).
Second, it can cause the prefetched audio and video {streams} to be severely unbalanced,
e.g., audio (resp. video) buffer underruns and hence leads to stalls, while there is a lot of prefetched {data for the corresponding} video content (resp. audio).

Although the audio/video rate adaptation is decided by the player at the client, we argue that the server needs to provide sufficient information to facilitate the decision. Determining a good set of audio and video combinations is not a trivial task. It depends on the nature of the specific content, device characteristics (e.g., screen size, sound system), and the business rules encoding the content provider's preferences for that content.
For instance, for music shows, the sound quality may be {relatively} more important than video quality, and hence it might be more desirable to combine high audio  tracks with low/medium video tracks; while for an action movie,
the desirable combinations may be the opposite.
The origin server knows the content information, client device types, and the business rules, and hence is at a better position for deciding the combinations.
Ideally, the server can determine a good  set of audio and video combinations beforehand, and pass the information to the player through the manifest file or other out-of-the-band mechanisms.

\mysubsection{High-level Methodology}
To understand the current practice for handling demuxed audio and video tracks, we examine both predominant ABR streaming protocols (DASH and HLS) and the treatment in three open-source popular players, specifically, ExoPlayer~\cite{exoplayer} and Shaka~\cite{shakaplayer} that are widely used in industry, and \texttt{dash.js}~\cite{dash-js} that showcases the DASH standard.
For each player, our focus is on understanding the  interaction between audio and video rate adaptation and their impact on each other.
Specifically,
we use a combination of source code analysis (all the three players are open-source players)  and controlled experiments
 (with additional instrumentation in the source code when necessary).
The combined approach  is necessary because
as a large-scale often multithreaded software system, the player source code contains a large number of interlocking components with complex interactions, and just code analysis by itself may not be sufficient to  understand the behavior of the system.
The controlled experiments therefore play a crucial role in profiling the overall behavior and also help us to confirm our understanding.
While our study focuses on these three players, our findings of specific issues (\S\ref{sec:behaviors}) in these players  have wide applicability to the services that use these players. Our  best practice recommendations (\S\ref{sec:suggestions}) are not player specific; we expect them to be widely applicable to any ABR streaming service.

\mysubsection{ABR Streaming Protocols} \label{sec:method-HLS-DASH}

DASH~\cite{dash-standard} and HLS~\cite{apple-hls} are the two predominant ABR streaming protocols. Both of them now support demuxed audio and video tracks.
%define the formats for audio and video tracks.
The DASH protocol defines an Adaptation Set as a set of interchangeable encoded versions of one or several media content components. For demuxed audio and video tracks, it defines one Adaptation Set for video tracks and another set for audio tracks.
{A \texttt{bandwidth} attribute is associated with each track (termed {\em Representation} in DASH) to declare the required bandwidth (which is close to the peak bitrate; see one example in Table~\ref{table:ytbitrateladder}).
In HLS, a top-level master playlist uses the {EXT-X-STREAM-INF} tag to specify an audio and video track combination (termed {\em Variant} in HLS). With HLS, one can specify all possible combinations or a subset thereof.
A \texttt{BANDWIDTH} attribute is associated with a video and audio track combination
to declare the aggregate bandwidth requirement of the combination (which is the sum of the peak bitrates of the audio and video tracks in the combination).
In the rest of the chapter, depending on the context, the phrase \emph{bandwidth requirement} refers to the \texttt{bandwidth} attribute in DASH or  \texttt{BANDWIDTH} attribute in HLS.
}

There are some key differences between DASH and HLS specifications in terms of how they treat demuxed audio and video tracks. These differences have important implications for both player {ABR logic design} and the resulting QoE.
One key difference is that DASH does not provide a mechanism for a content provider to specify only the desired subset of audio and video track combinations, while HLS provides an explicit mechanism to do so. {\em Therefore, DASH provides a player more freedom to select combinations {from the available audio and video tracks}, potentially making it more vulnerable to choosing undesirable combinations (see \S\ref{sec:behaviors}).}

Another key difference between DASH and HLS is that DASH specifies the bandwidth requirements of individual audio and video tracks, while HLS top-level manifest only supports specifying the aggregate bandwidth requirement of each audio and video track combination. This difference should be recognized by the player to avoid undesirable behaviors, as we shall see later.

\mysubsection{Players}
We focus on three widely used and open-source players, ExoPlayer (v2.10.2)~\cite{exoplayer}, Shaka Player (v2.5.1)~\cite{shakaplayer}, and \texttt{dash.js} player (v2.9.3)~\cite{dash-js}, and try to understand their behaviors towards audio and video rate adaptation using their latest versions. 
ExoPlayer is a widely used application level media player for Android platforms.
More than 140,000 applications in Google Play Store
use ExoPlayer to play media~\cite{exoplayeryt}. Shaka Player is an open-source JavaScript library for adaptive media, and has been used by more than 1,600 websites~\cite{shaka-statistics}. \texttt{dash.js} is an open-source JavaScript based player and is
 the reference player maintained by the DASH Industry Forum.
ExoPlayer and Shaka Player support both DASH and HLS streaming; \texttt{dash.js} only supports DASH.

	\mysection{Practice of Popular Players}\label{sec:behaviors}

\mysubsection{Experimental Setup} \label{sec:expe}
In the experiments below, we use a drama show downloaded from YouTube using  \texttt{youtube-dl}~\cite{youtube-dl}. It is around 5 minutes long, containing 6 video tracks
and 3 audio tracks. Table~\ref{table:ytbitrateladder} lists the average and peak bitrates of the video and audio tracks, as well as other key characteristics.
{The bitrate ladder in Table~\ref{table:ytbitrateladder} is commonly used by YouTube~\cite{youtubeladders}; the issues we point out below are related to the bitrate ladder (not particular to  the specific content), and hence are broadly applicable.}

\setlength{\arrayrulewidth}{1pt}
\begin{table}[h]
	\centering
	%\vspace{-.1in}
	\caption{Video and audio of a YouTube drama show.}
	%\vspace{-.1in}
	\begin{tabular}{llllll}\hline
		Audio/  & Average & Peak       & Declared & Audio channels,\\
		Video       & Bitrate  & Bitrate  & Bitrate for &  sampling rate \\  
		Track      &    (Kbps)  & (Kbps)  & DASH (Kbps)& Video resolution\\ \hline
		A1           & 128         & 134      &  128      & 2 channels, 44 kHz      \\ 
		A2           & 196         &  199      &  196     & 6 channels, 48 kHz  \\ 
		A3           &  384        & 391       &  384    & 6 channels, 48 kHz    \\ 
		V1           &  111        & 119      &  111      &    144p   \\ 
		V2           &  246       & 261       &  246      &    240p \\ 
		V3           &  362       & 641       &  473      &   360p  \\  
		V4           &  734       & 1190     &  914       &  480p  \\  
		V5           &  1421     & 2382     &  1852       &  720p    \\  
		V6           &  2728     & 4447     &   3746      & 1080p     \\ \hline
	\end{tabular}
	\vspace{-.2in}
	\label{table:ytbitrateladder}
\end{table}

We use the Bento4 {toolkit}~\cite{bento4} to create two sets of manifest files,  complying respectively with DASH and HLS standards.
For DASH, we create one manifest file, with the six video tracks and three audio tracks specified in two Adaptation Sets. {The declared bitrate for each audio/video track is shown in Table~\ref{table:ytbitrateladder}.}
For HLS, we create two manifest files. The first, $H_{all}$, specifies all 18 combinations of video and audio tracks\footnote{{While HLS provides a mechanism to specify a subset of video and audio combinations, it does not have an explicit recommendation that only carefully curated combinations should be specified. A content provider may choose to list all the combinations; we use $H_{all}$ to illustrate the issues that may arise from such practices.}}; the second, $H_{sub}$, specifies a subset of 6 combinations, i.e., V1+A1, V2+A1, V3+A2, V4+A2, V5+A3, V6+A3, where high quality video tracks are associated with high audio quality tracks, and vice versa.
The bitrates (both peak and average bitrates) for the combinations contained in these two manifest files are listed in Tables~\ref{table:va-all-combination} and \ref{table:va-sub-combination}.

Table~\ref{table:va-all-combination}  lists the full set of the 18 audio and video combinations  for the drama show in Table~\ref{table:ytbitrateladder}.  They are  the combinations listed in the HLS manifest file $H_{all}$.
For each combination, the peak bitrate  is the sum of the  peak bitrates of the audio and video tracks; the average bitrate is sum of their average bitrates. The combinations are placed in increasing order of the peak  bitrate.

Table~\ref{table:va-sub-combination} lists a subset of 6 audio and video combinations for the drama show in Table~\ref{table:ytbitrateladder}. They are the combinations listed in the HLS manifest file $H_{sub}$. For each combination, both the peak and average bitrates  are listed in the table.
\begin{table}[h]
	\centering
	\caption{Bitrates of the full set of audio and video combinations (used in HLS manifest file $H_{all}$).}
	\begin{tabular}{lllll}\hline
		Video/Audio   & Average Bitrate  & Peak Bitrate  \\
		Combination  &  (Kbps)        &  (Kbps) \\\hline
		V1 + A1 & 239 & 253\\
		V1 + A2 & 307 & 318\\
		V2 + A1 & 374 & 395\\
		V2 + A2 & 442 & 460\\
		V1 + A3 & 495 & 510\\
		V2 + A3 & 630 & 652\\
		V3 + A1 & 490 & 775\\
		V3 + A2 & 558 & 840\\
		V3 + A3 & 746 & 1032\\
		V4 + A1 & 862 & 1324\\
		V4 + A2 & 930 & 1389\\
		V4 + A3 & 1118 & 1581\\
		V5 + A1 & 1549 & 2516\\
		V5 + A2 & 1617 & 2581\\
		V5 + A3 & 1805 & 2773\\
		V6 + A1 & 2856 & 4581\\
		V6 + A2 & 2924 & 4646\\
		V6 + A3 & 3112 & 4838\\ \hline
	\end{tabular}
	\label{table:va-all-combination}
\end{table}

\begin{table}[h]
	\centering
	\caption{Bitrates of a subset of audio and video combinations (used in HLS manifest file $H_{sub}$).}
	\begin{tabular}{lllll}\hline
		Video/Audio   & Average Bitrate  & Peak Bitrate  \\
		Combination  &  (Kbps)        &  (Kbps) \\\hline
		V1+A1           &  239       & 253                    \\
		V2+A1           &  374        &  395                  \\
		V3+A2           &  558      & 840                    \\
		V4+A2           &  930         & 1389                    \\
		V5+A3           &  1805       & 2773                    \\
		V6+A3           &  3112       & 4838               \\ \hline
	\end{tabular}
	\label{table:va-sub-combination}
\end{table}

For controlled experiments, we set up a  HTTP server as the origin server.
The network bandwidths from the server to client are controlled by using \texttt{tc}~\cite{tc} at the server.
Our goal here is  to identify  demuxed audio-video related performance problems for each player, not to compare the performance across the multiple players. Therefore, we choose the experimental settings targeting the different issues of the different players. Whenever appropriate, we choose fixed network bandwidths,  since the behavior of a player is more easily understood under such bandwidth profiles.

\mysubsection{ExoPlayer}\label{subsec:exoplayer}

ExoPlayer adopted joint audio and video rate adaptation only very recently  (starting from version v2.10.0, released in May 2019). Prior to that,  it did not support simultaneous audio and video rate adaptation.
Specifically, for multiple demuxed video and audio tracks, it {selected} a fixed audio track {and used it throughout the session}
without any audio rate adaptation.
 In the following, we use ExoPlayer version v2.10.2, the latest version that is publicly available.

\textbf{DASH.}
Since a DASH manifest file does not restrict the combinations of audio and video tracks that can be selected,
ExoPlayer designs a specific logic that uses the per-track
{declared} bitrate in the manifest file to determine a subset of  combinations that
combines higher bitrate/quality audio tracks with higher bitrate/quality video tracks.
Specifically, for the audio and video tracks listed in Table~\ref{table:ytbitrateladder}, the resultant combinations, in increasing order of bandwidth requirement, are
V1+A1, V2+A1, V2+A2, V3+A2, V4+A2,  V4+A3, V5+A3, and V6+A3, where two adjacent combinations have either the same video or audio track.
We refer to these combinations as the \emph{predetermined} combinations by ExoPlayer.
The subsequent rate adaptation process \emph{only} considers these predetermined combinations.
Specifically, during rate adaptation, the player estimates the available network bandwidth by considering both video and audio downloading,
and conservatively assumes that the actual network bandwidth is $75\%$ of the estimated bandwidth. It then selects an audio and video combination out of the  predetermined combinations (based on the bitrate of each of the combinations, network bandwidth estimate, and the buffer status).

 The above treatment has the following limitation. The predetermined combinations by ExoPlayer, while may be reasonable for some bitrate ladder or content types, may not be suitable for others.
  As an example, for the above drama show, it may be desirable to allow V5+A2 (i.e., high quality video with medium quality audio), which is, however, not in the predetermined combinations,
  and hence will not be selected by ExoPlayer.

	We next further illustrate the above point using two experiments.
	In the first experiment,  we create an audio adaptation set with three audio tracks B1, B2 and B3 with the declared bitrate as 32, 64 and 128 Kbps, respectively. The video tracks are the same as listed in Table~\ref{table:ytbitrateladder}. In this case, the predetermined  combinations are V1+B1, V2+B1, V2+B2, V3+B2, V4+B2, V5+B2, V5+B3, and V6+B3.  The network bandwidth is fixed at 900 Kbps.
	Fig.~\ref{subfig:exodashlowaudio} plots the average bitrate of the selected tracks. We see that V3+B2 is selected, while
	V3+B3 would be a better choice, since it has higher audio quality, and its bandwidth requirement (601 Kbps) is below the available network bandwidth.
This more desirable combination  (V3+B3) is, however, not in the predetermined combinations, and hence will not be selected. In the second experiment, we switch the audio adaptation set to three audio tracks with relatively higher bitrate,
referred to as C1, C2 and C2 with the declared bitrate as 196, 384 and 768 Kbps, respectively. In this case, the predetermined combinations are V1+C1, V2+C1, V2+C2, V3+C2, V4+C2, V5+C2, V5+C3, and V6+C3. Fig.~\ref{subfig:exodashhighaudio} shows that  ExoPlayer selects V2+C2,  resulting in very low video quality and  high audio quality. The combination V3+C1 would be a better choice (V3+C1 has declared video and audio bitrates as 473 and 196 Kbps, respectively, versus 246 and 384 Kbps in V2+A2), which is however not in the predetermined combinations.

 \begin{figure}[t]
 	\centering
 	\subfigure[Selected bitrates  (experiment 1). \label{subfig:exodashlowaudio}]{%
 		\includegraphics[width=0.4\textwidth]{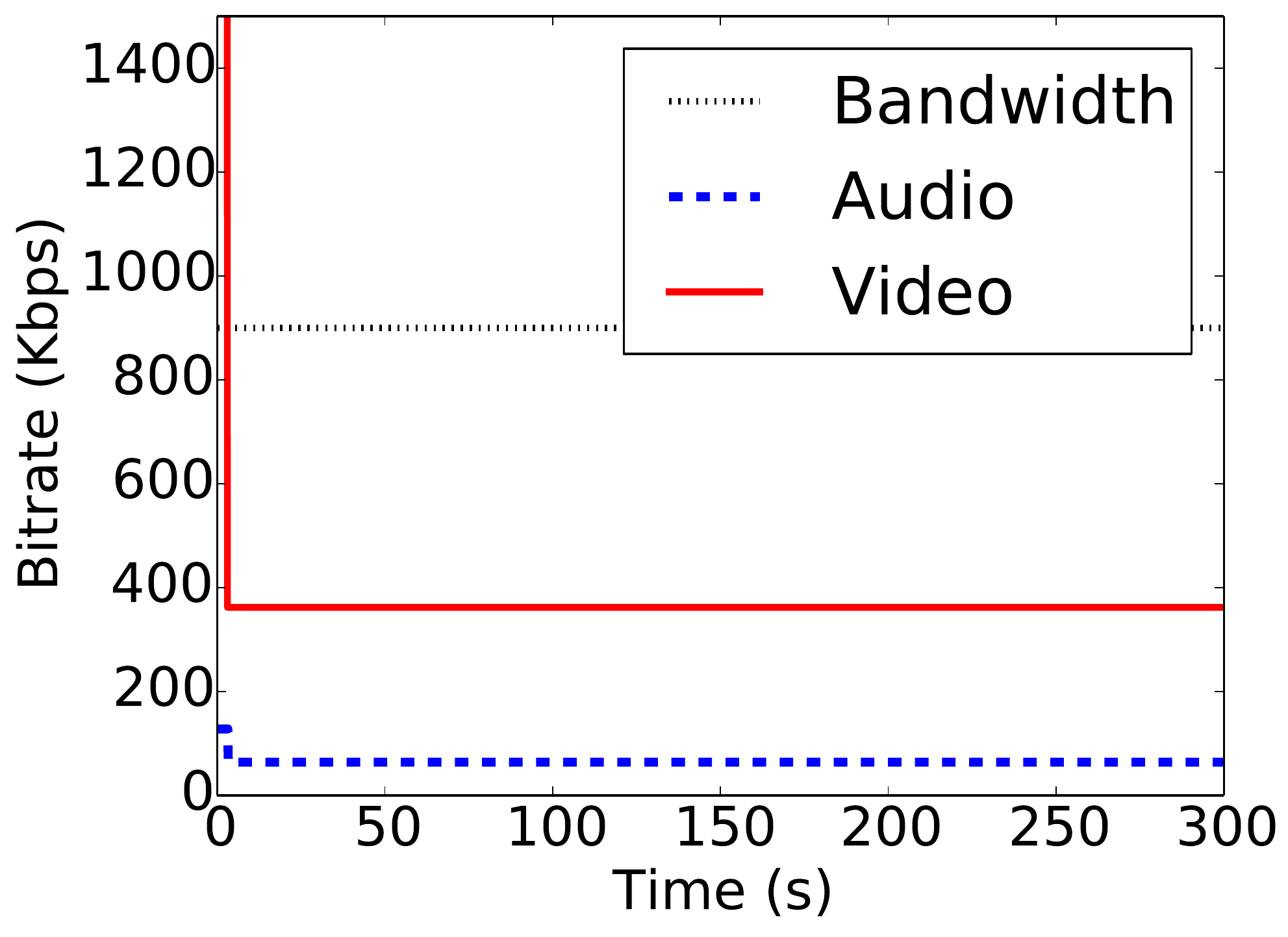}
 	}
 	\subfigure[Selected bitrates (experiment 2).\label{subfig:exodashhighaudio}]{%
 		\includegraphics[width=0.4\textwidth]{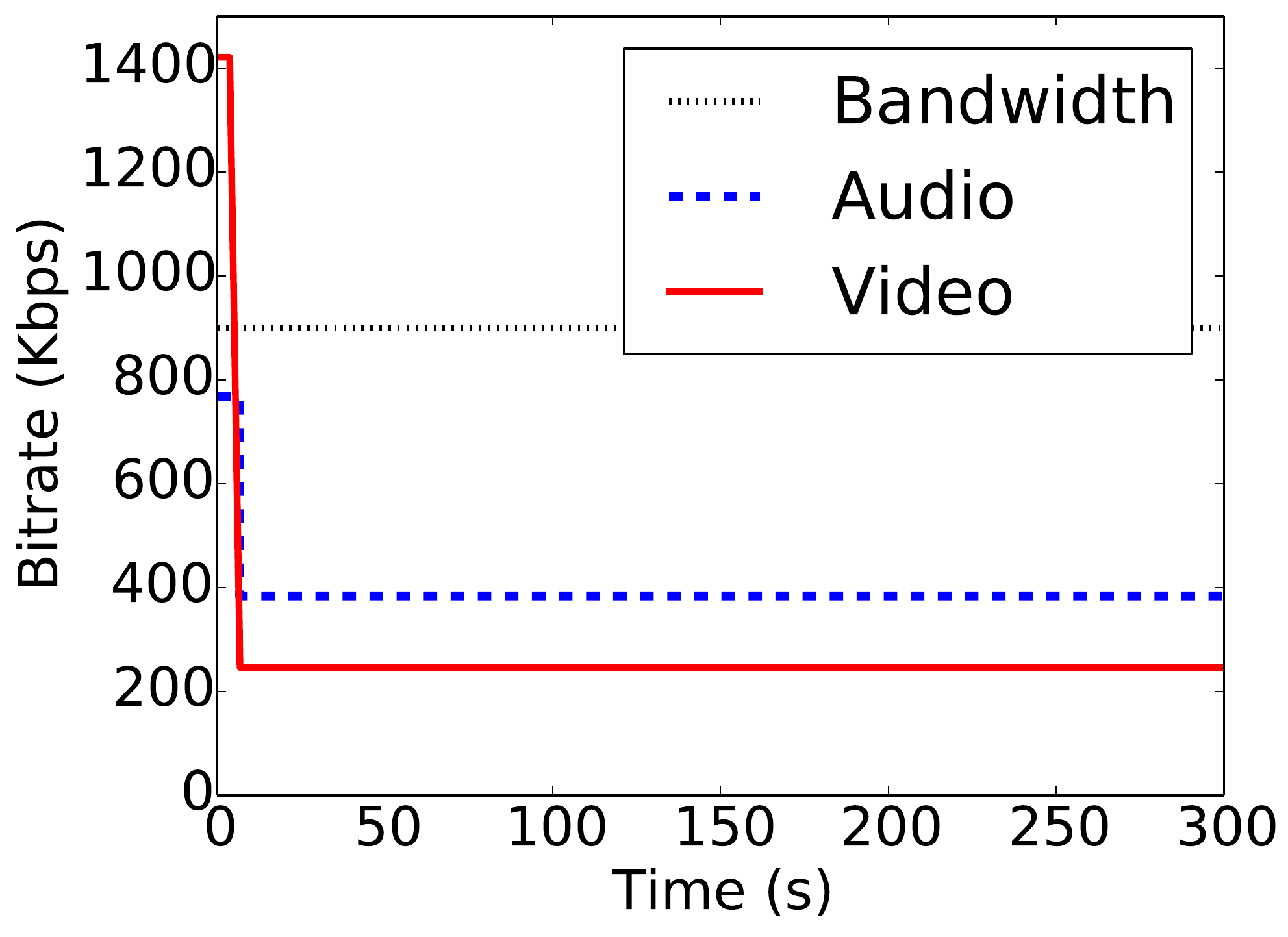}
 	}
 
 	\caption{ExoPlayer results (DASH).}
 	\label{fig:exoplayerDASH}
 \end{figure}

 While clearly undesirable, the root cause of the above problem  flows from a limitation in the DASH standard.
 As we mentioned in \S\ref{sec:method}, the server is better placed to  specify the desirable set of video and audio combinations.
 However, the DASH standard lacks such specifications.
 The DASH client, typically lacks detailed domain  knowledge of the content being streamed, and therefore is forced to  either (i) consider all possible combinations of video and audio track variants, or (ii) apply its own  policy to determine a subset of combinations (as in ExoPlayer).
 Both approaches have drawbacks. The former can lead to a player selecting
 an unsuitable combination for a  certain type of content, while the latter can potentially exclude desirable  combinations of audio and video tracks for a content.

\textbf{HLS.}  ExoPlayer uses the same rate adaptation code for both  DASH and HLS. In contrast to the behavior with DASH, for HLS, this unfortunately leads to selecting a fixed audio track.
In the following, we first illustrate this behavior using experiments and then explain the root cause.

In the first experiment, we specify a subset of audio and video combinations, $H_{sub}$, in the top-level master playlist;
the first audio track  in the manifest file is A3 (i.e., the highest bitrate/quality  audio track). The network bandwidth is set to be time-varying, with the average as 600 Kbps.
The evolution of the selected audio and video tracks is shown in  Fig.~\ref{subfig:exobitrate2}; the audio and video buffer levels over time are shown in  Fig.~\ref{subfig:exobuffer2}.
We note that ExoPlayer selects A3 throughout the playback, resulting in {\em 5 stall events and 36.9 seconds of rebuffering (marked as the shaded regions)}.  In addition, the  ABR selection has the effect of  {\em disobeying the subset of audio and video combinations specified in the HLS  manifest}  as it  selects some combinations (e.g., V1+A3) that are not in the specified subset.
In the second experiment, we modify the manifest file so that A1 (the lowest quality audio track) is listed as the first audio track and fix the network bandwidth to 5 Mbps. ExoPlayer selects A1 throughout the playback despite plenty of available network bandwidth (figures omitted), leading to unnecessarily poor audio QoE.

\begin{figure}[t]
	\centering
	\vspace{-.25in}
	\subfigure[Selected bitrates. \label{subfig:exobitrate2}]{%
		\includegraphics[width=0.4\textwidth]{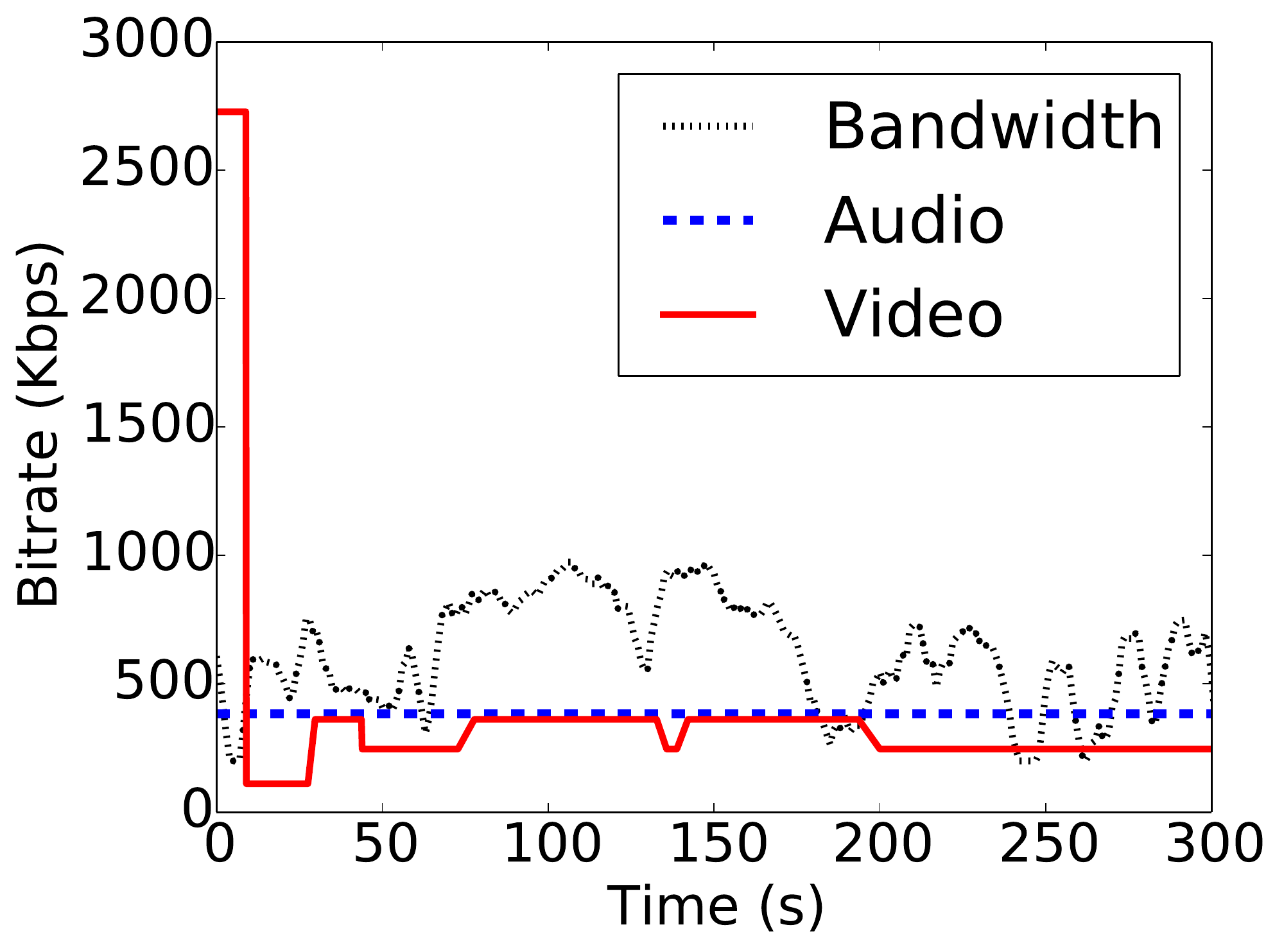}
	}
	\subfigure[Buffer levels.\label{subfig:exobuffer2}]{%
		\includegraphics[width=0.4\textwidth]{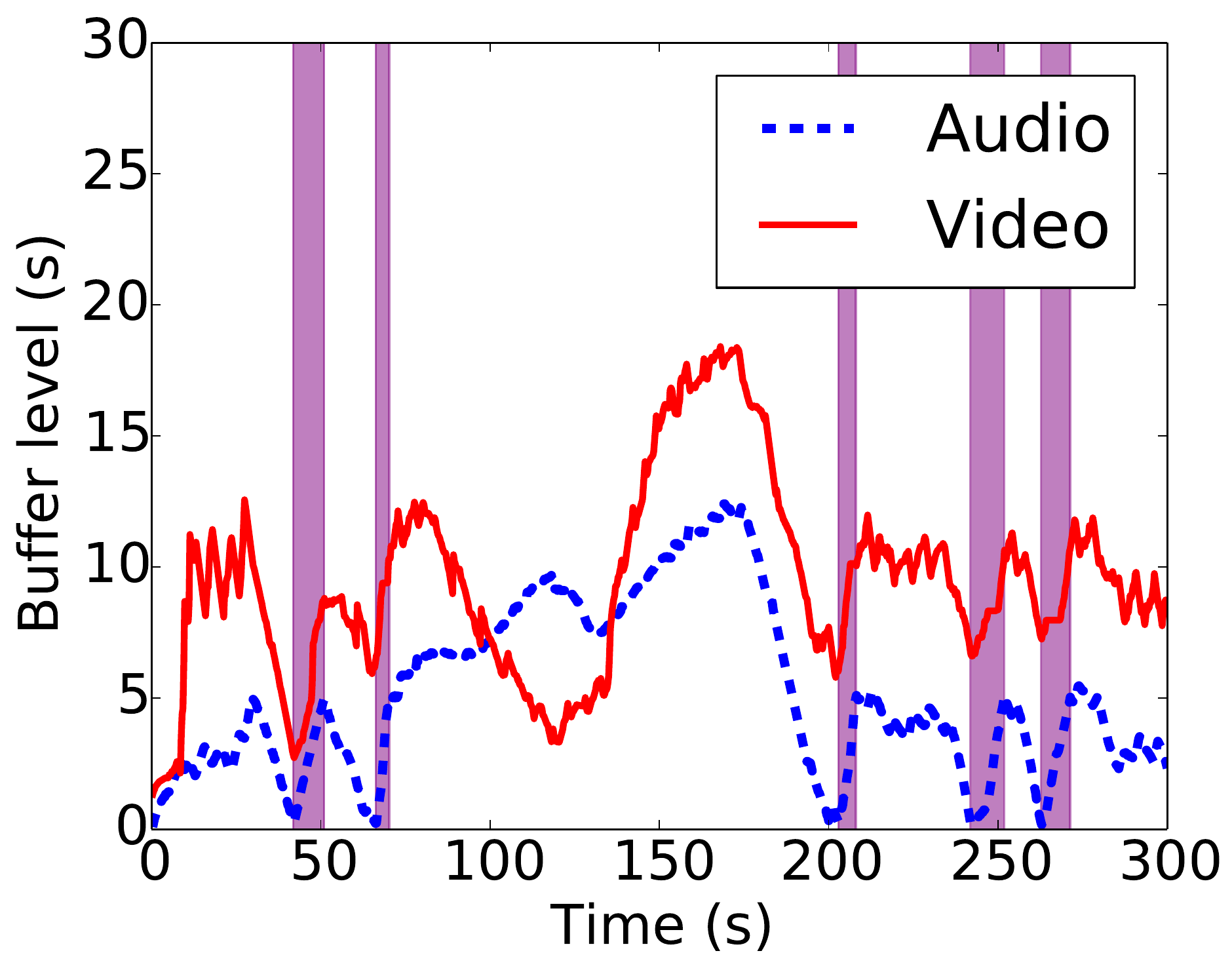}
	}
	\vspace{-.2in}
	\caption{ExoPlayer results (HLS).}
	\label{fig:exoplayer2}
	\vspace{-.05in}
\end{figure}

{What causes such behavior?
 Inspecting the ExoPlayer source code, we find that it is due to
 the lack of  specification of the individual audio and video track's  bitrate information in the top-level manifest file, which is needed by ExoPlayer to predetermine a subset of audio and video combinations.
 To accommodate the  lack of this information, ExoPlayer simply assumes that all the audio tracks have the same quality, thereby leading to a fixed audio track selection.
 For a video track, it uses the aggregate bitrate of the first variant in the top-level manifest file that contains this video track as its bitrate, which is clearly an overestimation. The overestimation will be even more severe when the order of the variants is not specified properly (i.e., the first variant for a video track contains the video track and the highest bitrate audio track).
 In \S\ref{sec:server-best-practice}, we outline practical solutions to this problem.
}

\mysubsection{Shaka Player} \label{subsec:shaka}

\begin{figure}[t]
	\centering
	\subfigure[Selected bitrates (experiment 1). \label{subfig:shakabitratecase1}]{%
		\includegraphics[width=0.4\textwidth]{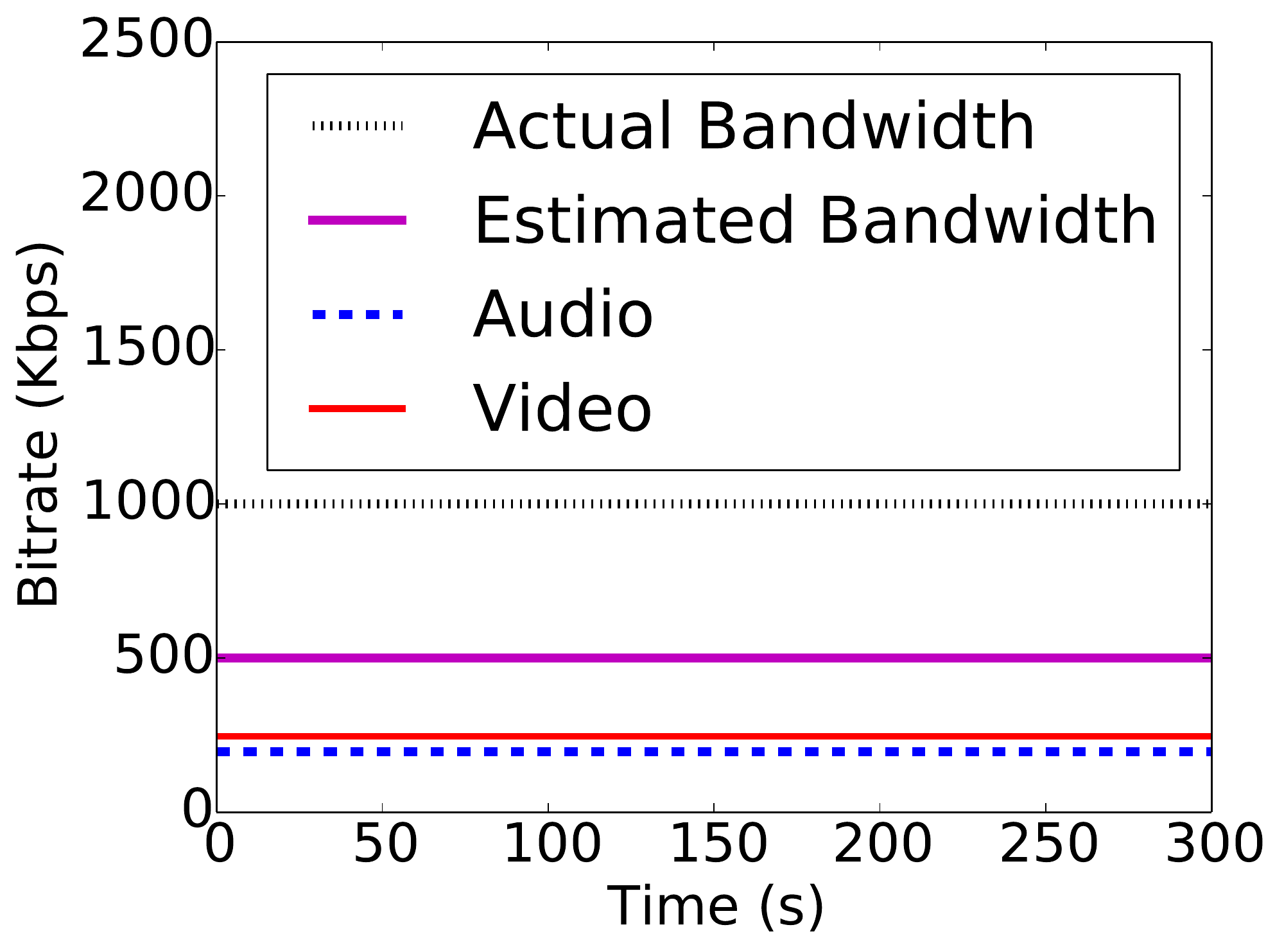}
	}
	\subfigure[Selected bitrates (experiment 2).  \label{subfig:shakabitratecase2}]{%
		\includegraphics[width=0.4\textwidth]{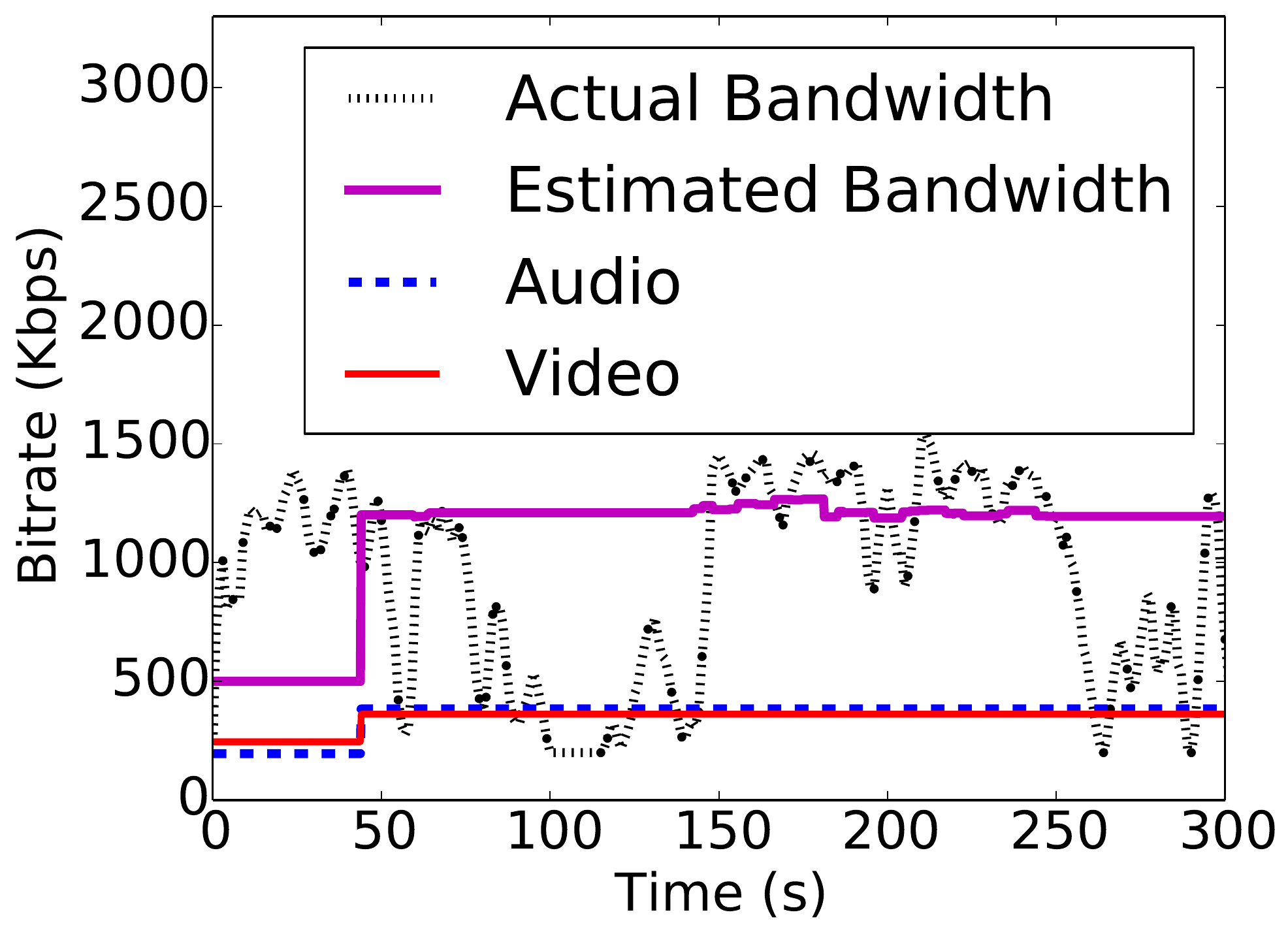}
	}
	\caption{Shaka Player results (HLS).}
	\label{fig:shakaplayer}

\end{figure}

\textbf{HLS.} We use two experiments with the HLS manifest file $H_{all}$ to understand the behavior of Shaka Player.
In the first experiment, the network bandwidth is set to a  constant  1 Mbps.
Fig.~\ref{subfig:shakabitratecase1} shows that the bandwidth estimated by Shaka  is a constant 500 Kbps, only half of the actual specified network bandwidth. As a result, 
V2+A2 (with aggregate peak bitrate of only 460 Kbps) is  selected. In the second experiment, the specified network bandwidth is dynamic (with the average as 600 Kbps, see Fig.~\ref{subfig:shakabitratecase2}).
In this case, Shaka first underestimates the network bandwidth, and then overestimates the bandwidth starting around 50 s.
Correspondingly, the selected video and audio tracks are initially low (V2+A2), and then overly high (V3+A3), leading to a total rebuffering of 39 s.
Inspecting the player code, we find that the above undesirable behaviors  are due to Shaka's network bandwidth estimation algorithm. 
Specifically,  Shaka uses past video and audio downloading throughputs as samples to estimate the bandwidth.
It treats video and audio downloading separately. While downloading a video track (the same applies to audio downloading), Shaka
considers each interval ($\delta=0.125$s), calculates the amount of data $d$ downloaded in that interval, and only counts the resultant throughput as a valid sample if $d \ge 16$ KB.
To estimate the overall available network bandwidth, it considers a set of samples $\{c_1, \ldots, c_n\}$, where $c_i$ is the throughput from either video or audio downloading, and sets the estimated bandwidth as an exponential weighted average of these samples.
For time intervals when the audio and video streams are being  concurrently downloaded over a shared network bottleneck link,  the above strategy will  severely underestimate  the  available  network bandwidth. 
In addition, the filtering rule (i.e., only consider samples with $d \ge 16$ KB) {itself} can lead to bandwidth underestimation or overestimation, depending on the network conditions. In the first experiment above, none of the throughput samples satisfies the filtering rule, and hence the default bandwidth estimation of 500 Kbps is used throughout the streaming session. In the second experiment, only the throughput samples under high network bandwidth satisfy the rule, while those under low network bandwidth are discarded, leading to significant bandwidth overestimation.

 Another issue is that Shaka uses a simple rate based adaptation scheme (i.e., selects the combination with the bandwidth requirement closest to the estimated bandwidth), which
can cause  the selected audio and video tracks to fluctuate frequently even if the bandwidth estimation is accurate.
The above fluctuation problem is more severe for the case of demuxed audio and video since 
a large number of audio and video combinations may have close bandwidth requirements.  
For example, suppose manifest file $H_{all}$ is used and the estimated network bandwidth varies between 300 to 700 Kbps.
Then the selected combinations can fluctuate among
V1+A2, V2+A1, V2+A2,  V1+A3 and V2+A3, with bandwidth requirements as 318, 395, 460, 510 and 652 Kbps, respectively.

\textbf{DASH.} Under DASH, since no combinations of audio and video tracks are specified in the manifest file, the player creates all the combinations of video and audio tracks when parsing the DASH manifest file. Therefore, the result is the same as that for HLS when using manifest file $H_{all}$.

\mysubsection{\texttt{dash.js} Player }

The default ABR algorithm used in \texttt{dash.js} is \texttt{DYNAMIC}~\cite{Spiteri2018bolae}, which switches between two schemes,  \texttt{THROUGHPUT} and \texttt{BOLA}. \texttt{THROUGHPUT} is a rate based approach that chooses tracks
whose declared peak bitrates are close to the estimated bandwidth.
% (obtained from recent throughput history). 
\texttt{BOLA} is a buffer based approach that optimizes a utility function~\cite{spiteri2016bola}.
% based on the buffer  level.
\texttt{DYNAMIC} starts by using \texttt{THROUGHPUT} rule. It switches to \texttt{BOLA} when the buffer level is above 12 s and \texttt{BOLA} selects a bitrate at least as high as that selected by \texttt{THROUGHPUT}; it switches back to \texttt{THROUGHPUT} if the buffer is less than 6 s and \texttt{BOLA} selects a bitrate lower than that selected by \texttt{THROUGHPUT}.

Examination of the source code of \texttt{dash.js} shows that  it utilizes \texttt{DYNAMIC} strategy  for both audio and video, and  performs rate adaptation for audio and video separately.
%, using the bandwidth estimation based on past audio and video downloading,
In addition, the bandwidth estimation for audio (video) is based on past audio (video) downloading only.
Fig.~\ref{fig:dashjsplayer}  plots the results when using a fixed network bandwidth of  700 Kbps.
%We see that selected video and audio combinations includes V3+A1, V3+A2, V2+A2, V2+A3, V3+A3, V2+A1, and V1+A1.
We see that the selected video and audio combinations includes V2+A3, V2+A2, V2+A3 and V3+A3.
Some of these combinations are clearly undesirable,  e.g., V2+A3. The combination, V3+A2, fits the network bandwidth profile (it actually has lower bandwidth requirement than V2+A3), while it has higher video quality and slightly lower audio quality than V2+A3, which might be more preferable from the user perspective for the drama show.
We further see from 
Fig.~\ref{fig:dashjsplayer}(b) that the buffer levels for audio and video can be unbalanced, which is undesirable since when one of them underruns, stalls will happen, even if there is a lot of content in the other buffer.

\begin{figure}[t]
	\centering
	\vspace{-.2in}
	\subfigure[Bitrate selections. \label{subfig:dashjsbitrate}]{%
		\includegraphics[width=0.4\textwidth]{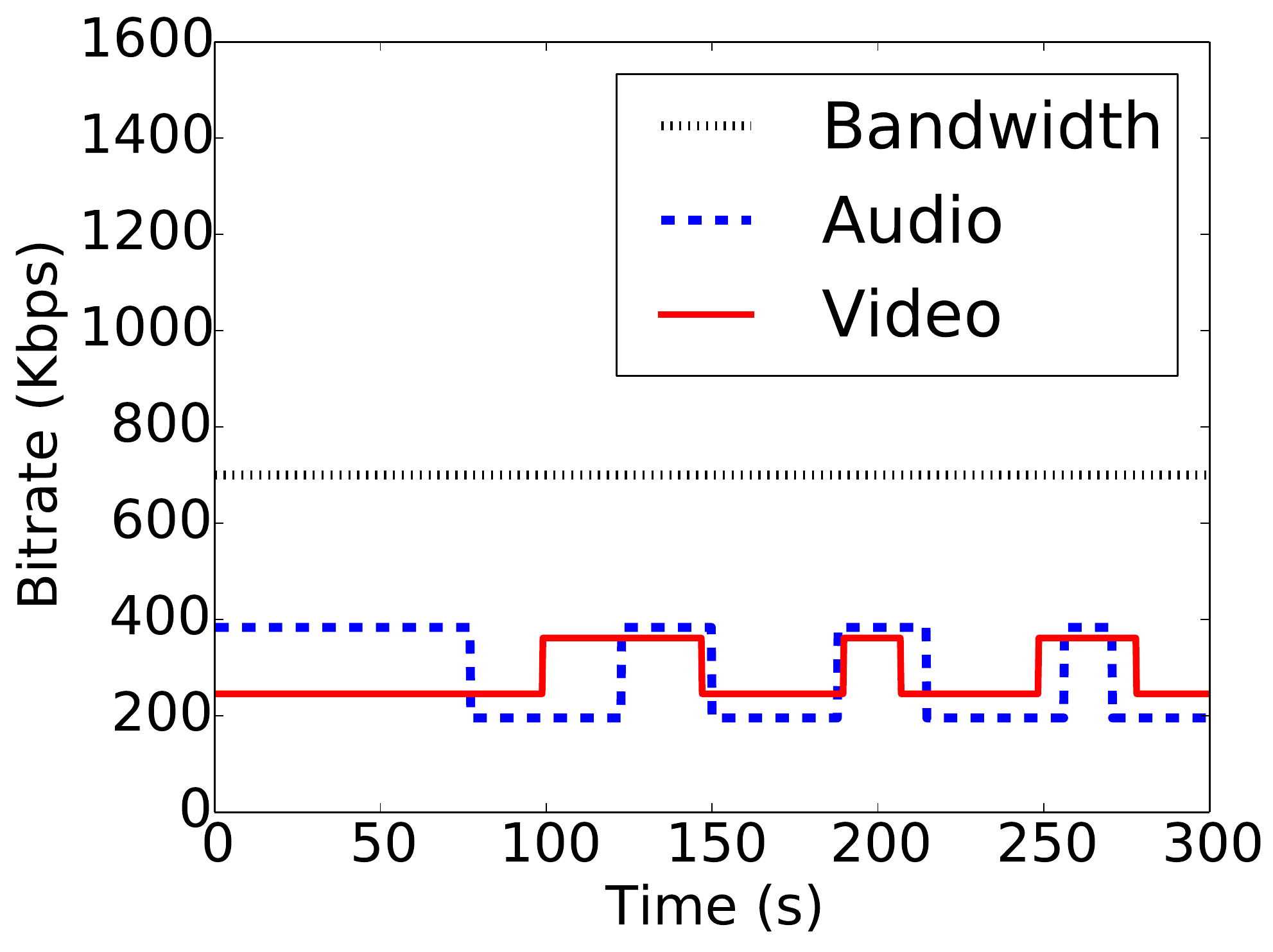}
	}
	\subfigure[Buffer levels.\label{subfig:dashjsbuffer}]{%
		\includegraphics[width=0.4\textwidth]{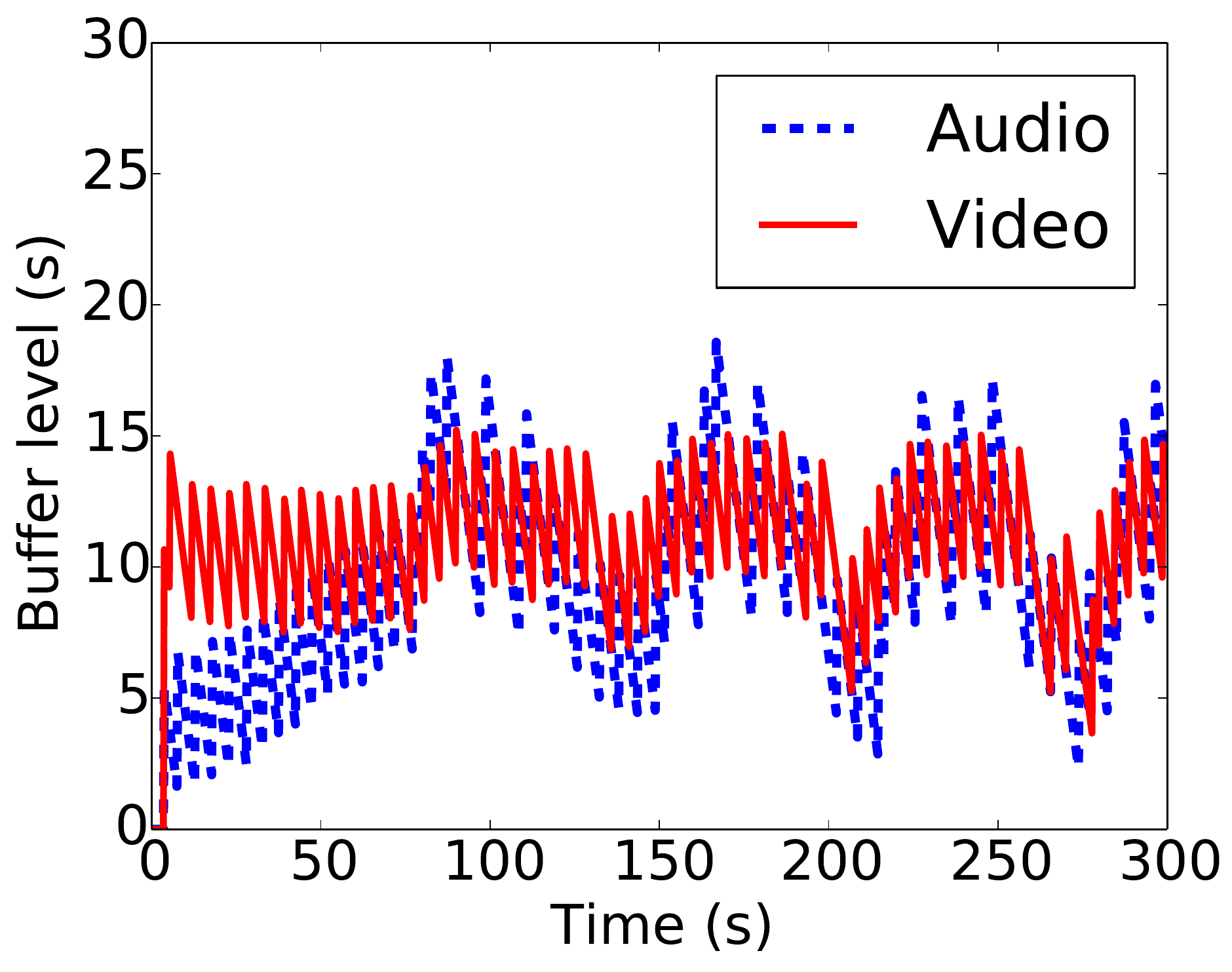}
	}
	\vspace{-.2in}
	\caption{dash.js results (DASH).}
	\label{fig:dashjsplayer}
	\vspace{-.35in}
\end{figure}

\mysubsection{Summary of Findings}

In summary, we observe the following problems with current ABR streaming protocols and player implementations.

\BULLET When a player does not perform rate adaptation for audio (e.g., ExoPlayer for HLS), it can lead to poor performance (e.g., a large amount of rebuffering). Therefore, it is desirable to perform rate adaptation for both audio and video.

\BULLET Some players select undesirable audio and video combinations, which can lead to poor viewing quality (e.g., the lowest video and highest audio tracks). The problem is more severe for DASH, which does not provide a mechanism to specify a subset of desirable audio and video combinations, making the player more vulnerable to selecting bad combinations.

\BULLET Some players make the rate adaptation decisions for audio and video completely independently (e.g.,  \texttt{dash.js}). Considering audio and video jointly is better than considering them independently. But considering them jointly in an overly {simplistic}  way is {also} not desirable (e.g., the practice in Shaka player can lead to frequent audio/video track changes).

\BULLET {Some players do not synchronize audio and video downloading explicitly (e.g.,  \texttt{dash.js}).}
This can lead to unbalanced content in video and audio buffers. Since we need both video and audio in the playback, having a lot more content in video/audio is not helpful. 
 Instead, it is more desirable to synchronize them on a finer granularity (e.g., on per chunk level as in ExoPlayer).
	
\BULLET  An HLS manifest can specify a subset of audio and video combinations, but some players do not conform to the manifest file. For example, ExoPlayer may select some bad combinations that are not in the manifest file.

	\mysection{Suggested Best Practices}\label{sec:suggestions}

Based on the analysis in \S\ref{sec:behaviors}, we suggest the following best practices for demuxed audio and video tracks for  the server side as well as the player side.

\mysubsection{Server Side Manifest Specifications} \label{sec:server-best-practice}

\textbf{Audio and video combinations.}
{\em The content provider should identify desirable combinations of audio and video tracks based on  content type and domain expertise, and specify these combinations in the manifest file.
 A player should only select chunks from the allowed combinations.}
This practice provides the following advantages: (i) allows the content owner to specify combinations that are suitable for the specific content (e.g., music video v.s. action movie) and devices (considering screen size, connected sound system, e.g., speaker or headphone),
%, (for example, the combinations that is suitable for PC is not necessary be good one for Phone.)},
and (ii)  simplifies the rate adaptation task for the player.
%;  and the player will not select an undesirable combination.
HLS already supports this  capability to include specific combinations in the manifest file. However it is  important for the content provider to utilize and  leverage this  capability appropriately, and not   specify all possible  combinations unless they are all desirable.
 Unlike HLS, DASH at present does not offer a way to list only specific allowed combinations in the manifest.
A practical short term workaround would be for the client to get this additional information from the server using HTTP. 
In the longer term, the DASH specification can be expanded to support this feature.

\smallskip
\noindent{\textbf{Audio and video bandwidth declaration.}}  {\em In the manifest file, it would be good to specify sufficient information about the aggregate bandwidth requirements of audio and video combinations, as well as the bandwidth requirements of individual audio/video tracks.} This is particularly important when audio and video are fetched over different network paths that have different network characteristics (e.g., when they are stored at different servers).
%In addition, having this information provides better support for a player that  conducts video and audio adaptation separately.
Currently, DASH specifies the bandwidth for each audio and video track, and the aggregate bandwidth requirement for audio and video combinations (if provided following the earlier suggestion) can be calculated from the individual tracks.
{
In HLS, the top-level master playlist \emph{only} specifies the aggregate bitrate for  each specified  audio and video combination. The bitrate of each individual audio/video track is not in the top-level master playlist, but can be obtained from the second-level media playlists as  follows (commercial players only use the information to identify the content  address   for an already selected chunk):
(i) When all the video/audio chunks are packaged into a single file,
the media playlists specify the EXT-X-BYTERANGE
information (i.e., the start and end
byte positions), which can be used to obtain the audio/video bitrate,
as pointed out in~\cite{QinVBR2018}.
(ii) When each video/audio chunk is packaged into an individual file,
no byte range information is provided in the media playlists. However, HLS recently added an optional EXT-X-BITRATE tag that
can be used to obtain per-chunk bitrate. We recommend that this option should  be
made mandatory. In addition, for both of the above two cases in HLS, since
the information required for computing the per-track bitrates is in the second-level manifest files, we suggest that the
player should download these files
% per-track manifests
and read the information before making rate adaptation decisions\footnote{We suggest avoiding the practice of ``lazy'' fetching, which only fetches
the playlist manifest file for a track when a chunk from that track
has been selected, at which point the player needs to know the address of that chunk.}.
A more robust longer term solution is to enhance the HLS specification so that the top-level master playlist directly provides per-track audio  and per-track video bitrate information.}
%}

\mysubsection{Player Side ABR Logic}

\textbf{Adopt audio rate adaptation.}  {\em For dynamic network bandwidth scenarios, it is important to perform audio rate adaptation.}  High quality audio tracks can have {similar or even} higher bitrates than  lower-rung video tracks. Audio rate adaptation is just as important as video rate adaptation
to avoid adverse impact on QoE (e.g., rebuffering as observed in \S\ref{subsec:exoplayer} or low audio quality).

\smallskip
\noindent{\textbf{Select only from allowed audio and video combinations.}}  {\em The ABR logic should only select  from the set of allowed audio and video combinations if provided by the server (e.g., via manifest file or certain out-of-band mechanism).}
As mentioned in \S\ref{sec:server-best-practice}, the allowed combinations reflect the desirable combinations between audio and video based on the content. Therefore, it is important for the client to select only from those combinations.

\smallskip
\noindent{\textbf{Joint adaptation of audio and video.}} {\em We suggest that the selection of the audio and video tracks for each chunk position (in playback order) be considered jointly.}
For both video and audio rate adaptation, it is desirable to satisfy conflicting goals in maximizing quality, minimizing stalls and minimizing quality variation.
Since the selection of audio and video is inherently coupled (so that good combinations of audio and video tracks are chosen), we recommend considering the combinations of audio and video (i.e., those specified in the manifest file if provided) while making rate adaptation decisions, instead of considering audio and video individually. The rate adaptation should be done carefully to avoid frequent changes in either audio or video tracks.

\smallskip
\noindent{\textbf{Maintain balance  between  audio and video prefetching.}}
{\em  It is desirable to keep the audio and video buffer levels (in seconds) balanced.}
% to avoid buffer underrun for either of them
This is because either empty audio or video buffer leads to stalls.
% -- it is no use having many audio chunks if the next video chunk is not there, and vice versa.
%This can be achieved by maintaining similar seconds of  audio and video in the prefetching buffer.
The balance can be achieved by synchronizing the duration of prefetched audio and video content at a fine granularity, e.g., at the chunk level or in terms of a small number of chunks.

	\mysection{Summary}\label{sec:conclusion}
In this chapter, we examined 
the  handling of 
ABR adaptation for demuxed video and audio tracks in predominant ABR streaming protocols (DASH and HLS), and in three popular ABR players.
We identified a number of limitations in existing practices that can adversely impact user QoE, and traced their root causes. We then proposed a number of best practices and principles for the server, ABR streaming protocols and client.
% for achieving good adaptation for demuxed audio and video tracks. 
We hope these findings will encourage further studies on this important topic.

\chapter{Conclusion and Future Work}
\label{chapter:conclusion}

In this dissertation, we first take a fresh look at PID-based bit rate adaptation
and
 design a framework called PIA (PID-control based ABR streaming) that strategically leverages PID control concepts and incorporates several novel strategies to account for the various requirements of ABR streaming. The evaluation results demonstrate that PIA outperforms state-of-the-art schemes in providing high average bitrate with significantly lower bitrate changes and stalls, while incurring very small runtime overhead. We further design PIA-E (PIA Enhanced), which improves the performance of PIA in the important initial playback phase.

As streaming providers have been moving towards using Variable Bitrate (VBR) encodings for the video content, we identify distinguishing characteristics of VBR encodings that impact user QoE and should be factored in any ABR adaptation decision and find that traditional ABR adaptation strategies designed for the CBR case are not adequate for VBR. We develop novel best practice design principles to guide ABR rate adaptation for VBR encodings. As a proof of concept, we design a novel and practical control-theoretic rate adaptation scheme, CAVA (Control-theoretic Adaption for VBR-based ABR streaming), incorporating these concepts. Extensive evaluations show that CAVA substantially outperforms existing state-of-the-art adaptation techniques.

We also explore quality-aware strategies for optimizing QoE while saving data. We analyze the underlying causes for suboptimal existing data saving practices and propose novel approaches to achieve better tradeoffs between video quality and data usage. The first approach is Chunk-Based Filtering (CBF), which can be retrofitted to any existing ABR scheme. The second approach is QUality-Aware Data-efficient streaming (QUAD), a holistic rate adaptation algorithm that is designed ground up. Our evaluations demonstrate that compared to the state of the art, the two proposed schemes achieve consistent video quality that is much closer to the user-specified target, lead to far more efficient data usage, and incur lower stalls.

We further examine the state of the art in the handling of demuxed audio and video tracks in predominant ABR protocols (DASH and HLS), as well as in real ABR client implementations in three popular players covering both browsers and mobile platforms. Combining experimental insights with code analysis, we shed light on a number of limitations in existing practices both in the protocols and the player implementations, which can cause undesirable behaviors such as stalls, selection of potentially undesirable combinations such as very low quality video with very high quality audio, etc. Based on our gained insights, we identify the underlying root causes of these issues, and propose a number of practical design best practices and principles  whose collective adoption will help avoid these issues and  lead to better QoE.

As future work, we plan to explore extending CAVA and its concepts to ABR streaming of live VBR encoded videos,
 develop data saving strategies that directly take account of a user’s data budget in ABR decisions.
Another direction is to further refine the suggested practices for ABR streaming with separate audio and video tracks, and design and implement rate adaptation schemes following the suggested practices.  
As cellular network infrastructure are upgrading from 4G/LTE to 5G, exploring video streaming over 5G is a promising
research direction based on our work.  Finally,  some of my previous works in SDN, quantum internet, quantum encryption and cybersecurity  ~\cite{ren2017,ren2018,li2019,lizhiWang2020,jiangweiwang2020,wenfeng2020,zhang2019enabling,tang2020quantum, wang2017,wang2020,chen2015,chen2020,jiang2021programmable,jiang2021quantum} inspire me that there might be a way to apply new approaches from other fields to video streaming study.

% As future work, we plan to explore the possibility of applying other machine learning mechanisms, and integrate even more lower layer information together with network layer information. We also plan to explore the possibility of applying our link bandwidth prediction technique to other network protocols and applications. For the video streaming adaptation, we plan to investigate other potential control models. 

%All the measurements in this paper are collected in one commercial LTE network. We believe that while different carriers may use different network configurations and their implementation of the LTE specification may differ, our approach is based on the fundamental mechanisms of LTE networks and is applicable to other carriers. One direction of future work is to evaluate our approach in other cellular networks. Another direction is to explore how to design upper-level transport protocols and applications using the predicted link bandwidth.

%People may doubts whether the prediction model described here can be directly applied to other networks. To answer this question, first we should note that the fundamental mechanism of LTE networks are the same. Even though different carriers may have slightly different network configuration, the fact that the performance is correlated to radio status should remain unchanged. Our scheme utilizes the regression tree model that can adjust the prediction model in real-time and adapt to different network (as described in Section~\ref{simu}.
%Still the evaluation of our model in other network carriers is will be our next step. 
\chapter{List of Publications during My PhD Study}

\begin{enumerate}
\item Y. Qin, C. Shende, C. Park, S. Sen, and B. Wang. “DataPlanner: Data-budget Driven Approach to Resource-efficient ABR Streaming,” Proceedings of ACM MMSys, 2021.  

\item Y. Qin, R. Jin, S. Hao, K. R. Pattipati, F. Qian, S. Sen, B. Wang, and C. Yue. “A Control Theoretic Approach to ABR Video Streaming: A Fresh Look at PID-based Rate Adaptation,” IEEE Transactions on Mobile Computing, vol. 19, no. 11, pp. 2505-2519, 1 Nov. 2020. 

\item Y. Qin, S. Sen, and B. Wang. “ABR Streaming with Separate Audio and Video Tracks: Measurements and Best Practices,” Proceedings of ACM CoNEXT, December 2019. 

\item Y. Qin, R. Jin, S. Hao, K. R. Pattipati, F. Qian, S. Sen, B. Wang, and C. Yue. “Quality-aware Strategies for Optimizing ABR Video Streaming QoE and Reducing Data Usage,” Proceedings of ACM MMSys, June 2019.

\item Y. Qin, R. Jin, S. Hao, K. R. Pattipati, F. Qian, S. Sen, B. Wang, and C. Yue. “ABR Streaming of VBR-encoded Videos: Characterization, Challenges, and Solutions,” Proceedings of ACM CoNEXT, December 2018. (Best Paper Award)

\item Y. Qin, R. Jin, S. Hao, K. R. Pattipati, F. Qian, S. Sen, B. Wang, and C. Yue. “A Control Theoretic Approach to ABR Video Streaming: A Fresh Look at PID-based Rate Adaptation,” Proceedings of IEEE INFOCOM, May 2017.

\item C. Yue. S. Sen, B. Wang, Y. Qin, and F. Qian, "Energy Considerations for ABR Video Streaming to Smartphones: Measurements, Models and Insights," Proceedings of ACM MMSys, June 2020.

\item C. Yue, R. Jin, K. Suh, Y. Qin, B. Wang, and W. Wei. “LinkForecast: Cellular Link Bandwidth Prediction in LTE Networks,” IEEE Transactions on Mobile Computing, vol. 17, no. 7, pp. 1582-1594, July 2018.

\item S. Chen, R. Morillo, Y. Qin, A. Russell, R. Jin, B. Wang, and S. Vasudevan, "Asynchronous Neighbor Discovery on Duty-Cycled Mobile Devices: Models and Schedules," in IEEE Transactions on Wireless Communications, vol. 19, no. 8, pp. 5204-5217, Aug. 2020

\item S. Chen, A. Russell, R. Jin, Y. Qin, B. Wang, and S. Vasudevan. “Asynchronous Neighbor Discovery on Duty-cycled Mobile Devices: Integer and Non-Integer Schedules,” Proceedings of ACM MobiHoc, June 2015.

\item Z. Jiang, Z. Tang, Y. Qin, C. Kang, and P. Zhang, “Quantum Internet for Resillient Electric Grids,” International Transactions on Electrical Energy Systems (2021): e12911.

\item Z. Jiang, Z. Tang, Y. Qin, and P. Zhang, “Programmable adaptive security scanning for networked microgrids”. Engineering, 2021.

\item Z. Tang, Y. Qin, Z. Jiang, W.O. Krawec, and P. Zhang, “Quantum-Secure Networked Microgrid”, IEEE Power \& Energy Society General Meeting, August 2020. (Best Paper Award) 

\item Z. Tang, Y. Qin, Z. Jiang, W. Krawec, and P. Zhang, “Quantum-Secure Microgrids,” IEEE Transactions on Power Systems, July, 2020.

\item W. Wan, M. A. Bragin, B. Yan, Y. Qin, J. Philhower, P. Zhang, and P. B. Luh, “Distributed and Asynchronous Active Fault Management for Networked Microgrids,” IEEE Transactions on Power System,  vol. 35, no. 5, pp. 3857-3868, Sept. 2020. 

\item L. Wang, Y. Qin, Z. Tang, and P. Zhang, “Software-defined microgrid control,” IEEE Open Access Journal of Power and Energy, vol. 7, pp. 173-182, 2020.

\item J. Wang, Y. Qin, Z. Tang, and P. Zhang, “Software-defined Cyber-Energy Secure Underwater Wireless Power Transfer,” IEEE Journal of Emerging and Selected Topics in Industrial Electronics, Oct. 2020. 

\item Y. Qin. “Resilient Networked Microgrids through Software Defined Networking,” Networked Microgrids, Cambridge University Press, 2021. (Book chapter)

\item Y. Li, Y. Qin, P. Zhang, and A. Herzberg. “SDN-Enabled Cyber-Physical Security in Networked Microgrids,” IEEE Transactions on Sustainable Energy, vol. 10, no. 3, pp. 1613-1622, 2018.

\item L. Ren, Y. Qin, Y. Li, P. Zhang, B. Wang, P. B. Luh, S. Han, T. Orekan, and T. Gong. “Enabling Resilient Distributed Power Sharing in Networked Microgrids through Software Defined Networking,” Applied Energy, vol. 56, no. 1, pp. 237-244, January 2018.

\item G. Wang, and Y. Qin. "MAC Protocols for Wireless Mesh Networks with Multi-beam Antennas: A Survey." In Future of Information and Communication Conference, pp. 117-142. Springer, Cham, 2019.

\item G. Wang, Y. Qin, and C. Chang, "Communication with Partial Noisy Feedback,"  IEEE Symposium on Computers and Communications (ISCC), Heraklion, 2017, pp. 602-607.

\item L. Ren, Y. Qin, B. Wang, P. Zhang, P. B. Luh, and R. Jin. “Enabling Resilient Microgrid through Programmable Network,” IEEE Transactions on Smart Grid, vol. 8, no. 6, November 2017.

\item L. Ren, Y. Qin, B. Wang, P. Zhang, P. B. Luh and R. Jin, "Enabling Resilient Microgrid Through Programmable Network," IEEE Power \& Energy Society General Meeting, Chicago, IL, 2017.

\item Y. Qin, L. Ren, B. Wang, P. Zhang and P. Luh, “SDN-enabled Highly Resilient and Efficient Microgrids,” GENI Engineering Conference 23, UIUC, June 15-18, 2015.

\item Y. Qin, L. Ren, B. Wang, P. Zhang and P. Luh, “Enabling Highly Resilient and Efficient Microgrids through Ultra-fast Programmable Networks,” US Ignite GENI Engineering Conference 22/, DC, March 23-26, 2015.

\item S. Sen,  S. Hao, Y. Qin, K.R. Pattipati, and B. Wang, ``Chunk-based filtering to optimize video streaming quality and data usage
,”  US Patent  App. 16/560,721, 2021. 

\item S. Sen, K.R. Pattipati, B. Wang and  Y. Qin, "Perceptual visual quality video chunk selection," US Patent 10917667, 2021.

\item S. Sen,  S. Hao, Y. Qin, K.R. Pattipati, and B. Wang, ``Differential Adaptive Bitrate Streaming Based on Scene Complexity,”  US Patent 10827181, 2020. 

\item S. Sen, S. Hao, Y. Qin, B. Wang, and K. R. Pattipati, “Methods, Systems, and Devices for Video Streaming Adaptation using Control Theoretic Approach,” US Patent 10362080, 2019.

\item S. Sen, S. Hao, Y. Qin, B. Wang, K. R. Pattipati, and R. Jin, "Apparatus, Storage Medium and Method for Adaptive Bitrate Streaming Adaptation of Variable Bitrate Encodings", US Patent 10728180, 2020.

\item P. Zhang, B. Wang, P. Luh, L. Ren, Y. Qin. “Enabling Resilient Microgrid Through Ultra-Fast Programmable Network,” US Patent 10257100, 2019.

\end{enumerate}

\newpage

\chapter*{Bibliography}
\addcontentsline{toc}{chapter}{Bibliography}

\bibliographystyle{uconn_unsrtnat}
\singlespacing\bibliography{ref-bing} 
\end{document}